\documentclass[pre,twocolumn,twoside,byrevtex,superscriptaddress,floatfix]{revtex4}

\renewcommand{\thetable}{\arabic{table}}

\usepackage{url}
\usepackage{graphicx,epsfig,verbatim,enumerate}
\usepackage{amssymb,amsmath}
\usepackage{ifthen}
\newboolean{twocolswitch}

\usepackage{color}

\definecolor{lightgrey}{rgb}{0.5,0.5,0.5}
\definecolor{gray}{rgb}{0.5,0.5,0.5}

\usepackage{rotating}
\usepackage{longtable}
\usepackage{array}

\newcommand{\tavg}[1]{\langle#1\rangle}

\newcommand{\sindex}[1]{}
\newcommand{\nindex}[1]{}

\newcommand{\www}[1]{\url{#1}}

\newcommand{\Req}[1]{Eq.~(\ref{#1})}

\newcommand{\PreserveBackslash}[1]{\let\temp=\\#1\let\\=\temp}
\newcommand{\PBS}[1]{\let\temp=\\#1\let\\=\temp}

\newcommand{\havg}[1]{h_{\rm avg}(#1)}
\newcommand{\havgsup}[1]{h_{\rm avg}^{#1}}

\newcommand{\havgfn}{h_{\rm avg}}
\newcommand{\havgfnamb}{h_{\rm avg}^{\rm (amb)}}
\newcommand{\havgfnnorm}{h_{\rm avg}^{\rm (norm)}}

\newcommand{\Neqnum}[1]{N^{\rm eq}_{#1}}
\newcommand{\Neqq}{\Neqnum{q}}

\newcommand{\Nsimpson}{N_{\rm S}}

\setboolean{twocolswitch}{true}

\newcommand{\revtexonly}[1]{#1}
\newcommand{\plainlatexonly}[1]{}

\begin{document}

\title{
  Temporal patterns of happiness and information in a global social network: \revtexonly{\\}
Hedonometrics and Twitter\\
\revtexonly{\small PLoS ONE, Vol 6, e26752, 2011}

}

\author{
\firstname{Peter Sheridan}
\surname{Dodds}
}
\email{peter.dodds@uvm.edu}

\author{
\firstname{Kameron Decker}
\surname{Harris}
}

\author{
\firstname{Isabel M.}
\surname{Kloumann}
}

\author{
\firstname{Catherine A.}
\surname{Bliss}
}

\author{
\firstname{Christopher M.}
\surname{Danforth}
}
\email{chris.danforth@uvm.edu}

\affiliation{
  Department of Mathematics and Statistics,
  Center for Complex Systems,
  \&
  the Vermont Advanced Computing Center,
  University of Vermont,
  Burlington,
  VT, 05401
}

\date{\today}

\begin{abstract}
   Individual happiness is a fundamental societal metric.
Normally measured through self-report, happiness
has often been indirectly characterized and overshadowed by 
more readily quantifiable economic indicators
such as gross domestic product.
Here, we examine expressions made on the 
online, global microblog and social networking service Twitter,
uncovering and explaining temporal variations in happiness and information levels
over timescales ranging from hours to years.
Our data set comprises over 46 billion words contained
in nearly 4.6 billion expressions posted over a 33 month span
by over 63 million unique users.
In measuring happiness, 
we construct a tunable, real-time, remote-sensing, and non-invasive, 
text-based hedonometer.
In building our metric, made available with this paper,
we conducted a survey to obtain happiness evaluations of over 10,000 individual words,
representing a tenfold size improvement over similar existing word sets.
Rather than being ad hoc, our word list is chosen solely by frequency of usage,
and we show how a highly robust and tunable metric can be constructed and defended.

\end{abstract}

\maketitle

\section*{Introduction}
\label{sec:twhap.intro}

One of the great modern scientific challenges we face
lies in understanding macroscale sociotechnical phenomena---i.e., 
the behavior of decentralized, networked systems 
inextricably involving people, information, and machine algorithms---such as global
economic crashes and the spreading of ideas and beliefs~\cite{hedstrom2006a}.
Accurate description through quantitative measurement is 
essential to the advancement of any scientific field,
and the shift from being data scarce to data rich
has revolutionized 
many areas~\cite{bell2009a,halevy2009a,hey2009a,collins2010a}
ranging 
from astronomy~\cite{sdss,lsst,stephens2008a} 
to ecology and biology~\cite{venter2004a}
to particle physics~\cite{lhc}.
For the social sciences, 
the now widespread usage of the Internet has led to
a collective, open recording of an enormous number
of transactions, interactions, and expressions, marking a clear transition
in our ability to quantitatively characterize, 
and thereby potentially understand, previously hidden as well as novel 
microscale mechanisms underlying sociotechnical systems~\cite{miller2011a}.

While there are undoubtedly limits to that which may 
eventually be quantified regarding human behavior,
recent studies have demonstrated a number of 
successful and diverse methodologies,
all impossible (if imaginable) prior to the Internet age.
Three examples relevant to public health,
markets, entertainment, history,
evolution of language and culture, and prediction 
are 
(1) 
Google's digitization of over 15 million books
and an initial analysis of 
the last two hundred years,
showing language usage changes,
censorship,
dynamics of fame,
and time compression of collective memory~\cite{michel2011a,googlebooks-ngrams};
(2) 
Google's Flu Trends~\cite{ginsberg2009a,choi2009a,goel2010a}
which allows for real-time monitoring of flu outbreaks
through the proxy of user search;
and 
(3) 
the accurate prediction of box office success based
on the rate of online mentions of individual movies~\cite{asur2010a} (see also~\cite{mishne2006e}).

Out of the many possibilities in
the `Big Data' age of social sciences,
we focus here on measuring, describing, and understanding
the well-being of large populations.
A measure of `societal happiness' is a crucial adjunct 
to traditional economic measures such
as gross domestic product and is of fundamental
scientific interest in its 
own right~\cite{layard2005a,gilbert2006a,lyubomirsky2007a,seaford2011a}.

Our overall objective is to use 
web-scale text analysis to 
remotely sense societal-scale levels
of happiness
using the singular source of the microblog and social networking
service Twitter.

Our contributions are both
methodological and observational.
First, our method for measuring the happiness of a given text, 
which we introduced in~\cite{dodds2009b}
and which we improve upon greatly in the present work,
entails word frequency distributions
combined with independently assessed
numerical estimates of the `happiness' of over 10,000 words
obtained using Amazon's Mechanical Turk~\cite{mechturk}.
We describe our method in full below and demonstrate its robustness.
We refer to our data set as 
`language assessment by Mechanical Turk 1.0',
which abbreviates as labMT 1.0,
and we provide all data as Supplementary Information
(Data Set S1).

Second, using Twitter as a data source,
we are able to explore happiness as a function of time, 
space, demographics, and network structure, with time being our focus here.
Twitter is extremely simple in nature, allowing users 
to place brief, text-only expressions online---`status updates' or `tweets'---that 
are no more than 140 characters in length.
As we will show, Twitter's framing tends to yield in-the-moment expressions
that reflect users' current experiences, making the service an ideal candidate
input signal for a real time societal `hedonometer'~\cite{edgeworth1881a}.

There is an important psychological distinction
between an individual's current, experiential happiness~\cite{killingsworth2010a}
and their longer term, 
reflective evaluation of their life~\cite{kahneman2005a},
and in using Twitter, our approach is tuned to the former kind.
Nevertheless, by following the written expressions
of individual users over long time periods,
we are potentially able to infer details of happiness
dynamics such as 
individual stability,
social correlation 
and contagion~\cite{fowler2008a},
and connections to well-being and health~\cite{kahneman2005a,layard2005a,seaford2011a}.

We further focus our present work on our essential findings 
regarding temporal variations in happiness 
including:
the overall time series;
regular cycles at the scale of days and weeks;
time series for subsets of tweets containing
specific keywords;
and
detailed comparisons between texts at the level of individual words.
We also compare happiness levels with measures of information content,
which we show are, in general, uncorrelated quantities (see~\ref{subsec:twhap.ambient}).
For information, as we explain 
below, we employ an estimate of lexical size
(or effective vocabulary size)
which is related to species diversity for ecological populations
and is derived from generalized entropy measures~\cite{jost2006a}.

Our methods and findings complement a number of related
efforts undertaken in recent years regarding happiness and well-being
including: 
large-scale surveys carried out by Gallup~\cite{gallup};
population-level happiness measurements carried out
by Facebook's internal data team~\cite{facebookgnh} and others~\cite{kramer2010a};
work focusing directly on sentiment detection
based on Twitter~\cite{oconnor2010a,mitrovic2010a,bollen2011c,bollen2011b,golder2011a};
and survey-based, psychological profiles as a function
of location, such as for the United States~\cite{rentfrow2008a}.
Our work also naturally builds on and shows consistency with 
earlier work on blogs~\cite{balog2006b,balog2006a,mishne2005a,mihalcea2006a,mishne2005b,mishne2006f,dodds2009b},
which in recent years have subsided due 
the ascent of Twitter and other services such as Facebook.

We structure our paper as follows:
in Sec.~\ref{sec:twhap.data}, we describe our data set;
in Secs.~\ref{sec:twhap.measurement1} and ~\ref{sec:twhap.measurement2},
we detail
our methods for measuring happiness and information content,
demonstrating in particular the robustness of our hedonometer
while uncovering some intriguing aspects
of the English language's emotional content;
in Sec.~\ref{sec:twhap.time},
we present and discuss the overall time series
for happiness and information;
in Secs.~\ref{sec:twhap.weeklycycle}
and~\ref{sec:twhap.dailycycle}, 
we
examine the average weekly and daily cycles
in detail;
in Sec.~\ref{sec:twhap.keywords},
we explore happiness and information time series 
for tweets containing keywords and short phrases;
and in Sec.~\ref{sec:twhap.conclusion},
we offer some concluding remarks.

\section{Description of data set}
\label{sec:twhap.data}

Since its inception, Twitter has provided various
kinds of dedicated data feeds for research purposes.
For the results we present here, we collected tweets 
over a three year period running from September 9, 2008 to 
September 18, 2011.
To the nearest million, our data set comprises 
46.076 billion words 
contained in
4.586 billion tweets  
posted by 
over 63 million individual users.
Up until November 6, 2010,
our collection represents approximately 8\% of all tweets 
posted to that point in time~\cite{gigatweet}.
A subsequent change in Twitter's message
numbering rendered such estimates more difficult,
but we can reasonably claim to have collected
over 5\% of all tweets.

Our rate of gathering tweets was not constant over time,
with regions of stability connected by short
periods of considerable fluctuations
(shown later in detail).
These changes were due to periodic alterations
in Twitter's feed mechanism as the company adjusted 
to increasing demand on their service~\cite{twitterapi}.
Twitter's tremendous growth in usage
and importance over this time frame
lead to several service outages,
and generated considerable technical issues for us
in handling and storing tweets.  
Nevertheless, we 
were able to amass a very large data set,
particularly so for one in the realm of social phenomena.
By August 31, 2011, we were receiving roughly 20 million tweets per day
(approximately 14,000 per minute),
and there were only a few days for which we did not record any data.

\begin{table}[tp!]
  \centering
  \begin{tabular}{|l|}
    \hline
    Tweet attributes:  \\
    \hline
    \hline
    Tweet text 
    \\
    Unique tweet ID
    \\
    Date and time tweet was posted$^\dagger$
    \\
    UTC offset (from GMT)
    \\
    User's location
    \\
    \hline 
    User ID
    \\
    Date and time user's account was created
    \\
    User's current follower count
    \\
    User's current friends count
    \\
    User's total number of tweets
    \\
    \hline
    In-reply-to tweet ID$^\ast$ 
    \\
    In-reply-to user ID$^\ast$ 
    \\    
    \hline
    Retweet (Y/N)
    \\
    \hline
  \end{tabular}
  \caption{
    List of key informational attributes accompanying each tweet.
    Information regarding the time of posting was altered ($\dagger$)
    on May 21, 2009 so that local time rather than Greenwich Mean Time (GMT)
    was reported.
    If a tweet is a reply to a previous tweet, the attributes
    also include those indicated by an asterisk:
    the ID of the specific tweet's and user's ID.
    Twitter initially issued tweets in XML format
    before moving to the JSON standard~\cite{twitterapi}.
    }
  \label{tab:twhap.dataformat}
\end{table}

Each tweet delivered by Twitter was accompanied
by a basic set of informational attributes;
we list the salient ones in Tab.~\ref{tab:twhap.dataformat}, and
summarize them briefly here.
First, for all tweets, we have a time stamp
referring to a single world clock running on 
US Eastern Standard Time;
and from May 21, 2009 onwards,
we also have local time.
Due to the importance of correcting for local time,
we focus much of our analysis on the time period
running from May 21, 2009 to December 31, 2010,
where we chose the end date as a clean stop point.

User location is available for some tweets
in the form of either current latitude and longitude,
as reported for example by a smartphone, 
or a static, free text entry of a home city along with state and country.
For measures of social interactions, we have a user's current follower 
and friend counts (but no information on who the followers and friends are),
and if a tweet is made in reply to another tweet, 
we also have the identifying number (ID) of the latter.
Finally, a `retweet' flag (`RT') indicates if a tweet is a rebroadcasting
of another tweet, encoding an important
kind of information spreading in the Twitter network.

Against the many benefits of using a data source such as Twitter,
there are a number of reasonable concerns to be raised,
notably representativeness.
First, in terms of basic sampling, 
tweets allocated to data feeds by Twitter
were effectively chosen at random from all tweets.  
Our observation of this apparent absence of bias
in no way dismisses the far stronger
issue that the full collection of tweets is a non-uniform
subsampling of all utterances made by a non-representative subpopulation
of all people~\cite{fox2009a,twitterdemog}.
While the demographic profile of individual Twitter users
does not match that of, say, the United States, where
the majority of users currently reside~\cite{smith2011a},
our interest is in finding suggestions of universal patterns.
Moreover, we note that like many other social networking services,
Twitter accommodates organizations as users, particularly
news services.  Twitter's user population is therefore a blend
of individuals, 
groups of individuals, 
organizations,
media outlets,
and
automated services such as bots~\cite{twitterusers}, representing a
kind of disaggregated, crowd-sourced media~\cite{kwak2010a}.
Thus, rather than analysing signals from a few news outlets,
which in theory represent and reflect the opinions and experiences
of many, we now have access to signals coming directly from
a vast number of individuals.  
Moreover, in our treatment,
tweets from, say, the New York Times or the White House are given
equal weight to those of any person-on-the-street.

In sum, we see two main arguments for pursuing the massive 
data stream of Twitter: 
(1) the potential for
describing universal human patterns, 
whether they be emotional, social, or otherwise;
and
(2) the current and growing importance of Twitter~\cite{congresstweets-nytimes} 
(surprising as that may be to critics of social media).

\begin{figure}[tb!]
  \centering
                \includegraphics[width=0.48\textwidth]{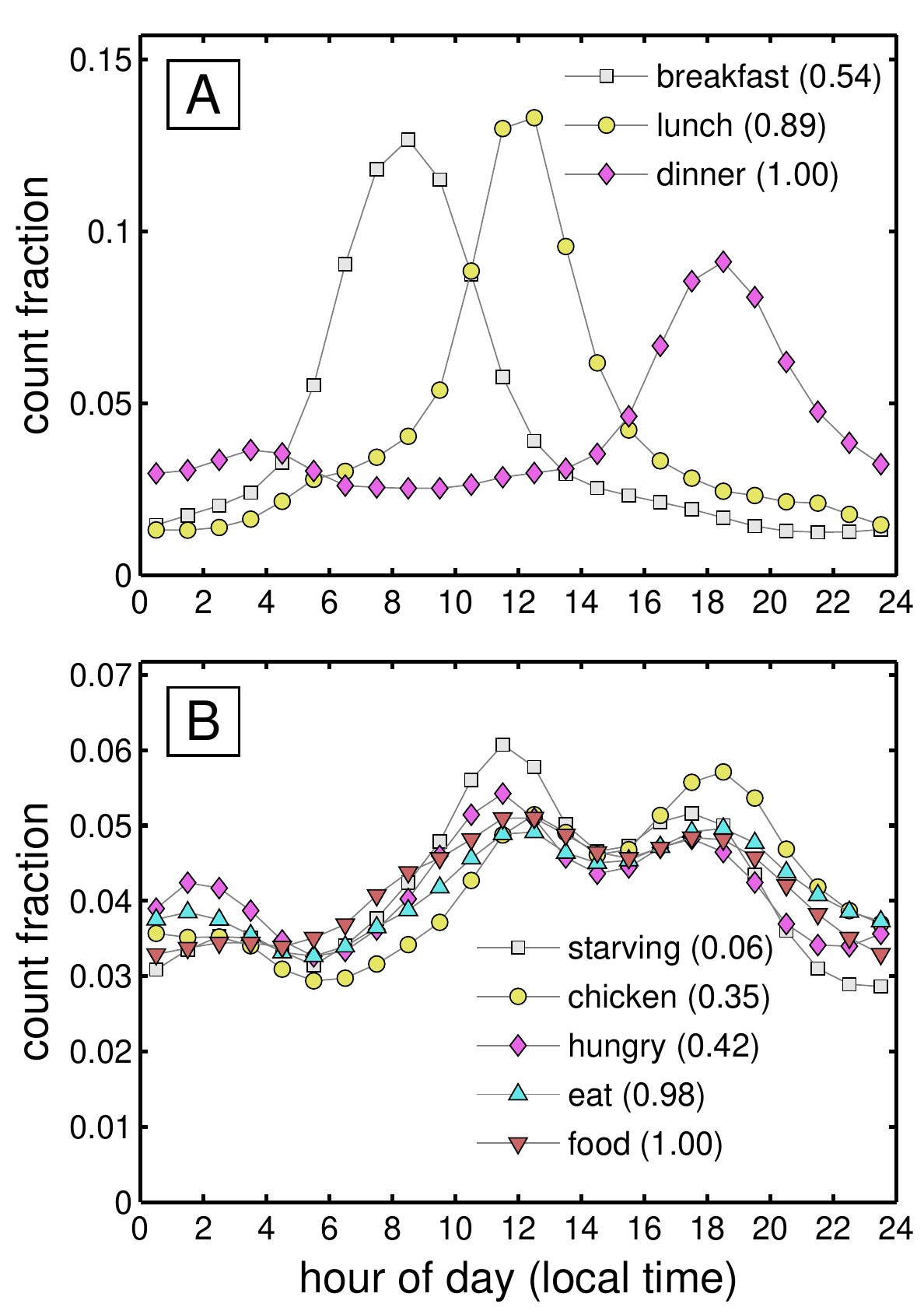}
  \caption{
    Daily trends for example sets of commonplace words
    appearing in tweets.  
    For purposes of comparison, each curve
    is normalized so that the count fraction represents
    the fraction of times a word is mentioned in a given hour
    relative to a day.
    The numbers in parentheses indicate the relative overall
    abundance normalized for each set of words by the most
    common word.
    Data for these plots is drawn from
    approximately 26.5 billion words
    collected from May 21, 2009 to December 31, 2010 inclusive,
    with the time of day adjusted to local time by Twitter
    from the former date onwards.
    The words `food' and `dinner'
    appeared a total of 2,994,745 (0.011\%)
    and 4,486,379 (0.016\%) times respectively.
  }
  \label{fig:twhap.daywordtrends}
\end{figure}

A preliminary glance at the data set shows
that the raw word content of tweets does appear to 
reflect people's current circumstances.
For example, Fig.~\ref{fig:twhap.daywordtrends}
shows normalized daily frequencies for two 
food-based sets of words, binned by hour of the day.
Fig.~\ref{fig:twhap.daywordtrends}A shows that,
as we would expect, the words `breakfast', `lunch', and `dinner'
respectively peak during the hours 8--9 am, 12--1 pm, and 6--7 pm.
In Fig.~\ref{fig:twhap.daywordtrends}B, we observe that
the words 
`starving', 
`chicken'
`hungry', 
`eat', 
and 
`food', 
all follow a similar cycle with three relative peaks, one around
midday, a smaller one before dinner, and another in the early morning.
These trends suggest more generally that words that are correlated conceptually
will be similarly congruent in their temporal patterns in tweets.
Other quotidian words follow equally reasonable trends:
the word `sunrise' peaks between 6 and 7 am, while `sunset' is most prominent around 6 pm;
and the daily high for `coffee' occurs between 8 and 9 am.
Regular cultural events also leave their imprint
with two examples from television being
`lost' (for the show `Lost')
and 
`idol' (for `American Idol') both sharply maximizing around 
their airing times in the evening.
Further evidence that everyday people are behind 
a large fraction of tweets
can be found in the prevalence of colloquial terms 
(e.g., `haha', `hahaha') and profanities, which we will
return to later.
Recent surveys also show
that approximately half of Twitter users 
engage with the service via mobile phones~\cite{smith2011a},
suggesting that individuals are often contributing tweets
from their current location.
Thus, while not statistically exhaustive,
we have reassuring, commonsensical support for the in-the-moment nature of tweets,
and we move on to our main descriptive focus: temporal patterns of societal happiness.

\section{A robust method for measuring emotional content}
\label{sec:twhap.measurement1}

\subsection{Algorithm for Hedonometer}
\label{subsec:twhap.algorithm}

We use a simple, fast method for measuring the happiness of texts
that hinges on two key components:
(1) human evaluations of the happiness of a set of individual words,
and 
(2) a naive algorithm for scaling up from
individual words to texts.
We substantially
improve here on the method introduced by 
two of the present authors in~\cite{dodds2009b} by
incorporating a tenfold larger word set for which
we have obtained happiness evaluations using Mechanical Turk~\cite{mechturk}.
As we demonstrate our, hedonometer exhibits an impressive
level of instrument robustness 
and a surprising property of tunability, 
similar in nature to a physical instrument such as a microscope.
For the algorithm, which is unchanged from~\cite{dodds2009b},
we first use a pattern-matching script to extract
the frequency of individual words in a given text $T$.
We then compute
the weighted average level of happiness for the text as
\begin{equation}
  \label{eq:twhap.htau}
  \havg{T}
  =
  \frac{\sum_{i=1}^{N} \havg{w_i} f_i}
  {\sum_{i=1}^{N} f_i}
  =
  \sum_{i=1}^{N} \havg{w_i} p_i,
\end{equation}
where $f_i$ is the frequency of the 
$i$th word $w_i$
for which we have an estimate of 
average happiness, $\havg{w_i}$,
and $p_i = f_i/\sum_{j=1}^{N} f_j$ is the corresponding normalized frequency.

For a single text, we would
naturally rank the $N$ unique words found in $T$ by decreasing frequency.
However, in wanting to rapidly compare in detail 
(e.g., at the level of individual words)
many pairs of massive texts assembled on the fly 
(e.g., by finding all tweets that contain a particular keyword),
it is useful to maintain a fixed, ordered list of words.
To do so, we took the most frequent 50,000 words from a large part of the overall Twitter
corpus (see Methods), as a standardized list,
and using this list, we then transformed texts into vectors of word frequencies.
The number 50,000 was chosen both for computational ease---a master 
list of all words appearing in our corpus would be too large---and 
the fact that various measures of information content (described below)
can be reliably computed.

\subsection{Word evaluations using Mechanical Turk}
\label{subsec:twhap.mechturk}

For human evaluations of happiness,
we used Amazon's Mechanical Turk~\cite{mechturk} 
to obtain ratings for individual words.
There are three main aspects to explain here: 
(1) how we created our initial word list, 
(2) the ratings procedure,
and
(3) how a requirement of robustness leads
us to using a tunable subset of words.
As per our introductory remarks, we will refer to
this data set as labMT 1.0 (Data Set S1).
We discuss the first two points in this section
and the third in the ensuing one.

We drew on four disparate text sources:
Twitter,
Google Books (English)~\cite{michel2011a,googlebooks-ngrams},
music lyrics (1960 to 2007)~\cite{dodds2009b},
and
the New York Times (1987 to 2007)~\cite{nytimescorpus2008a}.
For each corpus, we compiled word lists ordered 
by decreasing frequency of occurrence $f$, which is well known to follow
a power-law decay as a function of word rank $r$ for 
natural texts~\cite{zipf1949a}.
We merged the top 5,000 words from each source,
resulting in a composite set of 10,222 unique words.

By simply employing frequency as the measure of a word's importance,
we naturally achieve a number of goals:
(1) Precision: 
we have evaluations for as many words in a text as possible, given
cost restrictions (the number of unique `words' being tens of millions);
(2) Relevance: 
we  tailor our instrument to our focus of study;
and 
(3) Impartiality: 
we do not a priori decide
if a given word has emotional 
or meaningful content.
Our word set 
consequently involves multiple languages,
all parts of speech,
plurals, conjugations of verbs,
slang, abbreviations,
and emotionless, or neutral,
words such as `the' and `of'.

For the evaluations,
we asked users on Mechanical Turk to
rate how a given word made them feel
on a nine point integer scale,
obtaining 50 independent evaluations
per word.
We broke the overall assignment into 100 smaller
tasks of rating approximately 100 
randomly assigned words at a time.
We emphasized the scores 1, 3, 5, 7, and 9 
by stylized faces, representing a sad to happy spectrum.
Such five point scales are in widespread use on the web today
(e.g., Amazon) and would likely
be familiar with users.
The four intermediate scores of 2, 4, 6, 8 
allowed for fine tuning of assessments.
In using this scheme, we remained consistent
with the 1999 Affective Norms for English Words (ANEW) study 
by Bradley and Lang~\cite{bradley1999a}, 
the results of which we used 
in constructing our initial metric~\cite{dodds2009b}.

Some illustrative examples 
of average happiness we obtained
for individual words are:
\begin{itemize}
\item[] 
$\havg{\rm laughter}=8.50$,
\item[] 
$\havg{\rm food}=7.44$,
\item[] 
$\havg{\rm reunion}=6.96$,
\item[] 
$\havg{\rm truck}=5.48$,
\item[] 
$\havg{\rm the}=4.98$,
\item[] 
$\havg{\rm of}=4.94$,
\item[] 
$\havg{\rm vanity}=4.30$,
\item[] 
$\havg{\rm greed}=3.06$,
\item[] 
$\havg{\rm hate}=2.34$,
\item[] 
$\havg{\rm funeral}=2.10$,
\item[] 
and 
$\havg{\rm terrorist}=1.30$.
\end{itemize}
As this small sample indicates, 
we find the evaluations
are sensible with neutral words
averaging around 5.

Note that in analysing texts, we avoid stemming words,
i.e., conflating inflected words with their
root form, such as all conjugations of a specific verb.
For verbs in particular, by focusing on the most frequent words, we obtained
scores for those conjugations likely to appear in texts, obviating any need
for stemming.
Moreover, while we observe stemming works well in some cases
for happiness measures,
e.g., 
$\havg{\rm advance}$=6.58,
$\havg{\rm advanced}$=6.58,
and
$\havg{\rm advances}$=6.24,
it fails badly in others,
e.g., 
$\havg{\rm have}$=5.82
and
$\havg{\rm had}$=4.74;
$\havg{\rm arm}$=5.50
and
$\havg{\rm armed}$=3.84;
and
$\havg{\rm capture}$=4.18
and
$\havg{\rm captured}$=3.22.

In the Supplementary Information,
we provide happiness averages and standard deviations
for all 10,222 words, along with other information.

An immediate and reassuring sign
of the robustness of
the word happiness scores we
obtained via Mechanical Turk
is that our results agree very well
with that of the earlier ANEW study
which consisted of 1034 words~\cite{bradley1999a}
(Spearman's correlation coefficient 
$r_s = 0.944$ 
and $p$-value $< 10^{-10}$).
This adds to earlier suggestions of universality
in the form of a high correlation between
the ANEW study happiness scores
and those made by participants in Madrid
for a direct Spanish translation of
the ANEW study words~\cite{redondo2007a}.
Furthermore, the ANEW study involved students at the
University of Florida, a group evidently distinct 
from users on Mechanical Turk.

The ANEW study words were also
broadly chosen for their emotional and meaningful import
rather than usage frequency, and we show below 
that our larger frequency-based word set
affords a much greater coverage of texts.
(By coverage, we mean the percentage of words
in a text for which we have individual happiness estimates.)
Note that in the ANEW study and our earlier work~\cite{dodds2009b}, 
happiness
was referred to as psychological valence,
or simply valence, 
a standard terminology~\cite{osgood1957a}.

\begin{figure*}[tbp]
  \centering
    \includegraphics[width=\textwidth]{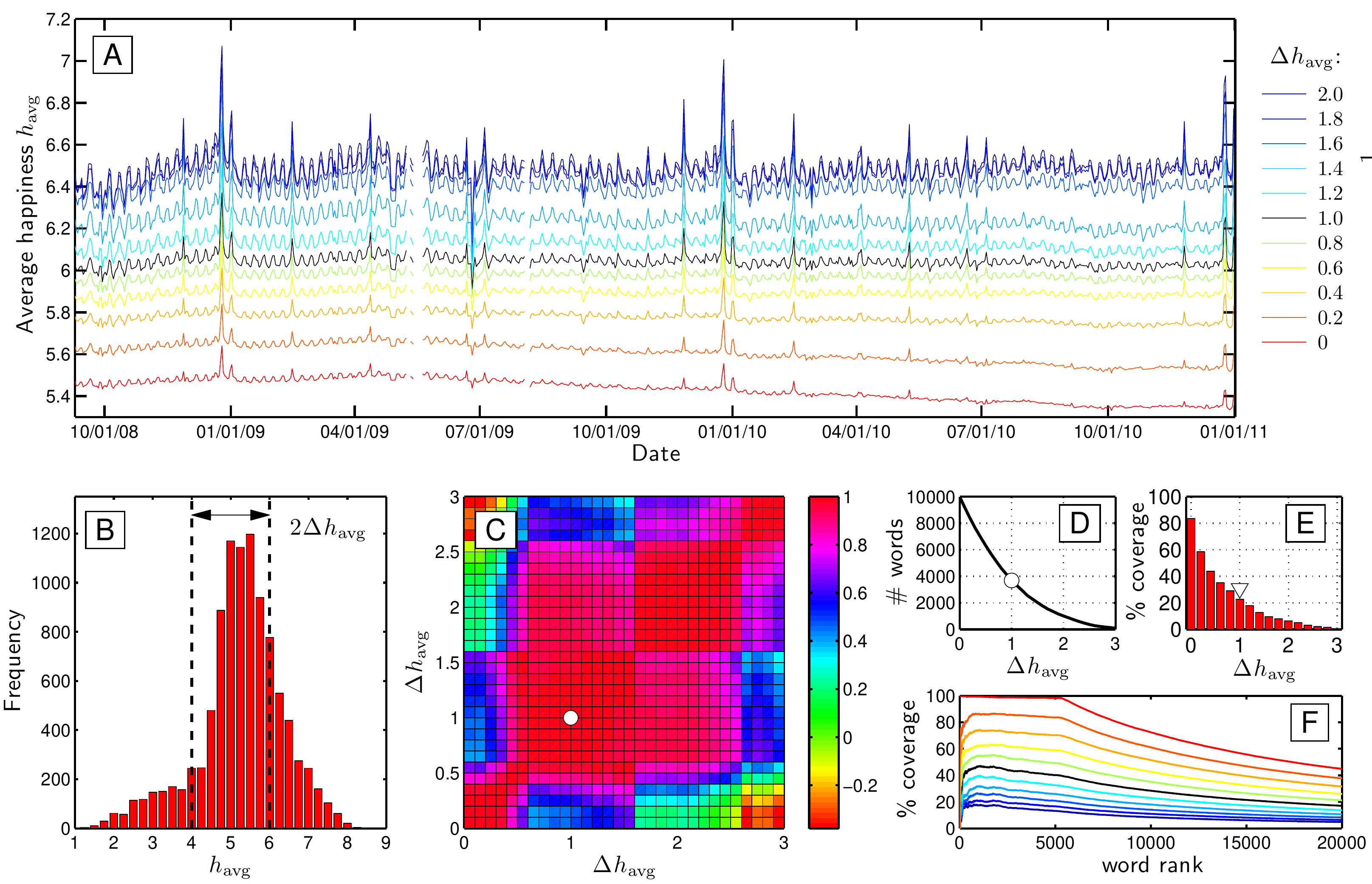}
    \caption{
      Demonstration of robustness and tunability of our text-based hedonometer,
      and reasoning for choice of a specific metric.
      To measure the happiness of a given text,
      we first compute frequencies of all words;
      we then create an overall happiness score, \Req{eq:twhap.htau},
      as a weighted average of subsets of 10,222 individual
      word happiness assessments on a 1 to 9 scale,
      obtained through Mechanical Turk 
      (see main text and Methods).
      In varying word sets by excluding stop words~\cite{wilbur1992a},
      we can systematically explore families of happiness metrics.
      In plot \textbf{A}, we show time series of average happiness for Twitter,
      binned by day, produced by different metrics.
      Each time series is generated by omitting
      words with $5-\Delta \havgfn < \havgfn < 5+\Delta \havgfn$ 
      as indicated in
      plot \textbf{B}, 
      which shows
      the overall distribution of average happiness of
      individual words.
      For $\Delta \havgfn = 0$ we use all words;
      as $\Delta \havgfn$ increases, we progressively
      remove words centered around the neutral evaluation of 5.
      Plot \textbf{C} provides a test for robustness through
      a pairwise comparison of all time series
      using Pearson's correlation coefficient.
      For $0.5 \le \Delta \havgfn \le 2.5$,
      the time series show very strong mutual agreement.
      We choose $\Delta \havgfn = 1$ 
      (black curve in \textbf{A} and \textbf{F},
      shown in \textbf{B},
      white symbols in \textbf{C}, \textbf{D}, and \textbf{E}) 
      for the present paper 
      because of its excellent correlation in output with 
      that of a wide range of $\Delta \havgfn$, and 
      for reasons concerning the following trade-offs.
      In \textbf{A}, we see that as the number of
      stop words increases, so does the variability of 
      the time series, suggesting an improvement
      in instrument sensitivity.
      However, at the same time, we lose coverage of texts.
      Plot \textbf{D} 
      first shows how the number of individual words for which
      we have evaluations decreases
      as $\Delta \havgfn$ increases.
      For $\Delta \havgfn = 1$, we have 3,686 individual words
      down from 10,222.
      Plot \textbf{E} next shows
      the percentage of the Twitter data set
      covered by each word list, accounting for word frequency;
      for $\Delta \havgfn = 1$,
      our metric uses 22.7\% of all words.
      Lastly, in plot \textbf{F} (which uses plot \textbf{A}'s legend), we show how coverage
      of words decreases with word rank.
      When $\Delta \havgfn=0$, we incorporate all
      low rank words, with a decline beginning
      at rank 5,000.  For $\Delta \havgfn > 0$,
      we see similar patterns with the maximum coverage
      declining; for $\Delta \havgfn=1$, we see a maximum
      coverage of approximately 50\%.
    }
  \label{fig:twhap.validation}
\end{figure*}

\subsection{Robustness and Refinement of Hedonometer}
\label{subsec:twhap.robustness}

We now show that our hedonometer can be
improved by considering the effects of
taking subsets of the overall list of 10,222 words.
Clearly, truly neutral words such as `the' and `of'
should be omitted, especially because of their
high relative abundance, thereby forming a list of excluded words
commonly referred to as stop words~\cite{wilbur1992a}.

Because we have filtered by frequency in
selecting our word list, we are able to determine
stop word lists in a principled way, leading
to a feature of tunability.
Here, we exclude words whose average happiness $\havgfn$ 
lies within $\Delta \havgfn$ of the neutral score of 5,
i.e.,
$5-\Delta \havgfn < \havgfn < 5+\Delta \havgfn$.
In other words, we remove all words lying
in a centered band of width $2\Delta \havgfn$ on our
happiness spectrum.  

We explore and demonstrate our hedonometer's behavior [\Req{eq:twhap.htau}]
with respect to different stop word lists by varying $\Delta \havgfn$,
with our main results and evidence displayed in
the six panels of Fig.~\ref{fig:twhap.validation}.
We will argue in particular that $\Delta \havgfn = 1$
yields a robust, sensitive, and informative hedonometer,
and this will be our choice for the remainder of the paper.
However, a range of values of $\Delta \havgfn$ will
also prove to be valid, meaning that $\Delta \havgfn$ is a tunable 
parameter.

As a test case and as shown in Fig.~\ref{fig:twhap.validation}A,
we focus on measuring the happiness time series for Twitter running
from September 9, 2008 to December 31, 2010, resolved at the level of days,
and  for $\Delta \havgfn = 0, 0.2, 0.4, \dots, 2.0$.
(Once we explain our selection of $\Delta \havgfn=1$,
black curve in~\ref{fig:twhap.validation}A,
we will return in the next section to study the overall
time series in detail.)
In Fig.~\ref{fig:twhap.validation}B, we show
a histogram of average happiness levels
for all 10,222 words, indicating the 
stop word selection for $\Delta \havgfn = 1$.
Several features are apparent: 
(1) the time series are broadly similar to the eye;
(2) as we expand the stop word list, the base line level
of happiness and size of fluctuations both increase;
(3) an overall downward trend apparent for 
small $\Delta \havgfn$ becomes less
pronounced as  $\Delta \havgfn$ increases;
and
(4) English words, as they appear in natural language,
are biased toward positivity,
a phenomenon we explore elsewhere~\cite{kloumann2011a}.
Note that point (4) explains point (2): the increasing relative abundance
of positive words leads to an inflation of overall happiness
as $\Delta \havgfn$ increases.

We quantify the similarity between time series
by computing Pearson's correlation coefficient for
each pair of time series with
$\Delta \havgfn = 0, 0.1, 0.2, \dots, 3.0$.
In Fig.~\ref{fig:twhap.validation}C,
we observe an impressively high correlation for all
pairs of time series with
$0.5 \lesssim \Delta \havgfn \lesssim 2.5$,
forming the central large square
(the white circle corresponds to $\Delta \havgfn = 1$).
For the range $\Delta \havgfn \lesssim 0.5$,
the resultant time series are internally consistent
but a clear break occurs with time series for $\Delta \havgfn \gtrsim 0.5$.
This transition appears to be due in part to the relative
increase of languages other than English on Twitter
since mid 2009, which we discuss later in 
Sec.~\ref{subsec:twhap.infocontent}.

The striking congruence for all time series
generated with $0.5 \lesssim \Delta \havgfn \lesssim 2.5$
suggests that we may use $\Delta \havgfn$ as a tuning
parameter, a remarkable consequence of the emotional
structure of the English language.
Larger values of
$\Delta \havgfn$ ($ \gtrsim 2.5$)
give us a higher resolution or sensitivity (the time series
fluctuate more)
but at a loss of overall word coverage leading
to a more brittle instrument.
This effect is reminiscent of increasing the 
contrast in an image, or edge detection.
More generally, we could choose any range of 
word happiness as a `lens' into a text's emotional
content.  For example, we could take words
with $7 < \havgfn \le 9$ to highlight the positive
elements of a text.  
Thus, as a practical instrument implemented online,
we would recommend the inclusion of
$\Delta \havgfn$ as a natural tuning parameter.

For the purposes of this paper, it is most useful
if we choose a specific value of $\Delta \havgfn$ in this range.
As we have indicated, we find $\Delta \havgfn = 1$ to be a suitable compromise
in balancing sensitivity versus robustness,
i.e., the ability to pick up variations across texts 
(requiring higher $\Delta \havgfn$)
versus text coverage
(requiring lower $\Delta \havgfn$).
In choosing $\Delta \havgfn = 1$, 
we are also safely above the transitional value
of $\Delta \havgfn \simeq 0.5$.

We support the robustness of our choice with evidence provided
in Figs.~\ref{fig:twhap.validation}D,
\ref{fig:twhap.validation}E,
and
\ref{fig:twhap.validation}F,
which together show how word coverage
declines with increasing $\Delta \havgfn$.
In Fig.~\ref{fig:twhap.validation}D,
we plot the number of unique words
left in our labMT 1.0 word list (Data Set S1)
as a function of $\Delta \havgfn$.
For $\Delta \havgfn = 1$, 3,686 unique words of the original
10,222 remain.  
The fraction of the Twitter corpus covered by these
3,686 word is approximately 23\% (Fig.~\ref{fig:twhap.validation}E).
By comparison, the ANEW study's 1,034 words
collectively cover only 3.7\% of the corpus,
typical of other
texts we have analysed such as blogs, books, 
and State of the Union Addresses~\cite{dodds2009b}.
This discrepancy in total coverage is 
again due to the ANEW word list's origin
being more to do with meaning than frequency.

Fig.~\ref{fig:twhap.validation}F shows
how our coverage of words in the Twitter
corpus decays as a function of frequency rank $r$.
For $\Delta \havgfn = 0$, our coverage is
complete out to $r=5,000$ where we begin
to miss words.  The same basic curve
is apparent for $\Delta \havgfn > 0$,
with a clear initial dip due to 
the exclusion of common neutral words.
For $\Delta \havgfn = 1$, we cover 
between 40 to 50\% for $r \le 5,000$.

As a final testament to the quality of our
hedonometer, we note that 
in an earlier version of the present paper~\cite{dodds2011a},
and prior to completing our word evaluation survey
using Mechanical Turk,
we used the ANEW study word list in all our analyses;
the interested reader will be able to make many direct comparisons
of figures and tables.
Broadly speaking, we find the same trends with our improved
word set, again speaking to the robustness of 
our instrument and indeed the English language.
In the manner of a true measuring instrument,
we obtain much greater resolution and fidelity
with the labMT 1.0 word list (Data Set S1), sharpening  
observations we made using the ANEW study,
and bringing new ones to light that were previously hidden.

\subsection{Limitations}
\label{subsec:limitations}

We address several key aspects and limitations of our measurement.
First, as with any sentiment analysis technique, our instrument
is fallible for smaller texts, 
especially at the scale of a typical sentence, where
ambiguity may render even human readers unable to judge
meaning or tone~\cite{lee2004a}.
Nevertheless, problems with small texts are not our 
concern, as our interest here is in dealing with 
and benefiting from very large data sets.

Second, we are also effectively extracting a happiness level as perceived by
a generic reader who sees only word frequency.
Indeed, our method is purposefully more
simplistic than traditional natural language processing (NLP) algorithms 
which attempt to infer meaning (e.g., OpinionFinder~\cite{choi2005a,choi2006a})
but suffer from a degree of inscrutability.
By ignoring the structure of a text, we are of course omitting
a great deal of content; nevertheless, we have shown 
using bootstrap-like approaches that
our method is sufficiently robust as to be meaningful for large
enough texts~\cite{dodds2009b}.

Third, we quantify only how people appear to others;
as should be obvious, our method cannot divine the internal emotional states
of specific individuals or populations.  
In attempting to truly understand a social system's potential dynamical
evolution, we would have to account for publicly 
hidden but accessible 
internal ranges and states of emotions, beliefs, etc.
However, a person's exhibited emotional tone,
now increasingly 
filtered through the signal-limiting medium of written interactions 
(e.g., status updates, emails, and text messages),
is that which other people evidently observe and react to.

Last, by using a simple kind of text analysis, we are able
to non-invasively, remotely sense the exhibited 
happiness of very large numbers of
people via their written, open, web-scale output.
Crucially, we do not ask people how happy they are, we merely observe
how they behave online.
As such, we avoid the many difficulties 
associated with self-report~\cite{list2006a,skitka2006a,kahneman2006a}.
We refer the reader to our initial work for more
discussion of our measurement technique~\cite{dodds2009b}.

\section{Measuring word diversity}
\label{sec:twhap.measurement2}

In quantifying a text's information content, we use 
concepts traditionally employed for
estimating species diversity in ecological studies~\cite{jost2006a}
which build on information theoretic approaches.
As we outline below, direct measures of information 
can be transformed into estimates of lexical size (or word diversity),
with the benefit that comparisons of the latter are more readily interpretable.

A first observation is that the sheer number of distinct words in a text
is not a good representation of lexical size.
Because natural texts generally exhibit highly skewed distributions of word frequencies,
such a measure discards much salient information, and 
moreover is difficult to estimate if a text is subsampled.

To arrive at a more useful and meaningful quantity,
we consider generalized entropy: $J_q = \sum_{i} p_i^q$
where, for a given text,
$p_i$ is the $i$th distinct word's normalized frequency of occurrence
and which we interpret as a probability.
In varying the parameter $q$, we tune the relative importance of common versus rare words, 
with large $q$ favoring common ones.

These generalized entropies can be seen as direct
measures of information but 
their values can be hard to immediately interpret.
To make comparisons between the information content of texts 
more understandable, if by adding an extra step,
we use these information measures to
compute an equivalent lexical size, $\Neqq$,
which is the number of words that would yield
the same information measure if all words appeared
with equal frequency~\cite{jost2006a}.  

We observe that the lexical sizes
$\Neqq$ for $q \gtrsim 1.5$
closely follow the same trends 
for the data we analyse here.
In therefore needing to show only one representative 
measure among the $\Neqq$, we choose 
$\Nsimpson=\Neqnum{2}$ based on Simpson's concentration $S = \sum_{i} p_i^2$,
corresponding to generalized entropy with $q=2$~\cite{simpson1949a}.
A simple calculation gives $\Nsimpson = 1/S$.
Simpson's concentration can be seen as
the probability that any two words chosen at random will be the same.
Simpson's concentration is also related to the Gini coefficient $G$,
which is often used to characterize income inequality,
as $S=1-G$.
For text analysis, $G$ represents the probability that two 
randomly chosen words are different.

Using $\Nsimpson = 1/S$ for lexical size
holds several theoretical and practical benefits:
(1) $S$ has the natural probablistic interpretation given above;
(2) The quantity $p_i^2$ decays sufficiently rapidly that we need
not be concerned about subsampling heavy tailed distributions
(see Methods);
and
(3) In comparing two texts, the contributions to $\Nsimpson$
due to changes in individual word frequencies combine linearly and thus
can be easily ranked.
From here on, we will focus on $\Nsimpson$ 
which we will refer to as a text's `Simpson lexical size.'

\section*{Results and Discussion}
\label{sec:twhap.results}

\section{Overall time dynamics of happiness and information}
\label{sec:twhap.time}
    
\revtexonly{\begin{turnpage}}
  \revtexonly{\begin{figure*}[tp!]}
    \plainlatexonly{\begin{figure}[tp!]}
                                                \centering
      \includegraphics[width=1.0\textheight]{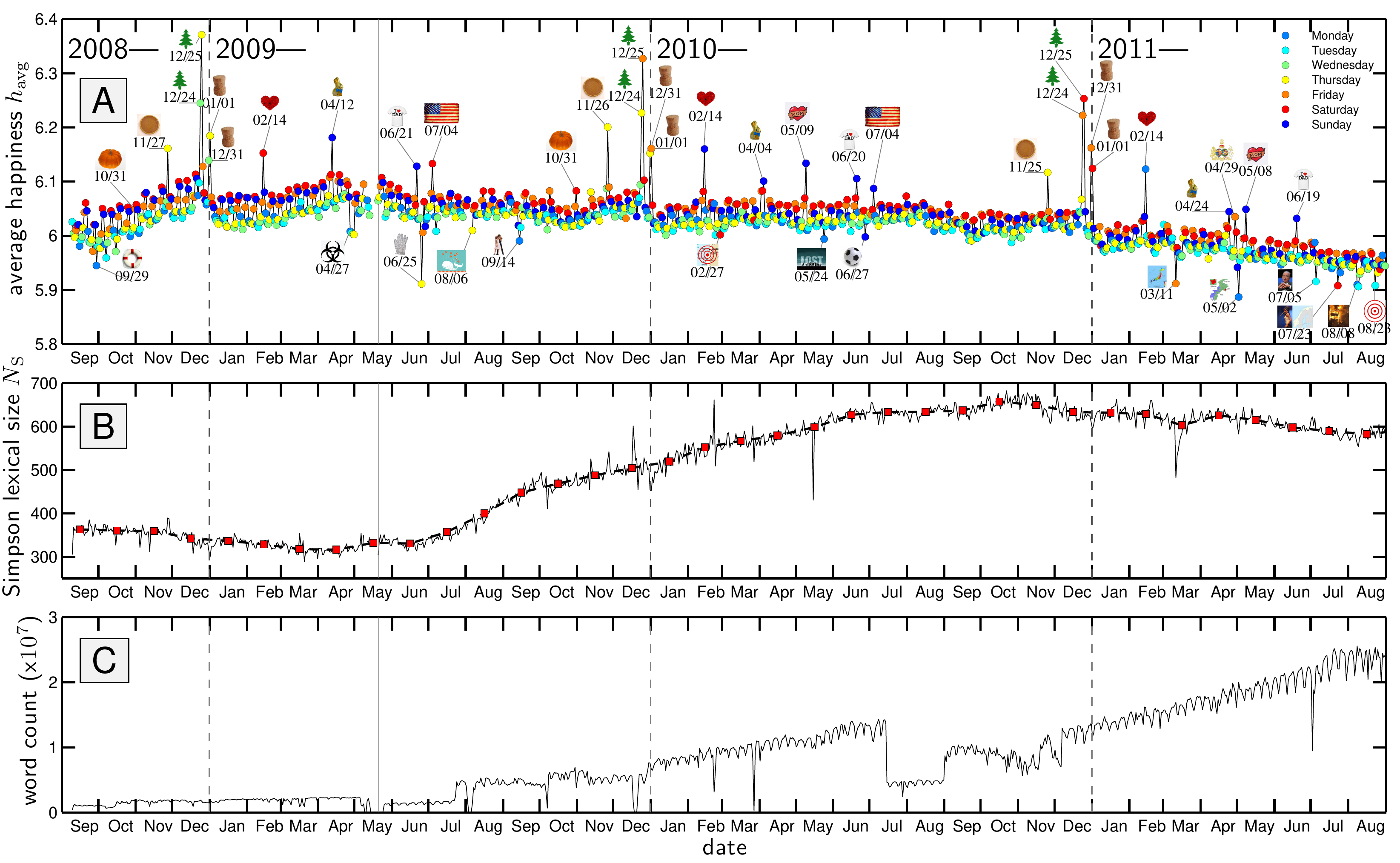} 
      \caption{
        Overall happiness, information, and count time series for all tweets averaged by individual day.
        \textbf{A.} 
        Average happiness measured
        over a three year period running from September 9, 2008 to August 31, 2011
        (see Sec.~\ref{sec:twhap.measurement2} for measurement explanation).
        A regular weekly cycle is clear with the red and blue of Saturday and Sunday
        typically the high points (examined further in Fig.~\ref{fig:twhap.twitter_timeseries_2a}).
        Post May 21, 2009 (indicated by a solid vertical line), 
        we use reported local time to assign tweets to particular dates.
        See also Figs.~\ref{fig:twhap.twitter_timeseries_zoomable_supp} and~\ref{fig:twhap.twitter_timeseries_simple_supp}.
        \textbf{B.}       
        Simpson lexical size $\Nsimpson$ as a function of date using Simpson's concentration as
        the base entropy measure (solid gray line; see Sec.~\ref{sec:twhap.measurement2}).
        The red squares with the dashed line show $\Nsimpson$ as a function of 
        calendar month.
        \textbf{C.} 
        The number of words extracted from all tweets as a function of date
        for which we used evaluations from Mechanical Turk.
        For both the happiness and Simpson lexical size plots, 
        we omit dates for which we have less than 1000 words with evaluations.
      }
      \label{fig:twhap.twitter_timeseries_1}
      \plainlatexonly{\end{figure}}
    \revtexonly{\end{figure*}}
\revtexonly{\end{turnpage}}

We observe a variety of temporal trends in 
happiness and information content
across timescales of hours, days, months, and years.
In Fig.~\ref{fig:twhap.twitter_timeseries_1}A
we present the average happiness time series
with tweets binned by day.
The accompanying plots,
Figs.~\ref{fig:twhap.twitter_timeseries_1}B
and~\ref{fig:twhap.twitter_timeseries_1}C,
show the Simpson lexical size $\Nsimpson$,
discussed in Sec.~\ref{subsec:twhap.infocontent},
and
the number of words for which we have evaluations
from Mechanical Turk (using $\Delta \havgfn = 1$).
We expect such a coarse-grained averaging to leave
only truly system wide signals, and
as we show later in Section~\ref{sec:twhap.keywords}, 
subsets of tweets
exhibit markedly different temporal trends.
In the Supplementary Information, 
we provide a zoomable, high resolution version,
Fig.~\ref{fig:twhap.twitter_timeseries_zoomable_supp},
as well as simpler plots of the time series only,
Fig.~\ref{fig:twhap.twitter_timeseries_simple_supp}.

Looking at the complete time series, 
we see that after a gradual
upward trend that ran from January to April, 2009,
the overall time series has shown a gradual downward
trend, accelerating somewhat over the first half of 2011.
We also see that average happiness gradually increased over the 
last months of 2008, 2009, and 2010, and dropped in January of
the ensuing years.
Moving down to timescales less than a month,
we see a clear weekly signal with
the peak generally 
occurring over the weekend, and the nadir on Monday and Tuesday
(c.f., \cite{balog2006a,mishne2006a,mihalcea2006a,kramer2010a,dodds2009b}).
We return to and examine the weekly cycle in detail in 
Sec.~\ref{sec:twhap.weeklycycle}.

\subsection{Outlier Dates}
\label{subsec:twhap.outlierdays}

At the scale of a day, we find a number of dates
which strongly deviate in their happiness levels 
from nearby dates, and we indicate these 
in Fig.~\ref{fig:twhap.twitter_timeseries_1}A.
We discuss positive and negative dates separately,
noting that anomalously positive days occur mainly on
annual religious, cultural, and national events,
whereas negative days 
typically arise from unexpected societal trauma
due for example to a natural disaster
or death of a celebrity.
(See~\cite{balog2006b} for similar, earlier work on blogs.)

In the following section, 
we look more closely at several dates, 
showing how individual words contribute
to their anomalous measurements.

For the outlying happy dates,
in 2008, 2009, and 2010, Christmas Day returned
the highest levels of happiness, followed by
Christmas Eve.
Other relatively positive dates include
New Year's Eve and Day,
Valentine's Day, Thanksgiving, Fourth of July, Easter Sunday,
Mother's Day, and Father's Day.
All of these observations are sensible,
and reflect a strong (though not universal) 
degree of social synchrony.
The spikes for Thanksgiving and Fourth of July reflects
the fact that while Twitter is a global service, the majority of users
still come from the United States~\cite{twitterdemog}.
The only singular, non-annual event to stand out as a positive day
was that of 
the Royal Wedding of Prince William and Catherine Middleton,
April 29, 2011.

Over the entire time span, we see substantial, system-wide drops in happiness
in response to a range of disparate events, both exogenous and endogenous
in nature. 
Working from the start of our time series, we first see
the  Bailout of the U.S.\ financial system, which induced
a multi-week depression in our time series.
The lowest point corresponds to Monday, September 29, 2008,
when the U.S.\ government agreed to an unprecedented purchase
of toxic assets in the form of mortgage backed securities.

Following the 2008 Bailout, we see the overall time series 
rebound well through the end of 2008, suffer the usual
post New Year's dip, and begin to rise again until
an extraordinary week long drop due to the 
onset of the 2009 swine flu or H1N1 pandemic.

The next decline occurred with Michael Jackson's death,
the largest single day drop we observed.
His memorial on July 7, 2009 induced another clear negative signal.
The death of actor Patrick Swayze on September 14, 2009
also left a discernible negative impact on the time series.
In between, Twitter itself was the victim of a large-scale
distributed denial of service attack, leading to an outage
of the service; upon resumption, tweets were noticeably
focused on this internal story.

Several natural disasters registered as 
days with relatively low happiness:
the February, 2010 Chilean earthquake,
the October, 2010 record size storm complex across
the U.S.,
and the
March, 2011 earthquake and tsunami 
which devastated Japan.

Reports of the killing of Osama Bin Laden on May 2, 2011
resulted in the day of the lowest happiness across
the entire time frame. 
And global sport left one identifiable drop:
the 4--1 victory of Germany over England in the 2010
Football World Cup.  Spain's ultimate victory in the
tournament was detectable in terms of word usage but
did not lead to a significant change in overall
happiness.

One arguably false finding of a cultural event
being negative
was the finale of the last
season of the highly rated television show `Lost',
marked by a drop in our time series on May 24, 2010,
and in part due to the word `lost'
having a low happiness score of $\havgfn$=2.76,
but also to an overall increase in negative words
on that date.

A number of these departures for specific dates
qualitatively match observations we made
in our earlier work
on blogs~\cite{dodds2009b},
though we make any comparison tentatively as
for blogs we focused on sentences written in
the first person containing a conjugation of the
verb `to feel'~\cite{harris2009a}.
For example, 
Christmas Eve and Day, New Year's Eve and Day, and 
Valentine's Day 
all exhibit jumps in happiness in both tweets and `I feel...' blog sentences.
Both time series also show a pronounced drop for Michael Jackson's death.
However, tweets did not register a similar lift as blogs for 
the US Presidential Election in 2008 and Inauguration Day, 2009,
while positive sentiment for both Mother's and Father's Day,
the Fourth of July, are much more evident in tweets.
Lastly, blogs typically showed drops for September 10 and/or 11 
that are largely absent in tweets, although relevant negative words
appear more frequently on those dates (e.g., `lost', `victims', and `tragedy').

\subsection{Word Shift Analysis}
\label{subsec:twhap.wordshift}

\begin{figure*}[tp!]
  \centering
    \includegraphics[width=\textwidth]{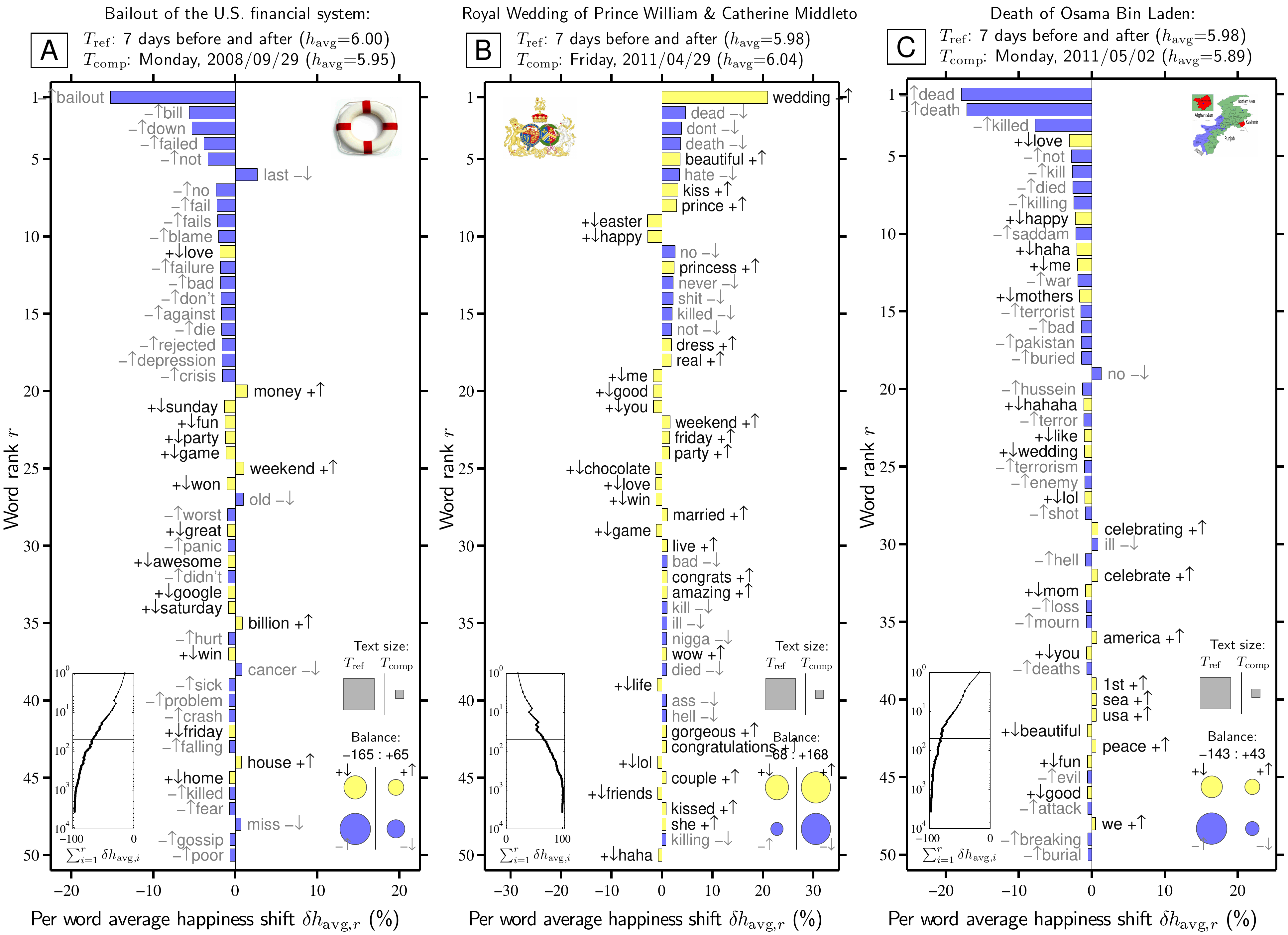}
  \caption{
      Word shift graph showing how changes in word frequencies
      produce spikes or dips in happiness for three example dates,
      relative to the 7 days before and 7 days after each date.
      Words are ranked by their 
      percentage contribution to the change
      in average happiness, $\delta h_{\rm avg,i}$.
      The background 14 days are set as the
      reference text ($T_{\rm ref}$) and 
      the individual dates
      as the comparison text ($T_{\rm comp}$).
      How individual words contribute to the shift
      is indicated by a pairing of two symbols:
      $+/-$ shows the word is more/less happy than $T_{\rm ref}$ as a whole,
      and 
      $\uparrow/\downarrow$ shows that the word is more/less
      relatively prevalent in $T_{\rm comp}$ than in $T_{\rm ref}$.
      Black and gray font additionally encode 
      the $+$ and $-$ distinction respectively.
      The left inset panel shows how the ranked 3,686 labMT 1.0 words (Data Set S1)
      combine in sum (word rank $r$ is shown on a log scale).
      The four circles in the bottom right show the total
      contribution of the four kinds of words 
      ($+$$\downarrow$,
      $+$$\uparrow$,
      \textcolor{gray}{$-$$\uparrow$},
      \textcolor{gray}{$-$$\downarrow$}).
      Relative text size is indicated by the areas
      of the gray squares.
      See Eqs.~\ref{eq:twhap.deltah} and \ref{eq:twhap.deltahfinal}
      and Sec.~\ref{subsec:twhap.wordshift} for complete details.
  }
  \label{fig:twhap.twitter_timeseries_wordshifts}
\end{figure*}

When comparing two or more texts
using a single summary statistic,
as we have here with average happiness,
we naturally need to look further into why
a given measure shows variation.
In Fig.~\ref{fig:twhap.twitter_timeseries_wordshifts}
we provide `word shift graphs'
for three outlier days
relative to the seven preceding and seven ensuing
days combined: 
the 2008 Bailout of the U.S.\ financial system,
the 2011 Royal Wedding,
and
Osama Bin Laden's death
(we include corresponding graphs for all 
identified outlier days 
in the Supplementary Information, 
Figs.~\ref{fig:twhap.interestingdates-supp001}--\ref{fig:twhap.interestingdates-supp046}).
We will use these word shift graphs,
which we introduced in~\cite{dodds2009b}
and improve upon here,
throughout the remainder of the paper
to illuminate how the 
difference between two texts' happiness levels arises
from changes in underlying word frequency.
In view of the utility of these graphs,
we take time now to describe and explain them in detail.

Consider two texts $T_{\rm ref}$ (for reference) 
and $T_{\rm comp}$ (for comparison)
with happiness scores $\havgsup{(\rm ref)}$ and $\havgsup{(\rm comp)}$.
If we wish to compare $T_{\rm comp}$ relative to $T_{\rm ref}$ then,
using \Req{eq:twhap.htau}, we can write
\begin{gather}
  \havgsup{(\rm comp)} - \havgsup{(\rm ref)}
  =
  \sum_{i=1}^{N} 
  \havg{w_i} 
  \left[
    p_i^{(\rm comp)} - p_i^{(\rm ref)}
  \right]
  \nonumber \\ 
  =
  \sum_{i=1}^{N} 
  \left[
    \havg{w_i} - \havgsup{(\rm ref)}
  \right]
  \left[
    p_i^{(\rm comp)} - p_i^{(\rm ref)}
  \right]
  \label{eq:twhap.deltah}
\end{gather}
where we have employed the fact that
$$
\sum_{i=1}^{N} 
\havgsup{(\rm ref)}
\left[
  p_i^{(\rm comp)} - p_i^{(\rm ref)}
\right]
=
\havgsup{(\rm ref)}
\sum_{i=1}^{N} 
\left[
  p_i^{(\rm comp)} - p_i^{(\rm ref)}
\right]
$$
$$
=
\havgsup{(\rm ref)}
(1-1)
=
0.
$$
In introducing the term $-\havgsup{(\rm ref)}$,
we are now able to make clear the contribution of the $i$th word
to the difference $\havgsup{(\rm comp)} - \havgsup{(\rm ref)}$.
From the form of \Req{eq:twhap.deltah},
we see that we need to consider two aspects
in determining the sign of the $i$th word's contribution:
\begin{enumerate}
\item Whether or not the $i$th word is on average 
happier than text $T_{\rm ref}$'s average, $\havgsup{(\rm ref)}$;
and
\item Whether or not the $i$th word is relatively
more abundant in text $T_{\rm comp}$ than in text $T_{\rm ref}$.
\end{enumerate}
We will signify a word's happiness relative
to text $T_{\rm ref}$ by $+$ (more happy) and $-$ (less happy),
and its relative abundance in text $T_{\rm comp}$ versus
text $T_{\rm ref}$ with $\uparrow$ (more prevalent) 
and $\downarrow$ (less prevalent).
Combining these two binary possibilities leads to four cases:
\begin{enumerate}
\item[\textcolor{black}{$+$$\uparrow$}:] 
  Increased usage of relatively positive words---If a word is 
  happier than text $T_{\rm ref}$ ($+$)
  and appears relatively more often in text $T_{\rm comp}$ ($\uparrow$),
  then the contribution to the
  difference $\havgsup{(\rm comp)} - \havgsup{(\rm ref)}$
  is positive;
\item[\textcolor{gray}{$-$$\downarrow$}:] 
  Decreased usage of relatively negative words---If a word is 
  less happy than text $T_{\rm ref}$ ($-$)
  and appears relatively less often in text $T_{\rm comp}$ ($\downarrow$), 
  then the contribution to the
  difference $\havgsup{(\rm comp)} - \havgsup{(\rm ref)}$
  is also positive;
\item[\textcolor{black}{$+$$\downarrow$}:]
  Decreased usage of relatively positive words---If a word is 
  happier than text $T_{\rm ref}$ ($+$)
  and appears relatively less often in text $T_{\rm comp}$ ($\downarrow$),
  then the contribution to the
  difference $\havgsup{(\rm comp)} - \havgsup{(\rm ref)}$
  is negative;
  and
\item[\textcolor{gray}{$-$$\uparrow$}:]
  Increased usage of relatively negative words---If a word is 
  less happy than text $T_{\rm ref}$ ($-$)
  and appears relatively more often in text $T_{\rm comp}$ ($\uparrow$), 
  then the contribution to the
  difference $\havgsup{(\rm comp)} - \havgsup{(\rm ref)}$
  is also negative.
\end{enumerate}

For the convenience of visualization,
we normalize the summands in~\Req{eq:twhap.deltah}
and convert to percentages to obtain:
\begin{eqnarray}
  \label{eq:twhap.deltahfinal}
  \lefteqn{\delta h_{\rm avg,i}  = } \nonumber \\
  &
  \frac{100}{
    \left|
      \havgsup{(\rm comp)} - \havgsup{(\rm ref)}
    \right|
    }
  \underbrace{
    \left[
      \havg{w_i} - \havgsup{(\rm ref)}
    \right]
  }_{+/-}
  \underbrace{
    \left[
      p_i^{(\rm comp)} - p_i^{(\rm ref)}
    \right]
    }_{\uparrow/\downarrow},
\end{eqnarray}
where $\sum_i \delta h_{\rm avg,i}= \pm 100$,
depending on the sign of the difference
in happiness between the two texts, $\havgsup{(\rm comp)} - \havgsup{(\rm ref)}$,
and where we have indicated the terms to which the symbols
$+/-$ and $\uparrow/\downarrow$ apply.
We call $\delta h_{\rm avg,i}$ the per word
happiness shift of the $i$th word.

Finally, in comparing two texts,
we rank words by their absolute contribution 
to the change in average happiness, $|\delta h_{\rm avg,i}|$,
from largest to smallest.
In doing so, we are able to make clear the 
most important words driving the separation of
two texts' emotional content.

With these definitions in hand,
we return to Fig.~\ref{fig:twhap.twitter_timeseries_wordshifts}
to complete our explanation of word shift graphs.
For brevity we will refer to these graphs
with the terms Bailout, Royal Wedding, and Bin Laden.

The primary element of our word shift graphs
is a central bar graph 
showing a desired number of highest ranked
labMT 1.0 words (Data Set S1)
as ordered 
by their absolute contribution to the change
in average happiness, $|\delta h_{\rm avg,i}|$.  
In Fig.~\ref{fig:twhap.twitter_timeseries_wordshifts},
the word shift graphs show the first 50 words for each date.
Bars corresponding to words that are more happy than the 
reference text $T_{\rm ref}$
are colored yellow, and less happy ones are colored blue.

In each graph in Fig.~\ref{fig:twhap.twitter_timeseries_wordshifts},
we see examples of each of the four ways
words can contribute to $\havgsup{(\rm comp)} - \havgsup{(\rm ref)}$.
For the Bailout,
both kinds of negative changes dominate
with 42 of the top 50 shifts,
including more of the relatively negative words
`bailout',
`bill',
`down', 
`no',
`not',
`fail',
`blame',
and
`panic'
(all \textcolor{gray}{$-$$\uparrow$}),
and less of the relatively positive words
`fun',
`party',
`game',
`awesome',
and
`home'
(all $+$$\downarrow$).
For the Bin Laden graph, 40 out of the first 
50 ranked words contribute to the overall drop
(bars on left).
The strongest decreases come
from `dead' and `death' and
these combine with more negativity found in
`killed',
`kill',
`died',
`killing',
`terrorist',
`buried',
and 
`Pakistan' (all \textcolor{gray}{$-$$\uparrow$}).

By contrast,
we see the happiness spike of the Royal Wedding is
due to higher prevalence of positive words
such as
`wedding',
`beautiful',
`kiss',
`prince',
`princess',
`dress',
and 
`gorgeous'
(all $+$$\uparrow$),
and a relative dearth of negative words such as 
`dead',
`death',
`hate',
`no',
`never',
and several profanities
(all \textcolor{gray}{$-$$\downarrow$}).

Beyond these dominant stories, our word shifts allow
us to make a number of supporting and clarifying observations.
First, since we have chosen to compare specific dates
to the surrounding 14 days, nearby anomalous events
appear in each word shift.  
For example, the Royal Wedding (2011/4/29)
has less `Easter' and `chocolate' because Easter occurred
five days earlier and less `dead' and `killed' 
because of Bin Laden's death three days later (2011/5/2).
The Bin Laden graph in turn shows
less `wedding', `happy', and `Mother's' (due to the Royal Wedding
and Mother's Day, 2011/5/8).
Other reference texts can be readily constructed 
for comparisons (e.g., tweets on all days or matching weekdays).
However, we find that the main words contributing to
word shifts reliably appear as we consider 
alternative, reasonable reference texts.

Second, in all text comparisons, we find some
words go against the main trend.
For example, 
we see more `money', `weekend', and `billion' 
(all \textcolor{black}{$+$$\uparrow$}),
and less `last' and `old'
(all \textcolor{gray}{$-$$\downarrow$})
for the Bailout word shift;
less `me', `good', and `haha' for the Royal Wedding
(all \textcolor{black}{$+$$\downarrow$});
and more 
`celebrating',
`America',
and
`USA'
for Bin Laden's death
(all \textcolor{black}{$+$$\uparrow$}).
Some shifts are genuinely at odds with the 
overall shift (e.g., `celebrating' for Bin Laden)
while others appear
due to our omission of context 
(e.g., the generally positive word
`money' was not being talked about in
a positive way during the Bailout).
In the case of the Bailout, our instrument overcomes its
inherent coarseness to yield intuitive overall measurements.
For Bin Laden's death, which would arguably be a positive moment
for many users of Twitter, the death of a profoundly negative character
results in word usage that appears, not unreasonably,
as a surge of negative emotion.  
Every reading on our hedonometer, anomalous or not,
and indeed that of any sentiment measurement,
must be validated by plain demonstration of
which words are most salient.

The three insets in the word shift graphs
of Fig.~\ref{fig:twhap.twitter_timeseries_wordshifts}
expand the story provided by the main bar charts in the following ways.
First and simplest is the pair of gray squares on the right which 
show, by their area, the relative sizes of the two texts, as measured by
the total number of 
labMT 1.0 words (Data Set S1)
(the absolute number of words is not indicated).  
For these comparisons, the ratio is therefore approximately 14:1.

Second, on the bottom left of each word shift graph, 
the inset line graph shows the cumulative sum
of the individual word contributions, $\sum_{i=1}^{r} \delta h_{\rm avg,i}$
as a function of $\log_{10} r$ where $r$ is word rank.
The graph shows how rapidly the word contributions converge to $\pm 100\%$
as we include all 3,686 words.  
The solid line marks 50 words, the number of words in the main panel.
We typically see that the first 1000 words
account for more than 99\% of the entire shift.

The third and final inset on the bottom right is a key one.
An increase in happiness may be due to the use of more positive 
words, an avoidance of negative words, or a combination of both,
and we need to quantify this in a simple way.
The inset's four circles show the relative total contributions of the four classes of words
to the overall shift in average happiness.  
For example, the area of the top right
(yellow) circle represents the sum of all contributions due to
relatively positive words that increase in frequency in $T_{\rm comp}$
with respect to $T_{\rm ref}$ ($+$$\uparrow$).  
We find that the sizes of these
circles are not always transparently connected to the top 50 words,
with smaller contributions combining over the full set of 3,686 words.  

The two numbers above the circles give the total percentage
change toward and away from the reference text's average happiness.
For the Bailout example, there is a drop in happiness of -165\% 
of $\havgsup{(\rm comp)} - \havgsup{(\rm ref)}$
due to less use of positive words, $+$$\downarrow$,
and more use of negative words, $-$$\uparrow$.
On the other side, 
more frequent positive words, $+$$\uparrow$,
and less frequent negative words, $-$$\downarrow$,
contribute to a rise in happiness equal to +65\% of 
$\havgsup{(\rm comp)} - \havgsup{(\rm ref)}$.
The two changes combine to give -100\% of
$\havgsup{(\rm comp)} - \havgsup{(\rm ref)}$.

For the Bailout and Bin Laden graphs,
we see similar overall patterns:
the more frequent use of negative words ($-$$\uparrow$)
dominates while
the less frequent use of positive words ($+$$\downarrow$)
is also substantive;
and we see the smaller countering effects of the other two
classes of words are about equal
($+$$\uparrow$ and $-$$\downarrow$).
For the Royal Wedding, the relative increase in happiness
of the day is equally due to more frequent use of positive words
and less frequent use of negative words
($+$$\uparrow$ and $-$$\downarrow$),
while very few negative words are more prevalent
($-$$\uparrow$).

\subsection{Information Content}
\label{subsec:twhap.infocontent}

To complete our analysis of the overall
time series, we turn to information content (Fig.~\ref{fig:twhap.twitter_timeseries_1}B).
We see a strong increase in Simpson lexical size $\Nsimpson$
climbing from approximately 300 to 700 words beginning around July, 2009.
(For $q > 1.5$, generalized word diversities all follow the same trajectory
with $\Neqq$ increasing as $q$ decreases.)
We also indicate in the same plot $\Nsimpson$ measured
at the scale of months (red squares). 
The smoothness
of the resulting curve shows that $\Nsimpson$ 
is unaffected by the two issues of missing data
and non-uniform sampling rates.
(Note that the month estimates of $\Nsimpson$ 
are computed from the word distribution for the
month and are not simply averages of daily
values of $\Nsimpson$.)

By examining shifts in word usage, 
we are able to attribute the more than doubling of $\Nsimpson$ to 
a strong relative increase in non-English languages,
notwithstanding the dramatic growth in English language tweets.
Recalling that the most
common words such as articles and prepositions figure
most strongly in the computation of the Simpson word diversity,
we see the dominant growth 
in Spanish (`que', `la', `y',`en', `el').
A few other example languages making headway
are Portuguese (`pra'), 
which also shares some
common words with Spanish,  and
Indonesian (`yg'). By contrast, English words appear relatively less
(including the word `Twitter')
while a minority of words move against the general
diversification by appearing more frequently,
with prominent examples being the abbreviations `RT' (for retweet) and `lol'
(for laugh out loud).

\section{Weekly cycle}
\label{sec:twhap.weeklycycle}

\subsection{Average Happiness of Weekdays}
\label{subsec:twhap.weeklycycle-happiness}

\begin{figure}[tp!]
              \centering
  \includegraphics[width=0.48\textwidth]{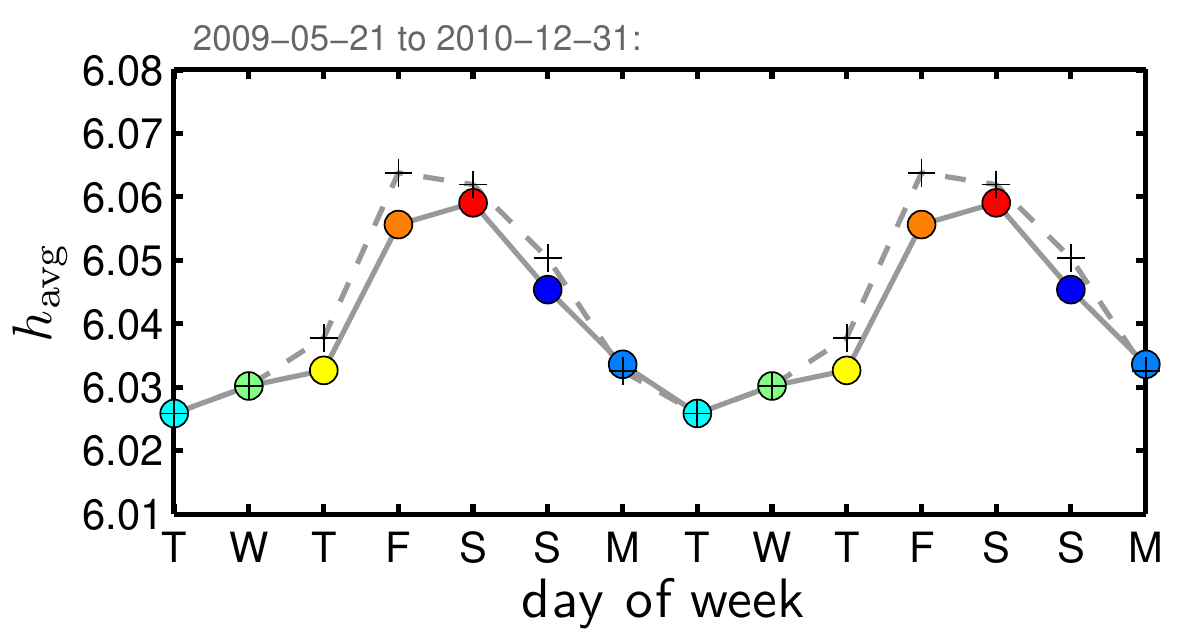}
  \caption{
    Average happiness as a function
    of day of the week for our complete data set.
    To make the average weekly cycle more clear, 
    we repeat the pattern for a second week.
    The crosses indicate happiness scores based on
    all data, while the filled circles show the results
    of removing
    the outlier days indicated 
    in Fig.~\ref{fig:twhap.twitter_timeseries_1}A.
    The colors for the days of the week match those used
    in Fig.~\ref{fig:twhap.twitter_timeseries_1}A.
    To circumvent the non-uniform sampling of tweets throughout time,
    we compute an average of averages: for example, we 
    find the average
    happiness for each Monday separately, and then average
    over these values, thereby giving equal weight to
    each Monday's score.
    We use data from May 21, 2009 to December 31, 2010, for which
    we have a local timestamp.
  }
  \label{fig:twhap.twitter_timeseries_2a}
\end{figure}

As we saw in Fig.~\ref{fig:twhap.twitter_timeseries_1}A,
a pronounced weekly cycle is present in the overall time series.
To reveal this feature more clearly, we compute average happiness
$\havgfn$  as a function of day of the week, Fig.~\ref{fig:twhap.twitter_timeseries_2a}.
Taking tweets for which we have local time information (May 21, 2009 onward), 
we show two curves, one for which we include
all data (crosses, dashed line), and one for which we exclude
the outlier days we identified 
in Fig.~\ref{fig:twhap.twitter_timeseries_1}A (labeled dates accompanied by icons).
Including outlier days yields a higher average happiness,
and the difference between the two curves is most pronounced
on Thursday, Friday, Saturday, and Sunday.
These discrepancies are explained by 
Thanksgiving (Thursday), and Christmas
Eve and Day and New Year's Eve and Day falling on
Thursday and Friday in 2009
and
Friday and Saturday in 2010,
as well as annual events such as Mother's Day
occurring on Sundays.

We take the reasonable step of focusing on 
the data with outlier days removed.
We see Saturday has the highest average happiness ($\havgfn \simeq 6.06$), 
closely followed by Friday and then Sunday.
From Saturday, we see a steady decline until
the weekly low occurs on Tuesday, which is then followed
by small increases on both Wednesday and Thursday ($\havgfn \simeq 6.03$).
We see a jump on Friday, leading back to the peak of Saturday.
Roughly similar patterns have been found in Gallup polls~\cite{gallup},
in Facebook by the company's internal research team~\cite{facebookgnh},
in binary sentiment analysis of tweets~\cite{golder2011a},
and in analyses of smaller collections of tweets~\cite{northeasterntwittermood}.
(In the last work and in contrast to our findings here for a data set
tenfold larger in size, Thursday evening was identified as the low point of the week.)

\begin{figure}[tp!]
      \centering
  \includegraphics[width=0.48\textwidth]{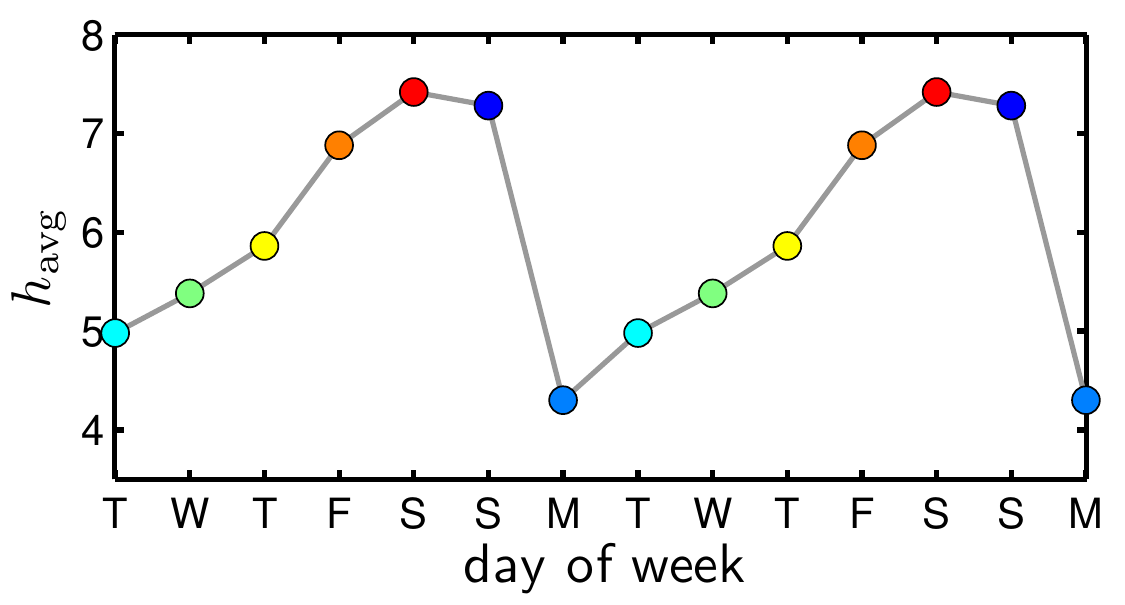}
  \caption{
    Evaluations of the individual days of the week 
    as isolated words using Mechanical Turk.
  }
  \label{fig:twhap.twitter_timeseries_2a_happy}
\end{figure}

While the weekend peak in the cycle conforms with everyday intuition,
the minimum on Tuesday goes against
standard notions of the Monday blues
with its back-to-work nature, and 
Wednesday's middle-of-the-week labeling as the 
work week's hump day~\cite{mihalcea2006a}.
To provide a quantitative comparison,
in Fig.~\ref{fig:twhap.twitter_timeseries_2a_happy},
we show how people's perception of days of the week
varies based on our Mechanical Turk study,
i.e., how people rate the words `Monday', `Tuesday', etc.,
when presented with them in a survey.
The overall pattern is similar in terms of ordering with
the exception of `Monday' being rated the lowest rather
than `Tuesday', and `Sunday' is rated above `Friday'.
The range of happiness is also much greater,
4.30 for `Monday' to 7.42 for `Saturday',
sensibly so since we are now considering evaluations
of individual words with no averaging over texts.
While people collectively have strong opinions
about the word `Monday', the reality, at least in terms of tweets,
is that Tuesday is the week's low point.

In our earlier work on blogs using 
the ANEW study word list~\cite{dodds2009b},
we saw a statistically significant
but much weaker cycle
for the days of the week;
the high and low days were Sunday
and Wednesday (see also~\cite{mihalcea2006a}).
The discrepancy appears to be due 
to the in-the-moment character of Twitter
versus the reflective one of blogs.

\begin{figure}[tp!]
                    \centering
  \includegraphics[width=0.48\textwidth]{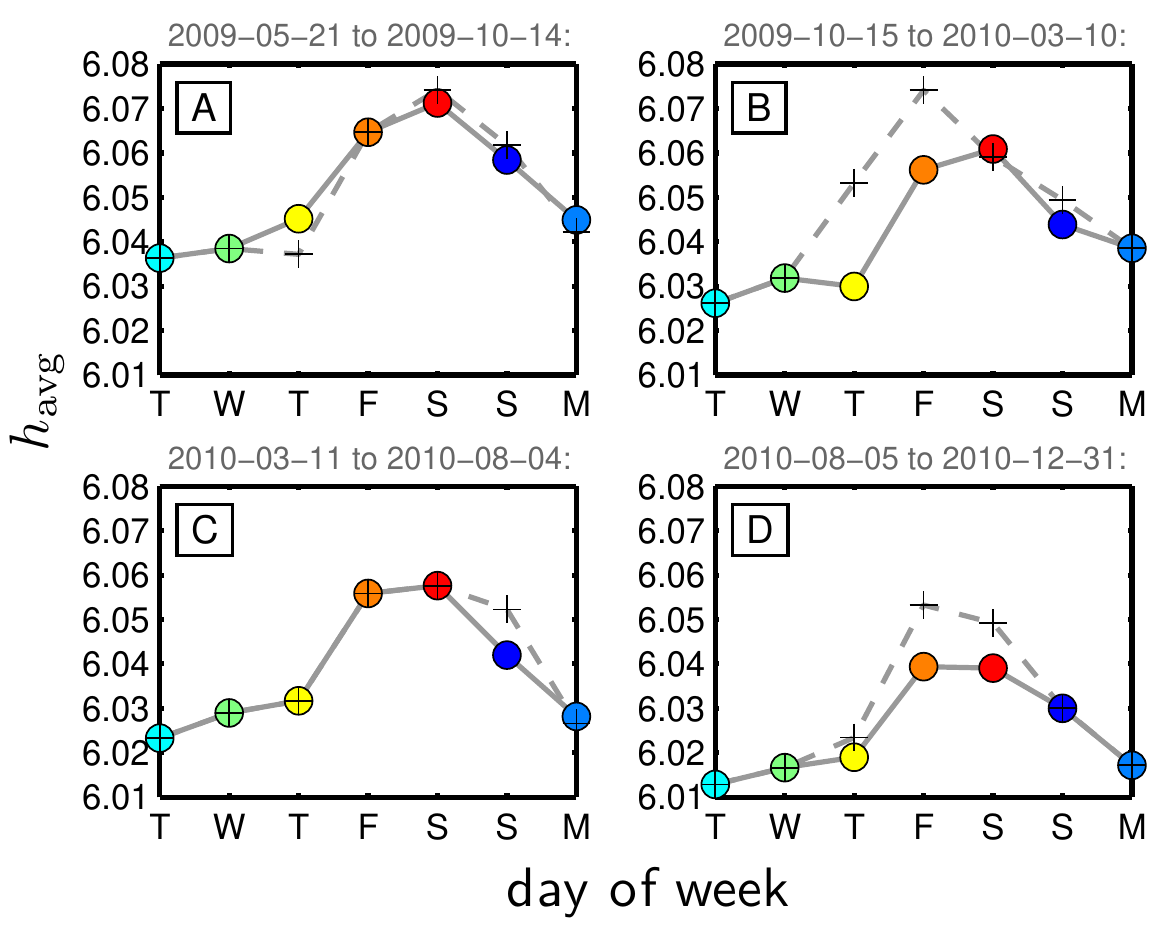}
  \caption{
    Average of daily average happiness for days of the week over
    four consecutive time periods of approximately five months duration each.
    As per Fig.~\ref{fig:twhap.twitter_timeseries_2a},
    crosses are based on all days, circles for days excluding
    outlier days marked in Fig.~\ref{fig:twhap.twitter_timeseries_1}.
    The vertical scale is the same
    in each plot and matches that used in Fig.~\ref{fig:twhap.twitter_timeseries_2a}.
  }
  \label{fig:twhap.dow-timechange}
\end{figure}

With any observed pattern, a fundamental issue is universality.
Is the three day midweek low followed by a peak around Saturday
a pattern we always see, given enough data?  
Further inspection of our Twitter data set shows 
a constancy in the weekly cycle occurring over time.
In Fig.~\ref{fig:twhap.dow-timechange}, we aggregate tweets
for days of the week for four time ranges, approximately equal in duration.
As before, we show the weekly pattern for all days (crosses, dashed curve)
and with outlier days marked in
Fig.~\ref{fig:twhap.twitter_timeseries_1}A removed (disks, solid curve).
The major differences we observe between
these two curves in the four panels
are predominantly explained as before by Christmas, New Year's, and Thanksgiving.
In terms of universality, we again see that Friday-Saturday-Sunday represents the peak
while Tuesday's level is the minimum in each period.
Only for Thursday in Fig.~\ref{fig:twhap.dow-timechange}B
do we see a change in the overall ordering of days.
Thus, we have some confidence that the overall weekly
cycle of happiness shown in Fig.~\ref{fig:twhap.twitter_timeseries_2a}
is a fair description of what appears to be a robust pattern
of users' expressed happiness.

\begin{figure}[tp!]
  \centering
          \includegraphics[width=0.48\textwidth]{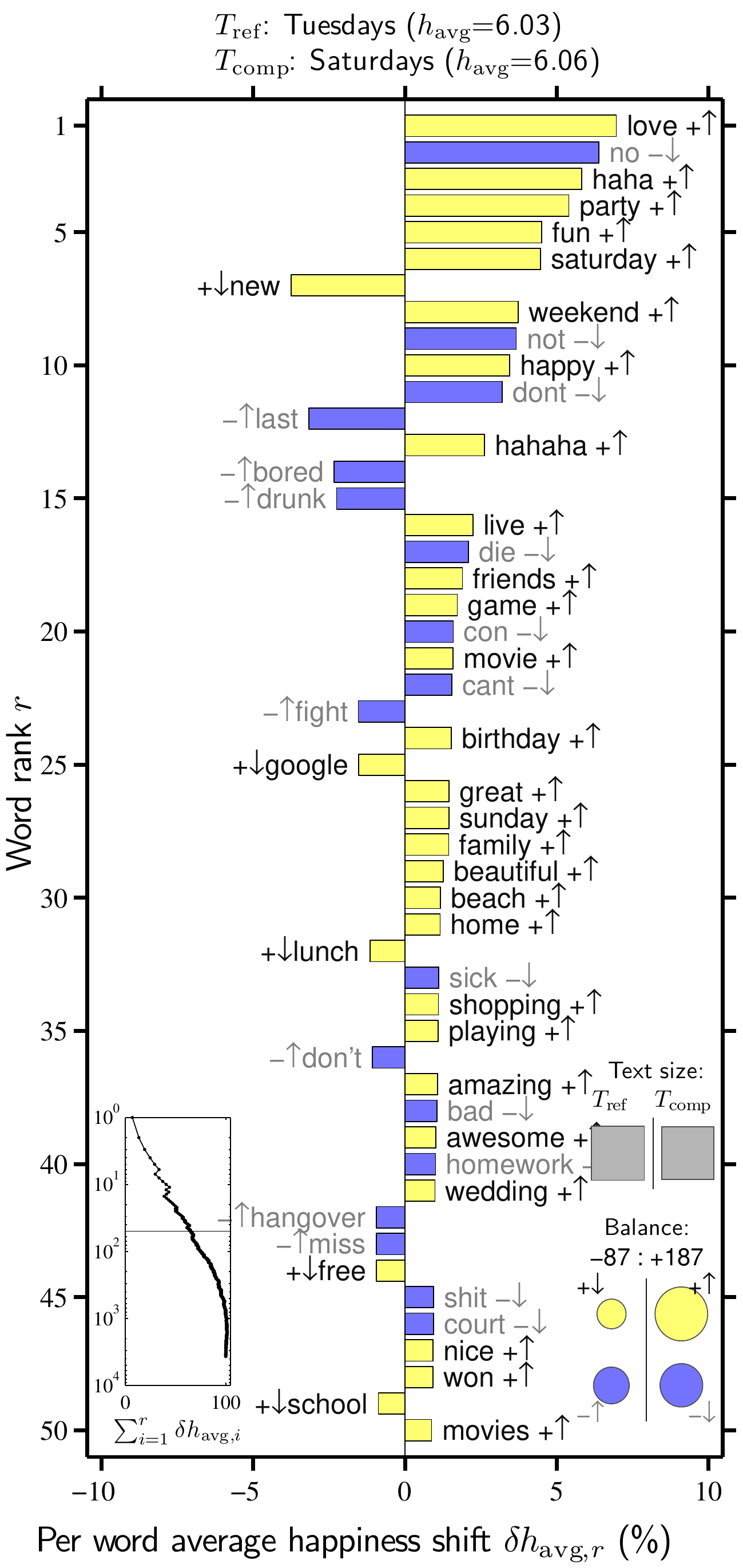}
  \caption{
    Word shift graph comparing Saturdays relative to Tuesdays.
    Each day of the week's word frequency distribution
    was generated by averaging normalized distributions
    for each instance of that week day in May 21, 2009 to December 31, 2010,
    with outlier dates removed.
    See Fig.~\ref{fig:twhap.twitter_timeseries_2b_supp} 
    in Supplementary Information for 
    word shifts based on alternate distributions.
      }
  \label{fig:twhap.twitter_timeseries_2b}
\end{figure}

\subsection{Word Shift Analysis}
\label{subsec:twhap.weeklycycle-wordshifts}

In Fig.~\ref{fig:twhap.twitter_timeseries_2b},
we present a word shift graph comparing tweets made on Saturdays relative
to those made on Tuesdays.
We created word frequency distributions 
for each day  by averaging
normalized distributions from
May 21, 2009 to December 31, 2010,
removing the outlier dates marked in 
Fig.~\ref{fig:twhap.twitter_timeseries_2a}A.
Alternate ways of creating the weekday distributions
do not change the word shifts appreciably
(See Fig.~\ref{fig:twhap.twitter_timeseries_2b_supp}
in Supplementary Information).
The two kinds of positive changes dominate
with 38 of the top 50 changes,
including more of 
`love', 
`haha', 
`party', 
`fun',
`Saturday',
`happy',
and
`hahaha'
(all $+$$\uparrow$),
and less of
`no',
`not',
`don't',
`can't',
`bad',
and
`homework'
(all \textcolor{gray}{$-$$\downarrow$}).
These changes are readily interpretable, with the weekend
involving more leisure and family time, and
a relative absence of work, school, and related concerns.
Words in the top 50 which move against the general trend are
the more prevalent, relatively negative words
`last',
`bored',
`drunk',
`fight',
and
`hangover'
(\textcolor{gray}{$-$$\uparrow$}),
and the less frequent positive words 
`new',
`google',
and
`lunch'
($+$$\downarrow$).
Thus while Saturdays may be on average happier than Tuesdays,
we also see evidence of boredom, fighting, and suffering
due to excessive drinking.

The insets of Fig.~\ref{fig:twhap.twitter_timeseries_2b}
provide further insight and information.
The gray squares indicate the word base for Tuesdays and 
Saturdays are of comparable size.
From the bottom left line graph, 
we see again that around
1000 words account for the shift
in average happiness between Tuesday and Saturday, and that
the first 50 words make up approximately 60\% of the shift.

The bottom right inset shows that
the overall positive shift from Tuesdays to Saturdays
is due to the more frequent use of positive words ($+$$\uparrow$),
and to a lesser extent,
the less frequent use of negative words ($-$$\downarrow$).
On the other side of the ledger, we see
a smaller total contribution of words going against 
the trend of happier Saturdays,
noting that the increased use of certain negative words ($-$$\uparrow$)
is slightly more appreciable in impact than 
the less frequent use of positive words ($+$$\downarrow$).

\begin{figure}[tp!]
  \centering
                  \includegraphics[width=0.48\textwidth]{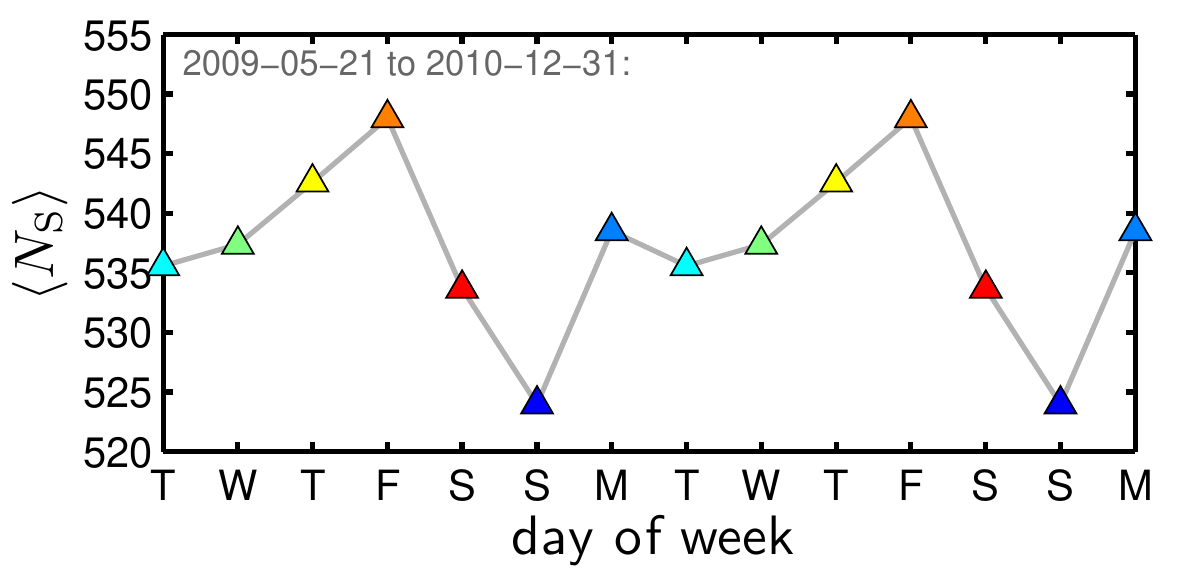}
      \caption{
    Simpson lexical size as a function
    of day of the week.
    We compute $\Nsimpson$ for individual dates
    Fig.~\ref{fig:twhap.twitter_timeseries_1}B,
    again excluding dates
    shown in Fig.~\ref{fig:twhap.twitter_timeseries_1}A,
    and then average these values.
    (See also Fig.~\ref{fig:twhap.twitter_timeseries_2a_simpson_supp}
    for the effects of alternate approaches.)
  }
  \label{fig:twhap.twitter_timeseries_2a_simpson}
\end{figure}

\subsection{Information Content}
\label{subsec:twhap.weeklycycle-info}

The average Simpson lexical size $\tavg{\Nsimpson}$ 
(Fig.~\ref{fig:twhap.twitter_timeseries_2a_simpson})
shows a pattern different 
to that of average happiness:
we observe
that a strong maximum appears on Friday
with a drop through the weekend to 
a distinct low on Sunday.
During the work week, Tuesday
presents a minor low, with a climb up
to Friday's high.
This pattern remains the same if 
we choose different averaging schemes
in generating a composite Simpson lexical size
(see also Fig.~\ref{fig:twhap.twitter_timeseries_2a_simpson_supp}).

To see further into these changes
between days,
we can generate word shift graphs for 
Simpson lexical size $\Nsimpson$.
These word shift graphs (not shown) are simpler
than those for average happiness
as they depend only on changes in word frequency.
Using the definition $\Nsimpson = 1/S = 1/\sum_{i=1}^{N} p_i^2$,
we obtain
\begin{eqnarray}
  \lefteqn{\Nsimpson^{(\rm{comp})} - \Nsimpson^{(\rm{ref})}
  = } \nonumber \\
  & 
  \frac{1}{
    S^{(\rm{comp})} S^{(\rm{ref})}
  }
  \sum_{i=1}^{N}
  \left(
    \left[
      p_i^{(\rm ref)} 
    \right]^2
    - 
    \left[
      p_i^{(\rm comp)}
    \right]^2
  \right).
\end{eqnarray}
We next define the individual percentage contribution in the shift in 
Simpson lexical size as
\begin{equation}
  \delta N_{{\rm{S}},i}  = 
  \frac{100}{
    \left|
      S^{(\rm ref)} - S^{(\rm comp)}
    \right|
    }
    \left(
      \left[
        p_i^{(\rm ref)} 
      \right]^2
      - 
      \left[
        p_i^{(\rm comp)}
      \right]^2
    \right),
    \label{eq:twhap.Nshift}
\end{equation}
where 
$\sum_{i} \delta N_{{\rm{S}},i} = \pm 100$ 
depending on the sign of
$S^{(\rm ref)} - S^{(\rm comp)}$.
Note that the reversal of the reference
and comparison elements in \Req{eq:twhap.Nshift} 
reflects the fact that any one word increasing
in frequency decreases overall diversity.
Further, no other diversity measure ($q \ne 2$)
allows for a linear superposition of contributions
such as we find in~\Req{eq:twhap.Nshift},
one of the reasons we provided earlier for choosing
a lexical size based on Simpson's concentration.

Using~\Req{eq:twhap.Nshift},
we find Friday's larger value of $\Nsimpson$
relative to Sunday's
can be attributed primarily to changes in 
the frequency of around 100 words.  
Most of these words
are those typically found at the start of a Zipf ranking
of a text, though their ordering is of interest.
A few words contributing the most to the shift
are 
`I',
`RT',
`you',
`me',
and 
`my'.
Decreases in the relative 
usage frequencies of personal pronouns
may suggest a shift in focus 
away from the self and toward the less
predictable, richer fare of Friday activities.
Words specific to Friday naturally appear
more frequently than on Sunday
serving to reduce Friday's Simpson lexical size.
Some examples include
`\#ff',
`follow',
`Friday',
`weekend',
and
`tonight'
(\#ff is an example of a hash tag,
in this case representing
a popular Friday custom of Twitter users 
recommending other users worth following).

\section{Daily cycle}
\label{sec:twhap.dailycycle}

\subsection{Average Happiness of Hours of the Day}
\label{subsec:twhap.dailycycle-happiness}

\begin{figure}[tp!]
  \centering
          \includegraphics[width=0.48\textwidth]{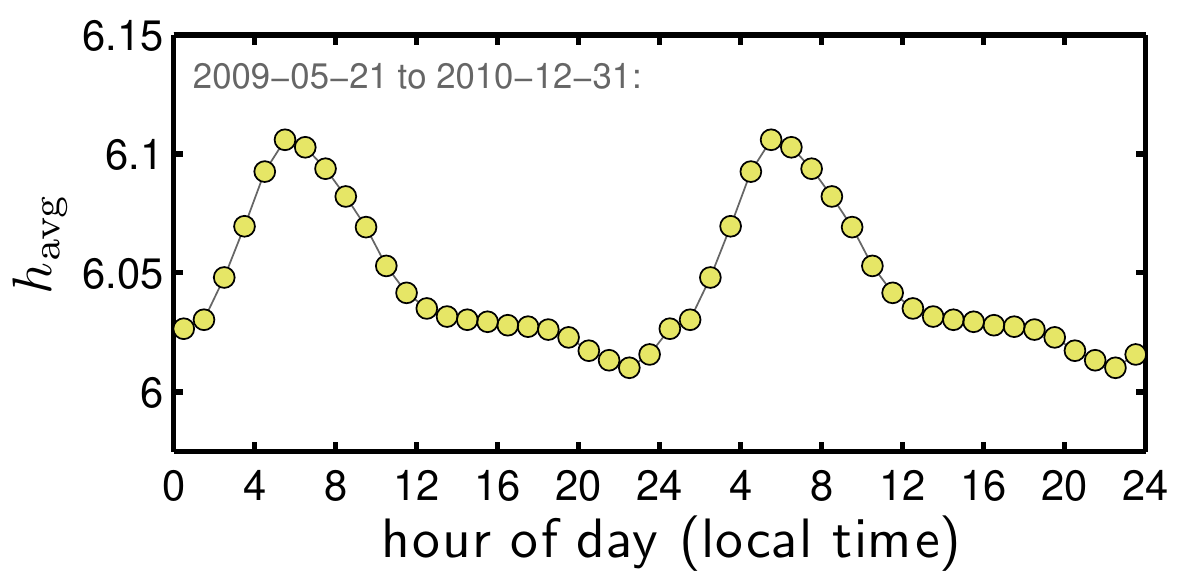}
  \caption{
    Average happiness level according to hour of the day,
    adjusted for local time.  
    As for days of the week in
    Fig.~\ref{fig:twhap.twitter_timeseries_2a},
    each data point represents
    an average of averages across days.
    The plot remains essentially unchanged if outlier dates
    marked in Fig.~\ref{fig:twhap.twitter_timeseries_1}A are
    excluded.  
    The maximum relative difference between
    the two plots is 0.08\%.
    The daily pattern of happiness in tweets
    shows more variation than we observed for the weekly cycle
    (Fig.~\ref{fig:twhap.twitter_timeseries_2a}),
    here ranging from 
    a low of $\havgfn \simeq 6.02$
    between 10 and 11 pm to
    a high of $\havgfn \simeq 6.12$
    between 5 and 6 am.
  }
  \label{fig:twhap.twitter_timeseries_daily}
\end{figure}

We next examine how average happiness levels change throughout
the day at the resolution of an hour. 
As shown in Fig.~\ref{fig:twhap.twitter_timeseries_daily},
the happiest hour of the day is 5 to 6 am,
after which we see a steep decline until midday
followed by a more gradual descent
to the on-average low of 10 to 11 pm,
and then a return to the daily peak through the night.
An afternoon low is consistent with self-reported moods;
Stone et al., in particular, observe a happiness dip in the afternoon~\cite{stone2006a},
though here we see negativity decreasing well into the night.
Our results are in contrast to some previous observations regarding blogs
and Facebook~\cite{mihalcea2006a,kramer2010a};
for example, Mihalcea and Liu~\cite{mihalcea2006a}
found a low occurring in the middle of the day
(part of their analysis involved the ANEW study word list).
The period 5--6 am marks 
`biological midnight' 
when, for example, body temperature is typically lowest
(see also~\cite{golder2011a}).
People after this point in time
are more likely to be rising for the day rather
than extending the previous one,
leading to a change in the 
kinds of mental states represented by active users.

\begin{figure}[tp!]
  \centering
                                  \includegraphics[width=0.48\textwidth]{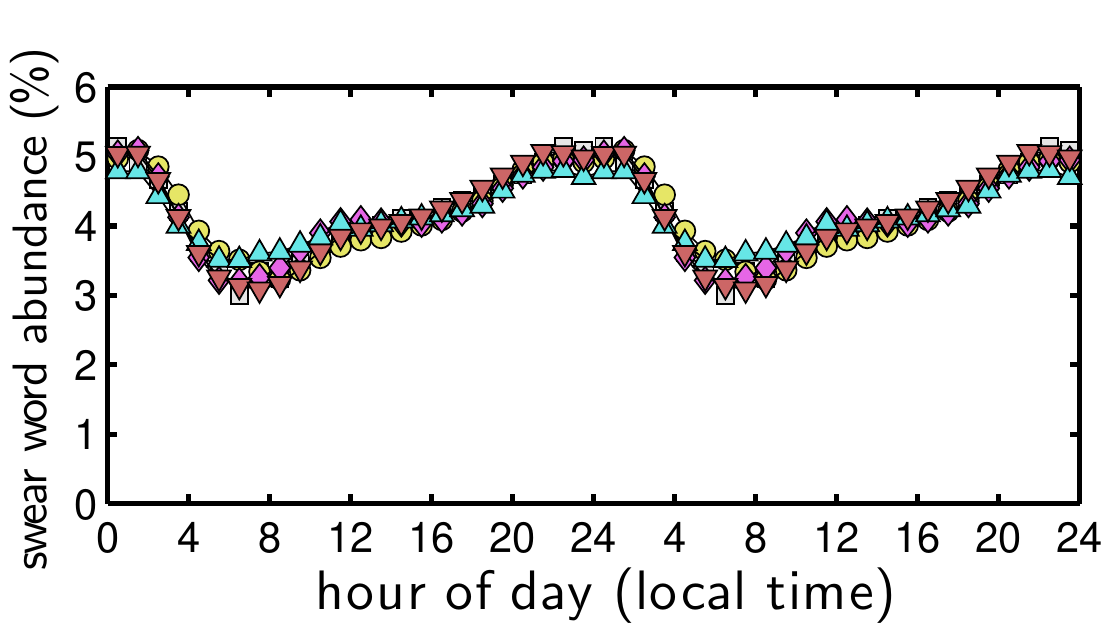}
  \caption{
    Normalized distributions of five example 
    common expletives as a function of hour of the day.
  }
  \label{fig:twhap.cursewords}
\end{figure}

We also find that usage rates of the
most common profanities are remarkably similar and
are roughly anticorrelated with the observed happiness cycle.
Fig.~\ref{fig:twhap.cursewords} shows the normalized
frequencies for five example profanities.
Cursing follows a sawtooth pattern with a maximum
occurring around 1 am, and the lowest 
relative usage of profanities matching up
with the daily early morning happiness peak between 5 and 6 am.
These patterns suggest a gradual, on-average, daily unraveling of the
human mind.

\begin{figure}[tp!]
  \centering
            \includegraphics[width=0.48\textwidth]{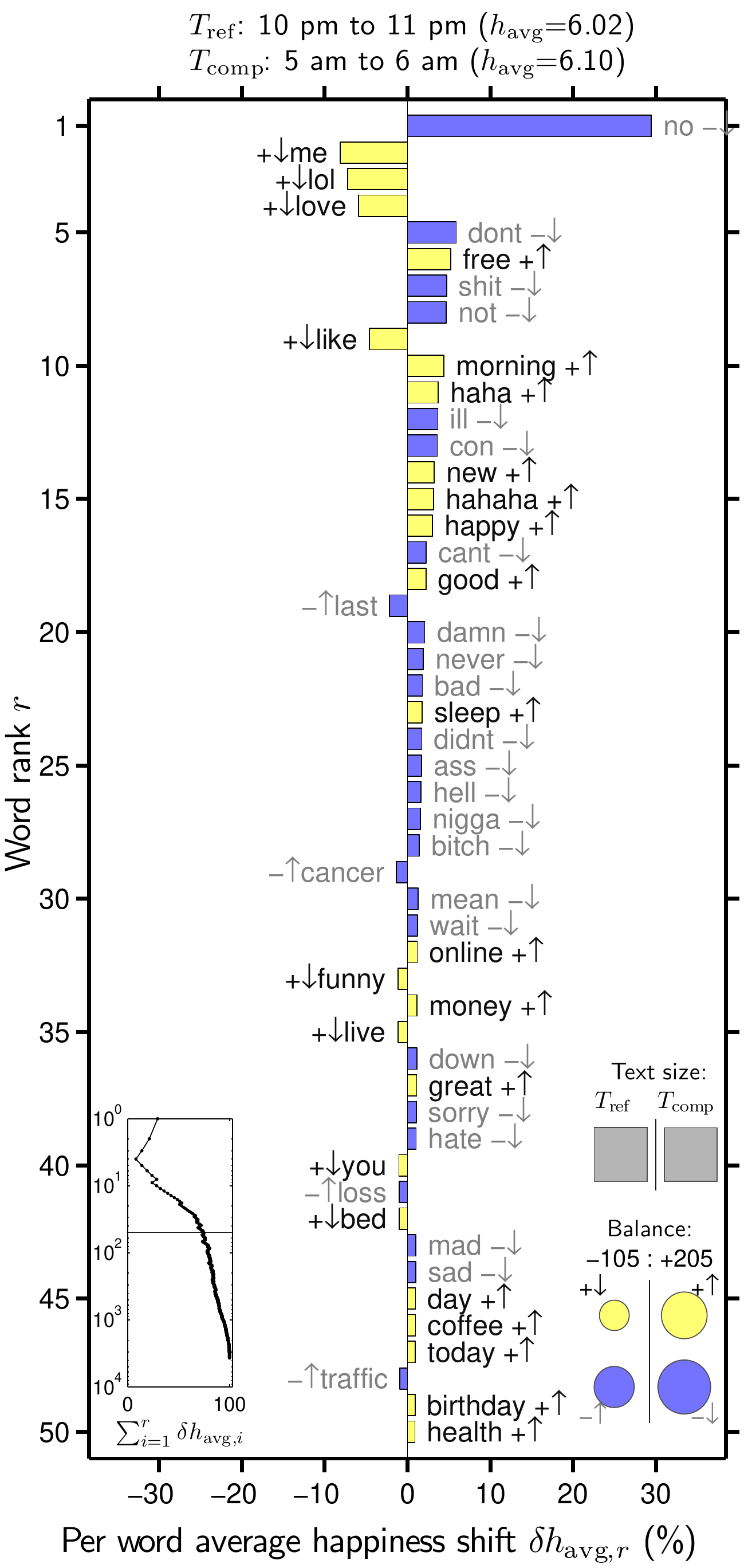}

  \caption{
    Word shift graph comparing the 
    happiest hour (5 am to 6 am)
    relative
    to the least happy hour (10 pm to 11 pm).
    Days given equal weighting with outlier dates removed.
    (See Fig.~\ref{fig:twhap.twitter_timeseries_daily_wsg_supp}
    in Supplementary Information for 
    word shifts based on alternate distributions.)
  }
  \label{fig:twhap.twitter_timeseries_daily_wsg}
\end{figure}

\subsection{Word Shift Analysis}
\label{subsec:twhap.dailycycle-happinesswordshift}

To give a deeper sense of the underlying
moods reflected in the low and high of the day,
we explore the word shift graph in
Fig.~\ref{fig:twhap.twitter_timeseries_daily_wsg},
comparing tweets made in the hours of 5 to 6 am and
10 to 11 pm.
For comparison, Fig.~\ref{fig:twhap.twitter_timeseries_daily_wsg_supp}
in Supplementary Information shows word shift graphs
under three averaging schemes.

The balance plot (bottom right inset)
shows that 5 to 6 am is happier because
of an overall preponderance of less abundant negative words 
and more abundant positive words, the former's
contribution marginally larger than the latter.
As the lower left inset cumulative plot shows, 
the first 50 words account for 
approximately 70\% of 
the total shift.
Thereafter, word shifts gradually bring
the overall difference up to 100\%, requiring
all words to do so.

The first few salient, relatively positive words more
abundant between 5 and 6 am
($+$$\uparrow$) are 
`free',
`morning' (likely appearing in good morning),
`haha',
`new',
`hahaha',
`happy',
and
`good'.
These are joined with 
decreases in negative word prevalences
(\textcolor{gray}{$-$$\downarrow$}) including
most strongly `no',
as well as
`don't',
`shit',
and
`not'.
Going against the overall trend 
are positive words used less often
and pointing
to a drop in social interactions,
such as 
`me',
`lol',
`love',
`like',
`funny',
and 
`you'
($+$$\downarrow$).
We also see more of
the early morning negative `traffic'
(\textcolor{gray}{$-$$\uparrow$}).
The word shift graph also holds suggestions
of automated tweets;
e.g., the word `cancer' may refer to the Zodiac sign.

\begin{figure}[tp!]
  \centering
                      \includegraphics[width=0.48\textwidth]{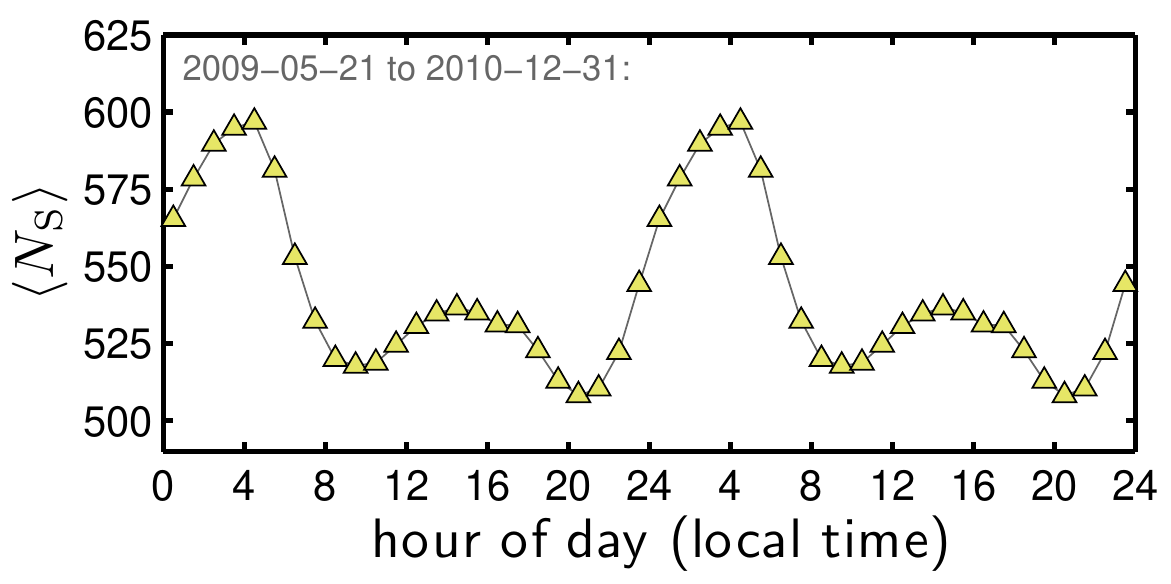} 
  \caption{
    Average Simpson lexical size $\Nsimpson$ for time of day, corrected according to local time,
    and computed for each day with outlier days removed, and then averaged
    across days.
    See also Fig.~\ref{fig:twhap.twitter_timeseries_daily_info_supp} in Supplementary Information
    for a demonstration of the robustness of the form of
    $\Nsimpson$ throughout the day under alternate averaging schemes.
  }
  \label{fig:twhap.twitter_timeseries_daily_info}
\end{figure}

\subsection{Information Content}
\label{subsec:twhap.dailycycle-info}

In Fig.~\ref{fig:twhap.twitter_timeseries_daily_info},
we show that average Simpson lexical size $\Nsimpson$ follows 
a daily cycle roughly similar in shape to average happiness.
The peak through the night is more pronounced
than for happiness, taking off around 9 pm, climbing until
5 to 6 am ($\Nsimpson\simeq 600$);
from there, $\Nsimpson$
drops rapidly to a local
minimum in the morning (9 to 10 am),
and then rises slightly to reach
a minor crest in the early afternoon
before slowly declining to the day's minimum
between 10 and 11 pm ($\Nsimpson\simeq 510$).
In examining the change in $\Nsimpson$
between the high at 5 to 6 am and the low in 10 to 11 pm, we see
the first few contributions by rank are
`I',
`a',
`the',
`de'
`me',
and
`que'
which appear less frequently between 5 and 6 am.
Most all other words making substantive contributions
are prepositions and pronouns.
The only word in the top 20 that becomes more
frequent and thus effects a decrease in $\Nsimpson$,
is the second ranked `RT'.
Tweets thus appear to be more rich and less predictable
during the night, with an apex near biological midnight.
Another potential explanation may involve automated tweets,
an analysis of which is beyond the scope of the present work.

Finally, we find that using alternate averaging schemes to create word frequency
distributions for hour of the day yields remarkably 
little variation in $\Nsimpson$
(see Fig.~\ref{fig:twhap.twitter_timeseries_daily_info_supp} in Supplementary Information).

\section{Happiness averages and dynamics for tweets containing keywords and phrases}
\label{sec:twhap.keywords}

We turn to our last area of focus: 
temporal happiness patterns for tweets
containing specific text elements.
We need not restrict ourselves to words,
considering also, for example, short phrases ($n$-grams),
dates, punctuation, emoticons, and phonemes.
We examine various collections of text elements,
ranging from long term importance
(`economy'),
to contemporary topics
(`Obama'),
to the everyday
(`today' and `!').
In doing so, we are effectively generating 
opinion polls regarding certain topics.  
Recent related work has explored
correlations between public opinion polls
and Twitter sentiment levels~\cite{oconnor2010a}, as well as
the use of emotional levels gleaned from Twitter 
to predict stock market behavior~\cite{bollen2011b}.
Here, we add to these findings by showing how certain happiness trends
based on keywords are clearly correlated with external events.
At the same time, we find many keyword-based trends are relatively
stable, and our interest turns to the average happiness level
which we do find to be highly variable across keywords.

\begin{figure}[tp!]
  \centering
    \includegraphics[width=0.48\textwidth]{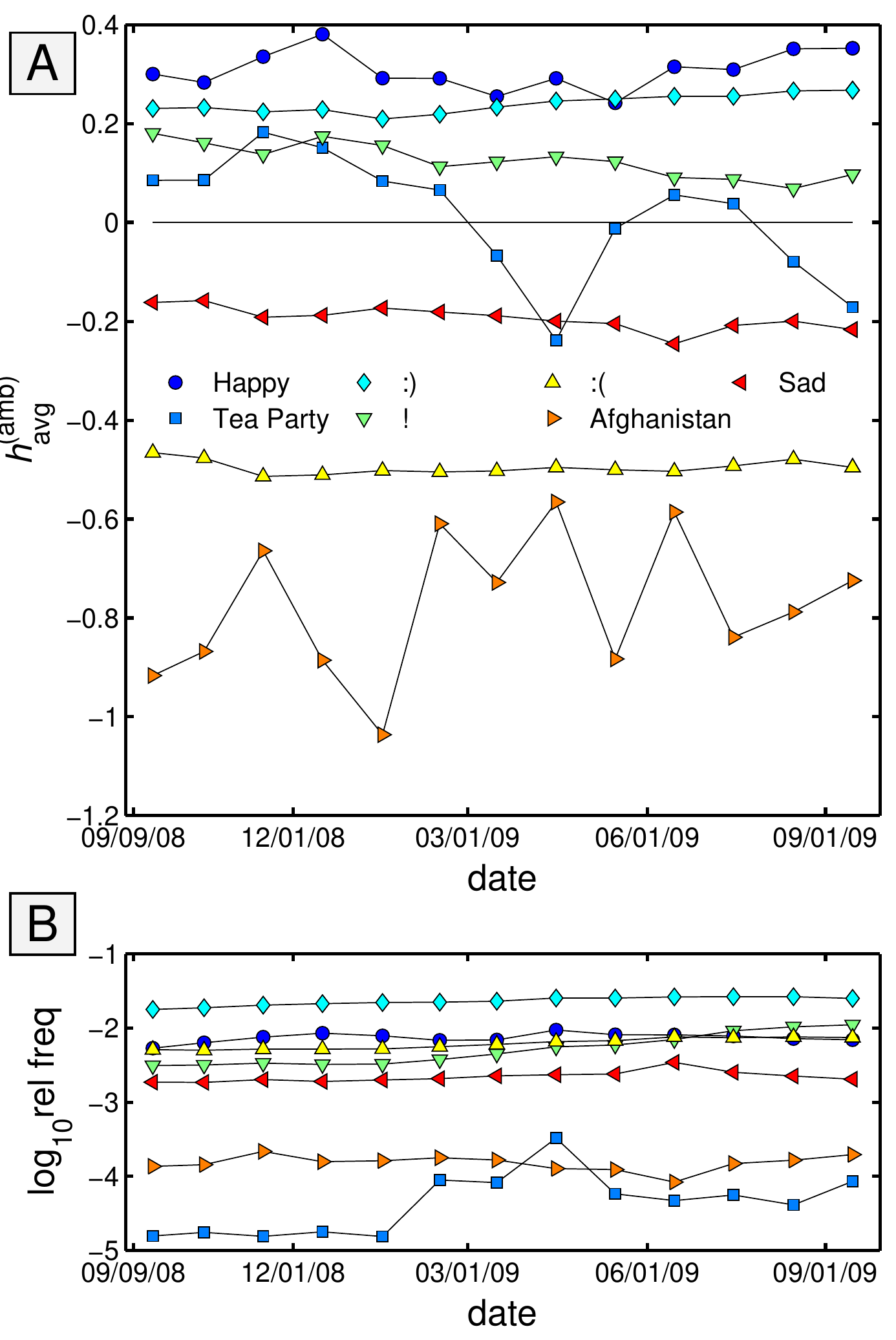}
  \caption{
    Ambient happiness $\havgfnamb$ and occurrence frequency time series for some
    illustrative text elements.
    \textbf{A.} 
    Ambient happiness is 
    the average happiness of all words found 
    co-occurring in tweets containing a given text element,
    with the background average happiness of all tweets removed
    (n.b., the text element's contribution is excluded).
    Binning is by calendar month and symbols are located
    at the center of each month.
    \textbf{B.} 
    Fraction of tweets containing text elements.
  }
  \label{fig:twhap.flatkeywordtimeseries}
\end{figure}

\subsection{Definition of Ambient Happiness}
\label{subsec:twhap.ambientdefn}

To facilitate comparisons, we now measure what we call
`normalized happiness' $\havgfnnorm$,
and `ambient happiness' $\havgfnamb$,
rather than absolute happiness $\havgfn$,
and which we define as follows.
For a given text element,
and a given pool of tweets (e.g., those falling in a specific month),
we first find all tweets containing the text element.
We measure the average happiness of the subset of tweets
in two ways: including the text element's own happiness
score for normalized happiness and excluding
it for ambient happiness.  
To create $\havgfnnorm$ and $\havgfnamb$,
we subtract the average happiness of 
all tweets in the pool.
In this way, we are able to separate out the effect of the text element, and 
can construct time series as the difference in happiness
between the text element time series and the overall
time series (Fig.~\ref{fig:twhap.twitter_timeseries_1}).

In Fig.~\ref{fig:twhap.flatkeywordtimeseries}A, we show
ambient happiness time series for seven example text elements,
chosen so as to exhibit both a range of happiness scores
and represent diverse topics and elements.
The lower plot in
Fig.~\ref{fig:twhap.flatkeywordtimeseries}B
shows the relative normalized frequency of tweets containing
each text element.
The trend for tweets containing the word `happy'
is to maintain a positive differential of approximately +0.3 to +0.4
above the overall average happiness time series.
By contrast, the counter of `sad' hovers around $-$0.2.
Words co-occurring with the emoticons `:)' and `:(' are strongly distinct
in terms of happiness with means near +0.25 and $-$0.5.
The exclamation point's
ambient happiness time series is a positive one
though clearly below that of `happy' and `:)',
and we see a slight downward trend toward a neutral score of 0.
Lastly, we show trends for two contemporary issues in the United
States, `Tea Party' and `Afghanistan'.  
Both phrases exhibit uneven signals,
with `Tea Party' reaching its lowest $\havgfnamb$ score
when its usage is most frequent.
`Afghanistan' is not surprisingly strongly
negative with ambient happiness scores consistently between
$-$1.1 and $-$0.6.

\subsection{Overall Ambient Happiness for Specific Tweets}
\label{subsec:twhap.ambient}

\begin{table*}[tp!]
  \centering
  \plainlatexonly{\begin{scriptsize}}
    \begin{tabular}{|l|c|l|l|}
\hline Word & $\havgfnamb$ & Total Tweets & $\havgfnnorm$  \\ \hline
1. happy & +0.430 & 1.65e+07 (13) & +1.104 (1) \\ 
\textcolor{lightgrey}{2. Christmas} & \textcolor{lightgrey}{+0.404} & \textcolor{lightgrey}{4.89e+06 (35)} & \textcolor{lightgrey}{+0.953 (3)} \\ 
3. vegan & +0.315 & 1.84e+05 (90) & -0.015 (46) \\ 
\textcolor{lightgrey}{4. :)} & \textcolor{lightgrey}{+0.274} & \textcolor{lightgrey}{1.04e+07 (20)} & \textcolor{lightgrey}{+0.630 (12)} \\ 
5. family & +0.251 & 5.01e+06 (32) & +0.716 (7) \\ 
\textcolor{lightgrey}{6. :-)} & \textcolor{lightgrey}{+0.228} & \textcolor{lightgrey}{1.67e+06 (60)} & \textcolor{lightgrey}{+0.560 (16)} \\ 
7. our & +0.207 & 1.41e+07 (16) & +0.159 (33) \\ 
\textcolor{lightgrey}{8. win} & \textcolor{lightgrey}{+0.204} & \textcolor{lightgrey}{7.98e+06 (26)} & \textcolor{lightgrey}{+0.924 (4)} \\ 
9. vacation & +0.200 & 9.35e+05 (67) & +0.817 (5) \\ 
\textcolor{lightgrey}{10. party} & \textcolor{lightgrey}{+0.170} & \textcolor{lightgrey}{6.44e+06 (29)} & \textcolor{lightgrey}{+0.679 (9)} \\ 
11. love & +0.164 & 4.67e+07 (6) & +0.977 (2) \\ 
\textcolor{lightgrey}{12. friends} & \textcolor{lightgrey}{+0.155} & \textcolor{lightgrey}{7.67e+06 (27)} & \textcolor{lightgrey}{+0.685 (8)} \\ 
13. hope & +0.149 & 1.18e+07 (18) & +0.515 (19) \\ 
\textcolor{lightgrey}{14. coffee} & \textcolor{lightgrey}{+0.147} & \textcolor{lightgrey}{2.80e+06 (46)} & \textcolor{lightgrey}{+0.518 (18)} \\ 
15. cash & +0.146 & 1.28e+06 (63) & +0.601 (14) \\ 
\textcolor{lightgrey}{16. sun} & \textcolor{lightgrey}{+0.144} & \textcolor{lightgrey}{2.39e+06 (52)} & \textcolor{lightgrey}{+0.737 (6)} \\ 
17. income & +0.137 & 5.10e+05 (76) & +0.621 (13) \\ 
\textcolor{lightgrey}{18. summer} & \textcolor{lightgrey}{+0.135} & \textcolor{lightgrey}{3.00e+06 (43)} & \textcolor{lightgrey}{+0.221 (29)} \\ 
19. church & +0.131 & 1.81e+06 (58) & -0.016 (47) \\ 
\textcolor{lightgrey}{20. Valentine} & \textcolor{lightgrey}{+0.127} & \textcolor{lightgrey}{2.47e+05 (84)} & \textcolor{lightgrey}{+0.593 (15)} \\ 
21. Stephen Colbert & +0.126 & 2.38e+04 (99) & +0.001 (45) \\ 
\textcolor{lightgrey}{22. USA} & \textcolor{lightgrey}{+0.113} & \textcolor{lightgrey}{2.16e+06 (54)} & \textcolor{lightgrey}{+0.325 (26)} \\ 
23. ! & +0.106 & 3.44e+06 (40) & +0.195 (31) \\ 
\textcolor{lightgrey}{24. winter} & \textcolor{lightgrey}{+0.101} & \textcolor{lightgrey}{1.26e+06 (64)} & \textcolor{lightgrey}{+0.050 (43)} \\ 
25. God & +0.099 & 8.58e+06 (25) & +0.468 (20) \\ 
\textcolor{lightgrey}{26. hot} & \textcolor{lightgrey}{+0.095} & \textcolor{lightgrey}{7.12e+06 (28)} & \textcolor{lightgrey}{-0.172 (54)} \\ 
27. ;) & +0.094 & 2.61e+06 (48) & +0.326 (25) \\ 
\textcolor{lightgrey}{28. Jesus} & \textcolor{lightgrey}{+0.094} & \textcolor{lightgrey}{2.03e+06 (56)} & \textcolor{lightgrey}{+0.247 (28)} \\ 
29. today & +0.092 & 2.56e+07 (9) & +0.126 (36) \\ 
\textcolor{lightgrey}{30. kiss} & \textcolor{lightgrey}{+0.072} & \textcolor{lightgrey}{1.70e+06 (59)} & \textcolor{lightgrey}{+0.632 (11)} \\ 
31. yes & +0.056 & 1.16e+07 (19) & +0.321 (27) \\ 
\textcolor{lightgrey}{32. tomorrow} & \textcolor{lightgrey}{+0.054} & \textcolor{lightgrey}{1.04e+07 (21)} & \textcolor{lightgrey}{+0.086 (38)} \\ 
33. you & +0.052 & 1.73e+08 (3) & +0.111 (37) \\ 
\textcolor{lightgrey}{34. heaven} & \textcolor{lightgrey}{+0.041} & \textcolor{lightgrey}{7.42e+05 (71)} & \textcolor{lightgrey}{+0.674 (10)} \\ 
35. ;-) & +0.041 & 9.39e+05 (66) & +0.395 (23) \\ 
\textcolor{lightgrey}{36. we} & \textcolor{lightgrey}{+0.035} & \textcolor{lightgrey}{3.91e+07 (7)} & \textcolor{lightgrey}{+0.146 (34)} \\ 
37. yesterday & +0.033 & 3.08e+06 (42) & -0.168 (53) \\ 
\textcolor{lightgrey}{38. dark} & \textcolor{lightgrey}{+0.031} & \textcolor{lightgrey}{1.58e+06 (61)} & \textcolor{lightgrey}{-0.766 (81)} \\ 
39. ? & +0.030 & 2.32e+06 (53) & -0.503 (68) \\ 
\textcolor{lightgrey}{40. RT} & \textcolor{lightgrey}{+0.028} & \textcolor{lightgrey}{3.39e+08 (1)} & \textcolor{lightgrey}{-0.443 (66)} \\ 
41. Michael Jackson & +0.018 & 8.26e+05 (70) & -0.213 (59) \\ 
\textcolor{lightgrey}{42. night} & \textcolor{lightgrey}{+0.014} & \textcolor{lightgrey}{1.71e+07 (12)} & \textcolor{lightgrey}{+0.074 (40)} \\ 
43. life & +0.012 & 1.40e+07 (17) & +0.422 (22) \\ 
\textcolor{lightgrey}{44. health} & \textcolor{lightgrey}{-0.000} & \textcolor{lightgrey}{2.58e+06 (50)} & \textcolor{lightgrey}{+0.447 (21)} \\ 
45. sex & -0.008 & 3.55e+06 (39) & +0.542 (17) \\ 
\textcolor{lightgrey}{46. work} & \textcolor{lightgrey}{-0.010} & \textcolor{lightgrey}{1.84e+07 (11)} & \textcolor{lightgrey}{-0.174 (56)} \\ 
47. girl & -0.010 & 1.01e+07 (22) & +0.331 (24) \\ 
\textcolor{lightgrey}{48. boy} & \textcolor{lightgrey}{-0.026} & \textcolor{lightgrey}{4.93e+06 (33)} & \textcolor{lightgrey}{+0.062 (41)} \\ 
49. I & -0.048 & 3.08e+08 (2) & -0.062 (49) \\ 
\textcolor{lightgrey}{50. commute} & \textcolor{lightgrey}{-0.048} & \textcolor{lightgrey}{9.01e+04 (94)} & \textcolor{lightgrey}{-0.206 (57)} \\ 
\hline
\end{tabular}\begin{tabular}{|l|c|l|l|}
\hline Word & $\havgfnamb$ & Total Tweets & $\havgfnnorm$  \\ \hline
51. snow & -0.051 & 2.60e+06 (49) & +0.083 (39) \\ 
\textcolor{lightgrey}{52. Jon Stewart} & \textcolor{lightgrey}{-0.052} & \textcolor{lightgrey}{5.21e+04 (97)} & \textcolor{lightgrey}{-0.024 (48)} \\ 
53. school & -0.056 & 9.26e+06 (24) & +0.050 (42) \\ 
\textcolor{lightgrey}{54. Lehman Brothers} & \textcolor{lightgrey}{-0.078} & \textcolor{lightgrey}{8.50e+03 (100)} & \textcolor{lightgrey}{-0.721 (79)} \\ 
55. them & -0.090 & 1.54e+07 (15) & -0.280 (60) \\ 
\textcolor{lightgrey}{56. right} & \textcolor{lightgrey}{-0.090} & \textcolor{lightgrey}{1.92e+07 (10)} & \textcolor{lightgrey}{+0.126 (35)} \\ 
57. woman & -0.115 & 2.54e+06 (51) & +0.202 (30) \\ 
\textcolor{lightgrey}{58. left} & \textcolor{lightgrey}{-0.118} & \textcolor{lightgrey}{4.89e+06 (34)} & \textcolor{lightgrey}{-0.383 (63)} \\ 
59. me & -0.119 & 1.44e+08 (4) & +0.160 (32) \\ 
\textcolor{lightgrey}{60. election} & \textcolor{lightgrey}{-0.127} & \textcolor{lightgrey}{5.60e+05 (75)} & \textcolor{lightgrey}{-0.306 (61)} \\ 
61. Sarah Palin & -0.128 & 2.26e+05 (87) & -0.681 (76) \\ 
\textcolor{lightgrey}{62. no} & \textcolor{lightgrey}{-0.132} & \textcolor{lightgrey}{9.51e+07 (5)} & \textcolor{lightgrey}{-1.415 (90)} \\ 
63. rain & -0.134 & 3.23e+06 (41) & +0.050 (44) \\ 
\textcolor{lightgrey}{64. climate} & \textcolor{lightgrey}{-0.135} & \textcolor{lightgrey}{3.64e+05 (80)} & \textcolor{lightgrey}{-0.160 (51)} \\ 
65. gay & -0.152 & 2.73e+06 (47) & -0.552 (72) \\ 
\textcolor{lightgrey}{66. lose} & \textcolor{lightgrey}{-0.157} & \textcolor{lightgrey}{2.06e+06 (55)} & \textcolor{lightgrey}{-1.181 (86)} \\ 
67. they & -0.159 & 2.74e+07 (8) & -0.208 (58) \\ 
\textcolor{lightgrey}{68. oil} & \textcolor{lightgrey}{-0.162} & \textcolor{lightgrey}{1.38e+06 (62)} & \textcolor{lightgrey}{-0.411 (65)} \\ 
69. cold & -0.162 & 3.67e+06 (36) & -0.546 (71) \\ 
\textcolor{lightgrey}{70. I feel} & \textcolor{lightgrey}{-0.173} & \textcolor{lightgrey}{5.17e+06 (31)} & \textcolor{lightgrey}{-0.129 (50)} \\ 
71. man & -0.175 & 1.59e+07 (14) & -0.163 (52) \\ 
\textcolor{lightgrey}{72. Republican} & \textcolor{lightgrey}{-0.181} & \textcolor{lightgrey}{2.30e+05 (86)} & \textcolor{lightgrey}{-0.539 (70)} \\ 
73. sad & -0.187 & 3.56e+06 (38) & -1.366 (89) \\ 
\textcolor{lightgrey}{74. gas} & \textcolor{lightgrey}{-0.193} & \textcolor{lightgrey}{1.02e+06 (65)} & \textcolor{lightgrey}{-0.471 (67)} \\ 
75. economy & -0.203 & 6.09e+05 (73) & -0.525 (69) \\ 
\textcolor{lightgrey}{76. Obama} & \textcolor{lightgrey}{-0.205} & \textcolor{lightgrey}{2.98e+06 (44)} & \textcolor{lightgrey}{-0.173 (55)} \\ 
77. Democrat & -0.226 & 9.32e+04 (93) & -0.384 (64) \\ 
\textcolor{lightgrey}{78. Congress} & \textcolor{lightgrey}{-0.231} & \textcolor{lightgrey}{3.92e+05 (79)} & \textcolor{lightgrey}{-0.580 (74)} \\ 
79. hell & -0.250 & 6.27e+06 (30) & -1.551 (96) \\ 
\textcolor{lightgrey}{80. sick} & \textcolor{lightgrey}{-0.262} & \textcolor{lightgrey}{3.58e+06 (37)} & \textcolor{lightgrey}{-1.630 (97)} \\ 
81. Muslim & -0.262 & 2.15e+05 (88) & -0.569 (73) \\ 
\textcolor{lightgrey}{82. war} & \textcolor{lightgrey}{-0.270} & \textcolor{lightgrey}{1.96e+06 (57)} & \textcolor{lightgrey}{-2.040 (100)} \\ 
83. Pope & -0.277 & 1.52e+05 (91) & -0.316 (62) \\ 
\textcolor{lightgrey}{84. hate} & \textcolor{lightgrey}{-0.282} & \textcolor{lightgrey}{9.65e+06 (23)} & \textcolor{lightgrey}{-1.520 (94)} \\ 
85. Glenn Beck & -0.282 & 1.14e+05 (92) & -0.776 (82) \\ 
\textcolor{lightgrey}{86. Islam} & \textcolor{lightgrey}{-0.299} & \textcolor{lightgrey}{1.87e+05 (89)} & \textcolor{lightgrey}{-0.710 (78)} \\ 
87. George Bush & -0.333 & 3.23e+04 (98) & -0.747 (80) \\ 
\textcolor{lightgrey}{88. Goldman Sachs} & \textcolor{lightgrey}{-0.337} & \textcolor{lightgrey}{5.27e+04 (96)} & \textcolor{lightgrey}{-0.984 (84)} \\ 
89. depressed & -0.339 & 2.81e+05 (82) & -1.541 (95) \\ 
\textcolor{lightgrey}{90. Senate} & \textcolor{lightgrey}{-0.340} & \textcolor{lightgrey}{4.48e+05 (78)} & \textcolor{lightgrey}{-0.601 (75)} \\ 
91. BP & -0.355 & 5.82e+05 (74) & -0.902 (83) \\ 
\textcolor{lightgrey}{92. gun} & \textcolor{lightgrey}{-0.367} & \textcolor{lightgrey}{6.81e+05 (72)} & \textcolor{lightgrey}{-1.476 (93)} \\ 
93. drugs & -0.382 & 5.10e+05 (77) & -1.452 (91) \\ 
\textcolor{lightgrey}{94. headache} & \textcolor{lightgrey}{-0.437} & \textcolor{lightgrey}{8.57e+05 (69)} & \textcolor{lightgrey}{-1.881 (98)} \\ 
95. :-( & -0.455 & 3.40e+05 (81) & -1.174 (85) \\ 
\textcolor{lightgrey}{96. :(} & \textcolor{lightgrey}{-0.472} & \textcolor{lightgrey}{2.89e+06 (45)} & \textcolor{lightgrey}{-1.288 (88)} \\ 
97. Afghanistan & -0.703 & 2.74e+05 (83) & -1.458 (92) \\ 
\textcolor{lightgrey}{98. mosque} & \textcolor{lightgrey}{-0.709} & \textcolor{lightgrey}{6.98e+04 (95)} & \textcolor{lightgrey}{-0.694 (77)} \\ 
99. flu & -0.735 & 9.01e+05 (68) & -1.912 (99) \\ 
\textcolor{lightgrey}{100. Iraq} & \textcolor{lightgrey}{-0.773} & \textcolor{lightgrey}{2.39e+05 (85)} & \textcolor{lightgrey}{-1.282 (87)} \\ 
\hline
\end{tabular}

  \plainlatexonly{\end{scriptsize}}
  \caption{
    Selection of 100 text elements ordered by 
    average ambient happiness $\havgfnamb$.
    The number of tweets and the value of normalized 
    happiness $\havgfnnorm$ (where the happiness value
    of the text element itself is included)
    are listed in the third and fourth columns, with the
    ranking of the text element according to these quantities
    shown in brackets.  
    For this list of text elements, we obtained additional happiness
    scores for phrases, punctuation, emoticons, etc.,
    using Mechanical Turk.
    All pattern matches with tweets
    were case-insensitive.
    Tab.~\ref{tab:twhap.supp-timeseriesE} in the Supplementary Information
    shows the same table sorted by normalized happiness $\havgfnnorm$.
  } 
  \label{tab:twhap.timeseriesA}
\end{table*}

\begin{table*}[tp!]
  \centering
  \plainlatexonly{\begin{scriptsize}}
    \begin{tabular}{|l|c|l|l|}
\hline Word & $\Nsimpson$ & Total Words & Frac top 50K \\ \hline
1. RT & 1019.5 & 4.751e+09 (1) & 0.653 (100) \\ 
\textcolor{lightgrey}{2. ?} & \textcolor{lightgrey}{662.1} & \textcolor{lightgrey}{2.608e+07 (58)} & \textcolor{lightgrey}{0.731 (98)}\\ 
3. ! & 621.1 & 3.682e+07 (50) & 0.742 (97) \\ 
\textcolor{lightgrey}{4. USA} & \textcolor{lightgrey}{501.5} & \textcolor{lightgrey}{3.150e+07 (54)} & \textcolor{lightgrey}{0.751 (94)}\\ 
5. no & 487.3 & 1.431e+09 (5) & 0.763 (93) \\ 
\textcolor{lightgrey}{6. ;-)} & \textcolor{lightgrey}{476.9} & \textcolor{lightgrey}{1.323e+07 (67)} & \textcolor{lightgrey}{0.75 (95)}\\ 
7. ;) & 389.2 & 3.379e+07 (52) & 0.791 (86) \\ 
\textcolor{lightgrey}{8. war} & \textcolor{lightgrey}{386.2} & \textcolor{lightgrey}{2.901e+07 (56)} & \textcolor{lightgrey}{0.785 (88)}\\ 
9. Goldman Sachs & 379.5 & 7.183e+05 (96) & 0.766 (92) \\ 
\textcolor{lightgrey}{10. gay} & \textcolor{lightgrey}{377.6} & \textcolor{lightgrey}{3.823e+07 (46)} & \textcolor{lightgrey}{0.823 (77)}\\ 
11. me & 368.4 & 2.136e+09 (4) & 0.829 (70) \\ 
\textcolor{lightgrey}{12. :-)} & \textcolor{lightgrey}{362.3} & \textcolor{lightgrey}{2.280e+07 (61)} & \textcolor{lightgrey}{0.773 (91)}\\ 
13. Islam & 355.2 & 2.776e+06 (89) & 0.678 (99) \\ 
\textcolor{lightgrey}{14. :)} & \textcolor{lightgrey}{347.1} & \textcolor{lightgrey}{1.313e+08 (24)} & \textcolor{lightgrey}{0.775 (90)}\\ 
15. Muslim & 343.9 & 3.327e+06 (86) & 0.779 (89) \\ 
\textcolor{lightgrey}{16. Michael Jackson} & \textcolor{lightgrey}{335.0} & \textcolor{lightgrey}{1.029e+07 (71)} & \textcolor{lightgrey}{0.803 (83)}\\ 
17. Obama & 325.8 & 4.412e+07 (43) & 0.825 (74) \\ 
\textcolor{lightgrey}{18. Lehman Brothers} & \textcolor{lightgrey}{324.5} & \textcolor{lightgrey}{1.161e+05 (100)} & \textcolor{lightgrey}{0.743 (96)}\\ 
19. :-( & 312.5 & 4.798e+06 (81) & 0.804 (82) \\ 
\textcolor{lightgrey}{20. health} & \textcolor{lightgrey}{312.4} & \textcolor{lightgrey}{3.817e+07 (47)} & \textcolor{lightgrey}{0.826 (72)}\\ 
21. gas & 311.8 & 1.580e+07 (65) & 0.822 (78) \\ 
\textcolor{lightgrey}{22. Jesus} & \textcolor{lightgrey}{311.4} & \textcolor{lightgrey}{3.011e+07 (55)} & \textcolor{lightgrey}{0.831 (69)}\\ 
23. :( & 304.5 & 3.802e+07 (48) & 0.798 (84) \\ 
\textcolor{lightgrey}{24. hot} & \textcolor{lightgrey}{298.3} & \textcolor{lightgrey}{9.826e+07 (28)} & \textcolor{lightgrey}{0.847 (46)}\\ 
25. cash & 298.0 & 1.909e+07 (63) & 0.832 (66) \\ 
\textcolor{lightgrey}{26. vegan} & \textcolor{lightgrey}{290.9} & \textcolor{lightgrey}{2.696e+06 (90)} & \textcolor{lightgrey}{0.845 (54)}\\ 
27. George Bush & 288.0 & 4.546e+05 (98) & 0.847 (48) \\ 
\textcolor{lightgrey}{28. BP} & \textcolor{lightgrey}{285.2} & \textcolor{lightgrey}{8.957e+06 (74)} & \textcolor{lightgrey}{0.791 (87)}\\ 
29. man & 283.3 & 2.333e+08 (15) & 0.845 (52) \\ 
\textcolor{lightgrey}{30. sex} & \textcolor{lightgrey}{276.2} & \textcolor{lightgrey}{5.186e+07 (37)} & \textcolor{lightgrey}{0.844 (57)}\\ 
31. Sarah Palin & 275.4 & 3.194e+06 (87) & 0.842 (58) \\ 
\textcolor{lightgrey}{32. we} & \textcolor{lightgrey}{272.4} & \textcolor{lightgrey}{6.434e+08 (6)} & \textcolor{lightgrey}{0.869 (34)}\\ 
33. flu & 270.8 & 1.279e+07 (68) & 0.826 (73) \\ 
\textcolor{lightgrey}{34. income} & \textcolor{lightgrey}{270.7} & \textcolor{lightgrey}{7.681e+06 (76)} & \textcolor{lightgrey}{0.835 (63)}\\ 
35. I & 269.8 & 4.590e+09 (2) & 0.881 (23) \\ 
\textcolor{lightgrey}{36. oil} & \textcolor{lightgrey}{267.1} & \textcolor{lightgrey}{2.147e+07 (62)} & \textcolor{lightgrey}{0.825 (75)}\\ 
37. Democrat & 262.4 & 1.469e+06 (94) & 0.832 (67) \\ 
\textcolor{lightgrey}{38. drugs} & \textcolor{lightgrey}{261.7} & \textcolor{lightgrey}{7.633e+06 (77)} & \textcolor{lightgrey}{0.862 (40)}\\ 
39. our & 257.6 & 2.394e+08 (14) & 0.869 (35) \\ 
\textcolor{lightgrey}{40. boy} & \textcolor{lightgrey}{256.7} & \textcolor{lightgrey}{7.174e+07 (33)} & \textcolor{lightgrey}{0.857 (42)}\\ 
41. Glenn Beck & 252.3 & 1.740e+06 (92) & 0.851 (44) \\ 
\textcolor{lightgrey}{42. Stephen Colbert} & \textcolor{lightgrey}{251.0} & \textcolor{lightgrey}{2.972e+05 (99)} & \textcolor{lightgrey}{0.844 (55)}\\ 
43. Valentine & 248.4 & 3.169e+06 (88) & 0.822 (79) \\ 
\textcolor{lightgrey}{44. party} & \textcolor{lightgrey}{242.9} & \textcolor{lightgrey}{9.466e+07 (29)} & \textcolor{lightgrey}{0.844 (56)}\\ 
45. gun & 241.9 & 1.030e+07 (70) & 0.836 (60) \\ 
\textcolor{lightgrey}{46. winter} & \textcolor{lightgrey}{240.2} & \textcolor{lightgrey}{1.871e+07 (64)} & \textcolor{lightgrey}{0.854 (43)}\\ 
47. Republican & 239.8 & 3.607e+06 (85) & 0.845 (53) \\ 
\textcolor{lightgrey}{48. they} & \textcolor{lightgrey}{239.8} & \textcolor{lightgrey}{4.749e+08 (8)} & \textcolor{lightgrey}{0.896 (8)}\\ 
49. you & 239.2 & 2.484e+09 (3) & 0.871 (33) \\ 
\textcolor{lightgrey}{50. Congress} & \textcolor{lightgrey}{236.8} & \textcolor{lightgrey}{6.221e+06 (79)} & \textcolor{lightgrey}{0.834 (64)}\\ 
\hline
\end{tabular}\begin{tabular}{|l|c|l|l|}
\hline Word & $\Nsimpson$ & Total Words & Frac top 50K \\ \hline
51. Iraq & 235.5 & 3.722e+06 (84) & 0.832 (68) \\ 
\textcolor{lightgrey}{52. Jon Stewart} & \textcolor{lightgrey}{234.9} & \textcolor{lightgrey}{7.053e+05 (97)} & \textcolor{lightgrey}{0.836 (62)}\\ 
53. Senate & 233.7 & 6.791e+06 (78) & 0.826 (71) \\ 
\textcolor{lightgrey}{54. happy} & \textcolor{lightgrey}{232.8} & \textcolor{lightgrey}{2.041e+08 (17)} & \textcolor{lightgrey}{0.834 (65)}\\ 
55. climate & 231.7 & 5.245e+06 (80) & 0.813 (81) \\ 
\textcolor{lightgrey}{56. yes} & \textcolor{lightgrey}{230.0} & \textcolor{lightgrey}{1.484e+08 (21)} & \textcolor{lightgrey}{0.846 (50)}\\ 
57. today & 225.3 & 3.802e+08 (9) & 0.883 (20) \\ 
\textcolor{lightgrey}{58. election} & \textcolor{lightgrey}{220.7} & \textcolor{lightgrey}{8.632e+06 (75)} & \textcolor{lightgrey}{0.847 (47)}\\ 
59. summer & 219.1 & 4.471e+07 (42) & 0.864 (39) \\ 
\textcolor{lightgrey}{60. Christmas} & \textcolor{lightgrey}{215.7} & \textcolor{lightgrey}{6.330e+07 (35)} & \textcolor{lightgrey}{0.862 (41)}\\ 
61. rain & 215.1 & 4.620e+07 (41) & 0.836 (61) \\ 
\textcolor{lightgrey}{62. girl} & \textcolor{lightgrey}{214.0} & \textcolor{lightgrey}{1.513e+08 (20)} & \textcolor{lightgrey}{0.873 (32)}\\ 
63. I feel & 214.0 & 7.141e+07 (34) & 0.901 (4) \\ 
\textcolor{lightgrey}{64. kiss} & \textcolor{lightgrey}{212.7} & \textcolor{lightgrey}{2.463e+07 (59)} & \textcolor{lightgrey}{0.845 (51)}\\ 
65. God & 211.6 & 1.298e+08 (25) & 0.884 (18) \\ 
\textcolor{lightgrey}{66. school} & \textcolor{lightgrey}{211.2} & \textcolor{lightgrey}{1.328e+08 (23)} & \textcolor{lightgrey}{0.88 (25)}\\ 
67. coffee & 209.1 & 3.926e+07 (45) & 0.878 (27) \\ 
\textcolor{lightgrey}{68. Afghanistan} & \textcolor{lightgrey}{208.8} & \textcolor{lightgrey}{3.898e+06 (83)} & \textcolor{lightgrey}{0.793 (85)}\\ 
69. heaven & 208.3 & 1.075e+07 (69) & 0.864 (38) \\ 
\textcolor{lightgrey}{70. left} & \textcolor{lightgrey}{207.8} & \textcolor{lightgrey}{8.017e+07 (31)} & \textcolor{lightgrey}{0.873 (31)}\\ 
71. family & 207.8 & 7.700e+07 (32) & 0.873 (30) \\ 
\textcolor{lightgrey}{72. them} & \textcolor{lightgrey}{205.1} & \textcolor{lightgrey}{2.672e+08 (12)} & \textcolor{lightgrey}{0.893 (9)}\\ 
73. sad & 203.6 & 5.482e+07 (36) & 0.886 (17) \\ 
\textcolor{lightgrey}{74. night} & \textcolor{lightgrey}{203.1} & \textcolor{lightgrey}{2.429e+08 (13)} & \textcolor{lightgrey}{0.883 (21)}\\ 
75. hell & 202.7 & 9.000e+07 (30) & 0.883 (19) \\ 
\textcolor{lightgrey}{76. mosque} & \textcolor{lightgrey}{198.3} & \textcolor{lightgrey}{1.081e+06 (95)} & \textcolor{lightgrey}{0.82 (80)}\\ 
77. tomorrow & 198.1 & 1.516e+08 (19) & 0.892 (11) \\ 
\textcolor{lightgrey}{78. friends} & \textcolor{lightgrey}{197.5} & \textcolor{lightgrey}{1.242e+08 (27)} & \textcolor{lightgrey}{0.886 (16)}\\ 
79. vacation & 197.1 & 1.341e+07 (66) & 0.876 (28) \\ 
\textcolor{lightgrey}{80. snow} & \textcolor{lightgrey}{195.6} & \textcolor{lightgrey}{3.698e+07 (49)} & \textcolor{lightgrey}{0.881 (22)}\\ 
81. yesterday & 192.7 & 5.003e+07 (39) & 0.887 (14) \\ 
\textcolor{lightgrey}{82. right} & \textcolor{lightgrey}{190.5} & \textcolor{lightgrey}{2.854e+08 (10)} & \textcolor{lightgrey}{0.887 (15)}\\ 
83. church & 189.1 & 2.668e+07 (57) & 0.879 (26) \\ 
\textcolor{lightgrey}{84. cold} & \textcolor{lightgrey}{188.4} & \textcolor{lightgrey}{5.116e+07 (38)} & \textcolor{lightgrey}{0.9 (5)}\\ 
85. lose & 187.2 & 3.335e+07 (53) & 0.881 (24) \\ 
\textcolor{lightgrey}{86. sick} & \textcolor{lightgrey}{186.6} & \textcolor{lightgrey}{4.985e+07 (40)} & \textcolor{lightgrey}{0.899 (6)}\\ 
87. economy & 186.5 & 9.512e+06 (73) & 0.847 (49) \\ 
\textcolor{lightgrey}{88. dark} & \textcolor{lightgrey}{186.1} & \textcolor{lightgrey}{2.403e+07 (60)} & \textcolor{lightgrey}{0.868 (36)}\\ 
89. Pope & 185.3 & 2.268e+06 (91) & 0.84 (59) \\ 
\textcolor{lightgrey}{90. win} & \textcolor{lightgrey}{185.1} & \textcolor{lightgrey}{1.261e+08 (26)} & \textcolor{lightgrey}{0.825 (76)}\\ 
91. life & 180.4 & 2.210e+08 (16) & 0.892 (10) \\ 
\textcolor{lightgrey}{92. woman} & \textcolor{lightgrey}{178.8} & \textcolor{lightgrey}{4.151e+07 (44)} & \textcolor{lightgrey}{0.874 (29)}\\ 
93. work & 178.3 & 2.791e+08 (11) & 0.898 (7) \\ 
\textcolor{lightgrey}{94. depressed} & \textcolor{lightgrey}{175.2} & \textcolor{lightgrey}{4.108e+06 (82)} & \textcolor{lightgrey}{0.906 (2)}\\ 
95. sun & 166.9 & 3.622e+07 (51) & 0.849 (45) \\ 
\textcolor{lightgrey}{96. commute} & \textcolor{lightgrey}{165.0} & \textcolor{lightgrey}{1.470e+06 (93)} & \textcolor{lightgrey}{0.887 (13)}\\ 
97. hope & 157.2 & 1.853e+08 (18) & 0.89 (12) \\ 
\textcolor{lightgrey}{98. love} & \textcolor{lightgrey}{149.9} & \textcolor{lightgrey}{6.409e+08 (7)} & \textcolor{lightgrey}{0.865 (37)}\\ 
99. headache & 126.7 & 1.005e+07 (72) & 0.907 (1) \\ 
\textcolor{lightgrey}{100. hate} & \textcolor{lightgrey}{106.5} & \textcolor{lightgrey}{1.382e+08 (22)} & \textcolor{lightgrey}{0.902 (3)}\\ 
\hline
\end{tabular}

  \plainlatexonly{\end{scriptsize}}
  \caption{
    The same keywords and text elements as listed in
    Tab.~\ref{tab:twhap.timeseriesA}
    sorted according to the
    Simpson lexical size $\Nsimpson$
    for all tweets containing them.
    Keywords themselves are not included
    in the calculation of $\Nsimpson$.
    The third and fourth columns
    show the total number of words (other than the keyword) used
    to measure $\Nsimpson$ and
    the fraction of these words that
    are in our fixed list of 50,000 words
    (the higher the better).
    The numbers in brackets give rankings.
  } 
  \label{tab:twhap.timeseriesB}
\end{table*}

We next examine a selection of 100 handpicked keywords and text elements.
As mentioned above, the ambient average happiness for
tweets containing many of these terms
are mostly stable over time, and 
in Tab.~\ref{tab:twhap.timeseriesA}
we show overall ambient average happiness $\havgfnamb$
for the list, sorted in descending order.
Our list is
in no way exhaustive; rather it contains 
political keywords (`Democrat' and `Republican'),
semantic differentials (`right' and `left'), 
terms relating to the economy  (`money' and `Goldman Sachs'),
families of related keywords (`Jon Stewart' and `Glenn Beck'),
personal pronouns,
emoticons,
and so on.
As such, the extremes (most and least happy words for example) are not to be 
presumed to remain so for larger sets of key words, 
and our main
interest is in making comparisons of related terms.
In Tab.~\ref{tab:twhap.timeseriesB}, we present
the same terms ordered according to Simpson lexical size $\Nsimpson$.
In computing each term's $\Nsimpson$, we exclude the term itself.
For ease of comparison, we include 
Tab.~\ref{tab:twhap.timeseriesA} reordered by
normalized happiness as Tab.~\ref{tab:twhap.supp-timeseriesE}
in Supplementary Information.

We observe many interesting patterns
and we invite the reader to explore the tables
beyond the observations we record here.
We begin with the highest and lowest
rankings of ambient happiness $\havgfnamb$,
for our list, finding them to be sensible.
The two top ranked words are
`happy' ($\havgfnamb$=$+$0.430)
and
`Christmas' ($\havgfnamb$=$+$0.404),
and
the last two are
`flu' ($\havgfnamb$=$-$0.735),
and 
`Iraq' ($\havgfnamb$=$-$0.773).
When we include the text element's score itself
(see Tab.~\ref{tab:twhap.supp-timeseriesE}),
the order shifts somewhat with
`happy' and `love' at the top
and 
`flu' and `war at the bottom.

An important finding is that
the average happiness
of text elements as assessed through Mechanical Turk
and their
ambient happiness
correlate very strongly
(Spearman's correlation coefficient $r_s=0.794$, $p \le 10^{-10}$),
as do 
ambient and normalized happiness 
(Spearman's correlation coefficient $r_s=0.984$, $p \le 10^{-10}$).
In terms of emotional content,
individual text elements 
therefore appear to be well connected to their contexts.
We caution again that this does not imply 
individual sentences will rigidly exhibit such structure,
but rather do so on average.

We nevertheless find some scores move substantially
when the text element's score is included;
For example, 
`vegan' ranks 3rd with $\havgfnamb$=$+$0.315,
and 46th with $\havgfnnorm$=$-$0.015;
`church' ranks 19th with $\havgfnamb$=$+$0.131,
and 47th with $\havgfnnorm$=$-$0.016;
and
`sex' ranks 45th with $\havgfnamb$=$-$0.008,
but rises to 17th with $\havgfnnorm$=$+$0.542.

For financial terms, 
we see tweets mentioning 
the dissolved firm of `Lehmann Brothers' 
and
`Goldman Sachs' 
are both negative (more so in the latter's case) 
while relatively high in 
lexical size
($\havgfnamb$=$-$0.078, $\havgfnnorm$=$-$0.721, $\Nsimpson$=324 and 
$\havgfnamb$=$-$0.337, $\havgfnnorm$=$-$0.984, $\Nsimpson$=379).
We see `economy' is pegged at the 
same somewhat negative level as political terms 
($\havgfnamb$=$-$0.203) but conversely returns
a low information level ($\Nsimpson$=186).
By contrast, the more personal term `cash'
appears in highly positive tweets with $\havgfnamb$=$+$0.146.

Tweets referring to United States politics are below average in happiness
with `Obama', `Sarah Palin', and `George Bush' 
registering $\havgfnamb$=$-$0.205, $-$0.128, and $-$0.333
($\havgfnnorm$=$-$0.173, $-$0.681, and $-$0.747).
At the same time, these political figures all 
correspond to large lexical sizes
($\Nsimpson$=326, 275, and 288 respectively).
A number of other political words also
fair poorly
such as `election' ($\havgfnamb$=$-$0.127, $\havgfnnorm$=$-$0.306),
`Senate' ($\havgfnamb$=$-$0.340, $\havgfnnorm$=$-$1.541),
and 
`Congress' ($\havgfnamb$=$-$0.231, $\havgfnnorm$=$-$0.580).
The ambient happiness for `Senate' is one rank lower than `depressed'
and one higher than `BP'.
`Republican' exceeds `Democrat' in ambient happiness
($\havgfnamb$=$-$0.181 versus $-$0.226)
but trails in information content
($\Nsimpson$=240 versus 262).

Tweets involving the word `war' rank high in information 
($\Nsimpson$=386)
and are unsurprisingly low in terms of happiness
($\havgfnamb$=$-$0.270, $\havgfnnorm$=$-$2.040).
The keywords `Muslim', `Islam', and `mosque'
also register some of the lower ambient
happiness scores: 
$\havgfnamb$=$-$$0.262$, $-$$0.299$, and $-$$0.709$.
($\havgfnnorm$=$-$$0.569$, $-$$0.710$, and $-$$0.694$).

Generally, personal pronouns tell a positive prosocial story
with `our' and `you' outranking `I' and `me' in happiness
($\havgfnamb$=$+$0.207 and +0.052 versus -0.048 and $-$0.119).
The least happy pronoun on our list is the easily
demonized `they' at $\havgfnamb$=$-$0.159.
However, tweets involving pronouns indicating self appear to be more information
rich in comparison with those pointing to others:
`me' and `we' rank 11th and 32nd ($\Nsimpson=368$ and 272), 
while `they' and `them' rank 48th and 72nd overall ($\Nsimpson=240$ and 205).

The ambient words in tweets 
containing `summer' are slightly happier than
those containing `winter' but are less diverse:
$\havgfnamb$=$+$0.135 and $\Nsimpson$=219
versus $\havgfnamb$=$+$0.101 and $\Nsimpson$=240.
Other semantic differentials show
reasonable differences.
Tweets with `hot' are happier
than those with `cold'
($\havgfnamb$=$-$0.095 versus $-$0.162).
The sequence `yesterday', `today', and `tomorrow'
suggests a preferential ordering of present, future, and past
with corresponding ambient happiness scores
of
$\havgfnamb$=$+$0.033, +0.092, and +0.054.

\begin{figure}[tp!]
  \centering
    \includegraphics[width=0.48\textwidth]{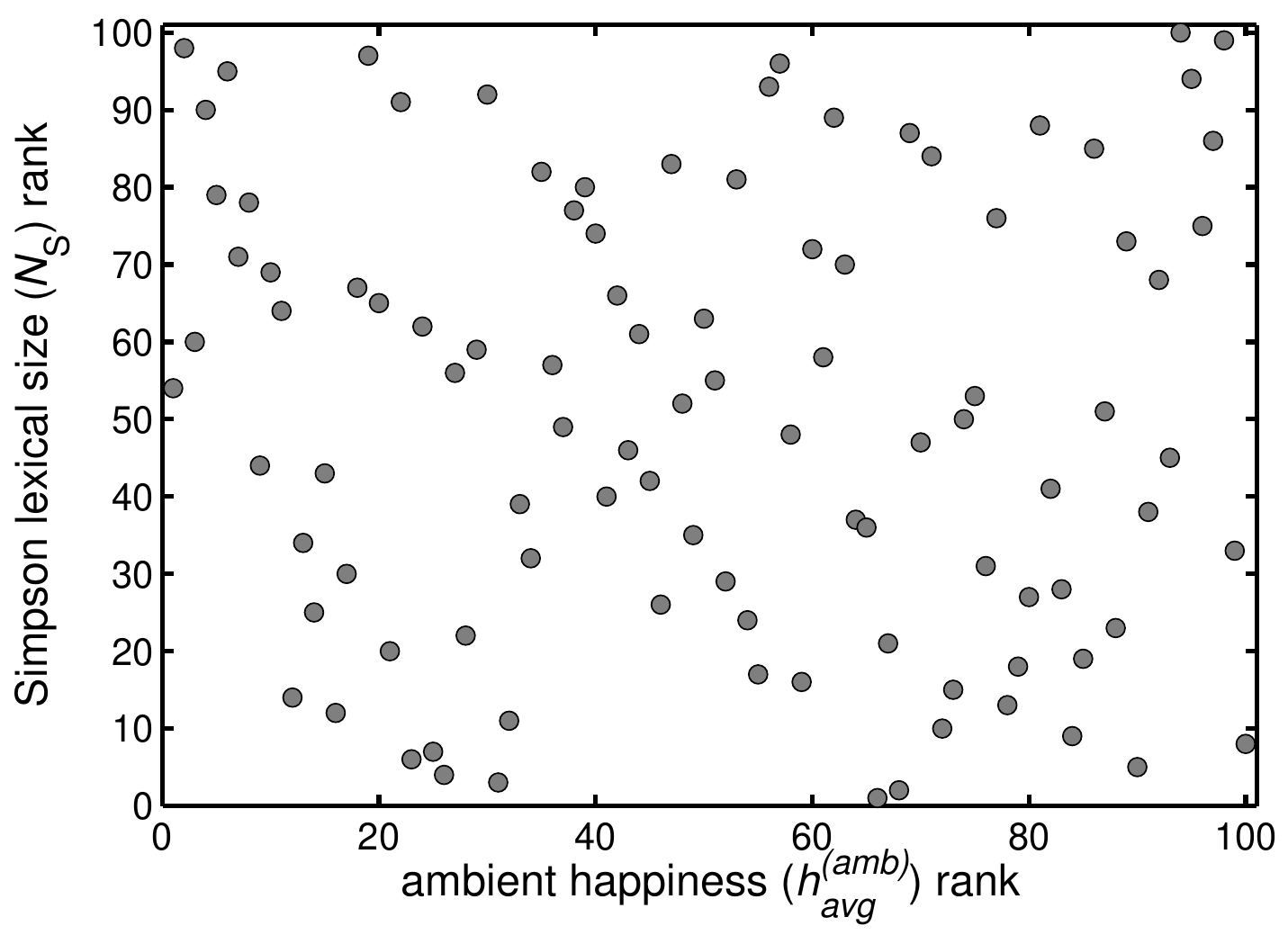}
  \caption{
    For the 100 keywords and text elements listed in 
    Tab.~\ref{tab:twhap.timeseriesA},
    a rank-rank plot
    of 
    Simpson lexical size $\Nsimpson$
    versus ambient happiness $\havgfnamb$.
    The two quantities show no 
    correlation with
    Spearman's correlation coefficient
    measuring
    $r_s = -0.038 $ ($p$-value $\simeq 0.71$).
  }
  \label{fig:twhap.valence_vs_information_ranks}
\end{figure}

Emoticons in increasing order of happiness
are 
\mbox{`:('},
\mbox{`:-('}, 
\mbox{`;-)'}, 
\mbox{`;)'},
\mbox{`:-)'},
and
\mbox{`:)'} with
$\havgfnamb$ spanning $-$0.472 to +0.274
(normalized happiness preserves the ordering
with the range increasing to -1.288 to +0.630).
In terms of increasing information level,
the order is
`:(', 
`:-(', 
`:)',
`:-)',
`;)',
and
`;-)'
with $\Nsimpson$ ranging from 305 to 477.
We see that happy emoticons correspond
to higher levels of both ambient happiness and information but
the ordering changes in a way that appears to reflect
a richness associated with cheekiness and mischief:
the two emoticons involving semi-colon
winks are third and fourth in terms of happiness but first
and second for information.

Tweets involving the `fake news' comedian Stephen Colbert are
both happier and of a higher information level 
than those concerning his senior colleague Jon Stewart 
($\havgfnamb$=$+$0.126 and $\Nsimpson$=251
versus
$\havgfnamb$=$-$0.052 and $\Nsimpson$=235).
By contrast, tweets mentioning 
Glenn Beck are lower in happiness
than both Colbert and Stewart
but comparable to Colbert in information content
($\havgfnamb$=$-$0.282 and $\Nsimpson$=252).

As noted above,
the exclamation point garners
a positive ambient happiness ($\havgfnamb$=$+$0.106),
and this is clearly above
the question mark's score of $\havgfnamb$=$-$0.030.
They have essentially equal values
for information content, ranking second (`?', $\Nsimpson$=662) 
and third (`!', $\Nsimpson$=621) overall.
These high values of $\Nsimpson$ are sensible
due to the versatility of punctuation,
and RT's top ranking reflects the 
diverse nature of status updates shared by users.

A reflection on the preceding survey suggests that groups of
related terms may possess positive, negative, or neutral correlation
between happiness and information content.
Overall, for our set of 100 keywords and text elements, 
we measure Spearman's correlation
coefficient as $r_s = -0.038 $ ($p$-value $\simeq 0.71$), 
indicating no correlation, a finding supported
visually in Fig.~\ref{fig:twhap.valence_vs_information_ranks}.
We thus have strong evidence that the two main quantities of interest
that we have studied in this paper are, generally speaking,
independent.
Several observations follow.
First of all, this independence
warrants further study for other texts
and, if possible, explanation.
Second, both quantities (or analogs)
should be reported in
any characterization of large-scale texts.
Third, for specific subfamilies of texts,
any finding of a statistically
and quantitatively significant
correlation between
happiness and lexical size 
is of interest and deserving
of further investigation.

\subsection{Analysis of Four Example Ambient Happiness Time Series}
\label{subsec:twhap.ambienttimeseries}

\begin{figure*}[tp!]
  \centering
    \includegraphics[width=\textwidth]{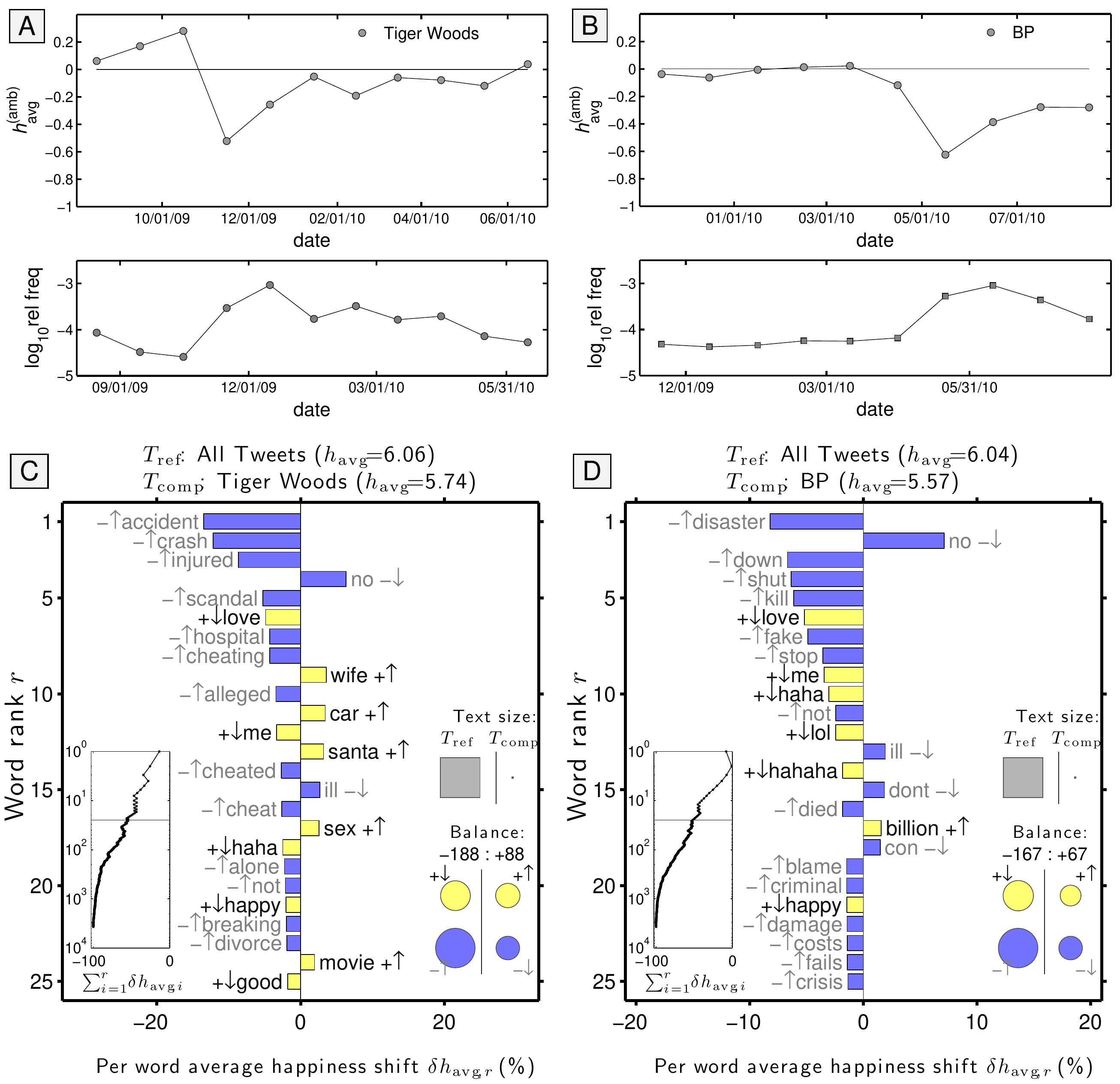}
  \caption{
    Ambient happiness time series and word shift graphs for tweets containing the keywords
    `Tiger Woods' and `BP'.  
    Ambient happiness of a keyword is $\havgfnamb$ for all words
    co-occurring in tweets containing that keyword, with the overall
    trend for all tweets subtracted.
    The word shift graphs are for tweets made during
    the worst month and the 
    ensuing one---November and December, 2009 for `Tiger Woods'
    and May and June, 2010 for `BP'.
  }
  \label{fig:twhap.fourtimeseries-1}
\end{figure*}

\begin{figure*}[tbp]
  \centering
    \includegraphics[width=\textwidth]{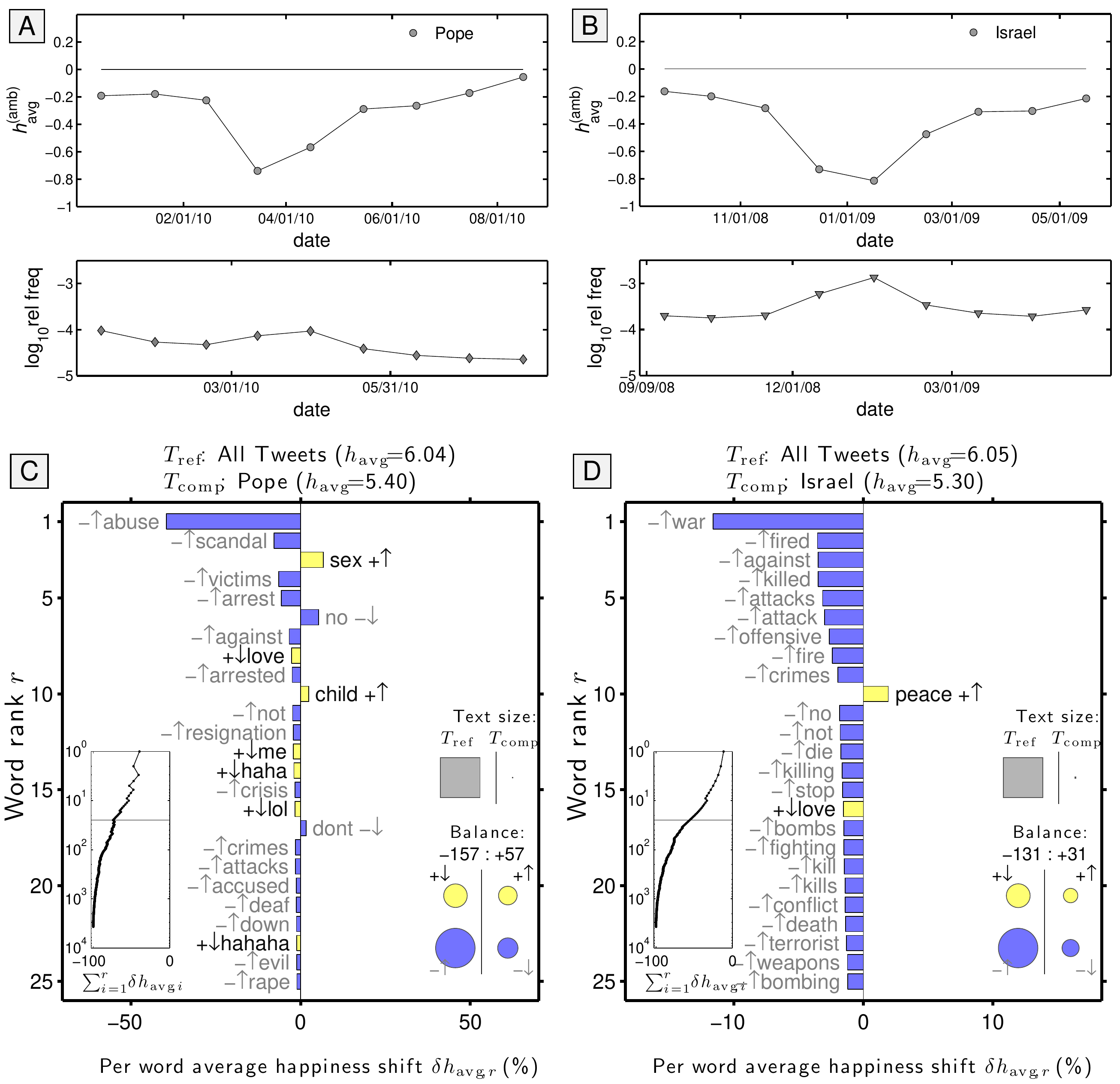}
  \caption{
    Time series and word shift graphs for tweets containing the keywords
    `Pope' and `Israel'.
    The word shift graphs are for the time periods
    March and April, 2010 for `Pope'
    and
    January and February, 2010 for `Israel.'
    See Fig.~\ref{fig:twhap.fourtimeseries-1}'s caption for more details.
  }
  \label{fig:twhap.fourtimeseries-2}
\end{figure*}

In Figs.~\ref{fig:twhap.fourtimeseries-1} 
and \ref{fig:twhap.fourtimeseries-2},
we present four ambient happiness time series for 
tweets containing the terms `Tiger Woods', `BP', `Pope', and `Israel'.  
For each example, we include word shift graphs that illuminate
the difference in word composition and tone for
the most extreme month and the following month in comparison to that of all tweets
during the same period.
All of these topics involve a negative event or events
leading to global media coverage.  

In Fig.~\ref{fig:twhap.fourtimeseries-1}A,
we show that the ambient happiness time series for
Tiger Woods drops abruptly in November, 2009
when his extramarital affairs famously became 
public after Woods crashed his car into a fire hydrant around Thanksgiving.
The National Enquirer had published a claim of 
infidelity a few days before,
and knowledge of Woods's manifold extra-marital relationships
were soon widely being reported in the general media.
Tweets concerning Woods, the world's longstanding number one golfer at the time,
dropped sharply in happiness level 
and then rebounded  over the next few months to a slightly
below average steady state.  
The jump in media coverage is reflected
in the number of tweets (middle plot).
As the word shift graph shows for November and December, 2009, 
negative words such as `accident', `crash', `scandal',
`hospital', and `divorce' pull the average happiness down below
the baseline.  The words `car' and `sex', in isolation
considered to be relatively happy words, here improved
$\havgfnamb$ for Woods,
showing one of the potential failings of our word-centric
approach.  
Nevertheless, the net effect is
clear and such microscopic errors are overcome for large enough texts.
Overall, of the four word types,
the largest contribution to the drop comes from
an increase in the use of negative words ($-$$\uparrow$).

In Fig.~\ref{fig:twhap.fourtimeseries-1}B, we see
the decline of British Petroleum's ambient happiness
following the April 20, 2010 explosion and collapse
of the deep sea drilling platform Deepwater Horizon
in the Gulf of Mexico.
The well proved to be extremely difficult to cap and
oil spewed into the Gulf for nearly three months.
In comparing tweets containing `BP' to
all tweets in May and June, 2010, we find a drop
in $\havgfnamb$ of $-$0.47
due to relative increases in words such as
`disaster',
`down',
`shut',
`kill',
`damage',
as well as
`blame',
`criminal',
and
`costs',
and decreases
in the appearance of
`love', 
`me',
`haha',
and 
`lol'.
Similar to the Tiger Woods word shift,
we see the more frequent use of 
relatively negative words ($-$$\uparrow$)
and the
less frequent use of more 
positive words ($+$$\downarrow$);
both contribute substantially
to the sharp decrease in average happiness.

In Fig.~\ref{fig:twhap.fourtimeseries-2}A,
we track the ambient happiness of the keyword `Pope'
over a nine month period starting with December, 2009.
While the relative frequency of tweets containing
`Pope' changes little, a clear minimum in
$\havgfnamb$ occurs in March, 2010.
The Catholic Church's long running child
molestation scandal was brought into even sharper
focus during this month, notably via a Papal apology
to the Irish church, and
the New York Times publishing documents
concerning Pope Benedict's past decisions
on child molestation cases, opening up
a highly charged dialogue between the media and the Vatican.
In the word shift graph, we see the nadir of March and April
arising from
the more frequent use of negative words
such as
`abuse',
`scandal',
`victims',
`arrest',
and
`resignation',
and the drop in positive words
such as 
`love',
`me',
and 
`haha'.
The increased use of the words 
`sex' ($\havgfn$=8.05)
and 
`child' ($\havgfn$=7.08)
in tweets containing `Pope' goes against the trend
(see remarks above for Tiger Woods).
The overall picture is similar to that 
for Tiger Woods and BP: 
the increase in negative words ($-$$\uparrow$)
is the main reason `Pope' tweets are far
below the average happiness level for March and April, 2010.

Our last example, Fig.~\ref{fig:twhap.fourtimeseries-2}B,
shows ambient happiness for tweets involving `Israel'
from September, 2008 through to May, 2009.  
The drop
in November and December reaching a minimum in January matches with the 
Gaza War, fought between Israel and Hamas.  
The increase in `Israel' tweets also captures
the increase in media reporting during
this conflict.  
In the top ranked 25 words contributing to the strong
decrease for January and February relative to the overall time series, 
we see the major changes primarily coming
from the more frequent use of negative words 
($-$$\uparrow$) such as 
`war',
`fire',
`kill',
`attack',
`bombs',
and
`conflict'.
Against this rather bleak sequence of 
negative word shifts, we may take some
solace in seeing the word `peace' 
appear more often ($+$$\uparrow$).
Once again, we see that the overall drop is
due largely to an increase in negative words
($-$$\uparrow$) and to a lesser extent
a decrease in positive words ($+$$\downarrow$).

\section{Concluding remarks}
\label{sec:twhap.conclusion}

In analysing temporal patterns of happiness
and information content for the very large
data set generated by Twitter thus far, we have been
able to uncover results ranging across many timescales
and topics.  The weekly and daily cycles in particular appear
to be robust and suggestive of universal forms,
accepting that the seven day week cycle is 
an historical and cultural artifact.  
With our greatly expanded word list as analysed using Mechanical Turk, labMT 1.0 (Data Set S1),
we believe we have provided a substantial methodological advance in the measurement
of sentiment in large-scale texts.
We hope that our tunable hedonometer and the associated words provided
in the Supplementary Information will be of use to other researchers.

An essential part of our comparative analyses is
the word shift graph, which we have primarily used
here for happiness.  These provide us with a detailed
view of why two texts differ based on changes
in word frequency.  These graphs, and their future
iterations, should be of use in a range of fields
where size distributions are compared through
summary statistics (e.g., understanding how species diversity
in ecological populations may differ as a result of 
changes in individual species abundances).

As we have described, the metadata accompanying
Twitter messages contains more information than time stamps.  
Future research will naturally address (and go beyond) 
geographic variations, particularly
for the United States; 
the change in expressions over time for individuals
and the possibility of correlation or contagion of
sentiment;
effects of popularity as measured by
follower count on users' expressions;
and the possibility of fine-scale emotional 
synchronization between individuals 
based on directed messages~\cite{bollen2011a}.
In terms of methodology, our hedonometer could be 
improved by incorporating happiness estimates for
common $n$-grams, e.g., 2-grams such as `child abuse' and `sex scandal' 
as well as negated sentiments such as `not happy'.
This would improve the reliability of our
happiness (and information) measures without
losing the transparency of our current approach,
and begin to address issues of words being
used in specific contexts and words having multiple meanings.
Language detection for tweets, recently added by Twitter
to their metadata, allows for language specific analyses.
The robustness of the measure we have used in our
present work further suggests that we should
be able to determine conversion
factors between the scores of
different text-based hedonometers.
For measures of information content, 
an improved handling of very long word lists, and
potential incorporation of $n$-grams, 
will allow us to use Shannon's entropy in future work.

As we have seen in both the work of others and ours,
Twitter and similar large-scale, online social networks
have thus far provided good evidence that 
scientifically interesting and meaningful patterns can be extracted
from these massive data sources of human behavior.
The extent to which small-scale patterns can be
elicited, e.g., for rare topics, also remains an open question,
as does the true generalizability to the broader population.
Whatever the case, Twitter is currently a substantial,
growing element
of the global media and is worth studying in its own right, just
as a study of newspapers would seem entirely valid.
And while current evidence suggests `instant polls' created by
remote-sensing text analysis methods are valid,
and that these instruments 
complement and may in some cases improve upon traditional surveys,
analysts will have to remain cognizant of
the ever present problem of users gaming
online expression systems to misinform.

Finally, the era of big data social sciences
has undoubtedly begun.  
Rather than being transformed or revolutionized
we feel the correct view is 
that the social sciences are expanding 
beyond a stable core to become
data-abundant fields.
In a data-abundant science, 
the challenge moves first to description
and pattern finding, with explanation
and experiments following.
Instead of first forming hypotheses,
we are forced to spend considerable
time and effort simply describing.
The approaches applicable for
a data-scarce science still remain of 
the same value 
but new, vast windows into social and psychological 
behaviour are now open, and new tools are
available and being developed to enable
us to take in the view.  

\section*{Methods}
\label{sec:twhap.methods}

We defined a word as
any contiguous set of characters bounded by white space and/or
a small set of punctuation characters.
We therefore included
all misspellings, words from any language used on Twitter, 
hyperlinks, etc.   
All pattern matches we made
were case-insensitive, and we did not perform
stemming (e.g., `love' and `loved' were counted separately).

The data feed from Twitter was provided 
in XML and JSON formats~\cite{twitterapi}.
Early on, the data feed contained
many repeated tweets,
and while the fraction
of duplicates dropped substantially over time,
we nevertheless were obliged to check for
and remove all such tweets.
(Due to these various changes, all measures
involving emoticons are derived from
the time series up until only November 9, 2009.)

In measuring and comparing information content, 
a computational difficulty with the Twitter data set
lies in accommodating the sheer number of distinct words.
We found 
approximately 230 million unique words (including URLs)
from a random sample of 25\% of the tweets in our database.
We determined that restricting our 
attention to a more manageable set
of the first 50,000 most frequent words
would be sufficient for highly accurate
estimates of generalized entropy $H_q$ with $q \gtrsim 1.5$,
and therefore Simpson's concentration $S$ when $q=2$.
We did not use Shannon's entropy~\cite{shannon1948a}
since it converges too slowly (akin to $q=1$) 
for the skew we observed in the Twitter word
frequency distribution.
Importantly, in fixing a list of words, we were able to account
for information content differences between texts
at the level of words.

Consequently, we recorded the frequencies for this specific set
of 50,000 words at the level of hours and days.
Note that we also always recorded the total number of words for
any particular subset of tweets,
so that our word probabilities were correctly normalized.

\begin{acknowledgments}
  The authors thank
B.~O'Connor for discussions 
and data sharing,
as well as
N.~Allgaier, 
T.~Gray,
J.~Harris, 
N.~Johnson,
P.~Lessard, 
L.~Mitchell,
E.~Pechenick,
M.~Pellon, 
A.~Reece,
N.~Riktor,
B.~Tivnan,
M.~Tretin,
and 
J.~Williams.

\end{acknowledgments}

\clearpage

\setcounter{page}{1}
\renewcommand{\thepage}{S\arabic{page}}
\renewcommand{\thefigure}{S\arabic{figure}}
\renewcommand{\thetable}{S\arabic{table}}
\setcounter{figure}{0}
\setcounter{table}{0}

\section*{Supplementary Material}
\label{sec:twhap.supp}

\begin{itemize}
\item 
  Supporting Figures and Table (this document).
\item 
  Data from Mechanical Turk study:\\
  labMT 1.0 = language assessment by Mechanical Turk 1.0.

  In the supplementary tab-delimited file named Data Set S1,
  we provide our set of 10,222 words,
  their average happiness evaluations according
  to users on Mechanical Turk, and other
  information as described below.

  Please cite the present paper when using
  this word set.  
  Within papers, we suggest using the abbreviation labMT 1.0
  when referencing this data set.

  The words are ordered according to average happiness (descending),
  and the file contains eight columns:
  \begin{enumerate}
  \item 
    word,
  \item 
    rank,
  \item 
    average happiness (50 user evalutions),
  \item 
    standard deviation of happiness,
  \item 
    Twitter rank,
  \item 
    Google Books rank,
  \item 
    New York Times rank,
  \item 
    Music Lyrics rank.
  \end{enumerate}

  The last four columns correspond to the ranking
  of a word by frequency of occurrence 
  in the top 5000 words for the specified
  corpus.  A double dash `\verb+--+' indicates a word was not
  found in the most frequent 5000 words for a corpus.

  Please see the main paper for more information
  regarding this data set.
\end{itemize}

\revtexonly{\begin{turnpage}}
  \revtexonly{\begin{figure*}[tp!]}
    \plainlatexonly{\begin{figure}[tp!]}
                                                \centering
      \includegraphics[width=1.0\textheight]{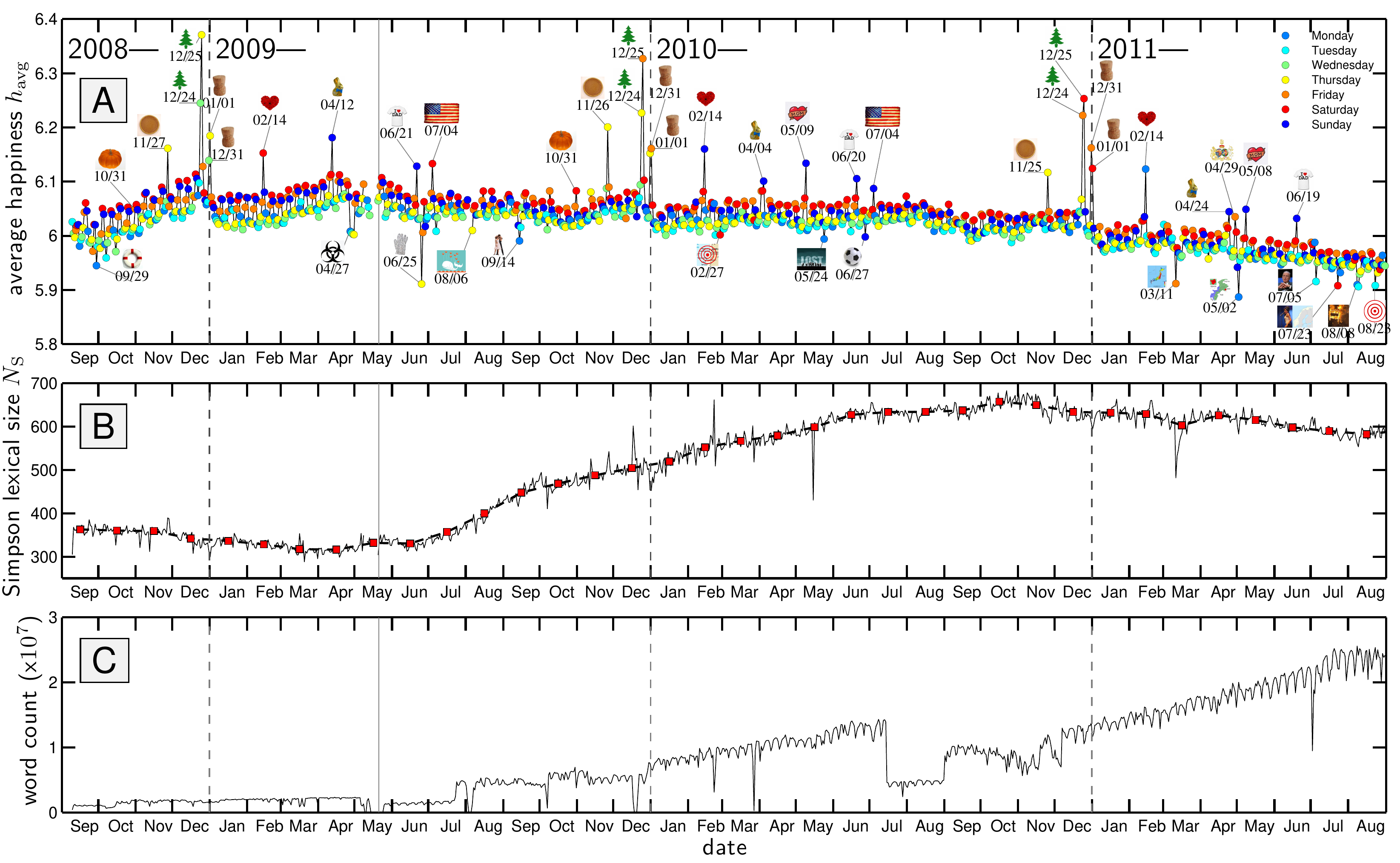} 
      \caption{
        High resolution, zoomable version of
        Fig.~\ref{fig:twhap.twitter_timeseries_1} in the main text.
      }
      \label{fig:twhap.twitter_timeseries_zoomable_supp}
      \plainlatexonly{\end{figure}}
    \revtexonly{\end{figure*}}
\revtexonly{\end{turnpage}}

\begin{figure*}[hp!]
  \centering
      \includegraphics[width=\textwidth]{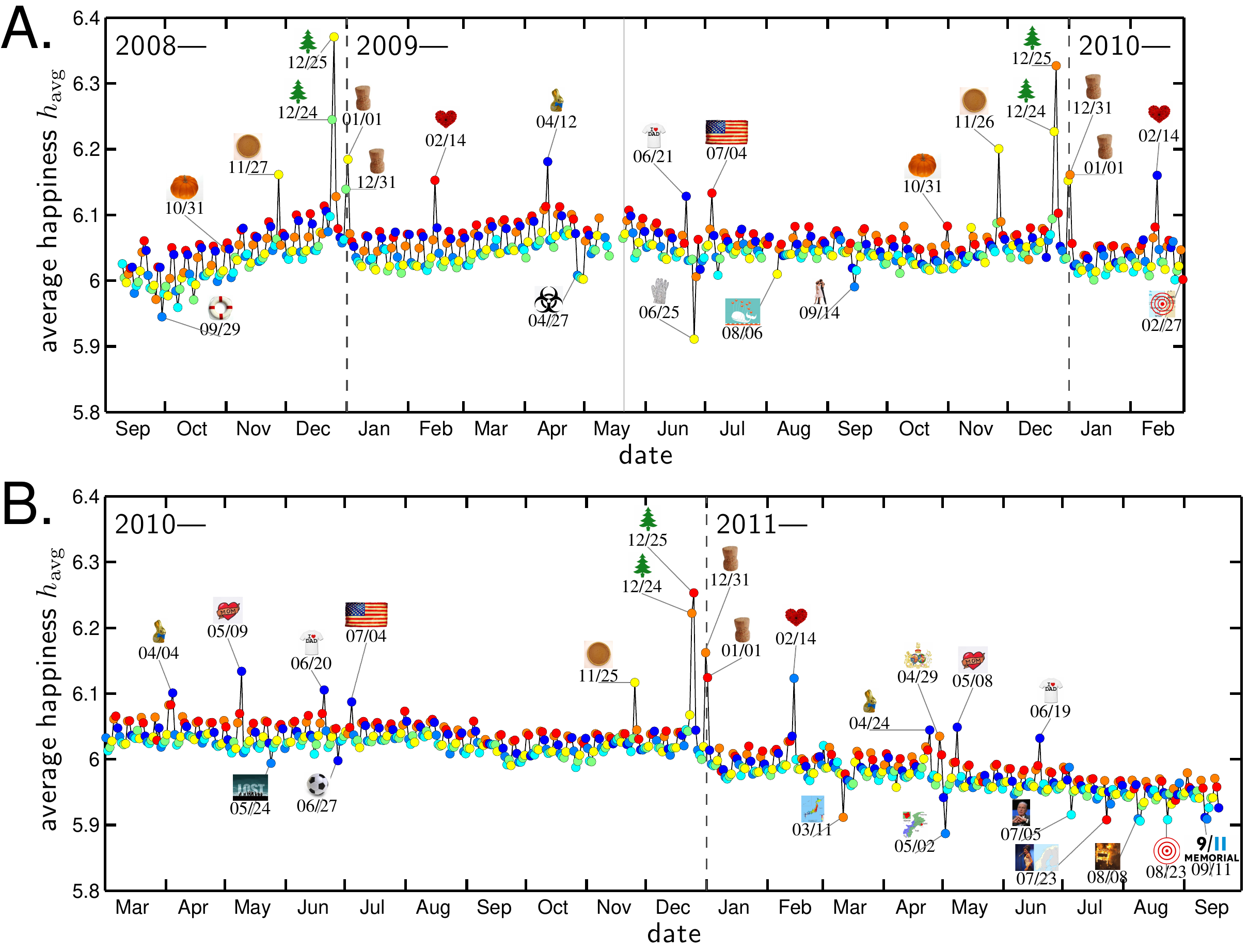}
  \caption{
    Simple average happiness time series plots.
    The time series is extended to
    include part of September, 2011, and 
    shows a drop corresponding to the tenth anniversary of the 9/11 terror attacks
    in the United States.
  }
  \label{fig:twhap.twitter_timeseries_simple_supp}
\end{figure*}

\clearpage

\begin{figure}[hbp!]
  \centering
                    \includegraphics[width=0.48\textwidth]{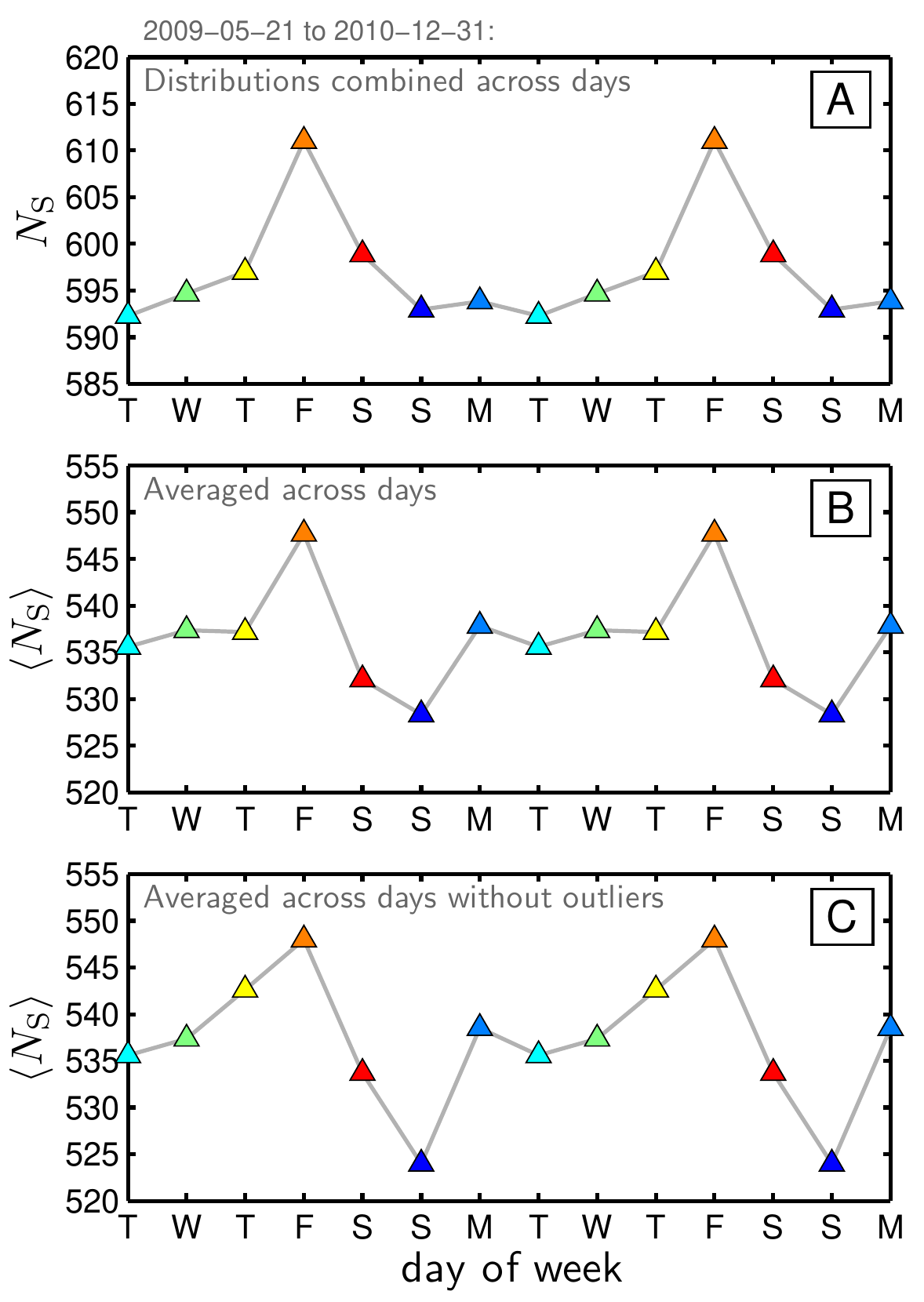}
  \caption{
    Simpson lexical size as a function
    of day of the week 
    using three different ways of creating distributions.
    Compare with Fig.~\ref{fig:twhap.twitter_timeseries_2a_simpson}.
  }
  \label{fig:twhap.twitter_timeseries_2a_simpson_supp}
\end{figure}

\begin{figure}[hbp!]
  \centering
                    \includegraphics[width=0.48\textwidth]{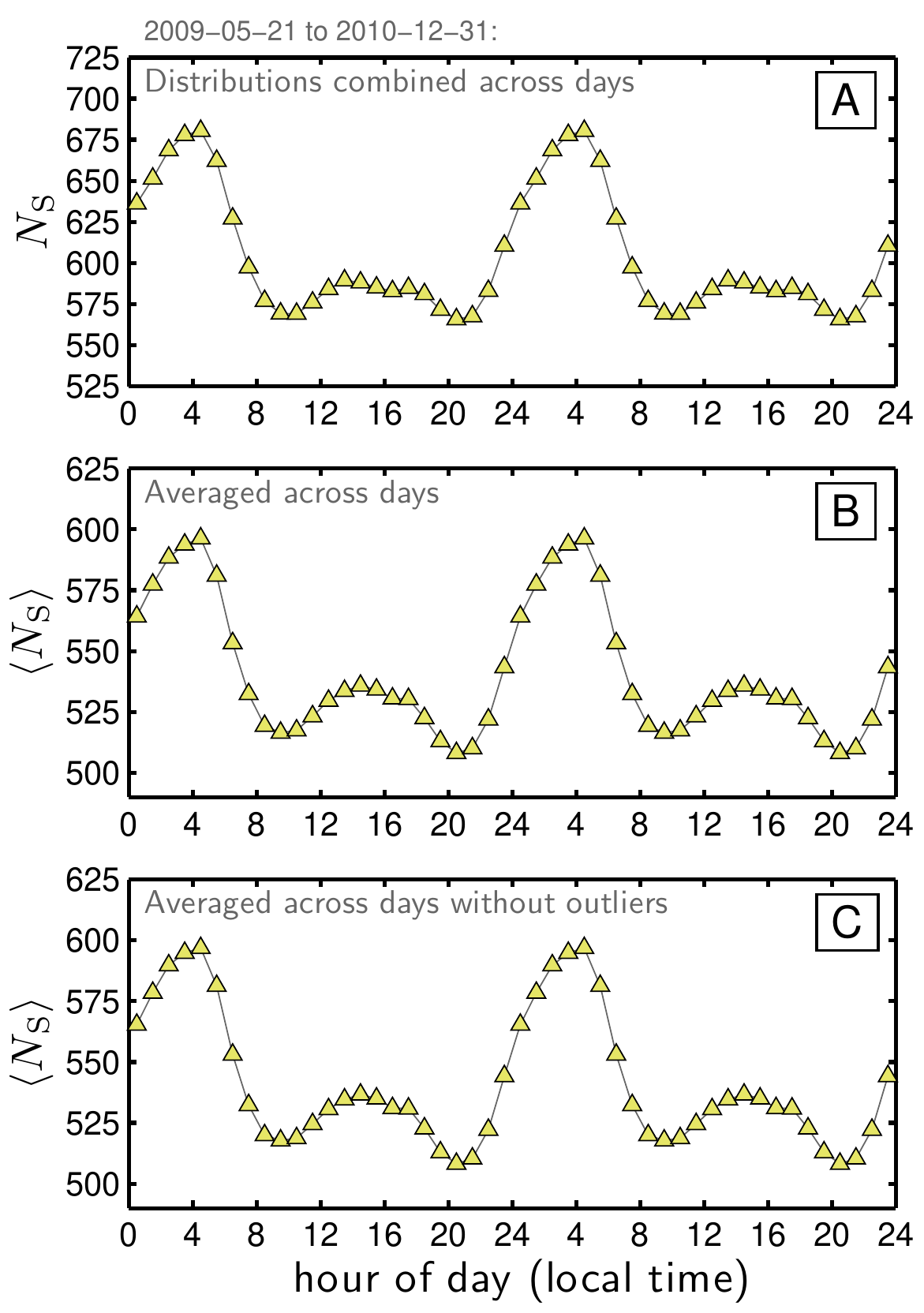} 
  \caption{
    Average Simpson lexical size $\Nsimpson$ for time of day, corrected according to local time,
    using three different ways of creating distributions.
    Compare with Fig.~\ref{fig:twhap.twitter_timeseries_daily_info}.
  }
  \label{fig:twhap.twitter_timeseries_daily_info_supp}
\end{figure}

\revtexonly{\begin{turnpage}}
  \revtexonly{\begin{figure*}[tp!]}
    \plainlatexonly{\begin{figure}[tp!]}
      \centering
                                                            \includegraphics[height=0.85\textwidth]{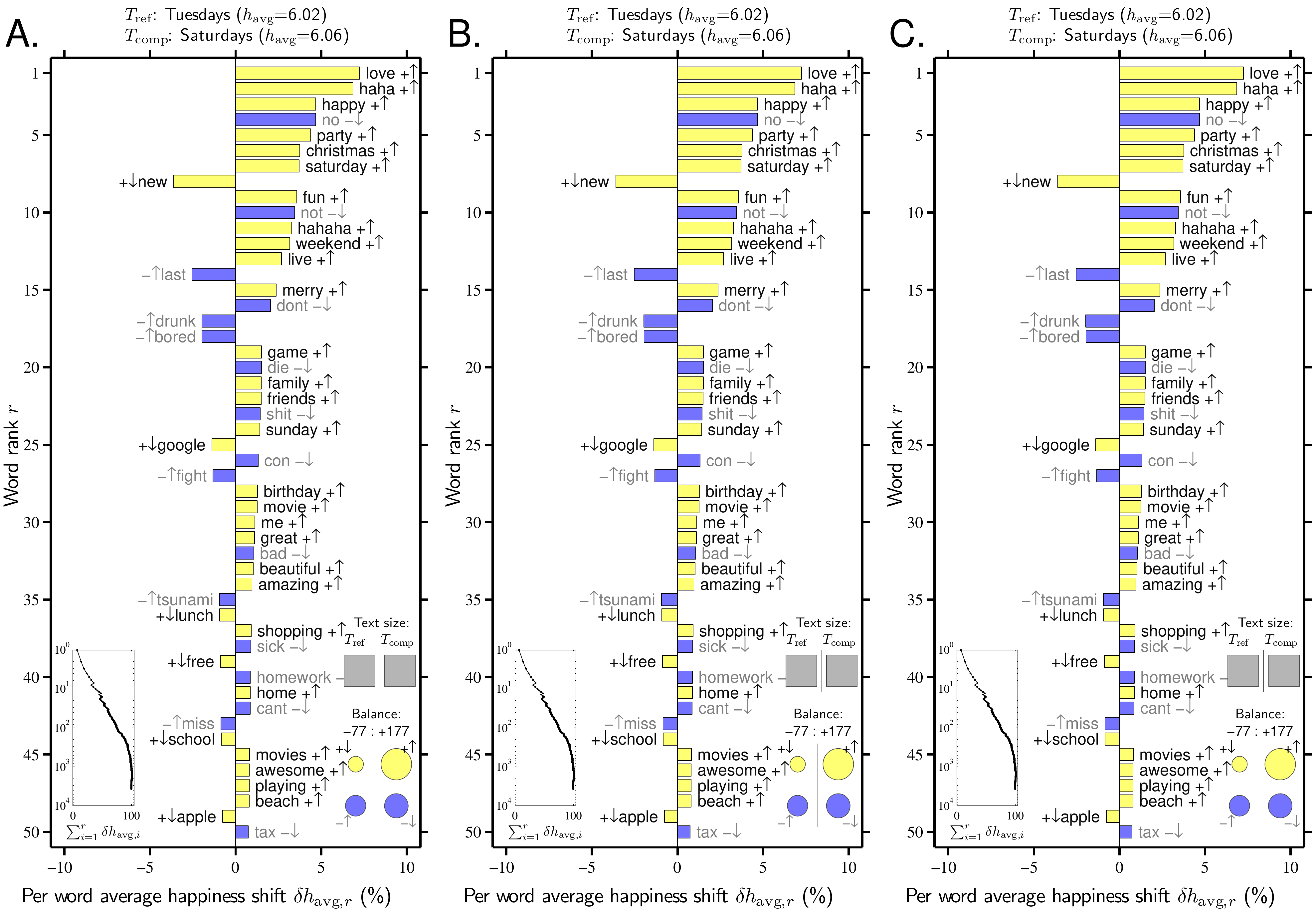}
      \caption{
        Word shift graph comparing the 
        Saturdays to Tuesdays
        using three approaches to generating the day word distributions.
        \textbf{A.} Days combined without regard to sampling frequency,
        \textbf{B.} Days given equal weighting,
        \textbf{C.} days given equal weighting with outlier dates removed.
        While words do move around the overall pattern remains similar.
        Compare with Fig.~\ref{fig:twhap.twitter_timeseries_2b}.
      }
      \label{fig:twhap.twitter_timeseries_2b_supp}
      \plainlatexonly{\end{figure}}
    \revtexonly{\end{figure*}}
  \revtexonly{\end{turnpage}}

\revtexonly{\begin{turnpage}}
  \revtexonly{\begin{figure*}[tp!]}
    \plainlatexonly{\begin{figure}[tp!]}
                              \includegraphics[height=0.85\textwidth]{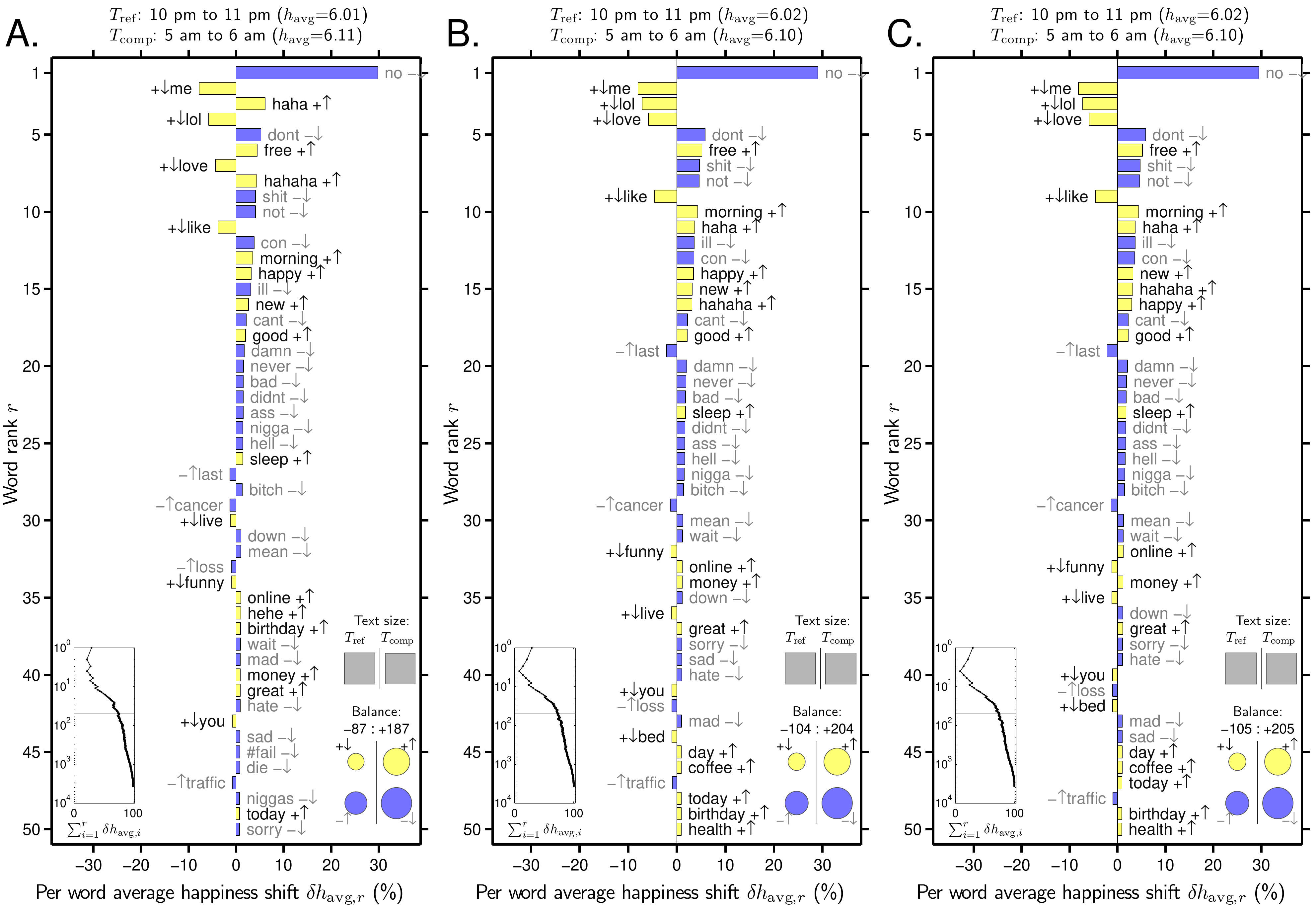}
      \caption{
        Word shift graph comparing the 
        happiest hour (5 am to 6 am)
        relative
        to the least happy hour (10 pm to 11 pm)
        using three approaches to generating the day word distributions.
        \textbf{A.} Days combined without regard to sampling frequency,
        \textbf{B.} days given equal weighting,
        \textbf{C.} days given equal weighting with outlier dates removed.
        Compare with Fig.~\ref{fig:twhap.twitter_timeseries_daily_wsg}
        and see Fig.~\ref{fig:twhap.twitter_timeseries_2b}
        and related text for further explanation.
      }
      \label{fig:twhap.twitter_timeseries_daily_wsg_supp}
      \plainlatexonly{\end{figure}}
    \revtexonly{\end{figure*}}
  \revtexonly{\end{turnpage}}

\begin{table*}[tp!]
  \centering
  \plainlatexonly{\begin{scriptsize}}
    \begin{tabular}{|l|c|l|l|}
\hline Word & $\havgfnnorm$ & Total Tweets & $\havgfnamb$  \\ \hline
1. happy & +1.104 & 1.65e+07 (13) & +0.430 (1) \\ 
\textcolor{lightgrey}{2. love} & \textcolor{lightgrey}{+0.977} & \textcolor{lightgrey}{4.67e+07 (6)} & \textcolor{lightgrey}{+0.164 (11)} \\ 
3. Christmas & +0.953 & 4.89e+06 (35) & +0.404 (2) \\ 
\textcolor{lightgrey}{4. win} & \textcolor{lightgrey}{+0.924} & \textcolor{lightgrey}{7.98e+06 (26)} & \textcolor{lightgrey}{+0.204 (8)} \\ 
5. vacation & +0.817 & 9.35e+05 (67) & +0.200 (9) \\ 
\textcolor{lightgrey}{6. sun} & \textcolor{lightgrey}{+0.737} & \textcolor{lightgrey}{2.39e+06 (52)} & \textcolor{lightgrey}{+0.144 (16)} \\ 
7. family & +0.716 & 5.01e+06 (32) & +0.251 (5) \\ 
\textcolor{lightgrey}{8. friends} & \textcolor{lightgrey}{+0.685} & \textcolor{lightgrey}{7.67e+06 (27)} & \textcolor{lightgrey}{+0.155 (12)} \\ 
9. party & +0.679 & 6.44e+06 (29) & +0.170 (10) \\ 
\textcolor{lightgrey}{10. heaven} & \textcolor{lightgrey}{+0.674} & \textcolor{lightgrey}{7.42e+05 (71)} & \textcolor{lightgrey}{+0.041 (34)} \\ 
11. kiss & +0.632 & 1.70e+06 (59) & +0.072 (30) \\ 
\textcolor{lightgrey}{12. :)} & \textcolor{lightgrey}{+0.630} & \textcolor{lightgrey}{1.04e+07 (20)} & \textcolor{lightgrey}{+0.274 (4)} \\ 
13. income & +0.621 & 5.10e+05 (76) & +0.137 (17) \\ 
\textcolor{lightgrey}{14. cash} & \textcolor{lightgrey}{+0.601} & \textcolor{lightgrey}{1.28e+06 (63)} & \textcolor{lightgrey}{+0.146 (15)} \\ 
15. Valentine & +0.593 & 2.47e+05 (84) & +0.127 (20) \\ 
\textcolor{lightgrey}{16. :-)} & \textcolor{lightgrey}{+0.560} & \textcolor{lightgrey}{1.67e+06 (60)} & \textcolor{lightgrey}{+0.228 (6)} \\ 
17. sex & +0.542 & 3.55e+06 (39) & -0.008 (45) \\ 
\textcolor{lightgrey}{18. coffee} & \textcolor{lightgrey}{+0.518} & \textcolor{lightgrey}{2.80e+06 (46)} & \textcolor{lightgrey}{+0.147 (14)} \\ 
19. hope & +0.515 & 1.18e+07 (18) & +0.149 (13) \\ 
\textcolor{lightgrey}{20. God} & \textcolor{lightgrey}{+0.468} & \textcolor{lightgrey}{8.58e+06 (25)} & \textcolor{lightgrey}{+0.099 (25)} \\ 
21. health & +0.447 & 2.58e+06 (50) & -0.000 (44) \\ 
\textcolor{lightgrey}{22. life} & \textcolor{lightgrey}{+0.422} & \textcolor{lightgrey}{1.40e+07 (17)} & \textcolor{lightgrey}{+0.012 (43)} \\ 
23. ;-) & +0.395 & 9.39e+05 (66) & +0.041 (35) \\ 
\textcolor{lightgrey}{24. girl} & \textcolor{lightgrey}{+0.331} & \textcolor{lightgrey}{1.01e+07 (22)} & \textcolor{lightgrey}{-0.010 (47)} \\ 
25. ;) & +0.326 & 2.61e+06 (48) & +0.094 (27) \\ 
\textcolor{lightgrey}{26. USA} & \textcolor{lightgrey}{+0.325} & \textcolor{lightgrey}{2.16e+06 (54)} & \textcolor{lightgrey}{+0.113 (22)} \\ 
27. yes & +0.321 & 1.16e+07 (19) & +0.056 (31) \\ 
\textcolor{lightgrey}{28. Jesus} & \textcolor{lightgrey}{+0.247} & \textcolor{lightgrey}{2.03e+06 (56)} & \textcolor{lightgrey}{+0.094 (28)} \\ 
29. summer & +0.221 & 3.00e+06 (43) & +0.135 (18) \\ 
\textcolor{lightgrey}{30. woman} & \textcolor{lightgrey}{+0.202} & \textcolor{lightgrey}{2.54e+06 (51)} & \textcolor{lightgrey}{-0.115 (57)} \\ 
31. ! & +0.195 & 3.44e+06 (40) & +0.106 (23) \\ 
\textcolor{lightgrey}{32. me} & \textcolor{lightgrey}{+0.160} & \textcolor{lightgrey}{1.44e+08 (4)} & \textcolor{lightgrey}{-0.119 (59)} \\ 
33. our & +0.159 & 1.41e+07 (16) & +0.207 (7) \\ 
\textcolor{lightgrey}{34. we} & \textcolor{lightgrey}{+0.146} & \textcolor{lightgrey}{3.91e+07 (7)} & \textcolor{lightgrey}{+0.035 (36)} \\ 
35. right & +0.126 & 1.92e+07 (10) & -0.090 (56) \\ 
\textcolor{lightgrey}{36. today} & \textcolor{lightgrey}{+0.126} & \textcolor{lightgrey}{2.56e+07 (9)} & \textcolor{lightgrey}{+0.092 (29)} \\ 
37. you & +0.111 & 1.73e+08 (3) & +0.052 (33) \\ 
\textcolor{lightgrey}{38. tomorrow} & \textcolor{lightgrey}{+0.086} & \textcolor{lightgrey}{1.04e+07 (21)} & \textcolor{lightgrey}{+0.054 (32)} \\ 
39. snow & +0.083 & 2.60e+06 (49) & -0.051 (51) \\ 
\textcolor{lightgrey}{40. night} & \textcolor{lightgrey}{+0.074} & \textcolor{lightgrey}{1.71e+07 (12)} & \textcolor{lightgrey}{+0.014 (42)} \\ 
41. boy & +0.062 & 4.93e+06 (33) & -0.026 (48) \\ 
\textcolor{lightgrey}{42. school} & \textcolor{lightgrey}{+0.050} & \textcolor{lightgrey}{9.26e+06 (24)} & \textcolor{lightgrey}{-0.056 (53)} \\ 
43. winter & +0.050 & 1.26e+06 (64) & +0.101 (24) \\ 
\textcolor{lightgrey}{44. rain} & \textcolor{lightgrey}{+0.050} & \textcolor{lightgrey}{3.23e+06 (41)} & \textcolor{lightgrey}{-0.134 (63)} \\ 
45. Stephen Colbert & +0.001 & 2.38e+04 (99) & +0.126 (21) \\ 
\textcolor{lightgrey}{46. vegan} & \textcolor{lightgrey}{-0.015} & \textcolor{lightgrey}{1.84e+05 (90)} & \textcolor{lightgrey}{+0.315 (3)} \\ 
47. church & -0.016 & 1.81e+06 (58) & +0.131 (19) \\ 
\textcolor{lightgrey}{48. Jon Stewart} & \textcolor{lightgrey}{-0.024} & \textcolor{lightgrey}{5.21e+04 (97)} & \textcolor{lightgrey}{-0.052 (52)} \\ 
49. I & -0.062 & 3.08e+08 (2) & -0.048 (49) \\ 
\textcolor{lightgrey}{50. I feel} & \textcolor{lightgrey}{-0.129} & \textcolor{lightgrey}{5.17e+06 (31)} & \textcolor{lightgrey}{-0.173 (70)} \\ 
\hline
\end{tabular}\begin{tabular}{|l|c|l|l|}
\hline Word & $\havgfnnorm$ & Total Tweets & $\havgfnamb$  \\ \hline
51. climate & -0.160 & 3.64e+05 (80) & -0.135 (64) \\ 
\textcolor{lightgrey}{52. man} & \textcolor{lightgrey}{-0.163} & \textcolor{lightgrey}{1.59e+07 (14)} & \textcolor{lightgrey}{-0.175 (71)} \\ 
53. yesterday & -0.168 & 3.08e+06 (42) & +0.033 (37) \\ 
\textcolor{lightgrey}{54. hot} & \textcolor{lightgrey}{-0.172} & \textcolor{lightgrey}{7.12e+06 (28)} & \textcolor{lightgrey}{+0.095 (26)} \\ 
55. Obama & -0.173 & 2.98e+06 (44) & -0.205 (76) \\ 
\textcolor{lightgrey}{56. work} & \textcolor{lightgrey}{-0.174} & \textcolor{lightgrey}{1.84e+07 (11)} & \textcolor{lightgrey}{-0.010 (46)} \\ 
57. commute & -0.206 & 9.01e+04 (94) & -0.048 (50) \\ 
\textcolor{lightgrey}{58. they} & \textcolor{lightgrey}{-0.208} & \textcolor{lightgrey}{2.74e+07 (8)} & \textcolor{lightgrey}{-0.159 (67)} \\ 
59. Michael Jackson & -0.213 & 8.26e+05 (70) & +0.018 (41) \\ 
\textcolor{lightgrey}{60. them} & \textcolor{lightgrey}{-0.280} & \textcolor{lightgrey}{1.54e+07 (15)} & \textcolor{lightgrey}{-0.090 (55)} \\ 
61. election & -0.306 & 5.60e+05 (75) & -0.127 (60) \\ 
\textcolor{lightgrey}{62. Pope} & \textcolor{lightgrey}{-0.316} & \textcolor{lightgrey}{1.52e+05 (91)} & \textcolor{lightgrey}{-0.277 (83)} \\ 
63. left & -0.383 & 4.89e+06 (34) & -0.118 (58) \\ 
\textcolor{lightgrey}{64. Democrat} & \textcolor{lightgrey}{-0.384} & \textcolor{lightgrey}{9.32e+04 (93)} & \textcolor{lightgrey}{-0.226 (77)} \\ 
65. oil & -0.411 & 1.38e+06 (62) & -0.162 (68) \\ 
\textcolor{lightgrey}{66. RT} & \textcolor{lightgrey}{-0.443} & \textcolor{lightgrey}{3.39e+08 (1)} & \textcolor{lightgrey}{+0.028 (40)} \\ 
67. gas & -0.471 & 1.02e+06 (65) & -0.193 (74) \\ 
\textcolor{lightgrey}{68. ?} & \textcolor{lightgrey}{-0.503} & \textcolor{lightgrey}{2.32e+06 (53)} & \textcolor{lightgrey}{+0.030 (39)} \\ 
69. economy & -0.525 & 6.09e+05 (73) & -0.203 (75) \\ 
\textcolor{lightgrey}{70. Republican} & \textcolor{lightgrey}{-0.539} & \textcolor{lightgrey}{2.30e+05 (86)} & \textcolor{lightgrey}{-0.181 (72)} \\ 
71. cold & -0.546 & 3.67e+06 (36) & -0.162 (69) \\ 
\textcolor{lightgrey}{72. gay} & \textcolor{lightgrey}{-0.552} & \textcolor{lightgrey}{2.73e+06 (47)} & \textcolor{lightgrey}{-0.152 (65)} \\ 
73. Muslim & -0.569 & 2.15e+05 (88) & -0.262 (81) \\ 
\textcolor{lightgrey}{74. Congress} & \textcolor{lightgrey}{-0.580} & \textcolor{lightgrey}{3.92e+05 (79)} & \textcolor{lightgrey}{-0.231 (78)} \\ 
75. Senate & -0.601 & 4.48e+05 (78) & -0.340 (90) \\ 
\textcolor{lightgrey}{76. Sarah Palin} & \textcolor{lightgrey}{-0.681} & \textcolor{lightgrey}{2.26e+05 (87)} & \textcolor{lightgrey}{-0.128 (61)} \\ 
77. mosque & -0.694 & 6.98e+04 (95) & -0.709 (98) \\ 
\textcolor{lightgrey}{78. Islam} & \textcolor{lightgrey}{-0.710} & \textcolor{lightgrey}{1.87e+05 (89)} & \textcolor{lightgrey}{-0.299 (86)} \\ 
79. Lehman Brothers & -0.721 & 8.50e+03 (100) & -0.078 (54) \\ 
\textcolor{lightgrey}{80. George Bush} & \textcolor{lightgrey}{-0.747} & \textcolor{lightgrey}{3.23e+04 (98)} & \textcolor{lightgrey}{-0.333 (87)} \\ 
81. dark & -0.766 & 1.58e+06 (61) & +0.031 (38) \\ 
\textcolor{lightgrey}{82. Glenn Beck} & \textcolor{lightgrey}{-0.776} & \textcolor{lightgrey}{1.14e+05 (92)} & \textcolor{lightgrey}{-0.282 (85)} \\ 
83. BP & -0.902 & 5.82e+05 (74) & -0.355 (91) \\ 
\textcolor{lightgrey}{84. Goldman Sachs} & \textcolor{lightgrey}{-0.984} & \textcolor{lightgrey}{5.27e+04 (96)} & \textcolor{lightgrey}{-0.337 (88)} \\ 
85. :-( & -1.174 & 3.40e+05 (81) & -0.455 (95) \\ 
\textcolor{lightgrey}{86. lose} & \textcolor{lightgrey}{-1.181} & \textcolor{lightgrey}{2.06e+06 (55)} & \textcolor{lightgrey}{-0.157 (66)} \\ 
87. Iraq & -1.282 & 2.39e+05 (85) & -0.773 (100) \\ 
\textcolor{lightgrey}{88. :(} & \textcolor{lightgrey}{-1.288} & \textcolor{lightgrey}{2.89e+06 (45)} & \textcolor{lightgrey}{-0.472 (96)} \\ 
89. sad & -1.366 & 3.56e+06 (38) & -0.187 (73) \\ 
\textcolor{lightgrey}{90. no} & \textcolor{lightgrey}{-1.415} & \textcolor{lightgrey}{9.51e+07 (5)} & \textcolor{lightgrey}{-0.132 (62)} \\ 
91. drugs & -1.452 & 5.10e+05 (77) & -0.382 (93) \\ 
\textcolor{lightgrey}{92. Afghanistan} & \textcolor{lightgrey}{-1.458} & \textcolor{lightgrey}{2.74e+05 (83)} & \textcolor{lightgrey}{-0.703 (97)} \\ 
93. gun & -1.476 & 6.81e+05 (72) & -0.367 (92) \\ 
\textcolor{lightgrey}{94. hate} & \textcolor{lightgrey}{-1.520} & \textcolor{lightgrey}{9.65e+06 (23)} & \textcolor{lightgrey}{-0.282 (84)} \\ 
95. depressed & -1.541 & 2.81e+05 (82) & -0.339 (89) \\ 
\textcolor{lightgrey}{96. hell} & \textcolor{lightgrey}{-1.551} & \textcolor{lightgrey}{6.27e+06 (30)} & \textcolor{lightgrey}{-0.250 (79)} \\ 
97. sick & -1.630 & 3.58e+06 (37) & -0.262 (80) \\ 
\textcolor{lightgrey}{98. headache} & \textcolor{lightgrey}{-1.881} & \textcolor{lightgrey}{8.57e+05 (69)} & \textcolor{lightgrey}{-0.437 (94)} \\ 
99. flu & -1.912 & 9.01e+05 (68) & -0.735 (99) \\ 
\textcolor{lightgrey}{100. war} & \textcolor{lightgrey}{-2.040} & \textcolor{lightgrey}{1.96e+06 (57)} & \textcolor{lightgrey}{-0.270 (82)} \\ 
\hline
\end{tabular}

  \plainlatexonly{\end{scriptsize}}
  \caption{
    The same selection of 100 keywords and text elements
    listed in the main text's Tab.~\ref{tab:twhap.timeseriesA},
    reordered by normalized happiness $\havgfnnorm$. 
  } 
  \label{tab:twhap.supp-timeseriesE}
\end{table*}

\clearpage

\begin{figure}[t]
\includegraphics[width=0.48\textwidth]{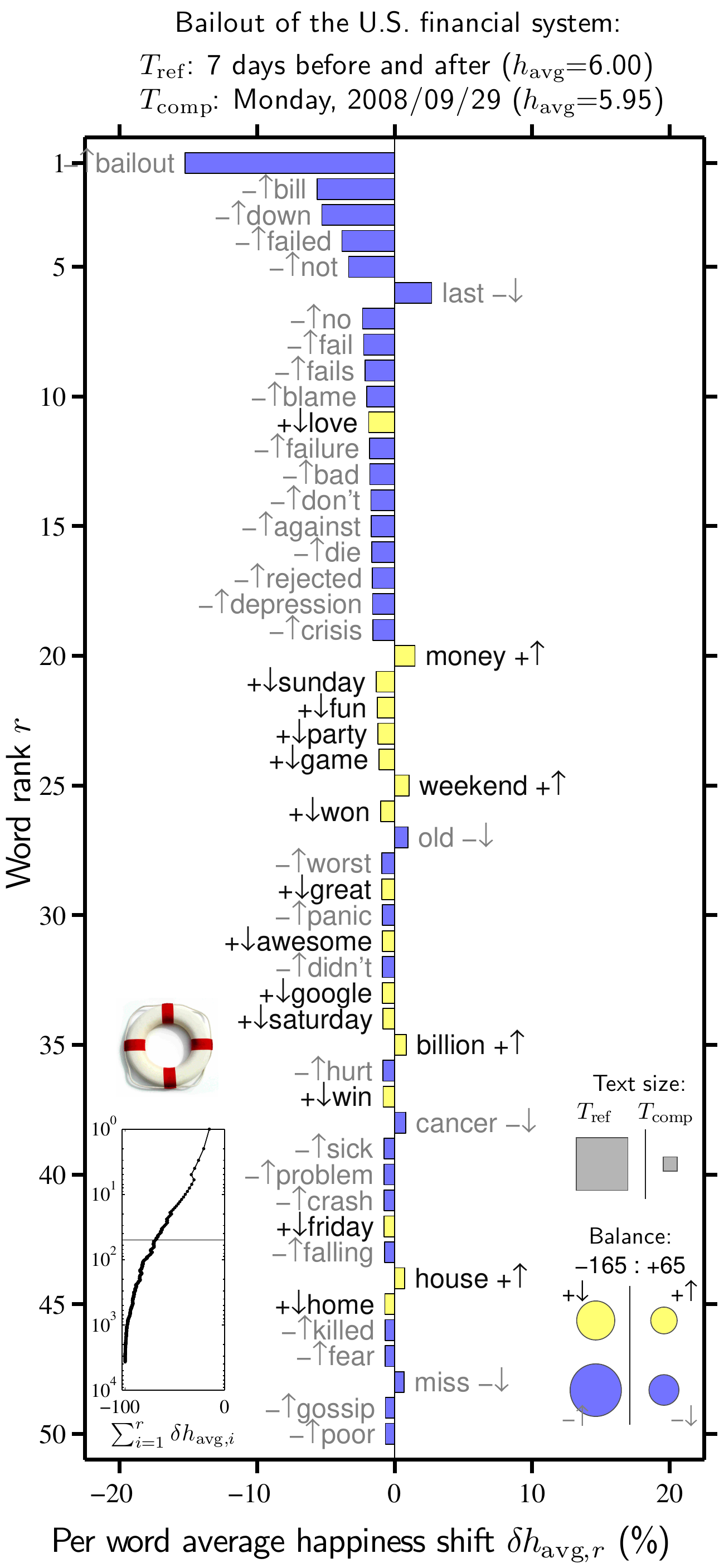}
\caption{
Word shift graph for Bailout of the U.S.~financial system, 2008/09/29, relative to 7 days before and 7 days after combined.
}
\label{fig:twhap.interestingdates-supp001}
\end{figure}

\begin{figure}[t]
\includegraphics[width=0.48\textwidth]{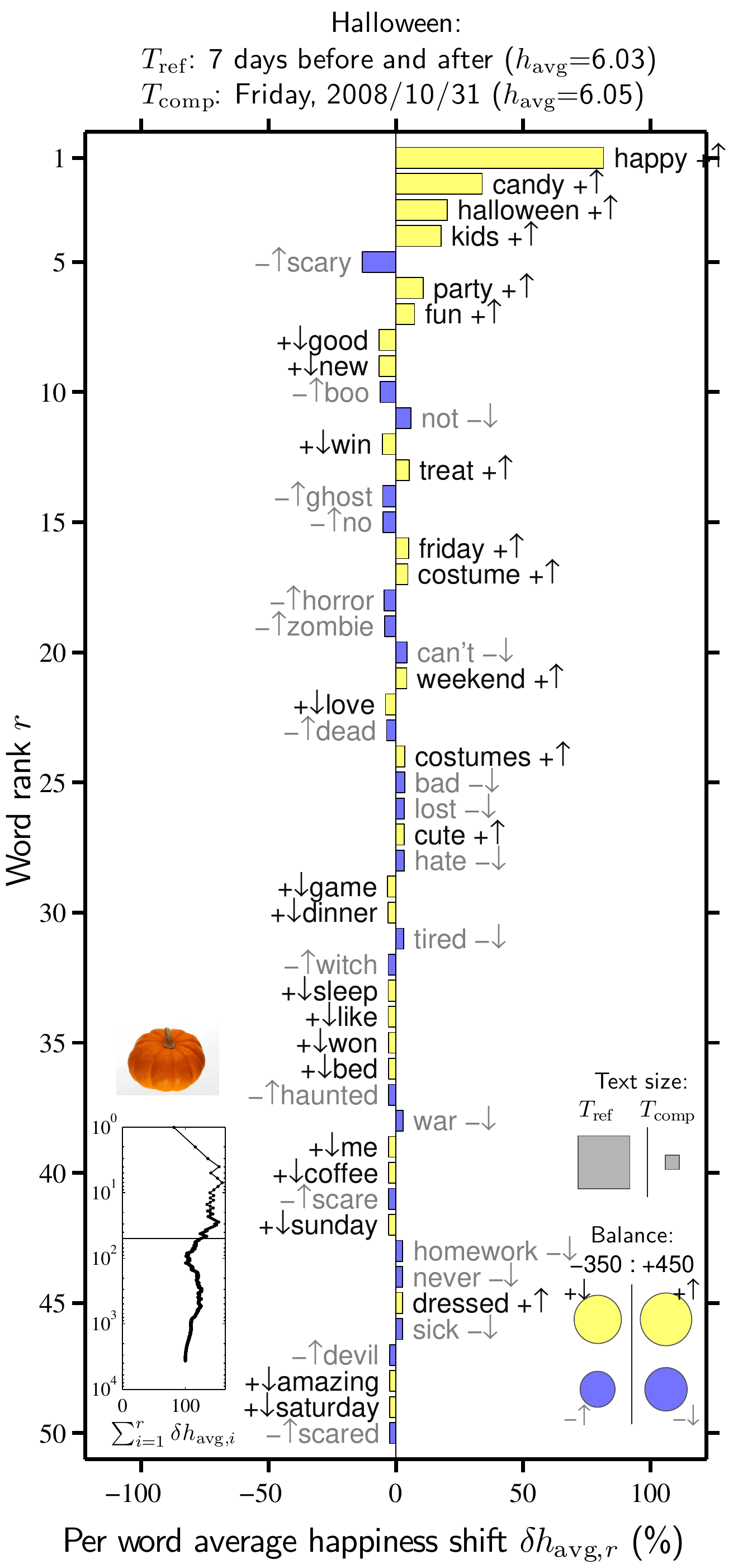}
\caption{
Word shift graph for Halloween, 2008/10/31, relative to 7 days before and 7 days after combined.
}
\label{fig:twhap.interestingdates-supp002}
\end{figure}

\clearpage
\begin{figure}[t]
\includegraphics[width=0.48\textwidth]{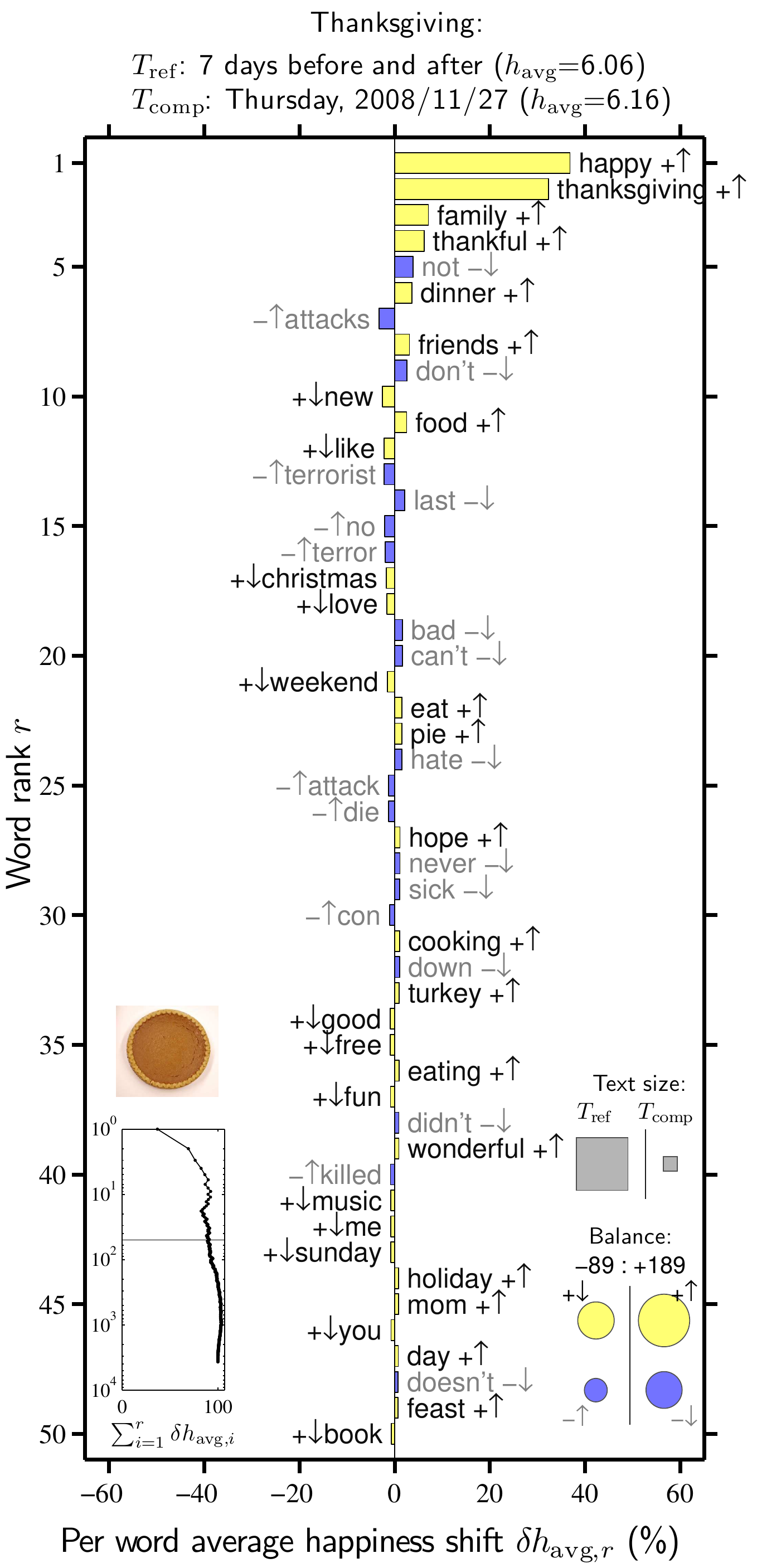}
\caption{
Word shift graph for Thanksgiving, 2008/11/27, relative to 7 days before and 7 days after combined.
}
\label{fig:twhap.interestingdates-supp003}
\end{figure}

\begin{figure}[t]
\includegraphics[width=0.48\textwidth]{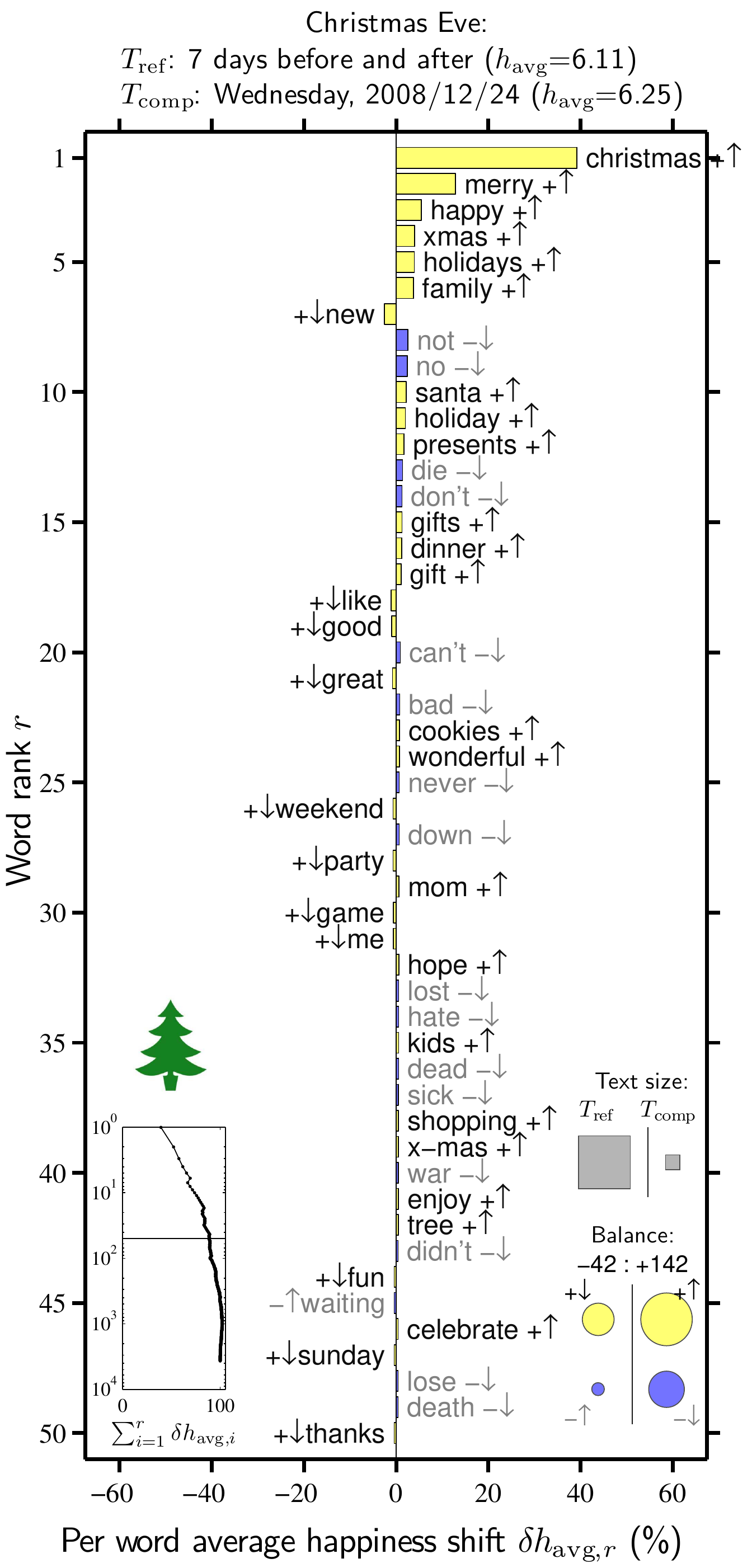}
\caption{
Word shift graph for Christmas Eve, 2008/12/24, relative to 7 days before and 7 days after combined.
}
\label{fig:twhap.interestingdates-supp004}
\end{figure}

\clearpage
\begin{figure}[t]
\includegraphics[width=0.48\textwidth]{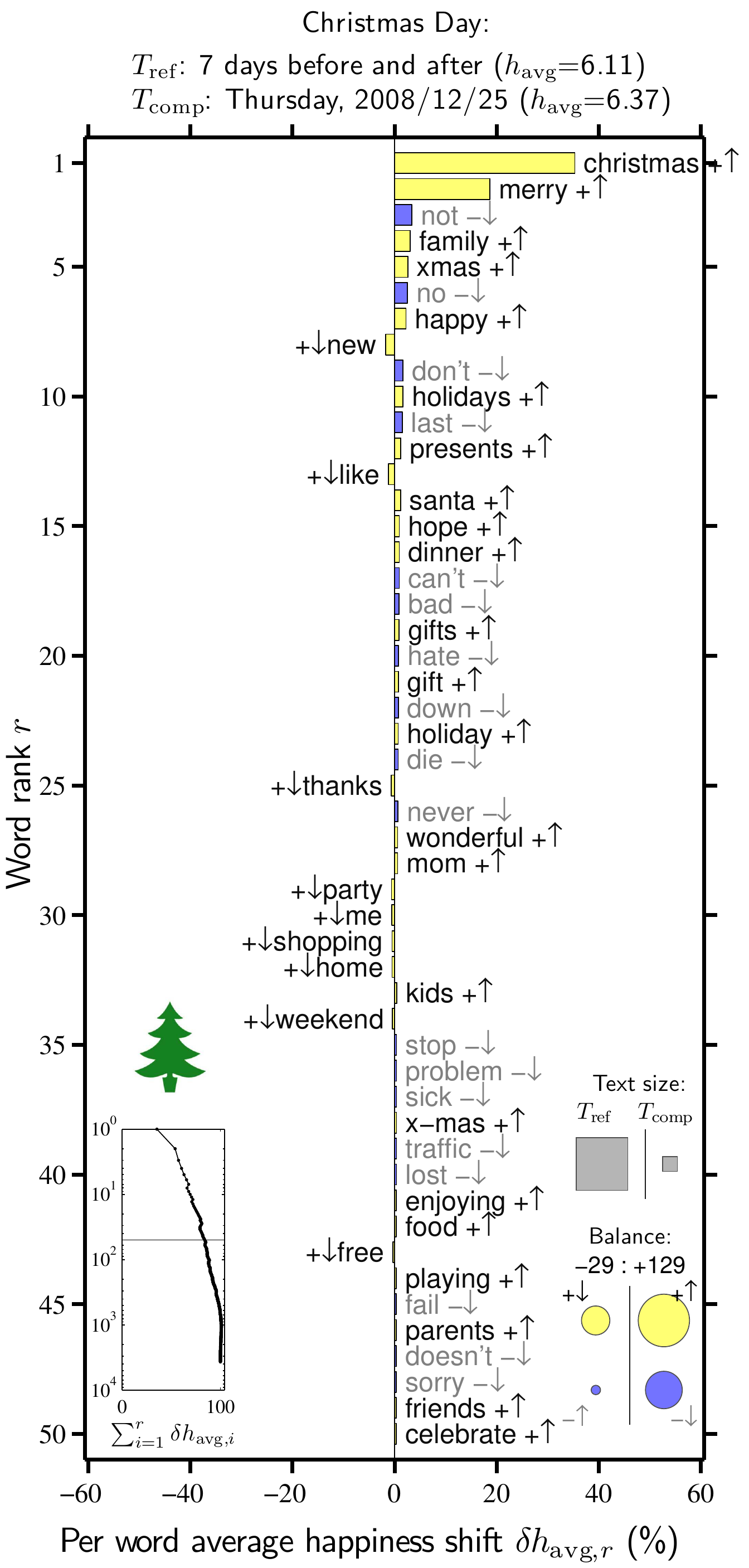}
\caption{
Word shift graph for Christmas Day, 2008/12/25, relative to 7 days before and 7 days after combined.
}
\label{fig:twhap.interestingdates-supp005}
\end{figure}

\begin{figure}[t]
\includegraphics[width=0.48\textwidth]{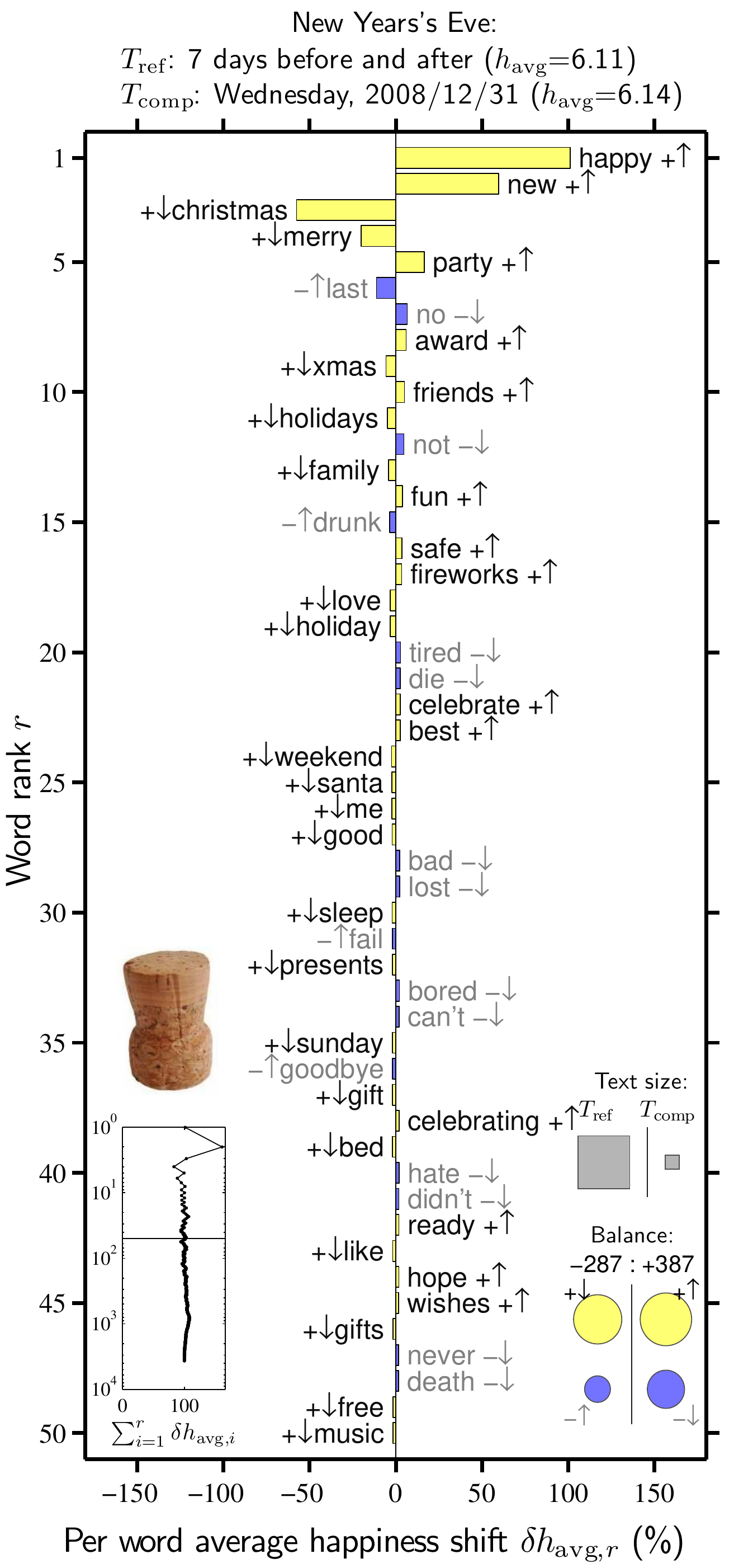}
\caption{
Word shift graph for New Years's Eve, 2008/12/31, relative to 7 days before and 7 days after combined.
}
\label{fig:twhap.interestingdates-supp006}
\end{figure}

\clearpage
\begin{figure}[t]
\includegraphics[width=0.48\textwidth]{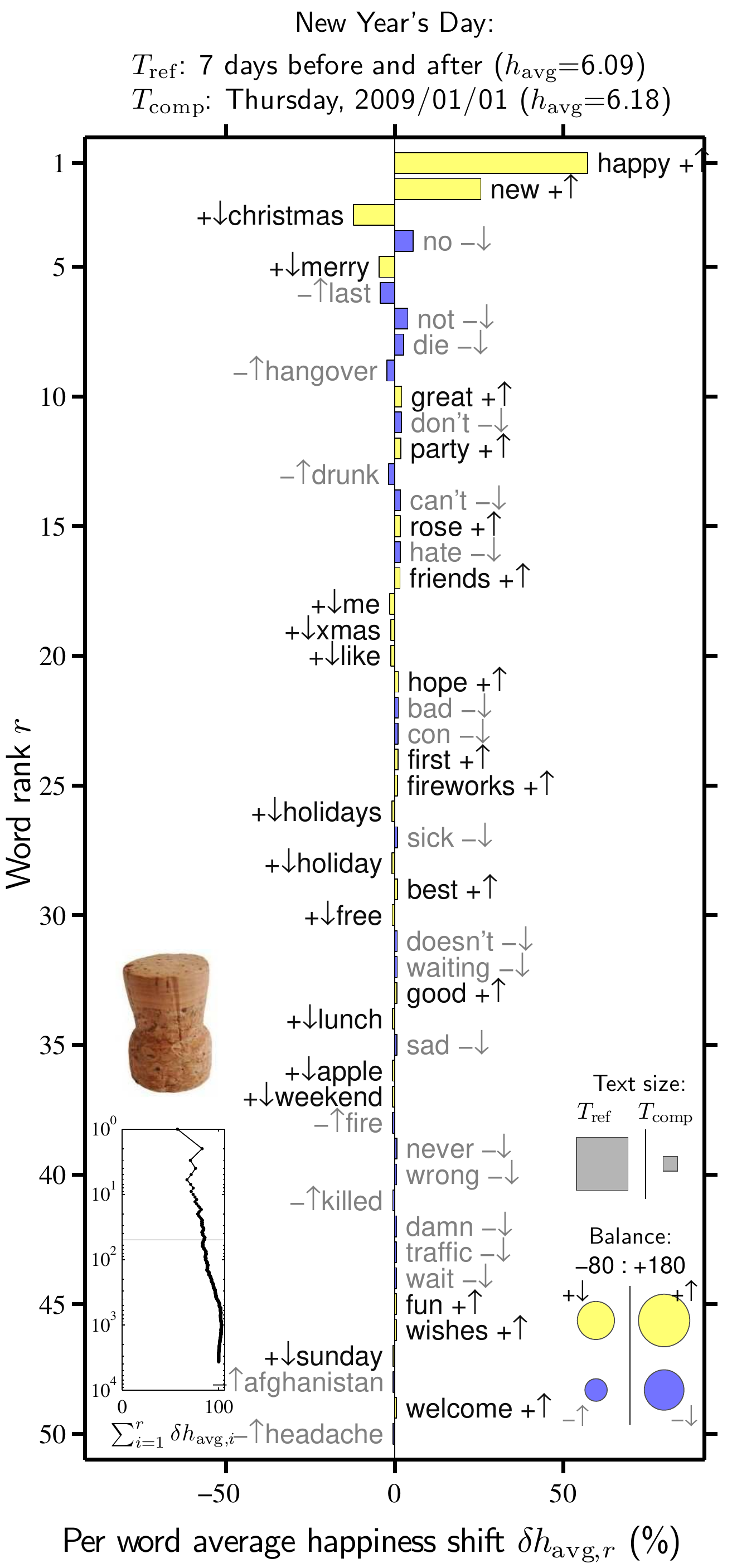}
\caption{
Word shift graph for New Year's Day, 2009/01/01, relative to 7 days before and 7 days after combined.
}
\label{fig:twhap.interestingdates-supp007}
\end{figure}

\begin{figure}[t]
\includegraphics[width=0.48\textwidth]{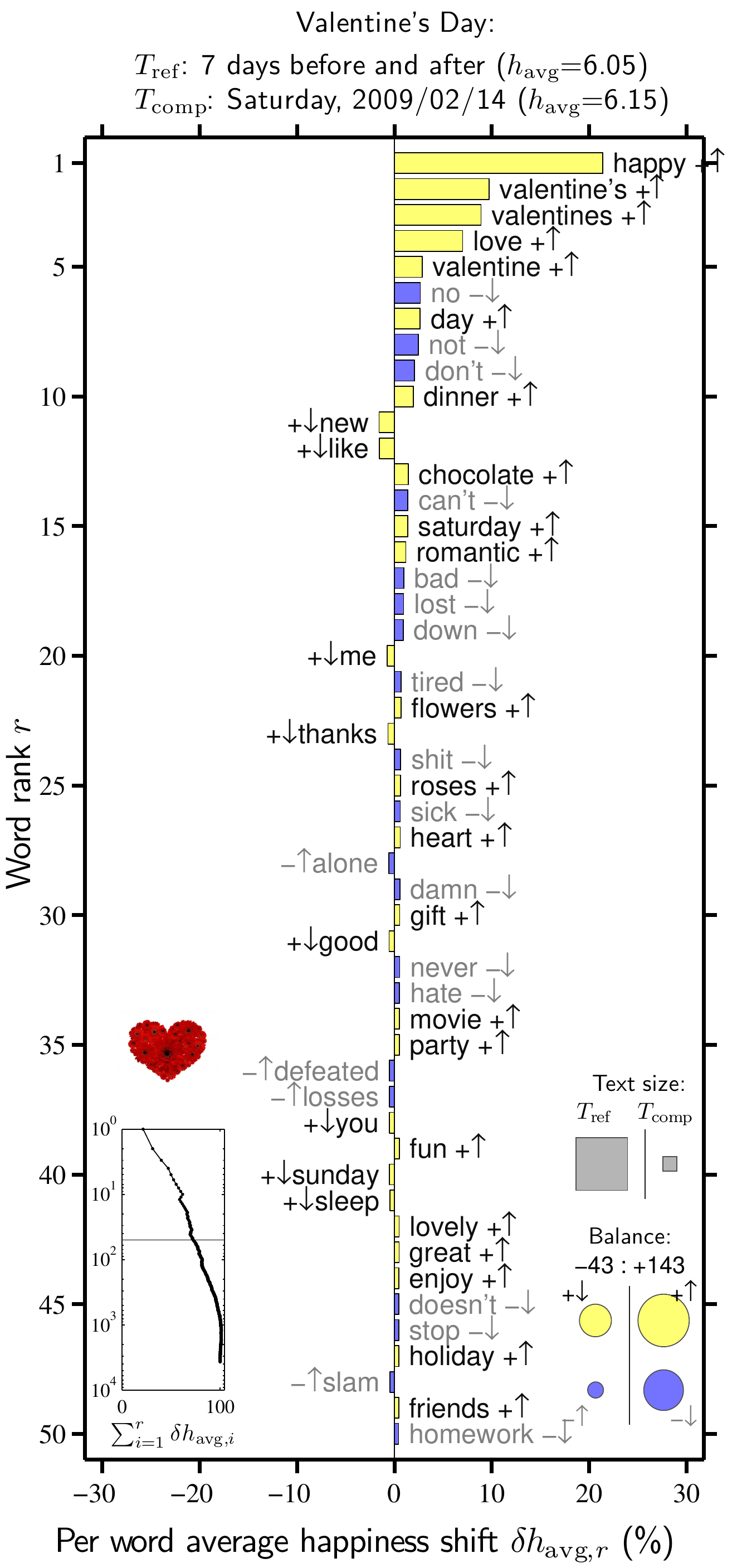}
\caption{
Word shift graph for Valentine's Day, 2009/02/14, relative to 7 days before and 7 days after combined.
}
\label{fig:twhap.interestingdates-supp008}
\end{figure}

\clearpage
\begin{figure}[t]
\includegraphics[width=0.48\textwidth]{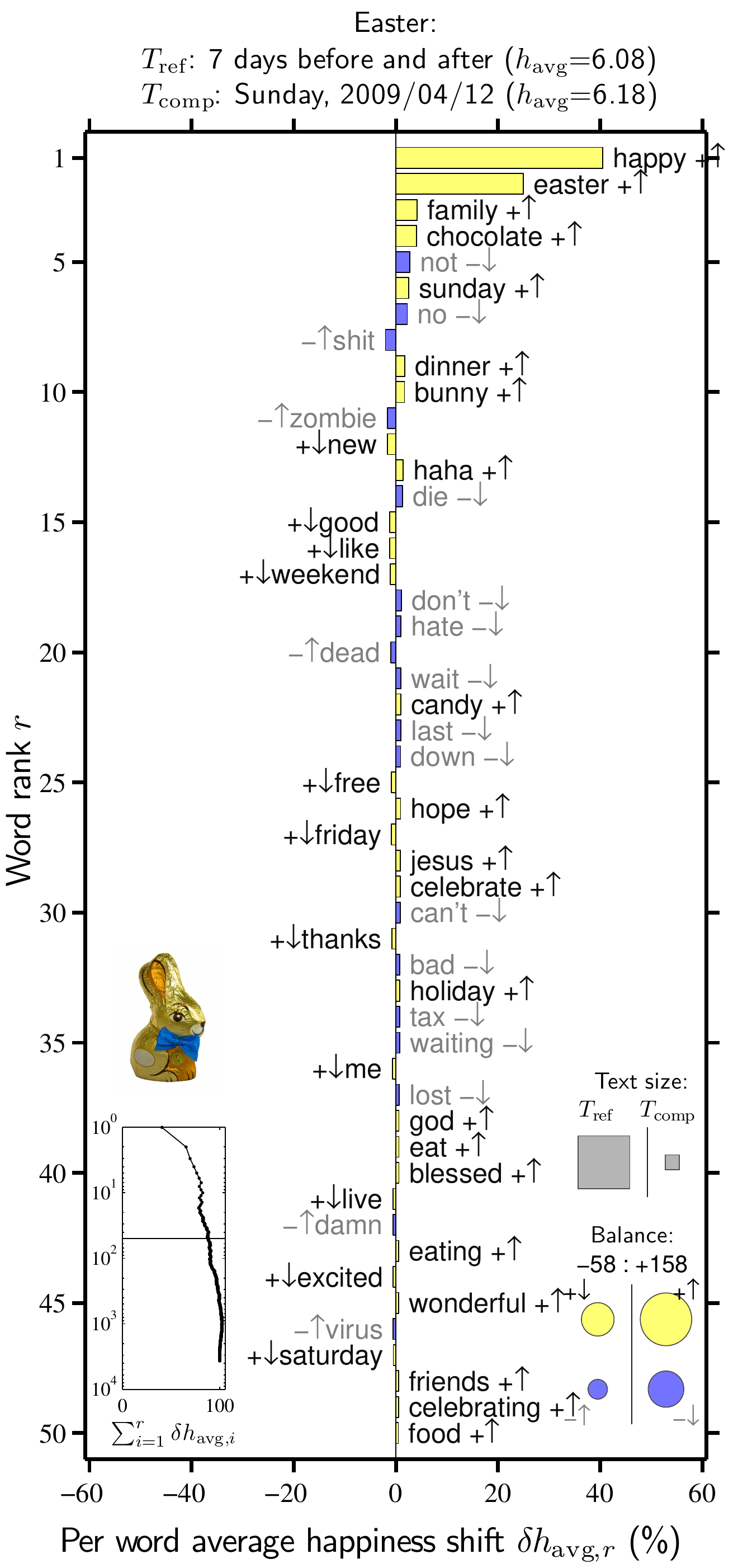}
\caption{
Word shift graph for Easter, 2009/04/12, relative to 7 days before and 7 days after combined.
}
\label{fig:twhap.interestingdates-supp009}
\end{figure}

\begin{figure}[t]
\includegraphics[width=0.48\textwidth]{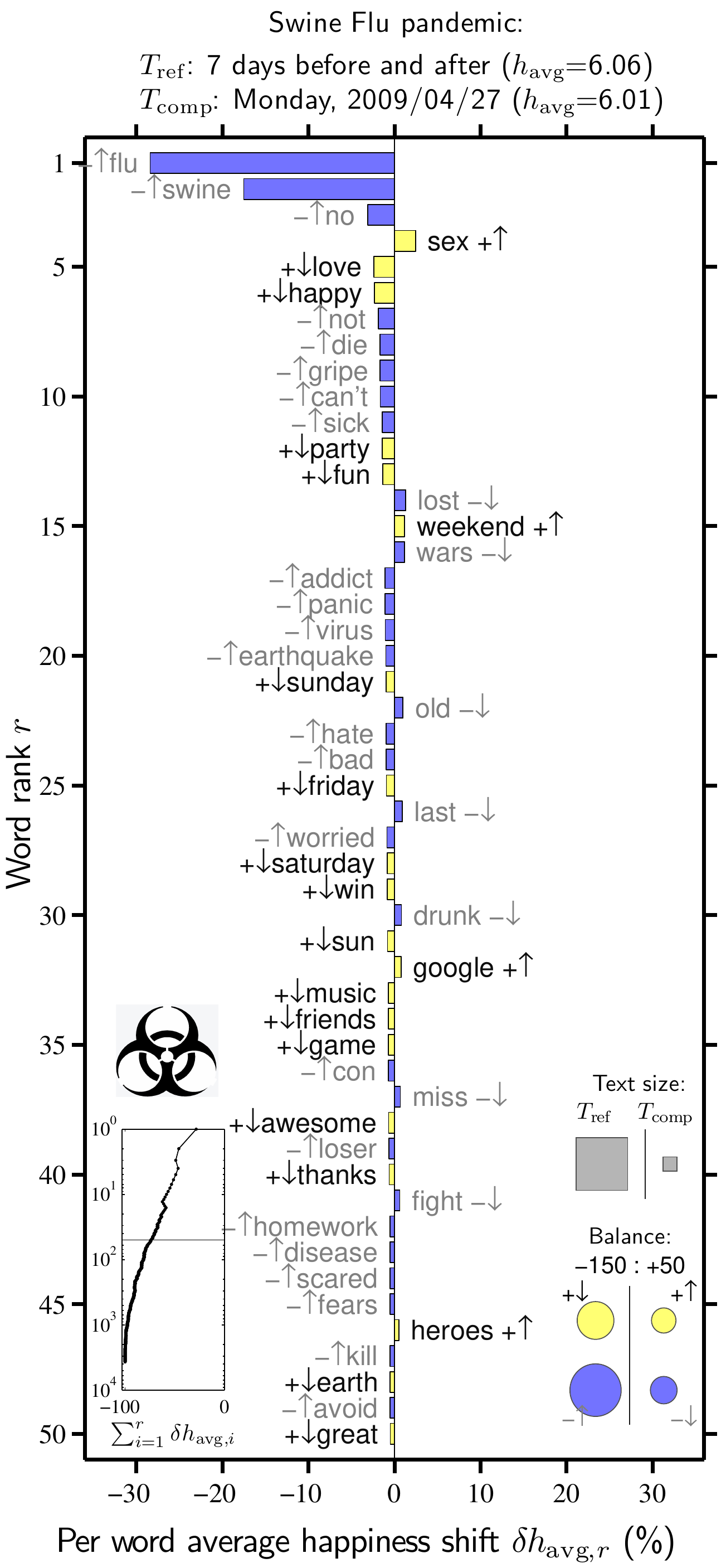}
\caption{
Word shift graph for Swine Flu pandemic, 2009/04/27, relative to 7 days before and 7 days after combined.
}
\label{fig:twhap.interestingdates-supp010}
\end{figure}

\clearpage
\begin{figure}[t]
\includegraphics[width=0.48\textwidth]{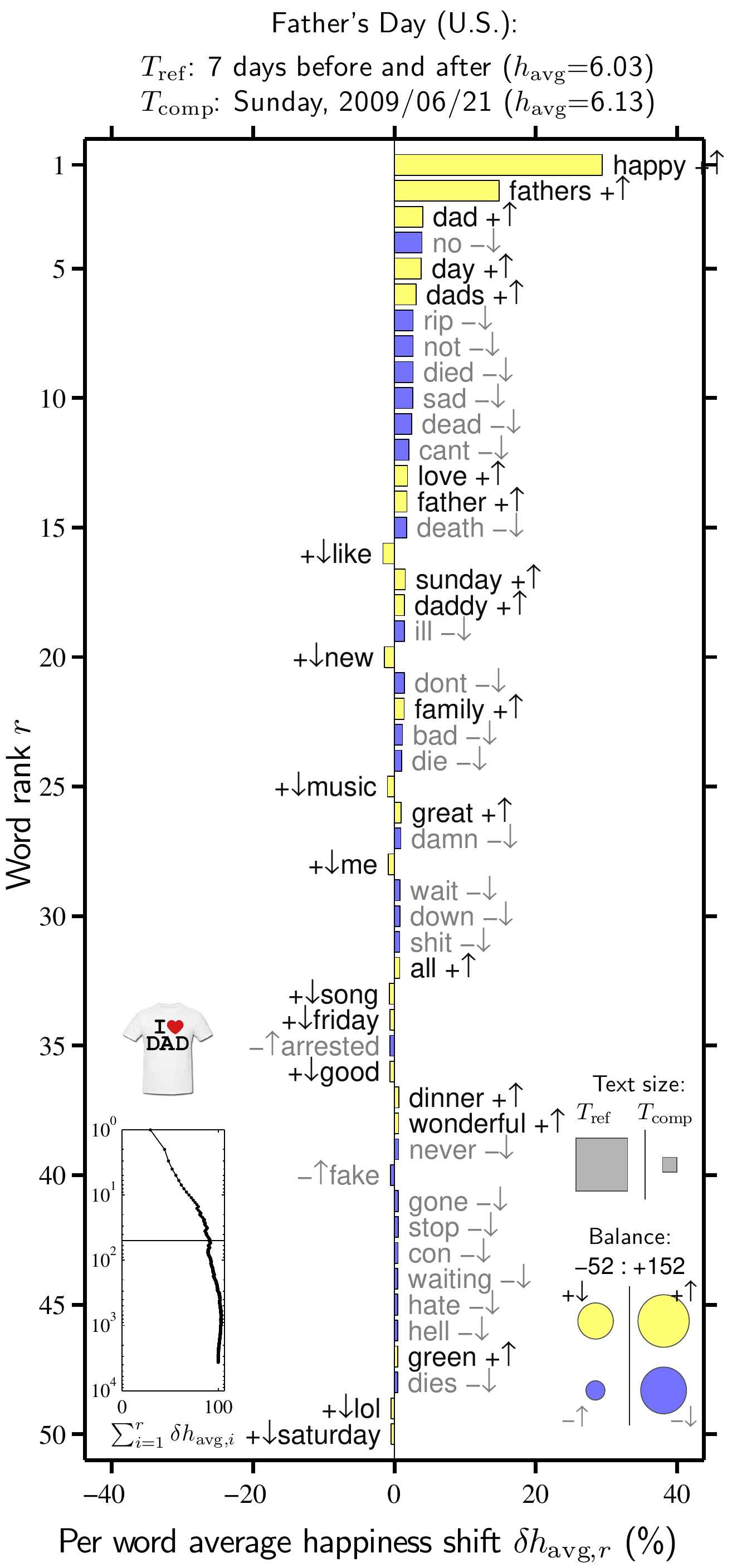}
\caption{
Word shift graph for Father's Day (U.S.), 2009/06/21, relative to 7 days before and 7 days after combined.
}
\label{fig:twhap.interestingdates-supp011}
\end{figure}

\begin{figure}[t]
\includegraphics[width=0.48\textwidth]{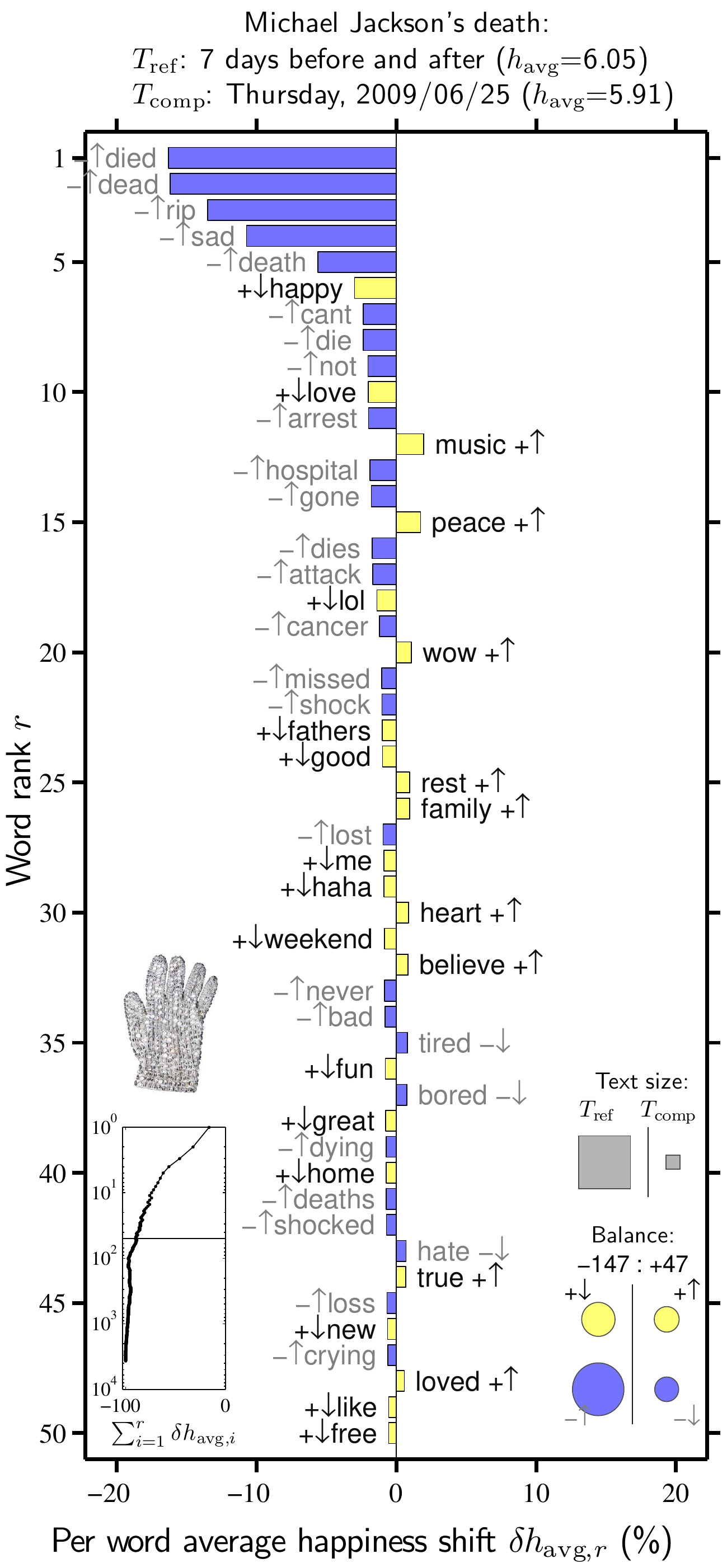}
\caption{
Word shift graph for Michael Jackson's death, 2009/06/25, relative to 7 days before and 7 days after combined.
}
\label{fig:twhap.interestingdates-supp012}
\end{figure}

\clearpage
\begin{figure}[t]
\includegraphics[width=0.48\textwidth]{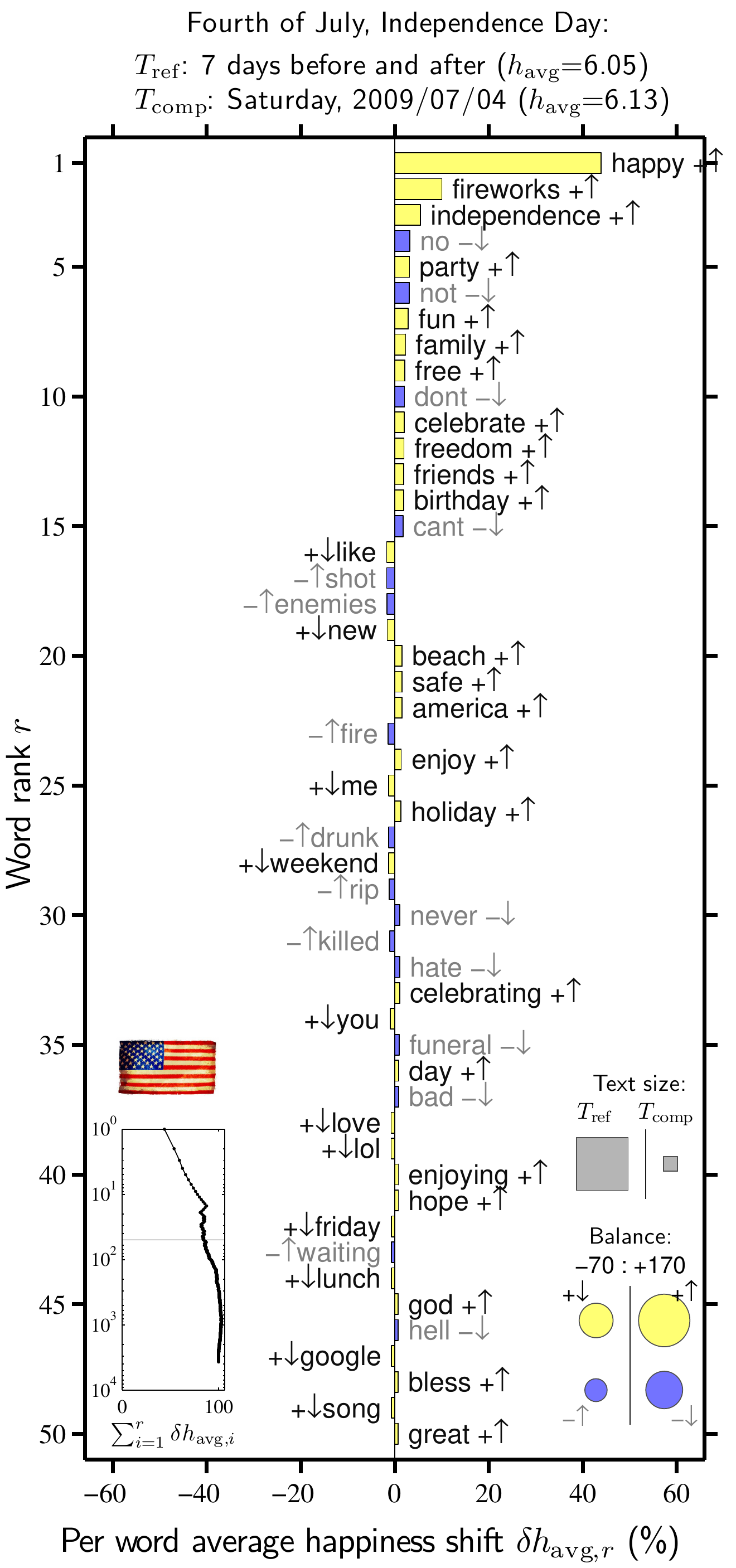}
\caption{
Word shift graph for Fourth of July, Independence Day, 2009/07/04, relative to 7 days before and 7 days after combined.
}
\label{fig:twhap.interestingdates-supp013}
\end{figure}

\begin{figure}[t]
\includegraphics[width=0.48\textwidth]{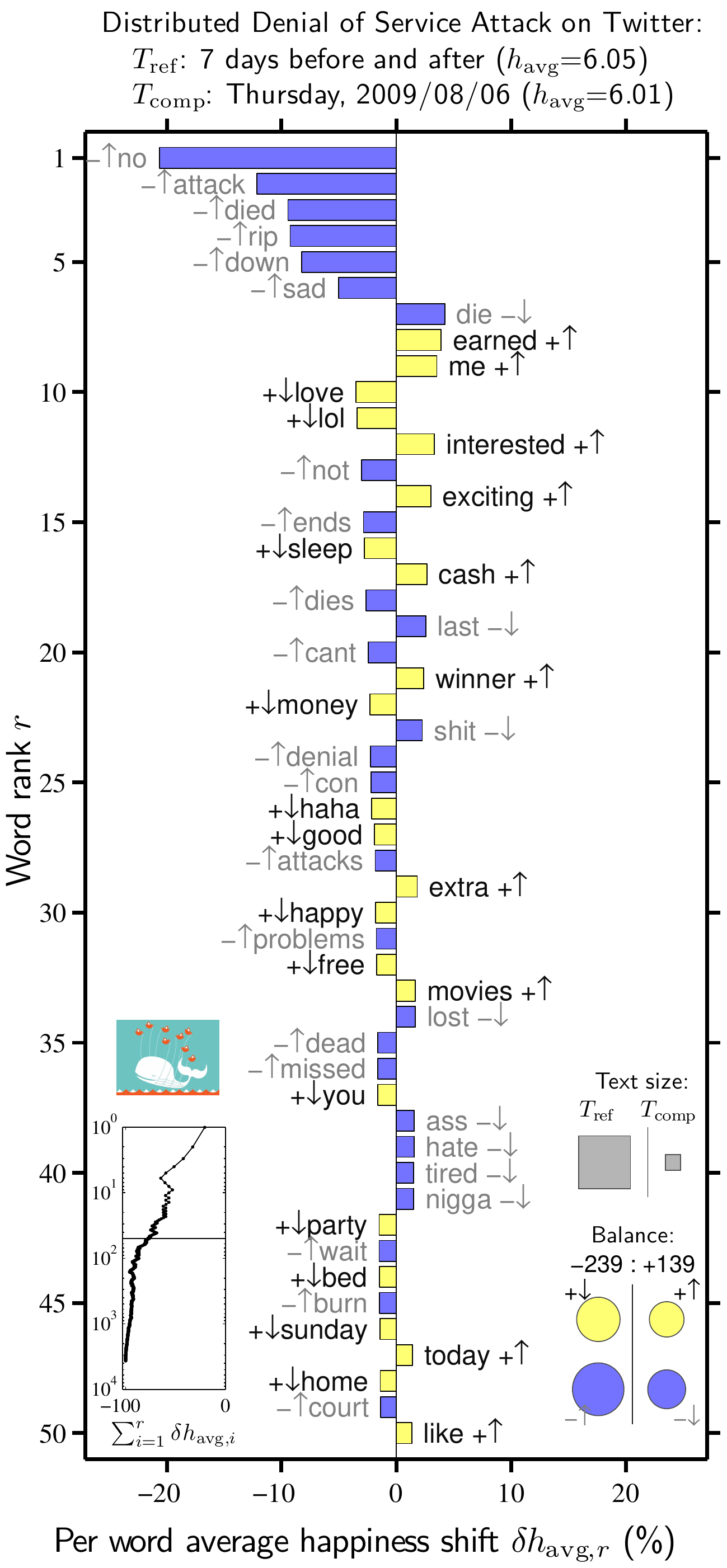}
\caption{
Word shift graph for Distributed Denial of Service Attack on Twitter, 2009/08/06, relative to 7 days before and 7 days after combined.
}
\label{fig:twhap.interestingdates-supp014}
\end{figure}

\clearpage
\begin{figure}[t]
\includegraphics[width=0.48\textwidth]{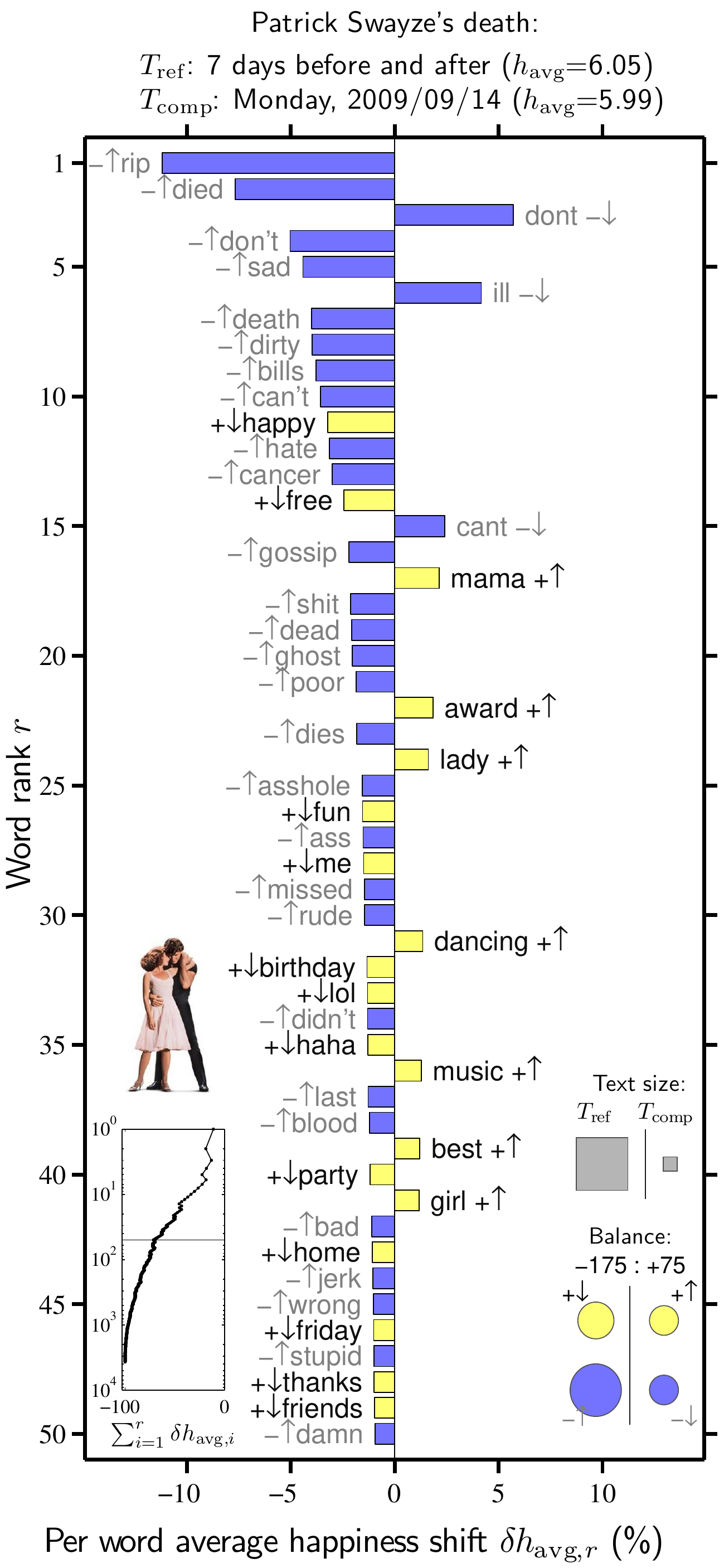}
\caption{
Word shift graph for Patrick Swayze's death, 2009/09/14, relative to 7 days before and 7 days after combined.
}
\label{fig:twhap.interestingdates-supp015}
\end{figure}

\begin{figure}[t]
\includegraphics[width=0.48\textwidth]{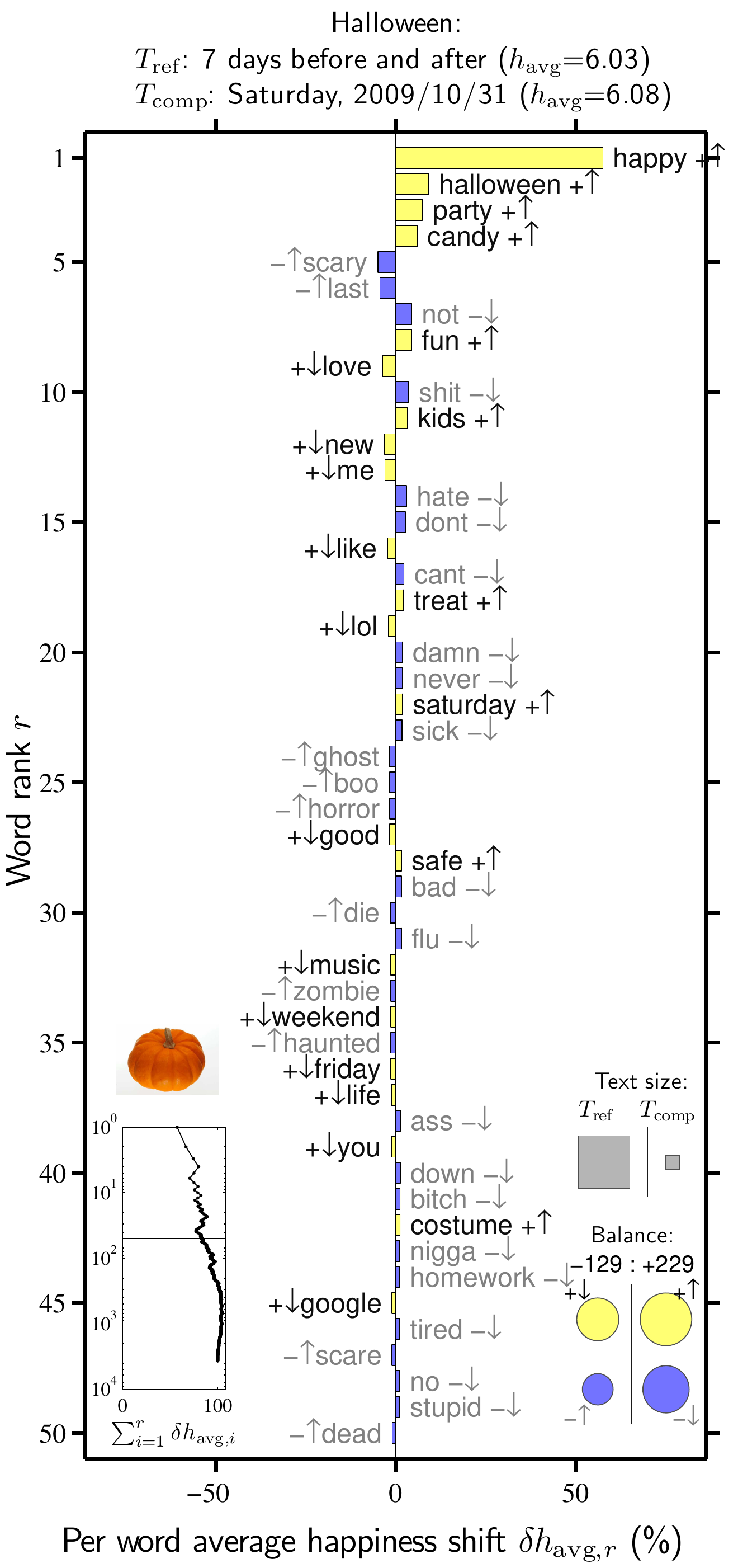}
\caption{
Word shift graph for Halloween, 2009/10/31, relative to 7 days before and 7 days after combined.
}
\label{fig:twhap.interestingdates-supp016}
\end{figure}

\clearpage
\begin{figure}[t]
\includegraphics[width=0.48\textwidth]{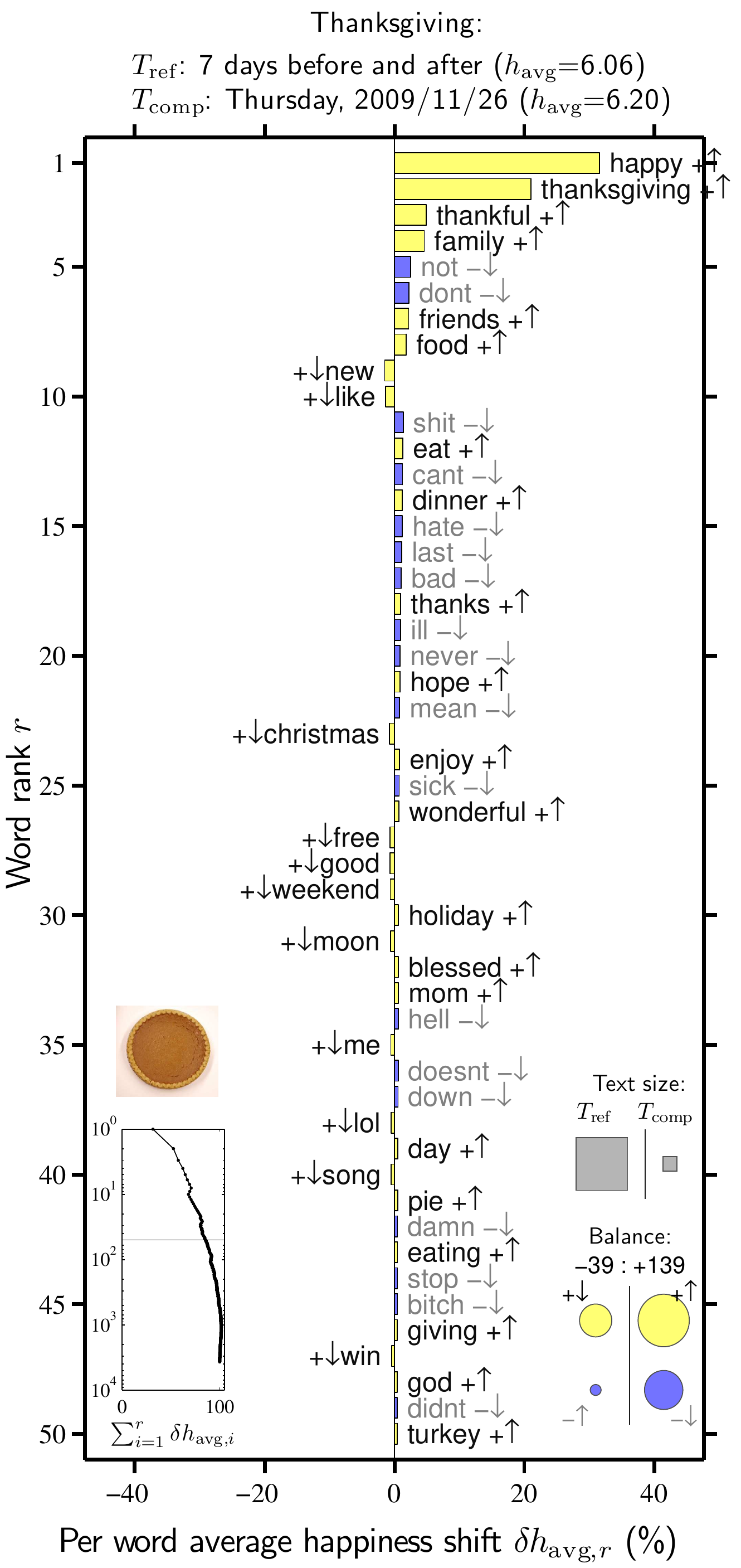}
\caption{
Word shift graph for Thanksgiving, 2009/11/26, relative to 7 days before and 7 days after combined.
}
\label{fig:twhap.interestingdates-supp017}
\end{figure}

\begin{figure}[t]
\includegraphics[width=0.48\textwidth]{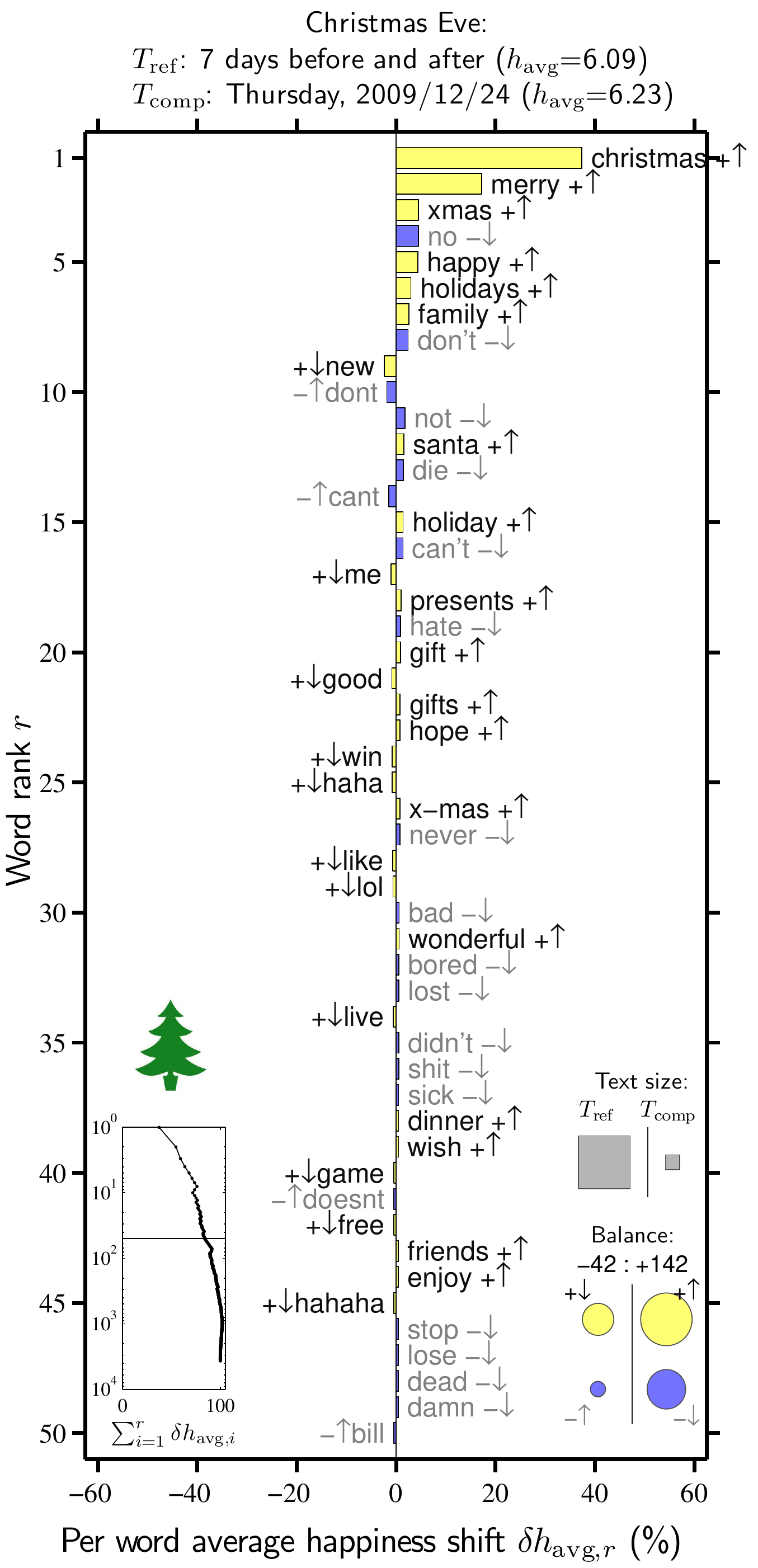}
\caption{
Word shift graph for Christmas Eve, 2009/12/24, relative to 7 days before and 7 days after combined.
}
\label{fig:twhap.interestingdates-supp018}
\end{figure}

\clearpage
\begin{figure}[t]
\includegraphics[width=0.48\textwidth]{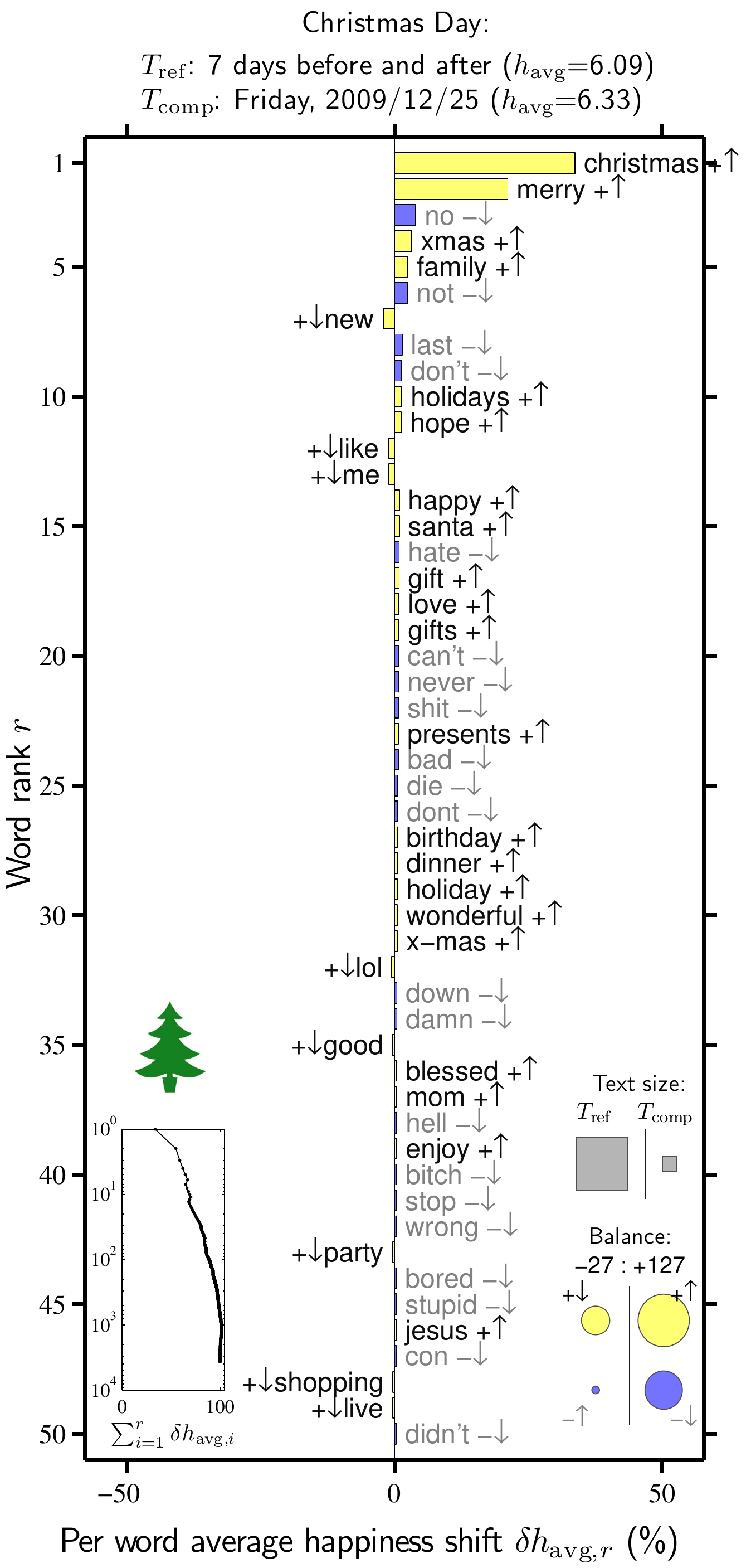}
\caption{
Word shift graph for Christmas Day, 2009/12/25, relative to 7 days before and 7 days after combined.
}
\label{fig:twhap.interestingdates-supp019}
\end{figure}

\begin{figure}[t]
\includegraphics[width=0.48\textwidth]{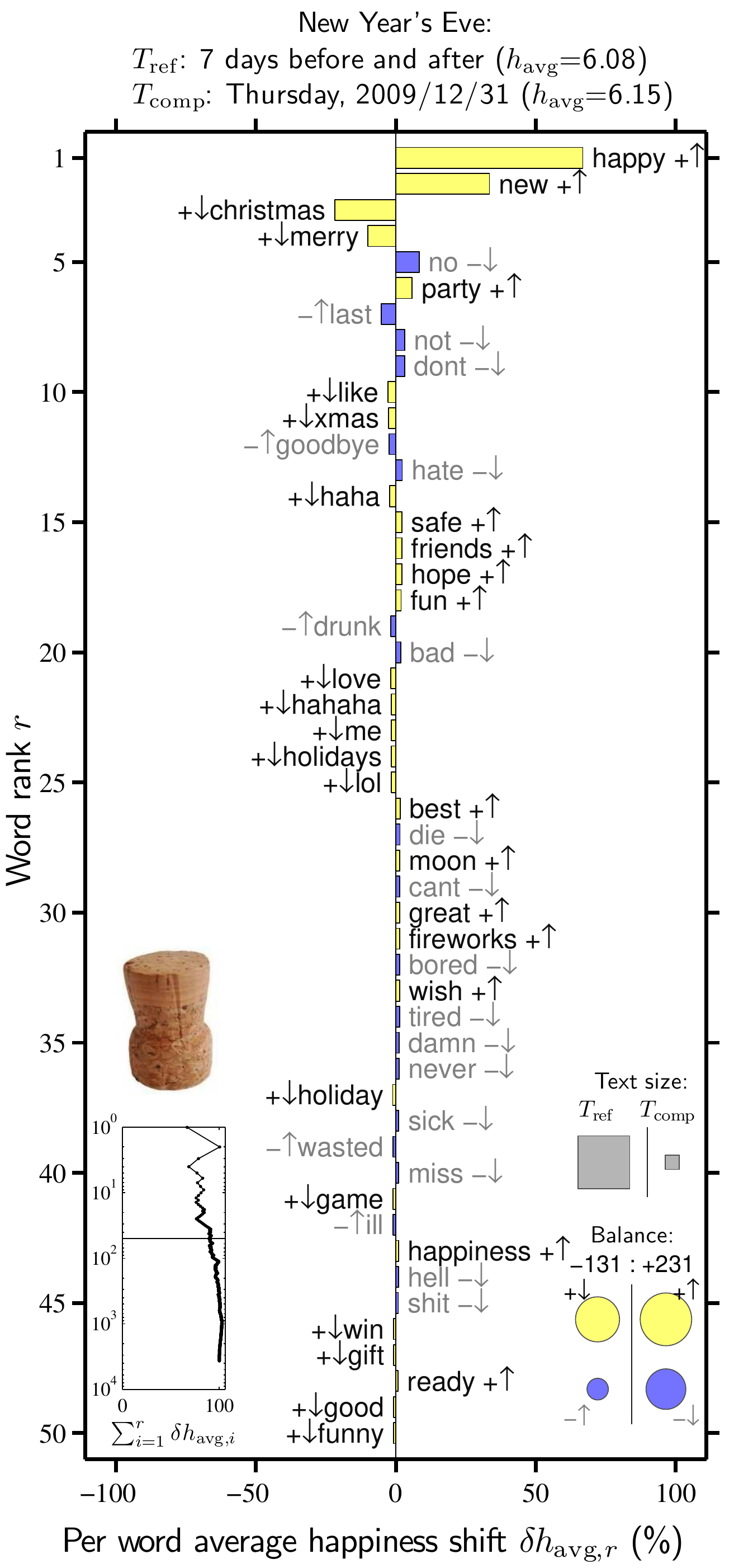}
\caption{
Word shift graph for New Year's Eve, 2009/12/31, relative to 7 days before and 7 days after combined.
}
\label{fig:twhap.interestingdates-supp020}
\end{figure}

\clearpage
\begin{figure}[t]
\includegraphics[width=0.48\textwidth]{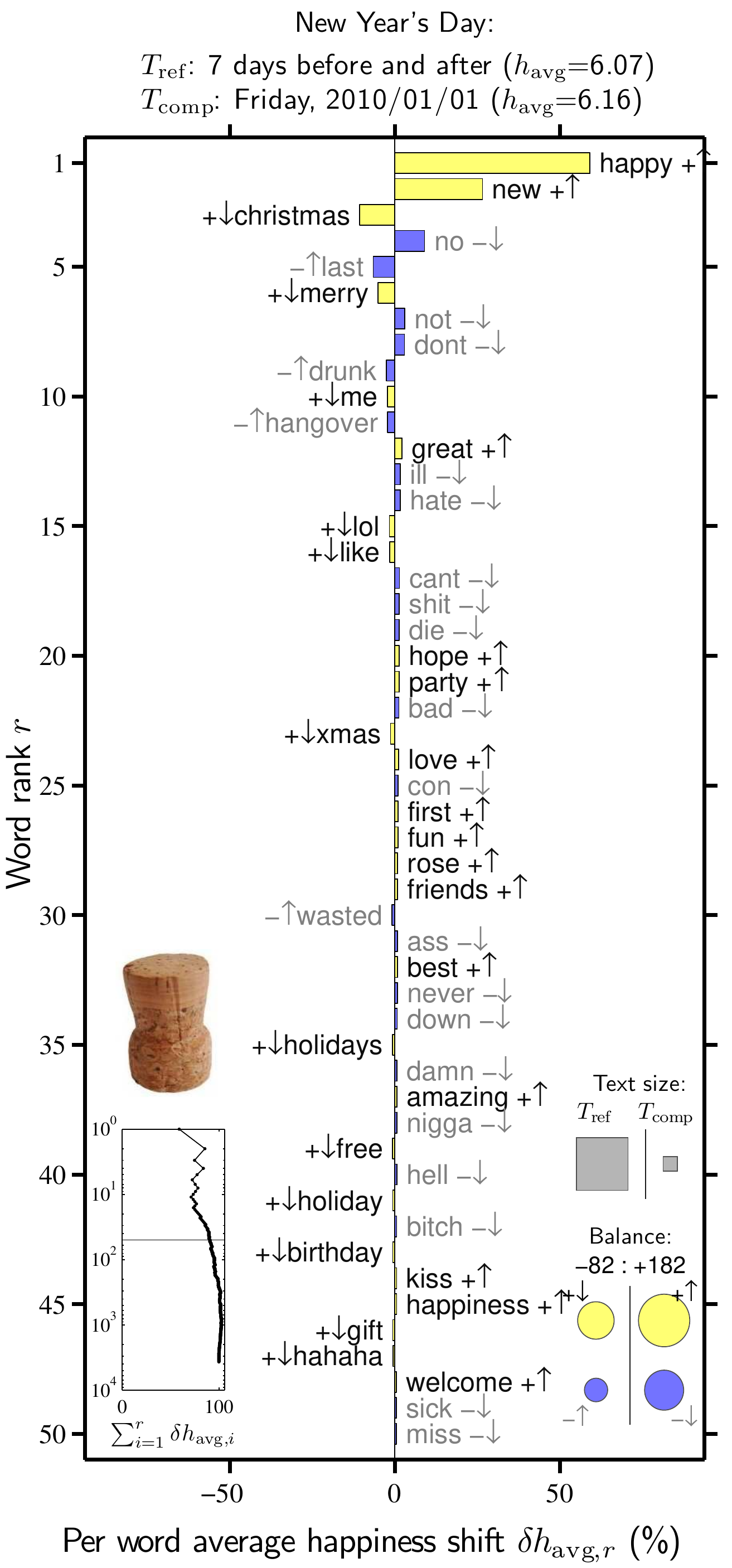}
\caption{
Word shift graph for New Year's Day, 2010/01/01, relative to 7 days before and 7 days after combined.
}
\label{fig:twhap.interestingdates-supp021}
\end{figure}

\begin{figure}[t]
\includegraphics[width=0.48\textwidth]{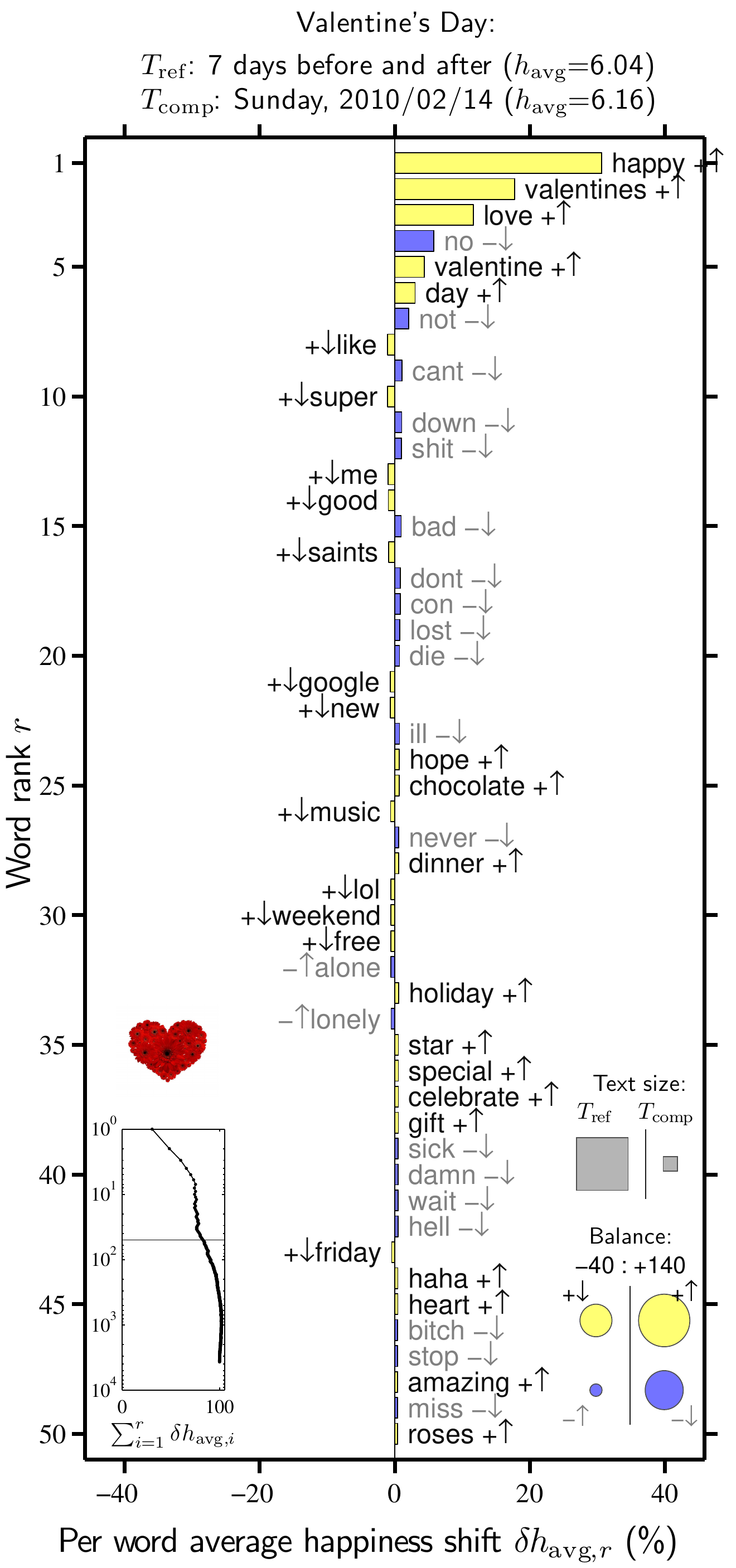}
\caption{
Word shift graph for Valentine's Day, 2010/02/14, relative to 7 days before and 7 days after combined.
}
\label{fig:twhap.interestingdates-supp022}
\end{figure}

\clearpage
\begin{figure}[t]
\includegraphics[width=0.48\textwidth]{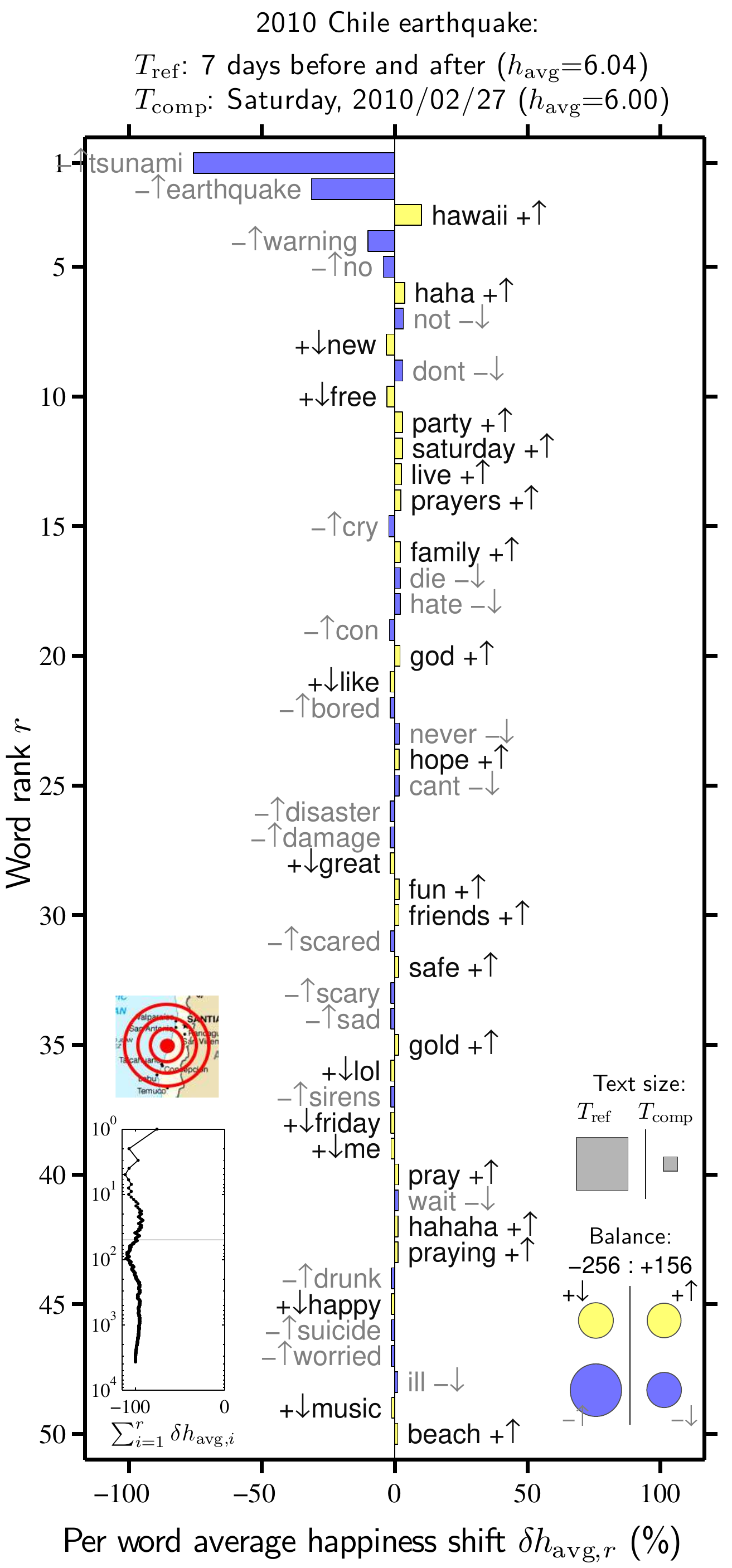}
\caption{
Word shift graph for 2010 Chile earthquake, 2010/02/27, relative to 7 days before and 7 days after combined.
}
\label{fig:twhap.interestingdates-supp023}
\end{figure}

\begin{figure}[t]
\includegraphics[width=0.48\textwidth]{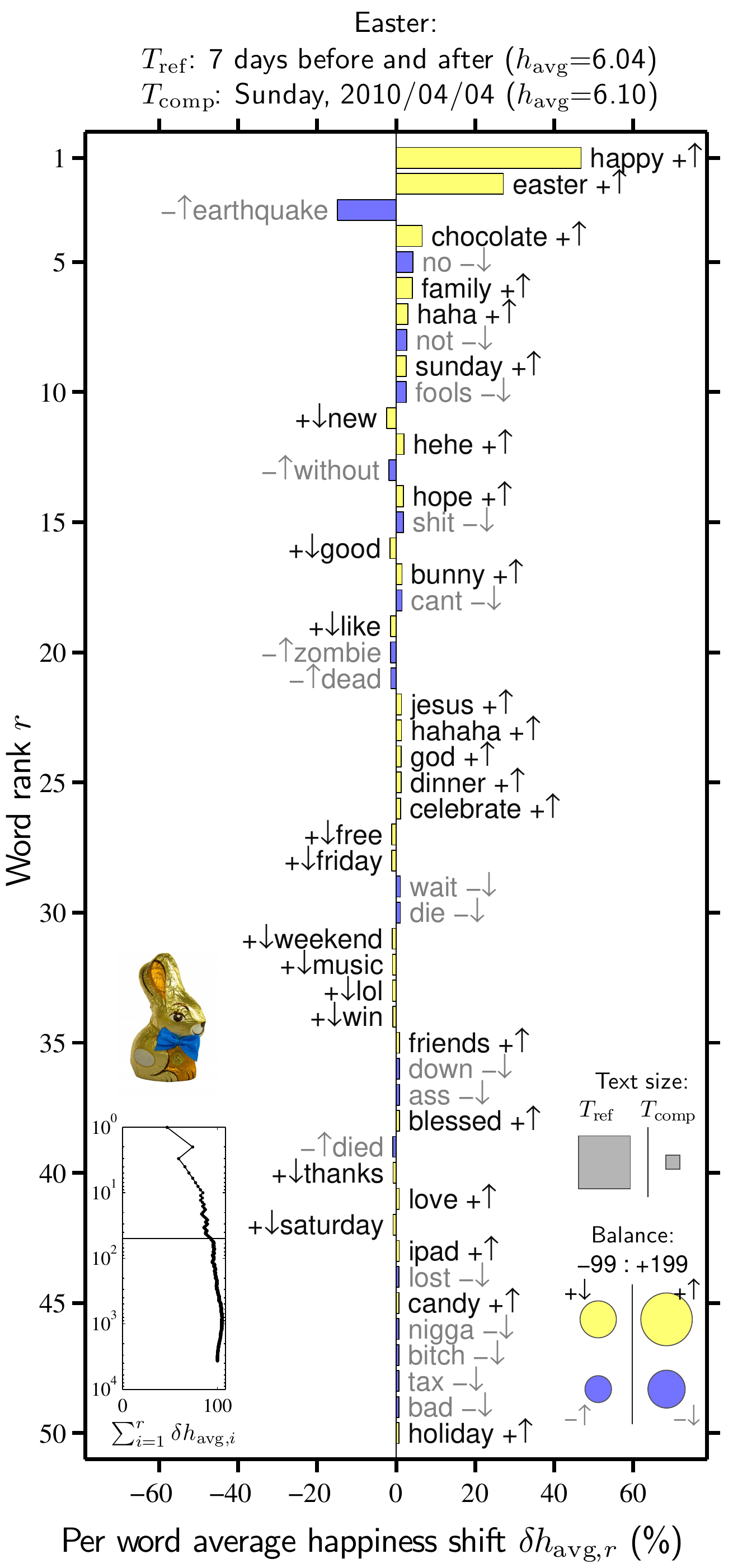}
\caption{
Word shift graph for Easter, 2010/04/04, relative to 7 days before and 7 days after combined.
}
\label{fig:twhap.interestingdates-supp024}
\end{figure}

\clearpage
\begin{figure}[t]
\includegraphics[width=0.48\textwidth]{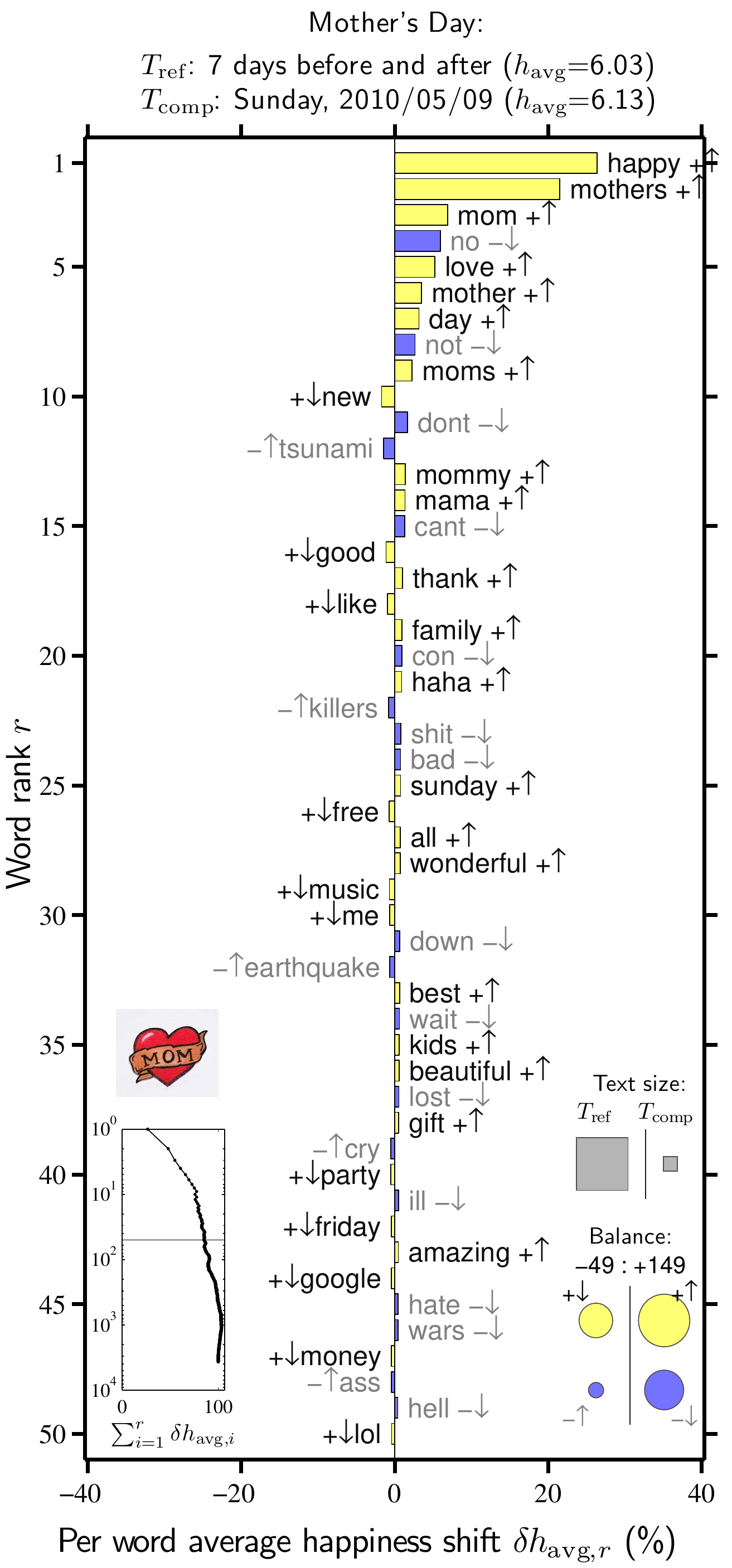}
\caption{
Word shift graph for Mother's Day, 2010/05/09, relative to 7 days before and 7 days after combined.
}
\label{fig:twhap.interestingdates-supp025}
\end{figure}

\begin{figure}[t]
\includegraphics[width=0.48\textwidth]{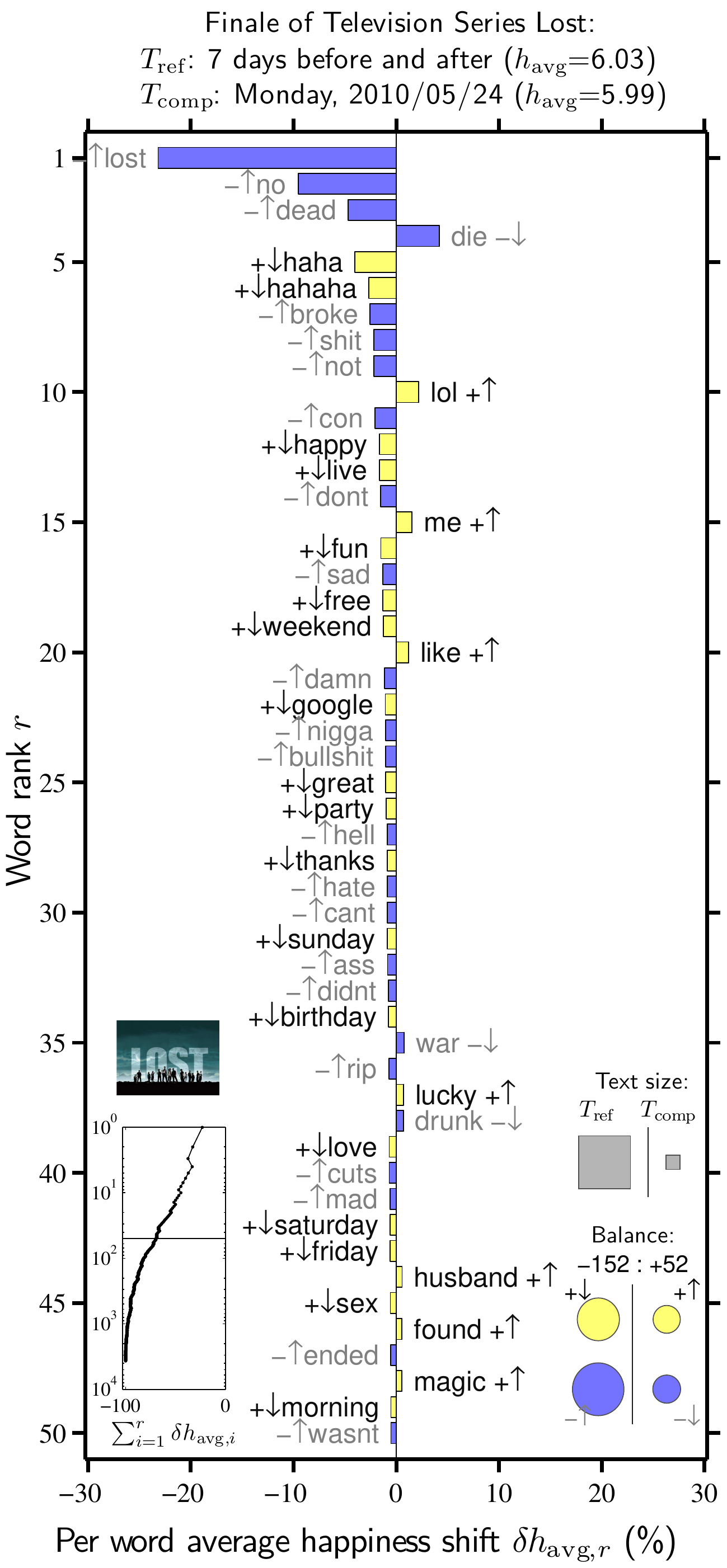}
\caption{
Word shift graph for Finale of Television Series Lost, 2010/05/24, relative to 7 days before and 7 days after combined.
}
\label{fig:twhap.interestingdates-supp026}
\end{figure}

\clearpage
\begin{figure}[t]
\includegraphics[width=0.48\textwidth]{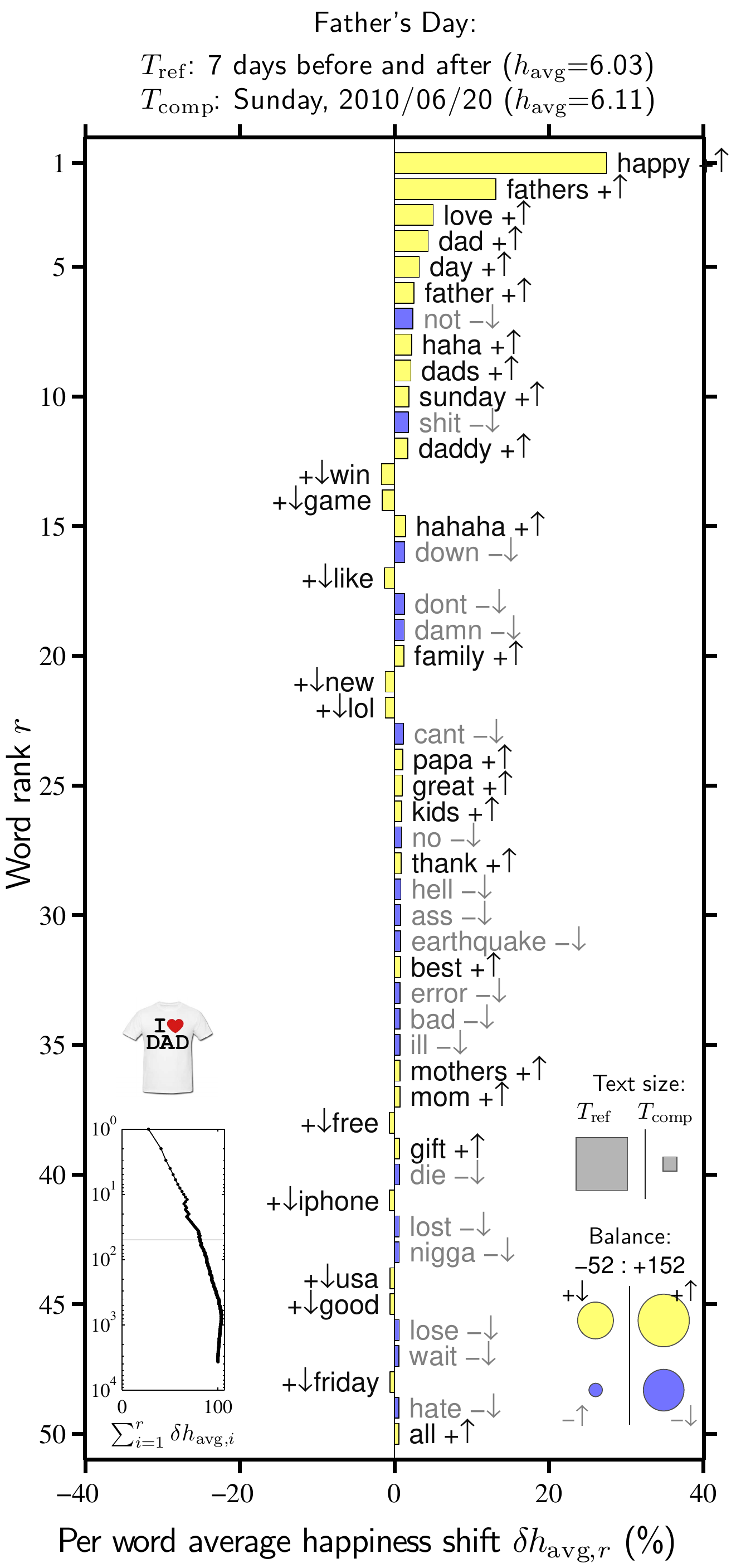}
\caption{
Word shift graph for Father's Day, 2010/06/20, relative to 7 days before and 7 days after combined.
}
\label{fig:twhap.interestingdates-supp027}
\end{figure}

\begin{figure}[t]
\includegraphics[width=0.48\textwidth]{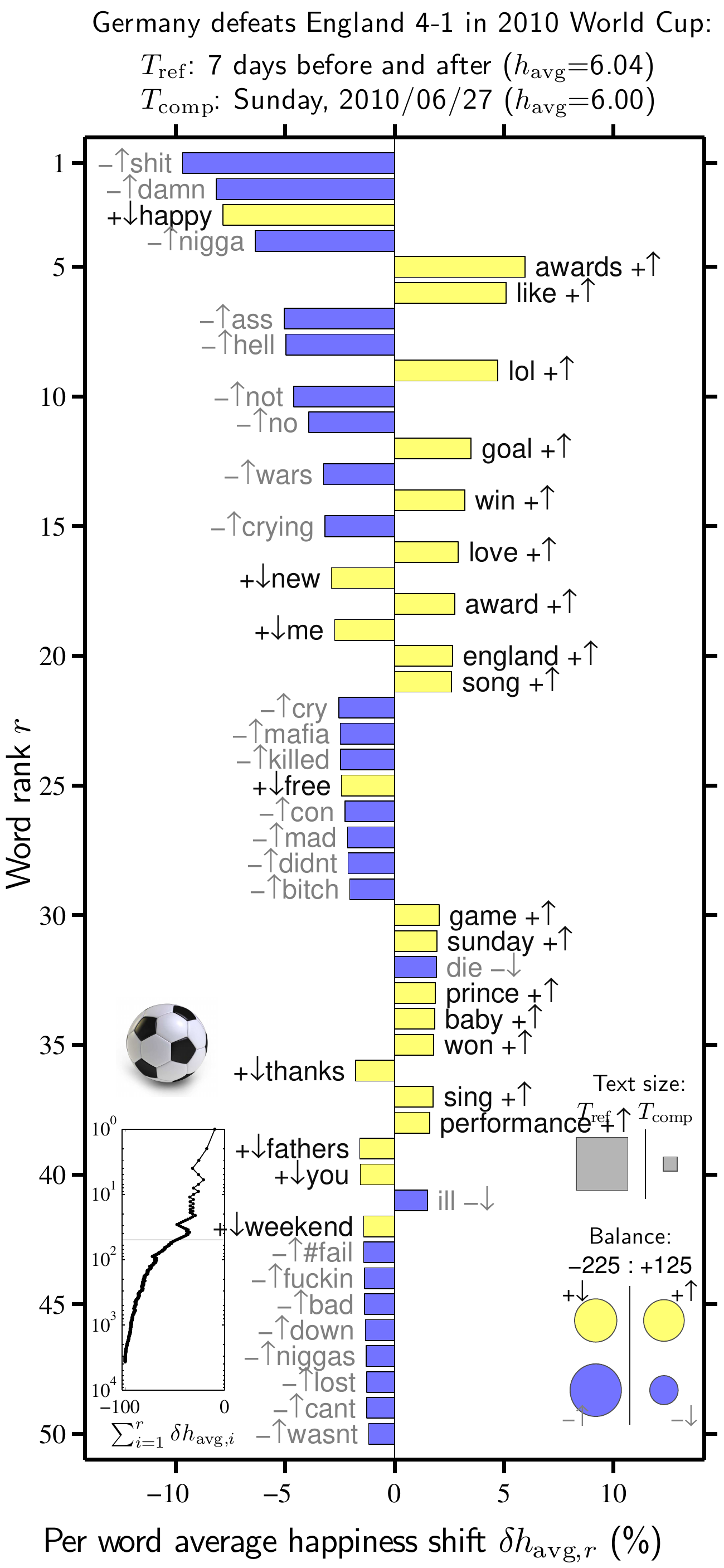}
\caption{
Word shift graph for Germany defeats England 4-1 in 2010 World Cup, 2010/06/27, relative to 7 days before and 7 days after combined.
}
\label{fig:twhap.interestingdates-supp028}
\end{figure}

\clearpage
\begin{figure}[t]
\includegraphics[width=0.48\textwidth]{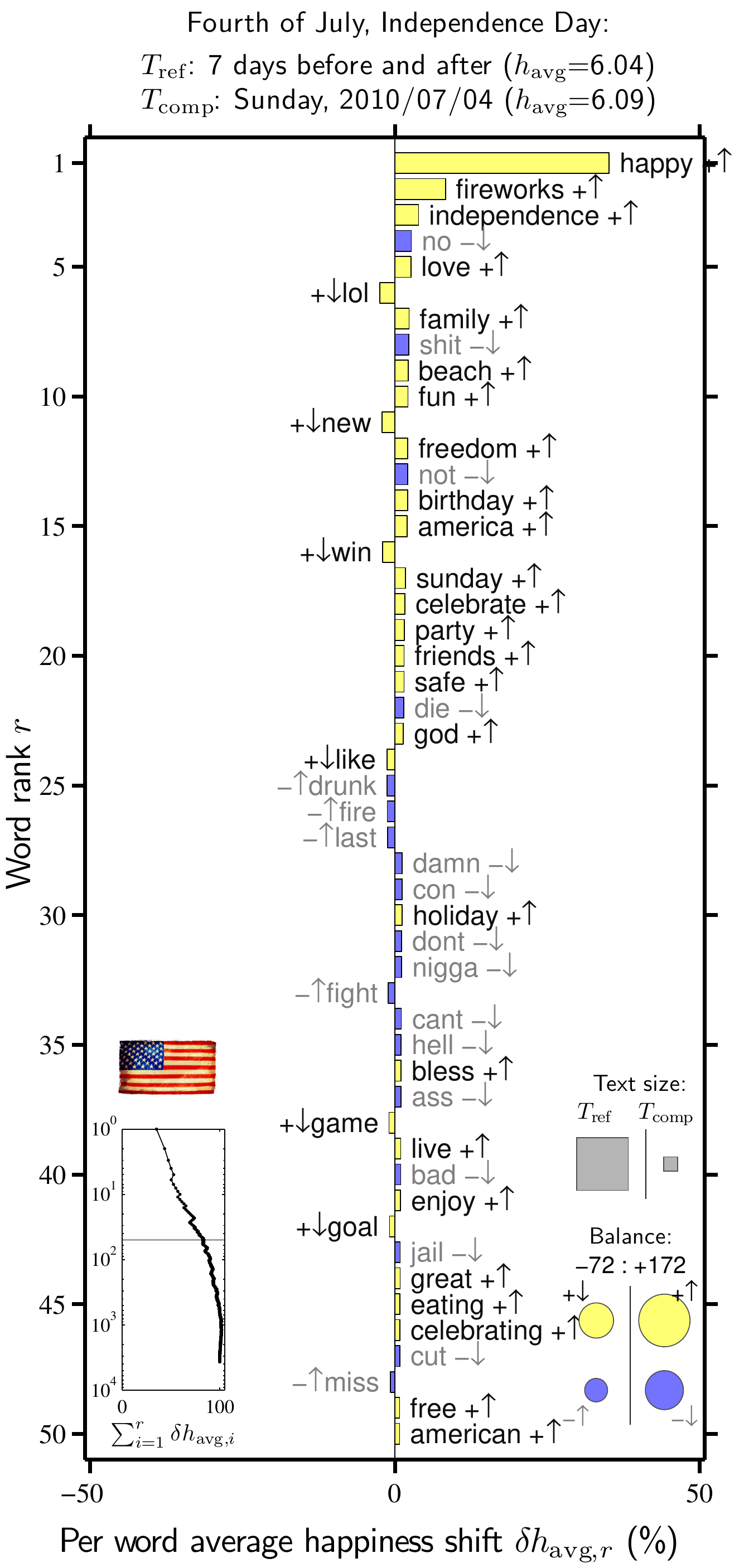}
\caption{
Word shift graph for Fourth of July, Independence Day, 2010/07/04, relative to 7 days before and 7 days after combined.
}
\label{fig:twhap.interestingdates-supp029}
\end{figure}

\begin{figure}[t]
\includegraphics[width=0.48\textwidth]{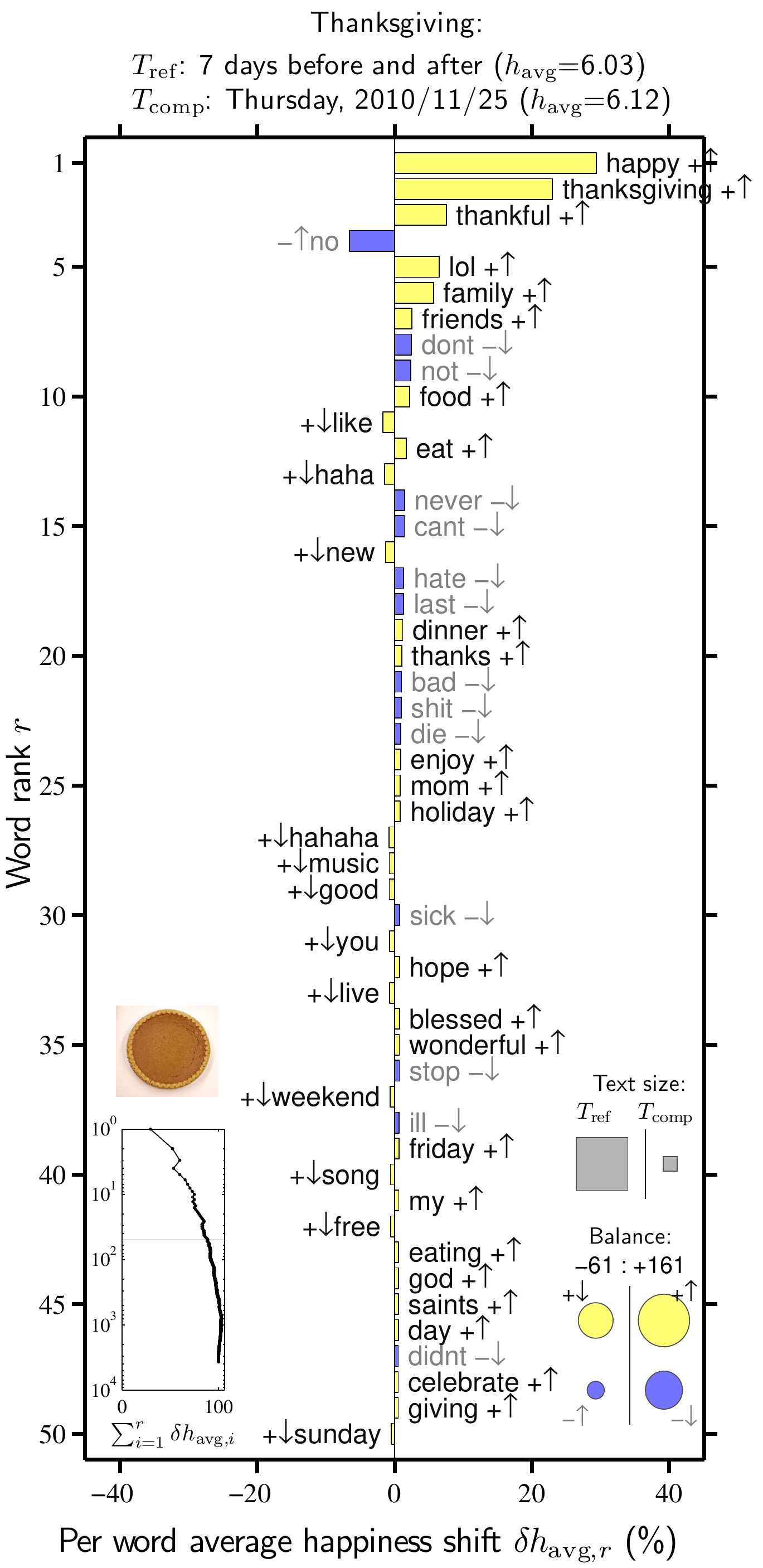}
\caption{
Word shift graph for Thanksgiving, 2010/11/25, relative to 7 days before and 7 days after combined.
}
\label{fig:twhap.interestingdates-supp030}
\end{figure}

\clearpage
\begin{figure}[t]
\includegraphics[width=0.48\textwidth]{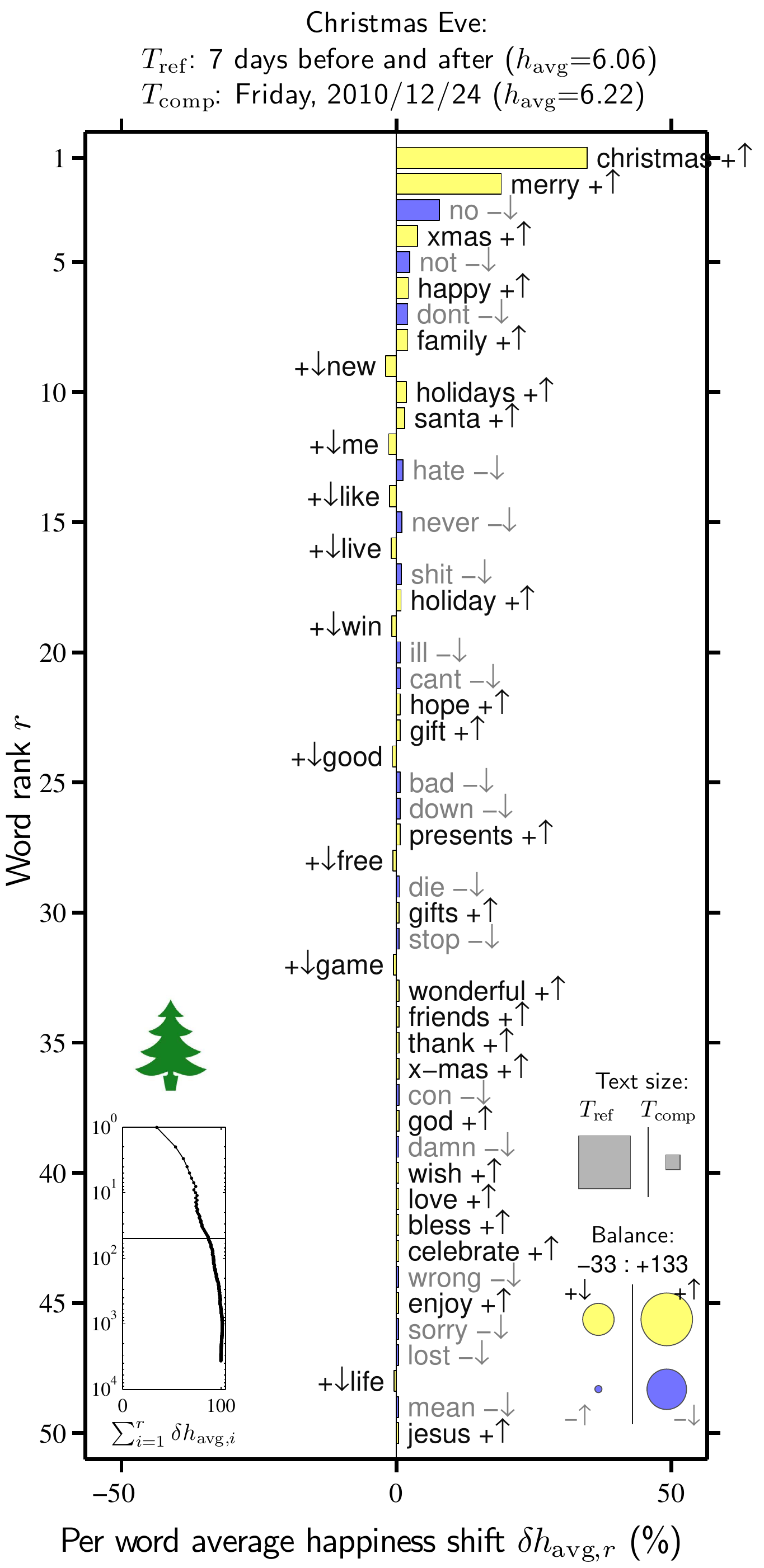}
\caption{
Word shift graph for Christmas Eve, 2010/12/24, relative to 7 days before and 7 days after combined.
}
\label{fig:twhap.interestingdates-supp031}
\end{figure}

\begin{figure}[t]
\includegraphics[width=0.48\textwidth]{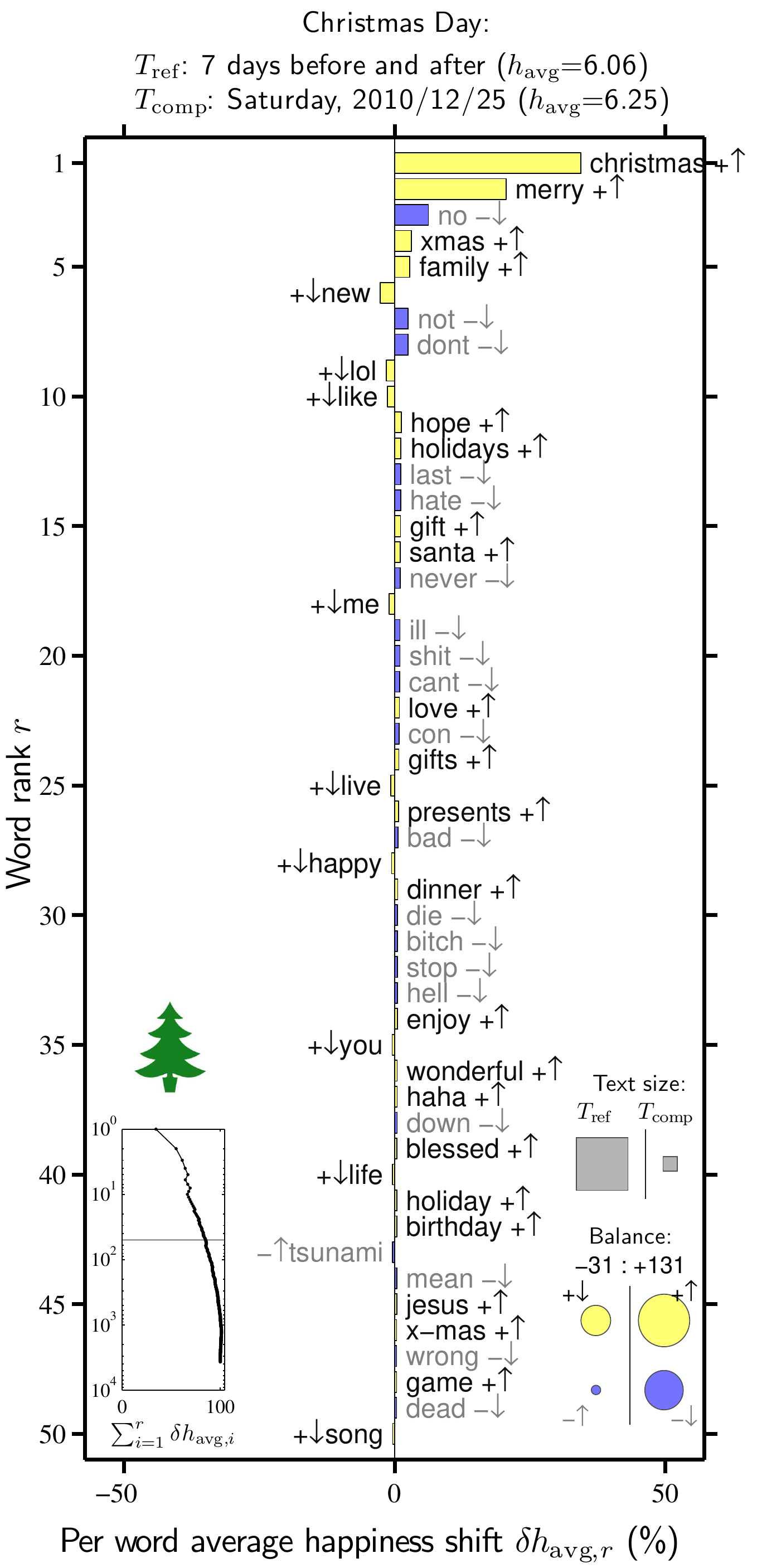}
\caption{
Word shift graph for Christmas Day, 2010/12/25, relative to 7 days before and 7 days after combined.
}
\label{fig:twhap.interestingdates-supp032}
\end{figure}

\clearpage
\begin{figure}[t]
\includegraphics[width=0.48\textwidth]{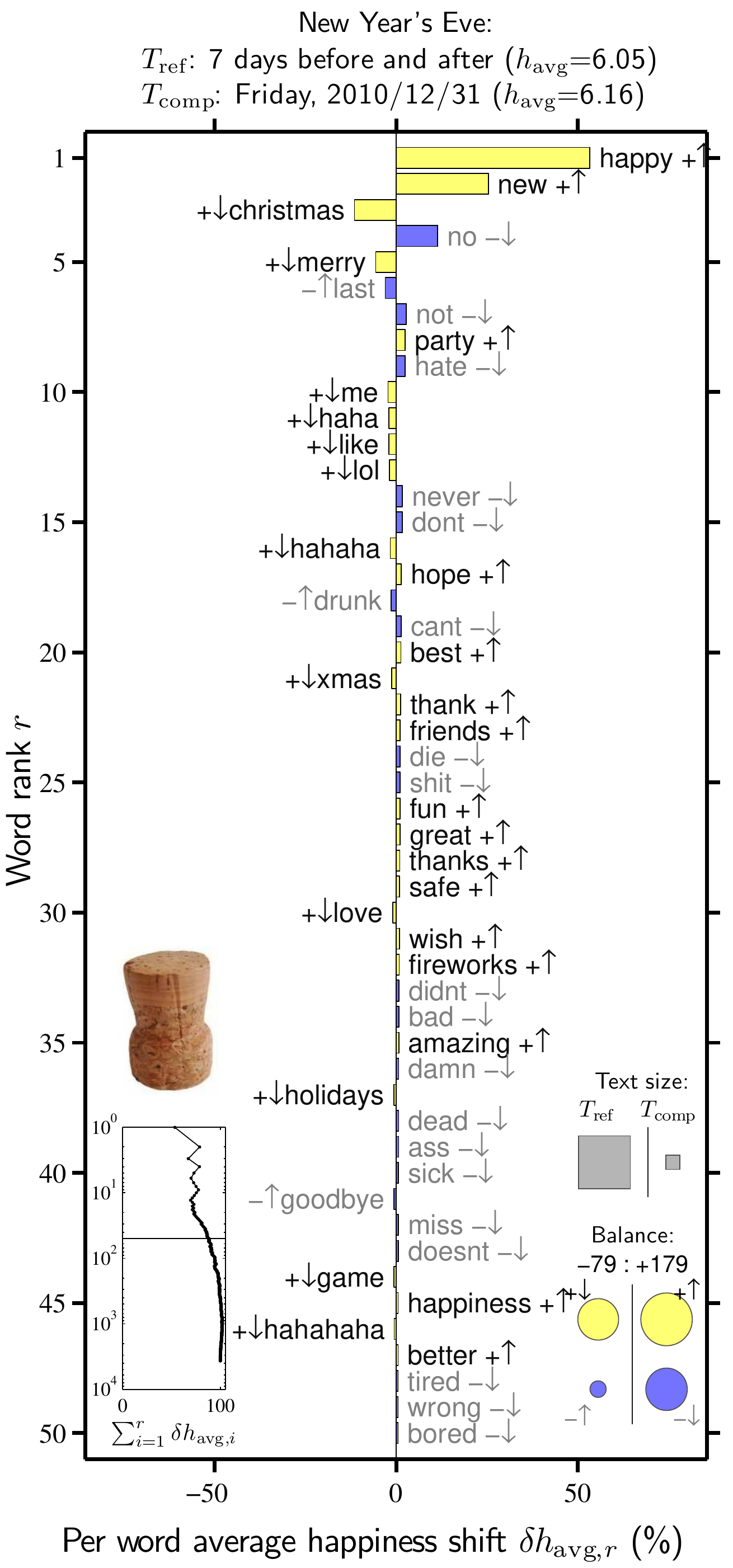}
\caption{
Word shift graph for New Year's Eve, 2010/12/31, relative to 7 days before and 7 days after combined.
}
\label{fig:twhap.interestingdates-supp033}
\end{figure}

\begin{figure}[t]
\includegraphics[width=0.48\textwidth]{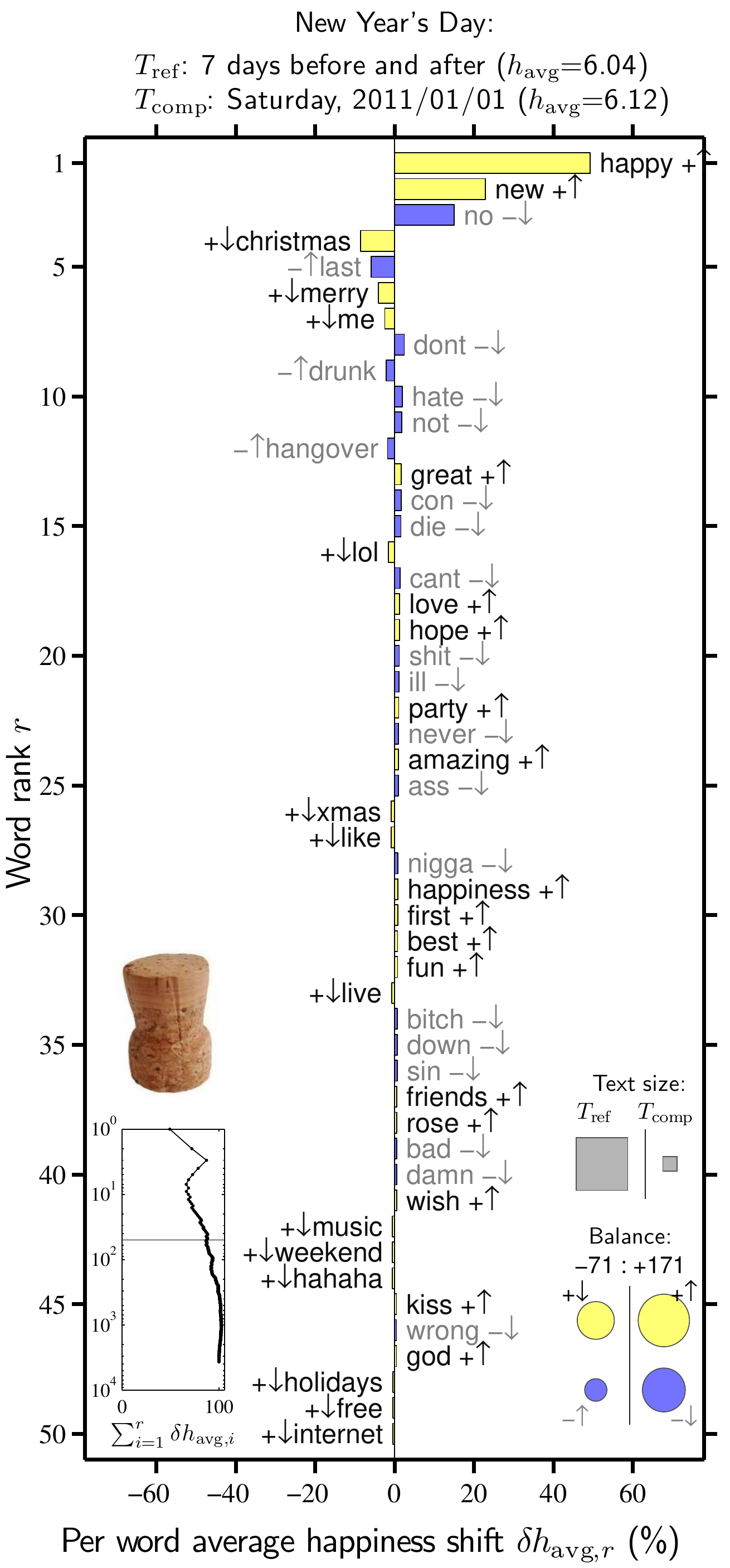}
\caption{
Word shift graph for New Year's Day, 2011/01/01, relative to 7 days before and 7 days after combined.
}
\label{fig:twhap.interestingdates-supp034}
\end{figure}

\clearpage
\begin{figure}[t]
\includegraphics[width=0.48\textwidth]{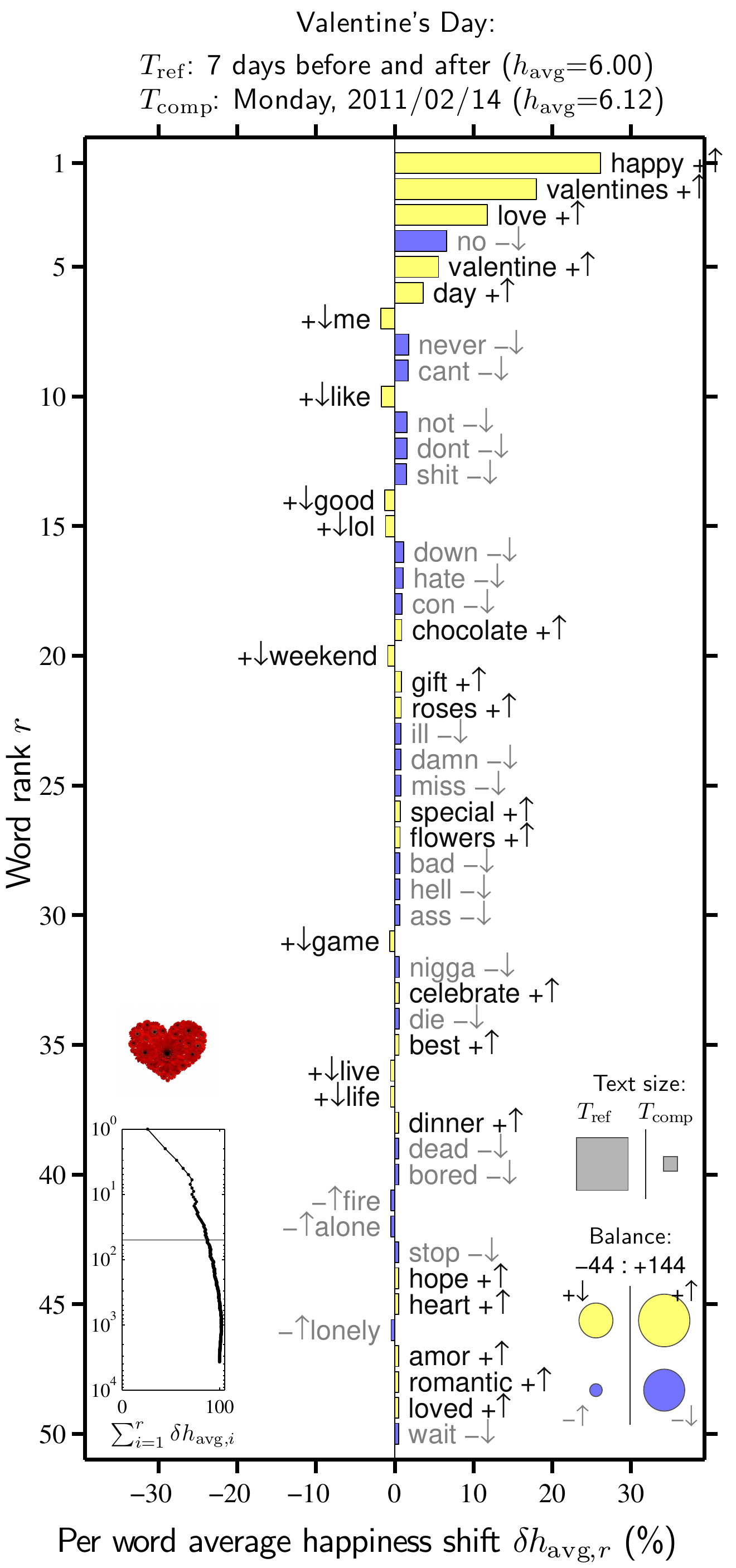}
\caption{
Word shift graph for Valentine's Day, 2011/02/14, relative to 7 days before and 7 days after combined.
}
\label{fig:twhap.interestingdates-supp035}
\end{figure}

\begin{figure}[t]
\includegraphics[width=0.48\textwidth]{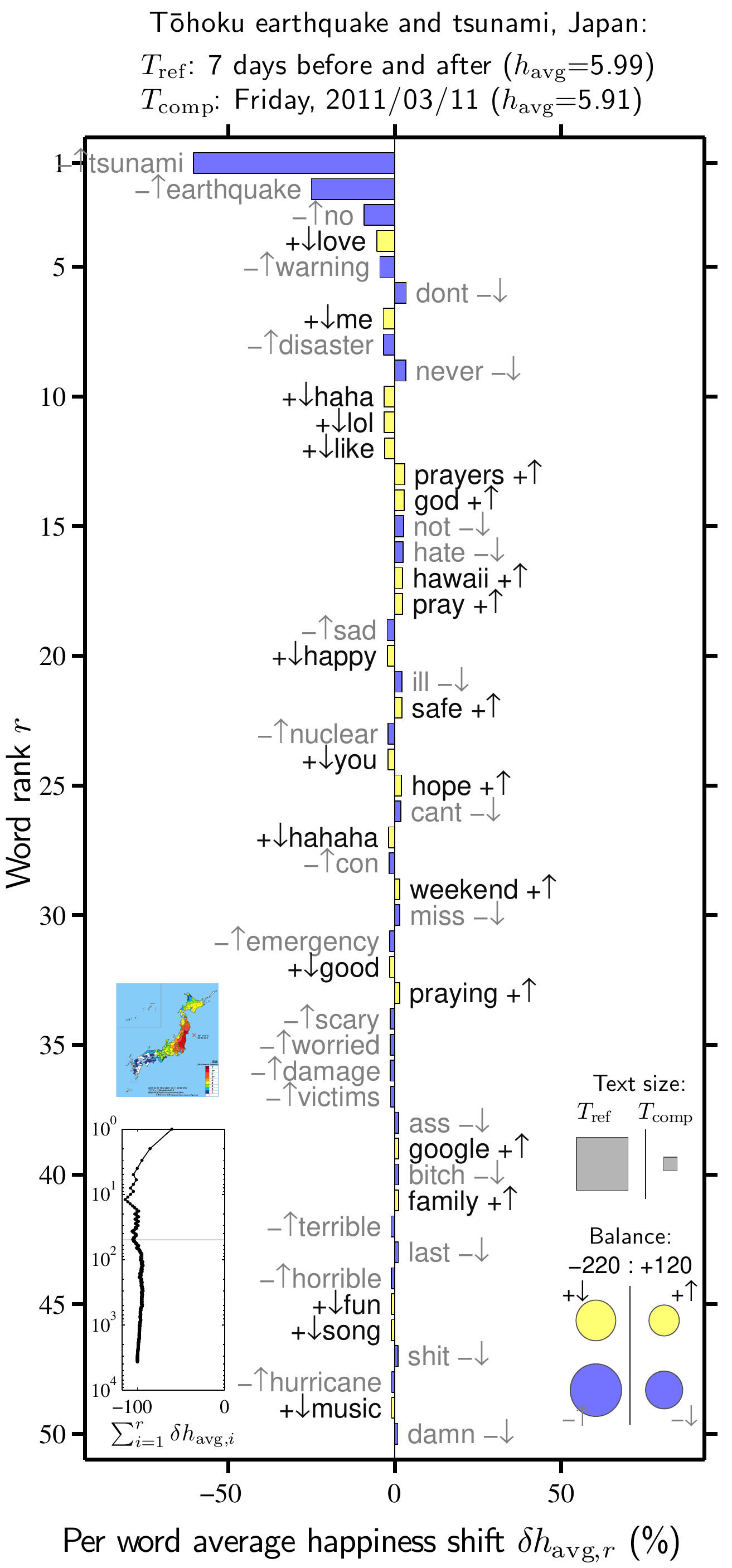}
\caption{
Word shift graph for T\={o}hoku earthquake and tsunami, Japan, 2011/03/11, relative to 7 days before and 7 days after combined.
}
\label{fig:twhap.interestingdates-supp036}
\end{figure}

\clearpage
\begin{figure}[t]
\includegraphics[width=0.48\textwidth]{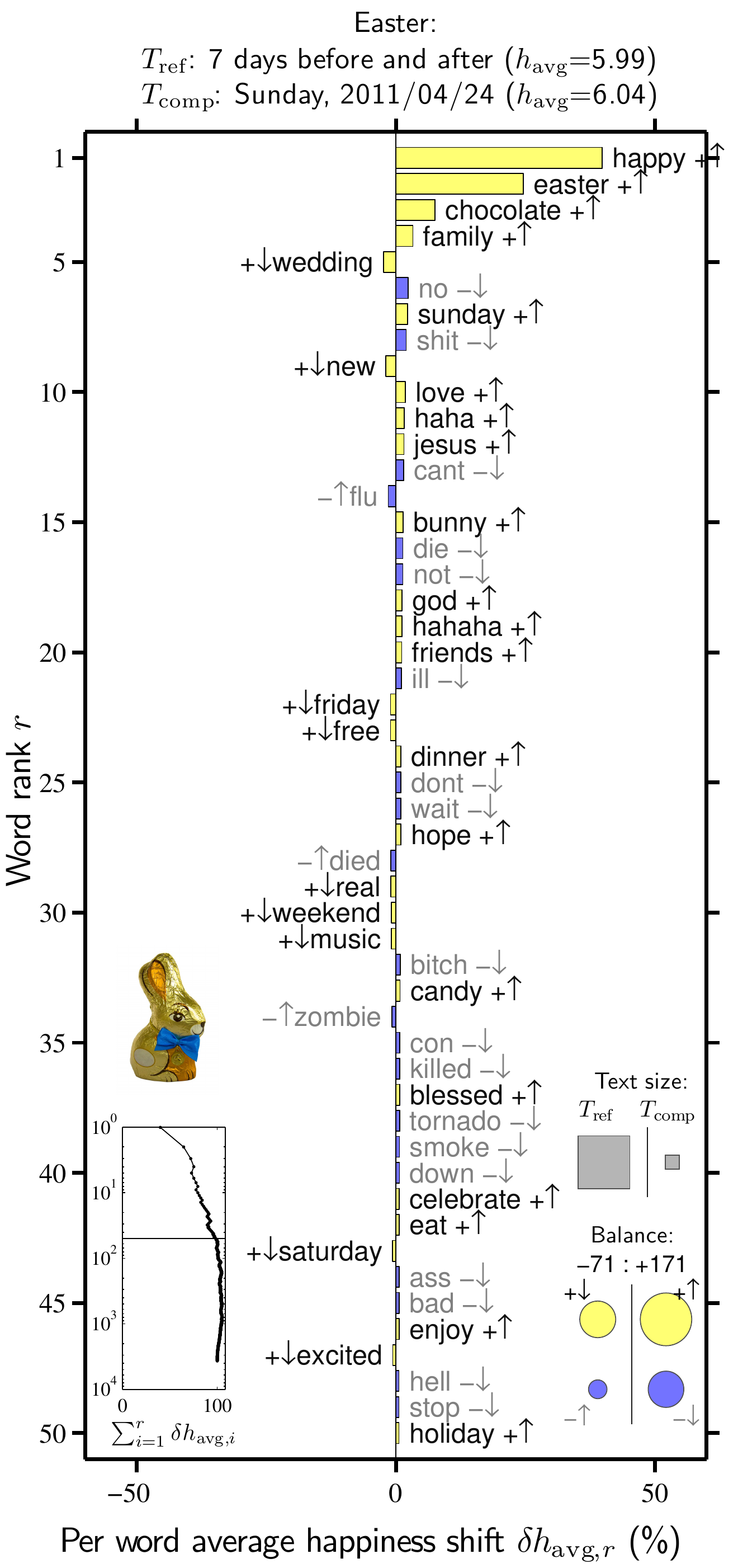}
\caption{
Word shift graph for Easter, 2011/04/24, relative to 7 days before and 7 days after combined.
}
\label{fig:twhap.interestingdates-supp037}
\end{figure}

\begin{figure}[t]
\includegraphics[width=0.48\textwidth]{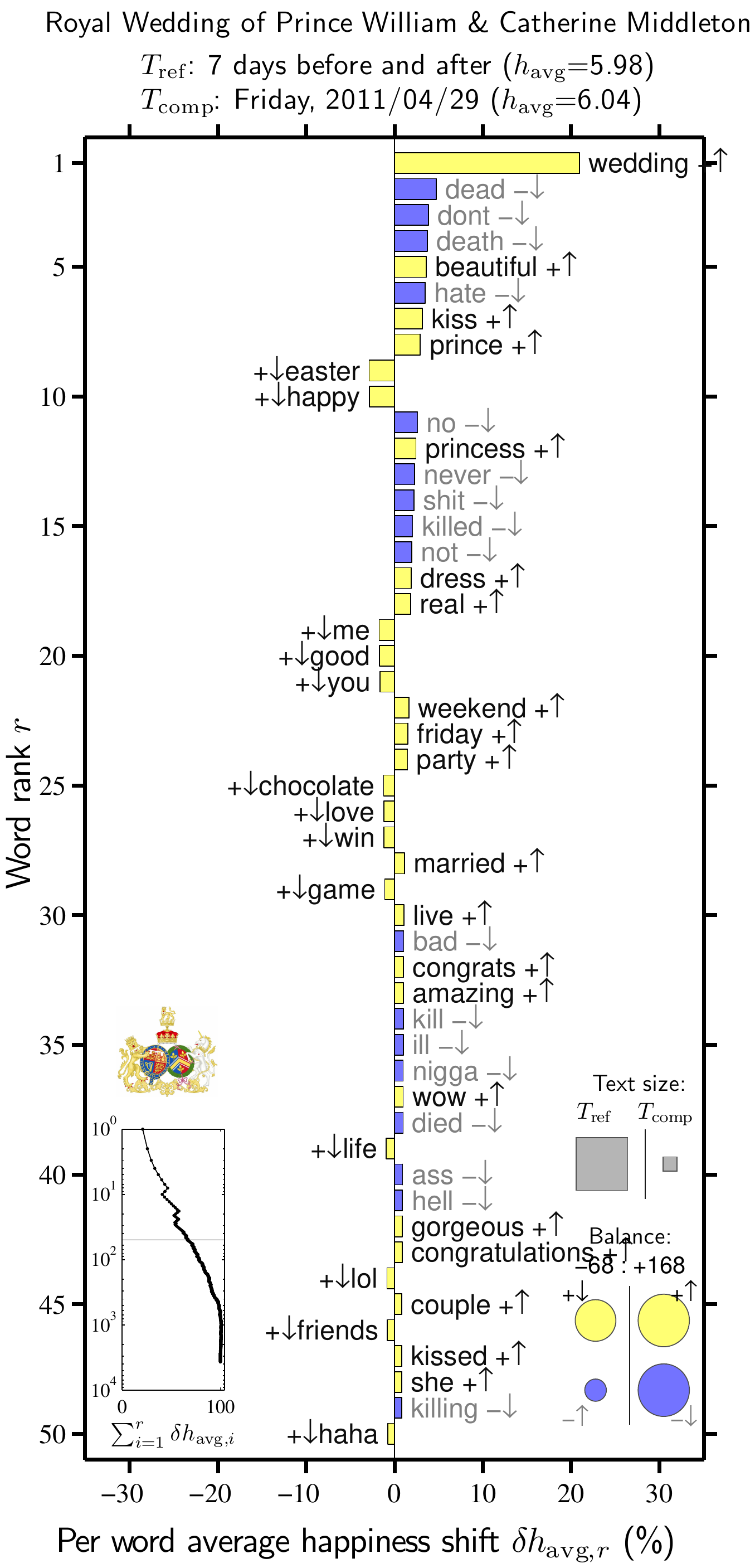}
\caption{
Word shift graph for Royal Wedding of Prince William \& Catherine Middleton, 2011/04/29, relative to 7 days before and 7 days after combined.
}
\label{fig:twhap.interestingdates-supp038}
\end{figure}

\clearpage
\begin{figure}[t]
\includegraphics[width=0.48\textwidth]{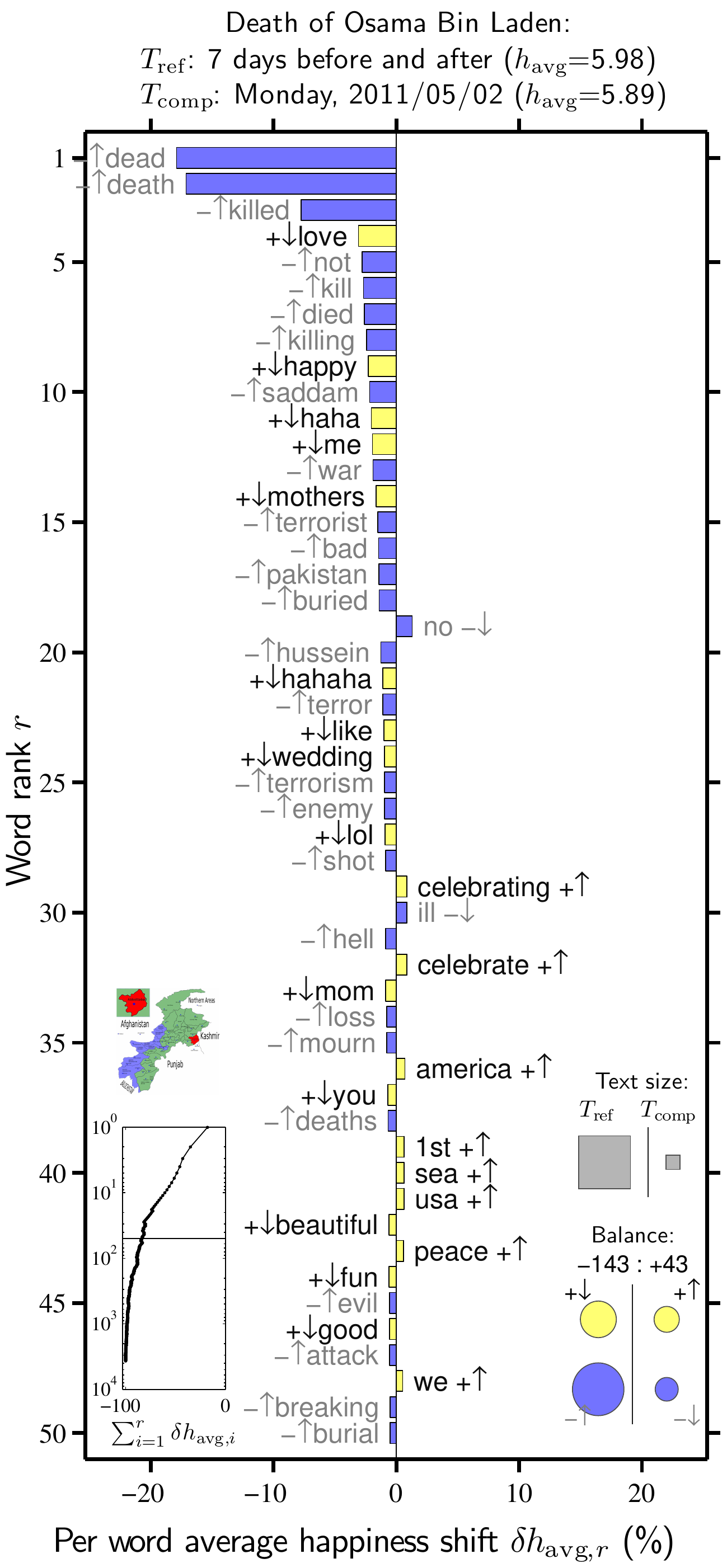}
\caption{
Word shift graph for Death of Osama Bin Laden, 2011/05/02, relative to 7 days before and 7 days after combined.
}
\label{fig:twhap.interestingdates-supp039}
\end{figure}

\begin{figure}[t]
\includegraphics[width=0.48\textwidth]{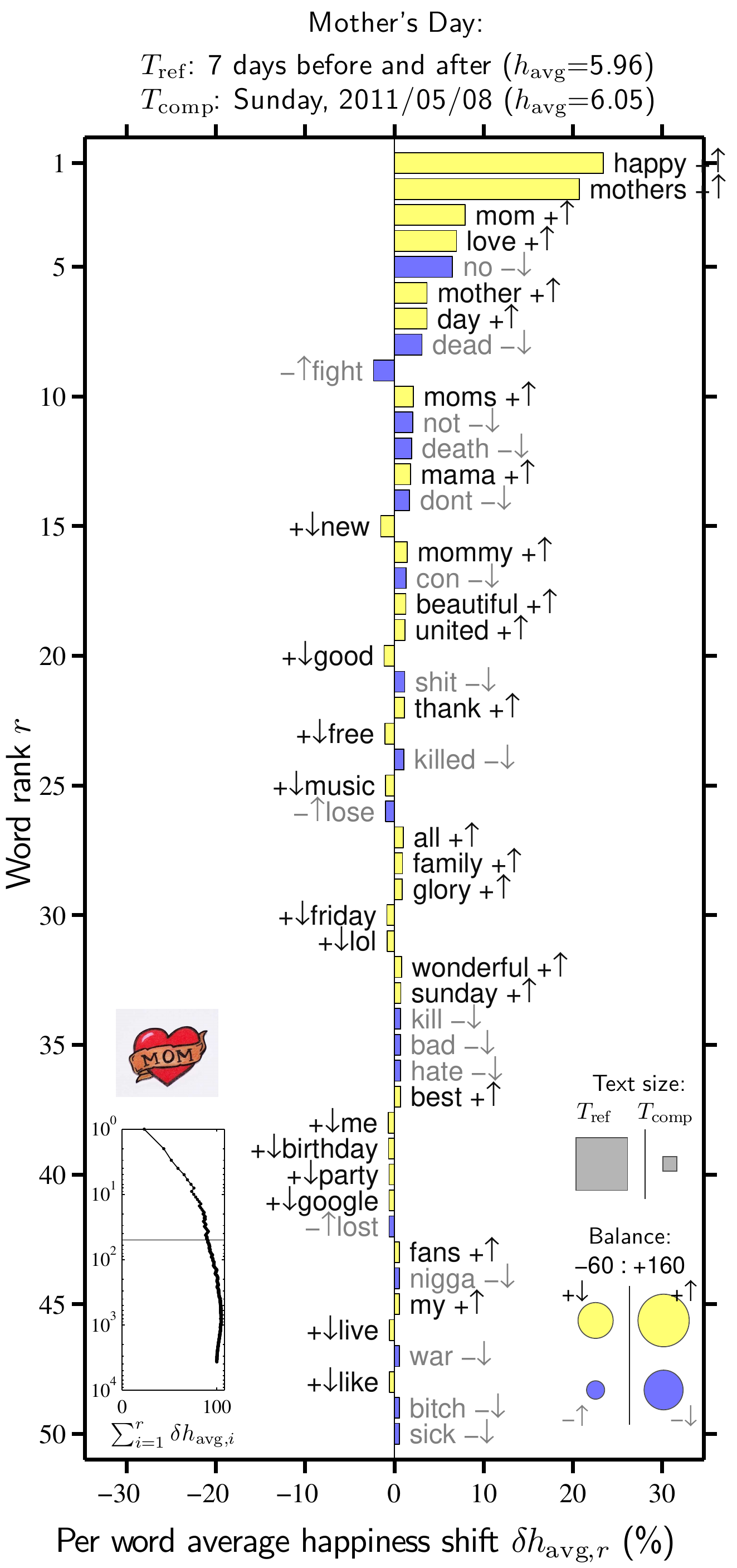}
\caption{
Word shift graph for Mother's Day, 2011/05/08, relative to 7 days before and 7 days after combined.
}
\label{fig:twhap.interestingdates-supp040}
\end{figure}

\clearpage
\begin{figure}[t]
\includegraphics[width=0.48\textwidth]{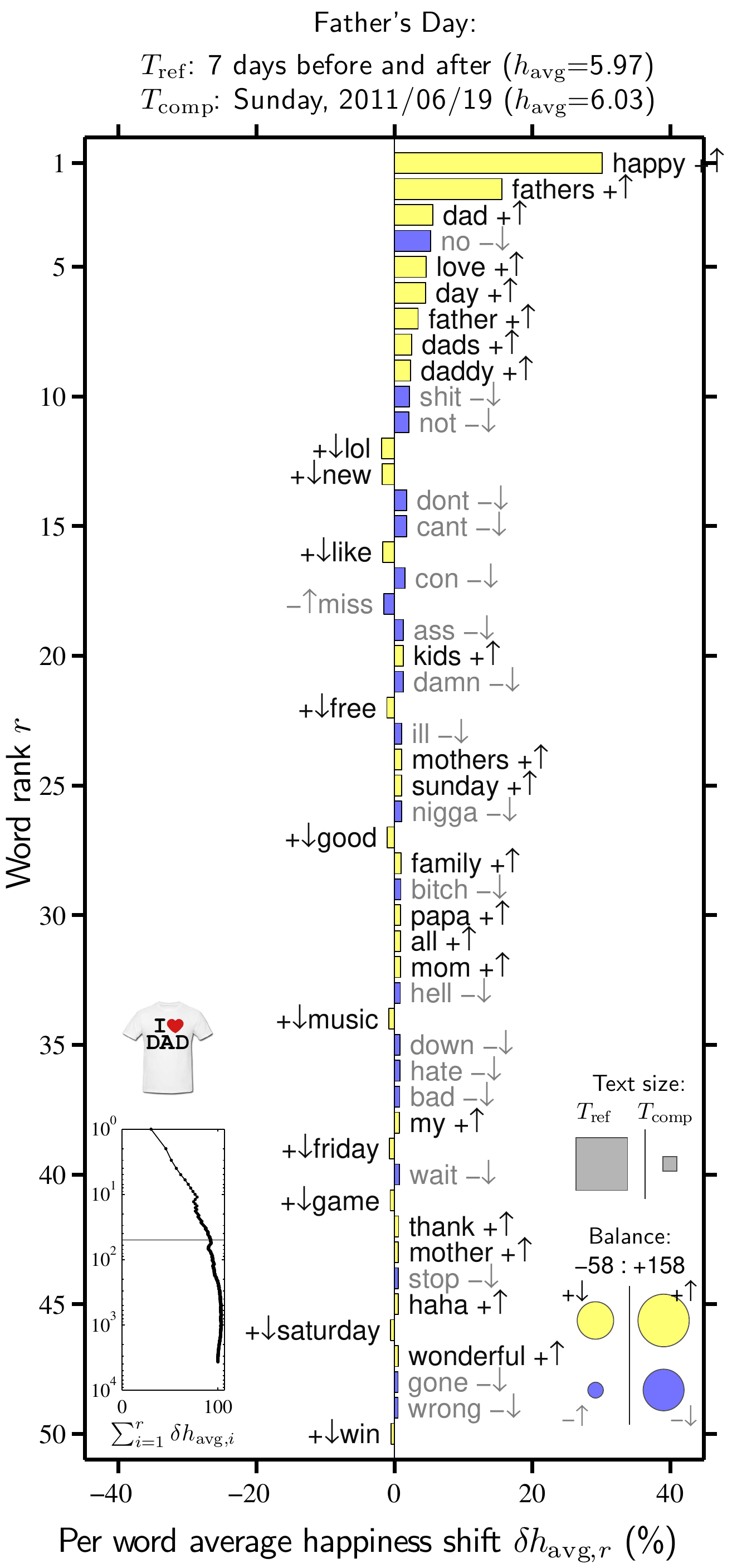}
\caption{
Word shift graph for Father's Day, 2011/06/19, relative to 7 days before and 7 days after combined.
}
\label{fig:twhap.interestingdates-supp041}
\end{figure}

\begin{figure}[t]
\includegraphics[width=0.48\textwidth]{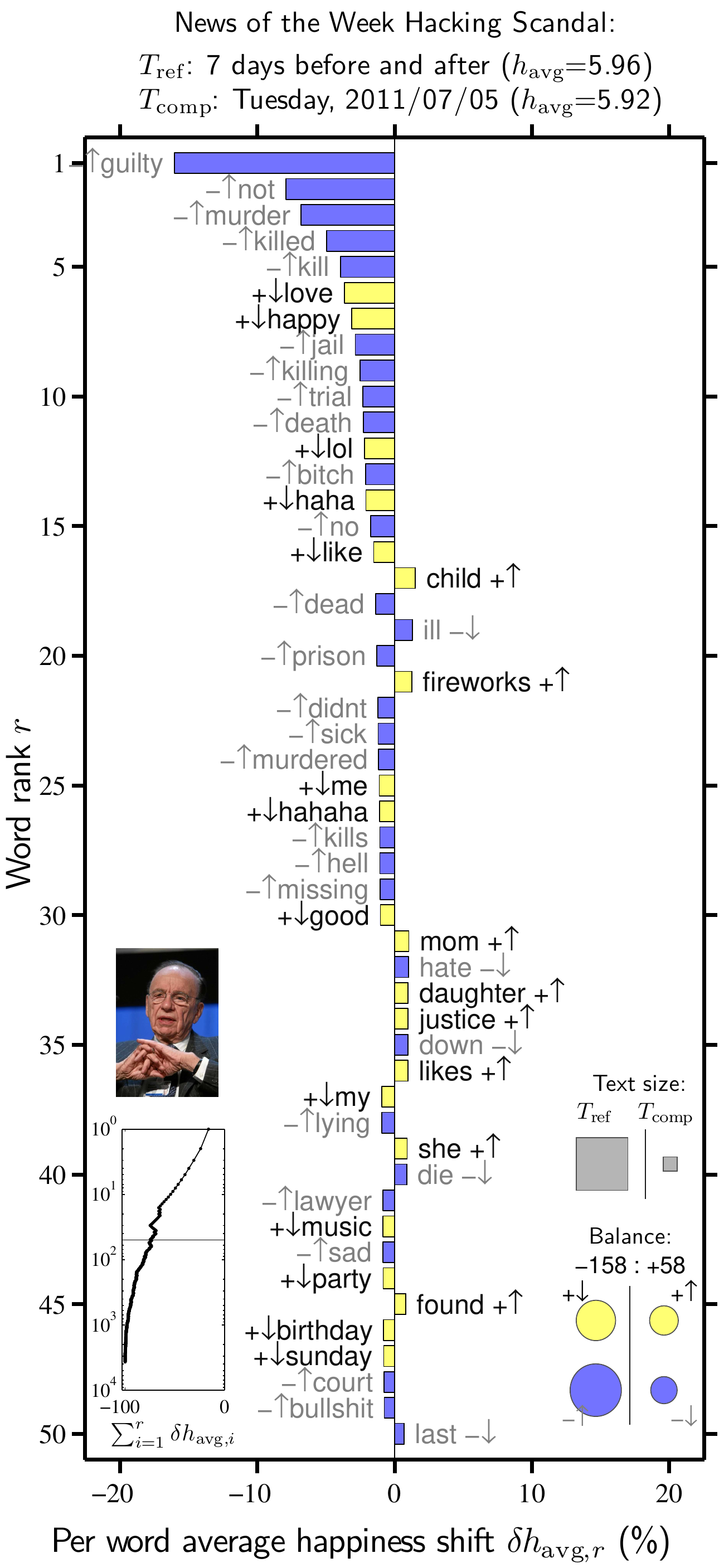}
\caption{
Word shift graph for News of the Week Hacking Scandal, 2011/07/05, relative to 7 days before and 7 days after combined.
}
\label{fig:twhap.interestingdates-supp042}
\end{figure}

\clearpage
\begin{figure}[t]
\includegraphics[width=0.48\textwidth]{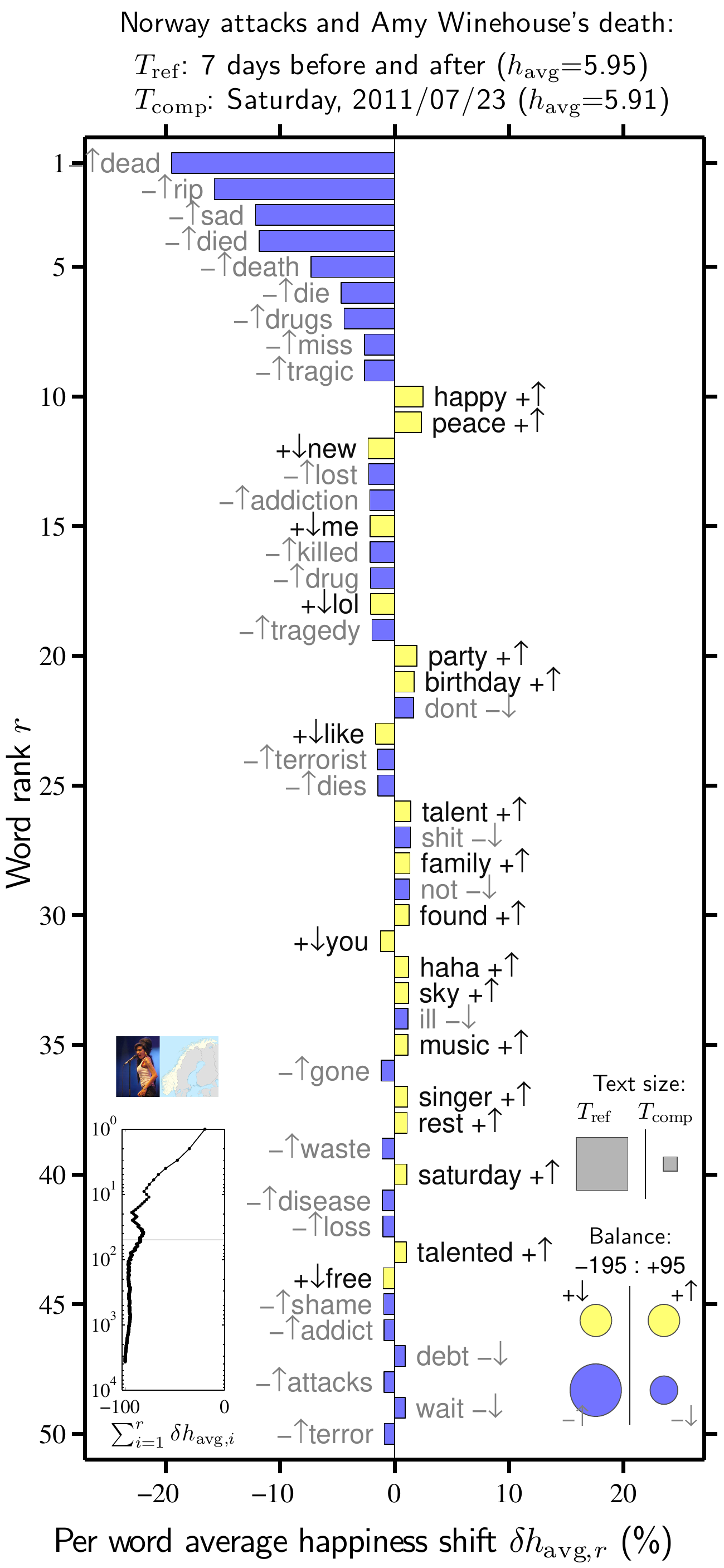}
\caption{
Word shift graph for Norway attacks and Amy Winehouse's death, 2011/07/23, relative to 7 days before and 7 days after combined.
}
\label{fig:twhap.interestingdates-supp043}
\end{figure}

\begin{figure}[t]
\includegraphics[width=0.48\textwidth]{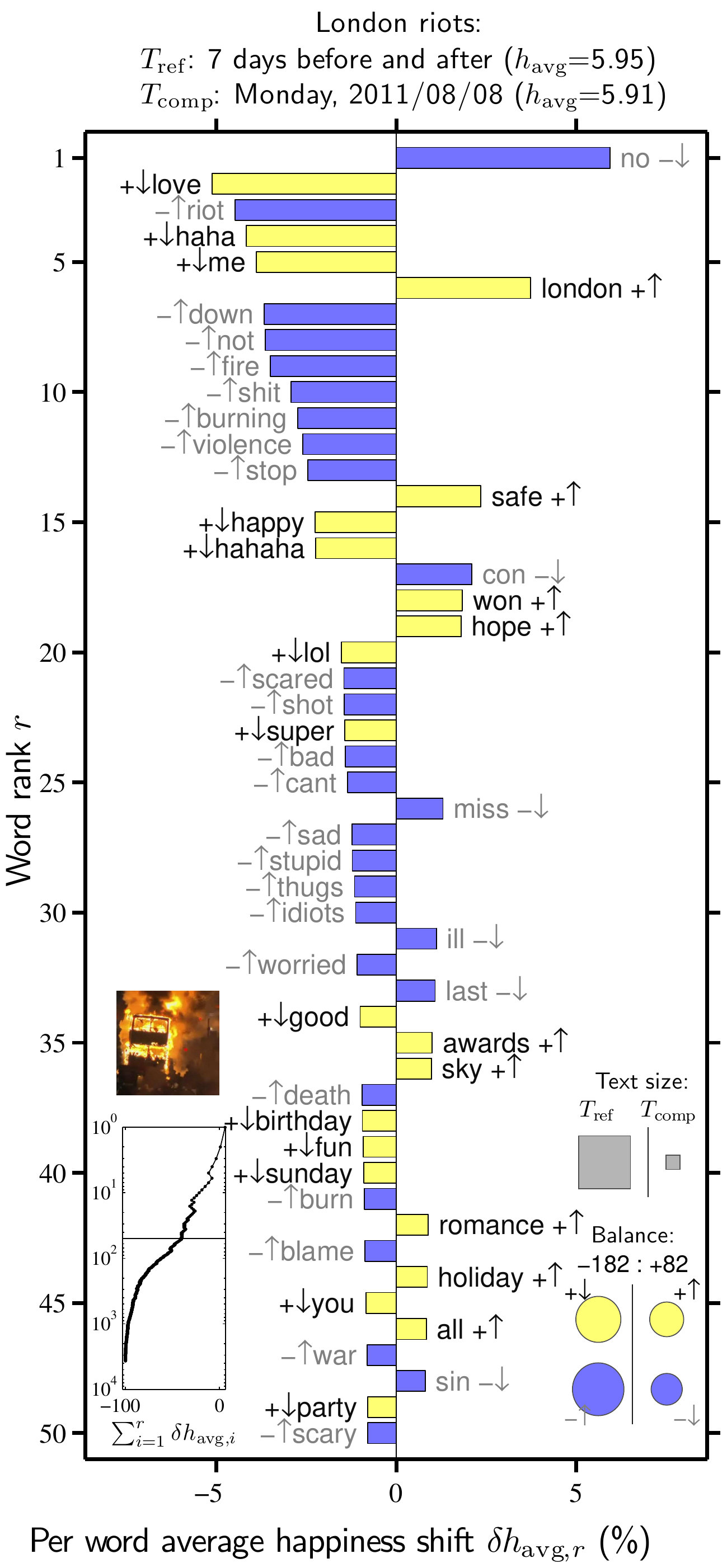}
\caption{
Word shift graph for London riots, 2011/08/08, relative to 7 days before and 7 days after combined.
}
\label{fig:twhap.interestingdates-supp044}
\end{figure}

\clearpage
\begin{figure}[t]
\includegraphics[width=0.48\textwidth]{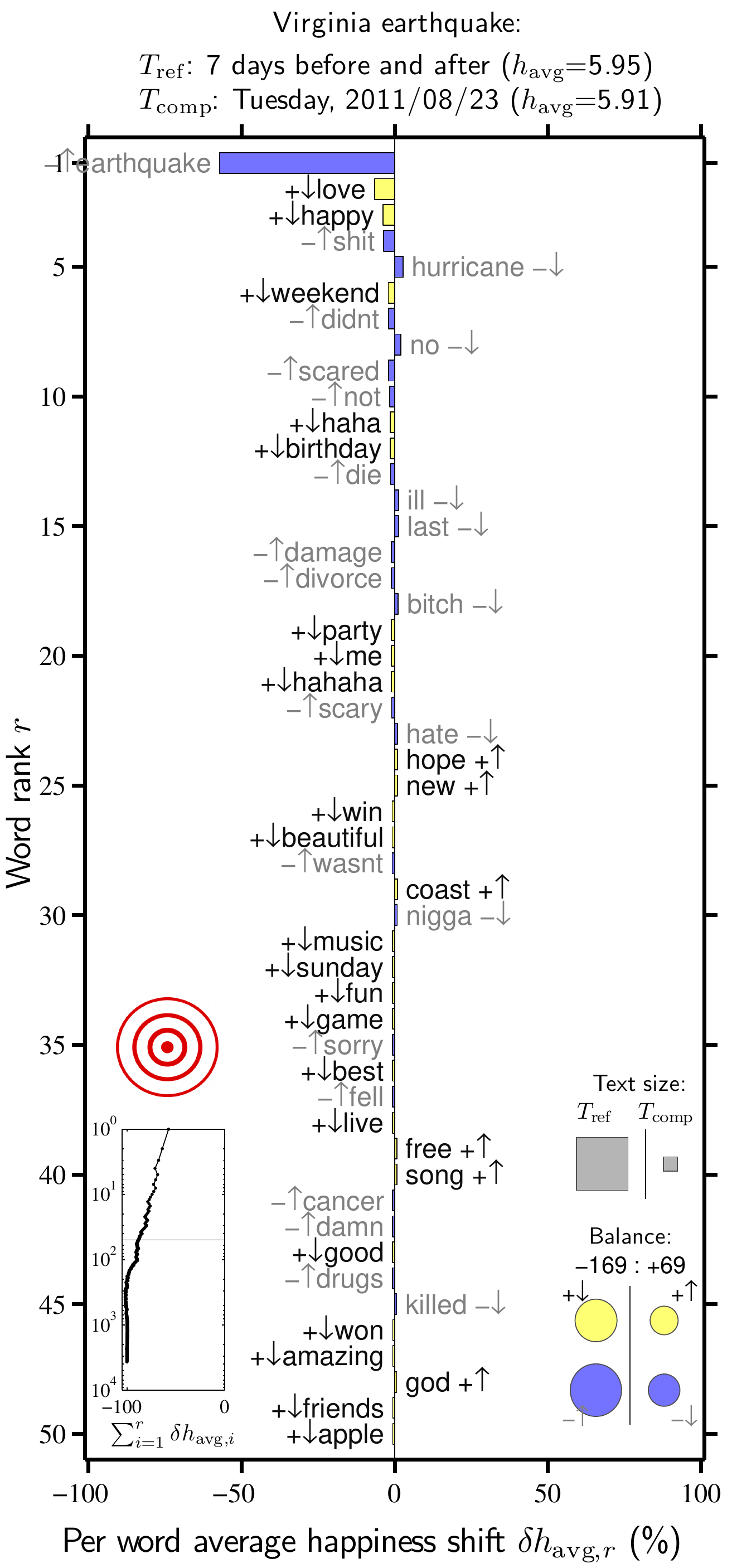}
\caption{
Word shift graph for Virginia earthquake, 2011/08/23, relative to 7 days before and 7 days after combined.
}
\label{fig:twhap.interestingdates-supp045}
\end{figure}

\begin{figure}[t]
\includegraphics[width=0.48\textwidth]{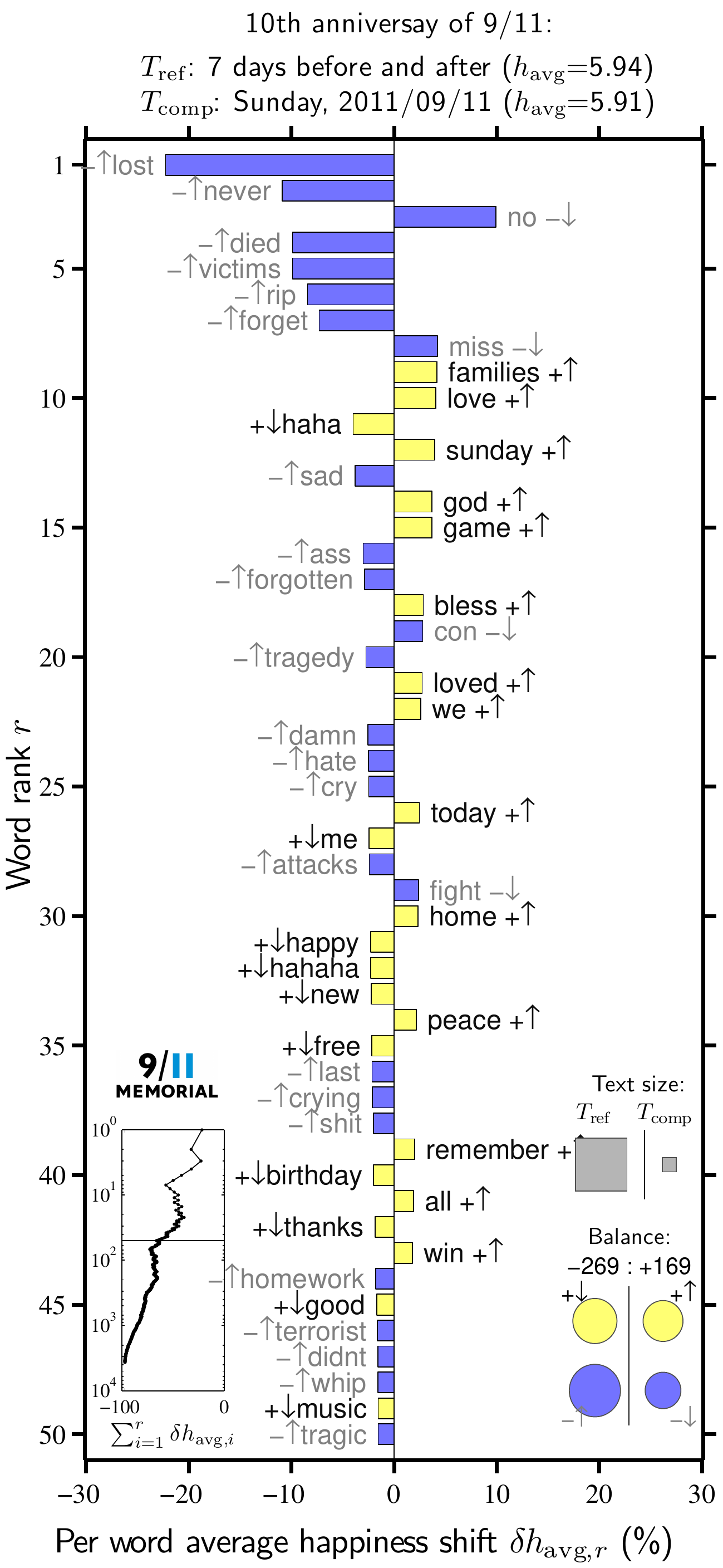}
\caption{
Word shift graph for 10th anniversay of 9/11, 2011/09/11, relative to 7 days before and 7 days after combined.
}
\label{fig:twhap.interestingdates-supp046}
\end{figure}

\clearpage


\begin{thebibliography}{73}
\expandafter\ifx\csname natexlab\endcsname\relax\def\natexlab#1{#1}\fi
\expandafter\ifx\csname bibnamefont\endcsname\relax
  \def\bibnamefont#1{#1}\fi
\expandafter\ifx\csname bibfnamefont\endcsname\relax
  \def\bibfnamefont#1{#1}\fi
\expandafter\ifx\csname citenamefont\endcsname\relax
  \def\citenamefont#1{#1}\fi
\expandafter\ifx\csname url\endcsname\relax
  \def\url#1{\texttt{#1}}\fi
\expandafter\ifx\csname urlprefix\endcsname\relax\def\urlprefix{URL }\fi
\providecommand{\bibinfo}[2]{#2}
\providecommand{\eprint}[2][]{\url{#2}}

\bibitem[{\citenamefont{Hedstr\"{o}m}(2006)}]{hedstrom2006a}
\bibinfo{author}{\bibfnamefont{P.}~\bibnamefont{Hedstr\"{o}m}},
  \bibinfo{journal}{Science} \textbf{\bibinfo{volume}{311}},
  \bibinfo{pages}{786} (\bibinfo{year}{2006}).

\bibitem[{\citenamefont{Bell et~al.}(2009)\citenamefont{Bell, Hey, and
  Szalay}}]{bell2009a}
\bibinfo{author}{\bibfnamefont{G.}~\bibnamefont{Bell}},
  \bibinfo{author}{\bibfnamefont{T.}~\bibnamefont{Hey}}, \bibnamefont{and}
  \bibinfo{author}{\bibfnamefont{A.}~\bibnamefont{Szalay}},
  \bibinfo{journal}{Science} \textbf{\bibinfo{volume}{323}},
  \bibinfo{pages}{1297} (\bibinfo{year}{2009}).

\bibitem[{\citenamefont{Halevy et~al.}(2009)\citenamefont{Halevy, Norvig, and
  Pereira}}]{halevy2009a}
\bibinfo{author}{\bibfnamefont{A.}~\bibnamefont{Halevy}},
  \bibinfo{author}{\bibfnamefont{P.}~\bibnamefont{Norvig}}, \bibnamefont{and}
  \bibinfo{author}{\bibfnamefont{F.}~\bibnamefont{Pereira}},
  \bibinfo{journal}{IEEE Intelligent Systems} \textbf{\bibinfo{volume}{24}},
  \bibinfo{pages}{8} (\bibinfo{year}{2009}).

\bibitem[{\citenamefont{Hey et~al.}(2009)\citenamefont{Hey, Tansley, and
  Tolle}}]{hey2009a}
\bibinfo{editor}{\bibfnamefont{T.}~\bibnamefont{Hey}},
  \bibinfo{editor}{\bibfnamefont{S.}~\bibnamefont{Tansley}}, \bibnamefont{and}
  \bibinfo{editor}{\bibfnamefont{K.}~\bibnamefont{Tolle}}, eds.,
  \emph{\bibinfo{title}{The Fourth Paradigm: Data-Intensive Scientific
  Discovery}} (\bibinfo{publisher}{Microsoft Research},
  \bibinfo{address}{Redmond, WA}, \bibinfo{year}{2009}).

\bibitem[{\citenamefont{Collins}(2010)}]{collins2010a}
\bibinfo{author}{\bibfnamefont{J.~P.} \bibnamefont{Collins}},
  \bibinfo{journal}{Science Magazine} \textbf{\bibinfo{volume}{327}},
  \bibinfo{pages}{1455} (\bibinfo{year}{2010}).

\bibitem[{sds()}]{sdss}
\bibinfo{note}{Sloan Digital Sky Survey. Available at
  \url{http://www.sdss.org/}. Accessed October 24, 2011.}

\bibitem[{lss()}]{lsst}
\bibinfo{note}{Large Synoptic Survey Telescope, \url{http://www.lsst.org/lsst}.
  Accessed October 24, 2011.}

\bibitem[{\citenamefont{Stephens}(2008)}]{stephens2008a}
\bibinfo{author}{\bibfnamefont{M.}~\bibnamefont{Stephens}},
  \emph{\bibinfo{title}{Mapping the universe at 30 terabytes a night: Jeff
  kantor, on building and managing a 150 petabyte database}},
  \bibinfo{howpublished}{The Register} (\bibinfo{year}{2008}).

\bibitem[{\citenamefont{Venter et~al.}(2004)\citenamefont{Venter, Remington,
  Heidelberg, Halpern, Rusch, Eisen, Wu, Paulsen, Nelson, Nelson
  et~al.}}]{venter2004a}
\bibinfo{author}{\bibfnamefont{J.~C.} \bibnamefont{Venter}},
  \bibinfo{author}{\bibfnamefont{K.}~\bibnamefont{Remington}},
  \bibinfo{author}{\bibfnamefont{J.~F.} \bibnamefont{Heidelberg}},
  \bibinfo{author}{\bibfnamefont{A.~L.} \bibnamefont{Halpern}},
  \bibinfo{author}{\bibfnamefont{D.}~\bibnamefont{Rusch}},
  \bibinfo{author}{\bibfnamefont{J.~A.} \bibnamefont{Eisen}},
  \bibinfo{author}{\bibfnamefont{D.}~\bibnamefont{Wu}},
  \bibinfo{author}{\bibfnamefont{I.}~\bibnamefont{Paulsen}},
  \bibinfo{author}{\bibfnamefont{K.~E.} \bibnamefont{Nelson}},
  \bibinfo{author}{\bibfnamefont{W.}~\bibnamefont{Nelson}},
  \bibnamefont{et~al.}, \bibinfo{journal}{Science}
  \textbf{\bibinfo{volume}{304}}, \bibinfo{pages}{66} (\bibinfo{year}{2004}).

\bibitem[{lhc()}]{lhc}
\bibinfo{note}{Large Hadron Collider. Available at
  \url{http://lhc.web.cern.ch/lhc/}}.

\bibitem[{\citenamefont{Miller}(2011)}]{miller2011a}
\bibinfo{author}{\bibfnamefont{G.}~\bibnamefont{Miller}},
  \bibinfo{journal}{Science Magazine} \textbf{\bibinfo{volume}{333}},
  \bibinfo{pages}{1814} (\bibinfo{year}{2011}).

\bibitem[{goo()}]{googlebooks-ngrams}
\bibinfo{note}{Google Labs ngram viewer. Availabe at
  \url{http://ngrams.googlelabs.com/}. Accessed October 24, 2011}.

\bibitem[{\citenamefont{Michel et~al.}(2011)\citenamefont{Michel, Shen, Aiden,
  Veres, Gray, {The Google Books Team}, Pickett, Hoiberg, Clancy, Norvig
  et~al.}}]{michel2011a}
\bibinfo{author}{\bibfnamefont{J.-B.} \bibnamefont{Michel}},
  \bibinfo{author}{\bibfnamefont{Y.~K.} \bibnamefont{Shen}},
  \bibinfo{author}{\bibfnamefont{A.~P.} \bibnamefont{Aiden}},
  \bibinfo{author}{\bibfnamefont{A.}~\bibnamefont{Veres}},
  \bibinfo{author}{\bibfnamefont{M.~K.} \bibnamefont{Gray}},
  \bibinfo{author}{\bibnamefont{{The Google Books Team}}},
  \bibinfo{author}{\bibfnamefont{J.~P.} \bibnamefont{Pickett}},
  \bibinfo{author}{\bibfnamefont{D.}~\bibnamefont{Hoiberg}},
  \bibinfo{author}{\bibfnamefont{D.}~\bibnamefont{Clancy}},
  \bibinfo{author}{\bibfnamefont{P.}~\bibnamefont{Norvig}},
  \bibnamefont{et~al.}, \bibinfo{journal}{Science Magazine}
  \textbf{\bibinfo{volume}{331}}, \bibinfo{pages}{176} (\bibinfo{year}{2011}).

\bibitem[{\citenamefont{Goel et~al.}(2010)\citenamefont{Goel, Hofman, Lahaie,
  Pennock, and Watts}}]{goel2010a}
\bibinfo{author}{\bibfnamefont{S.}~\bibnamefont{Goel}},
  \bibinfo{author}{\bibfnamefont{J.~M.} \bibnamefont{Hofman}},
  \bibinfo{author}{\bibfnamefont{S.}~\bibnamefont{Lahaie}},
  \bibinfo{author}{\bibfnamefont{D.~M.} \bibnamefont{Pennock}},
  \bibnamefont{and} \bibinfo{author}{\bibfnamefont{D.~J.} \bibnamefont{Watts}},
  \bibinfo{journal}{Proc. Natl. Acad. Sci.} \textbf{\bibinfo{volume}{107}},
  \bibinfo{pages}{17486} (\bibinfo{year}{2010}).

\bibitem[{\citenamefont{Ginsberg et~al.}(2009)\citenamefont{Ginsberg, Mohebbi,
  Patel, Brammer, Smolinski, and Brilliant}}]{ginsberg2009a}
\bibinfo{author}{\bibfnamefont{J.}~\bibnamefont{Ginsberg}},
  \bibinfo{author}{\bibfnamefont{M.~H.} \bibnamefont{Mohebbi}},
  \bibinfo{author}{\bibfnamefont{R.~S.} \bibnamefont{Patel}},
  \bibinfo{author}{\bibfnamefont{L.}~\bibnamefont{Brammer}},
  \bibinfo{author}{\bibfnamefont{M.~S.} \bibnamefont{Smolinski}},
  \bibnamefont{and}
  \bibinfo{author}{\bibfnamefont{L.}~\bibnamefont{Brilliant}},
  \bibinfo{journal}{Nature} \textbf{\bibinfo{volume}{459}},
  \bibinfo{pages}{1012} (\bibinfo{year}{2009}).

\bibitem[{\citenamefont{Choi and Varian}(2009)}]{choi2009a}
\bibinfo{author}{\bibfnamefont{H.}~\bibnamefont{Choi}} \bibnamefont{and}
  \bibinfo{author}{\bibfnamefont{H.}~\bibnamefont{Varian}},
  \bibinfo{type}{Tech. Rep.}, \bibinfo{institution}{Google Inc.}
  (\bibinfo{year}{2009}).

\bibitem[{\citenamefont{Asur and Huberman}(2010)}]{asur2010a}
\bibinfo{author}{\bibfnamefont{S.}~\bibnamefont{Asur}} \bibnamefont{and}
  \bibinfo{author}{\bibfnamefont{B.~A.} \bibnamefont{Huberman}}, in
  \emph{\bibinfo{booktitle}{Proceedings of the 2010 IEEE/WIC/ACM International
  Conference on Web Intelligence and Intelligent Agent Technology - Volume 01}}
  (\bibinfo{publisher}{IEEE Computer Society}, \bibinfo{address}{Washington,
  DC, USA}, \bibinfo{year}{2010}), WI-IAT '10, pp. \bibinfo{pages}{492--499}.

\bibitem[{\citenamefont{Mishne and Glance}(2006)}]{mishne2006e}
\bibinfo{author}{\bibfnamefont{G.}~\bibnamefont{Mishne}} \bibnamefont{and}
  \bibinfo{author}{\bibfnamefont{N.}~\bibnamefont{Glance}},
  \bibinfo{journal}{AAAI 2006 Spring Symposium on Computational Approaches to
  Analysing Weblogs}  (\bibinfo{year}{2006}).

\bibitem[{\citenamefont{Layard}(2005)}]{layard2005a}
\bibinfo{author}{\bibfnamefont{R.}~\bibnamefont{Layard}},
  \emph{\bibinfo{title}{Happiness}} (\bibinfo{publisher}{The Penguin Press},
  \bibinfo{address}{London}, \bibinfo{year}{2005}).

\bibitem[{\citenamefont{Gilbert}(2006)}]{gilbert2006a}
\bibinfo{author}{\bibfnamefont{D.}~\bibnamefont{Gilbert}},
  \emph{\bibinfo{title}{Stumbling on Happiness}} (\bibinfo{publisher}{Knopf},
  \bibinfo{address}{New York}, \bibinfo{year}{2006}).

\bibitem[{\citenamefont{Lyubomirsky}(2007)}]{lyubomirsky2007a}
\bibinfo{author}{\bibfnamefont{S.}~\bibnamefont{Lyubomirsky}},
  \emph{\bibinfo{title}{The How of Happiness}} (\bibinfo{publisher}{The Penguin
  Press}, \bibinfo{address}{New York}, \bibinfo{year}{2007}).

\bibitem[{\citenamefont{Seaford}(2011)}]{seaford2011a}
\bibinfo{author}{\bibfnamefont{C.}~\bibnamefont{Seaford}},
  \bibinfo{journal}{Nature} \textbf{\bibinfo{volume}{477}},
  \bibinfo{pages}{532} (\bibinfo{year}{2011}).

\bibitem[{\citenamefont{Dodds and Danforth}(2009)}]{dodds2009b}
\bibinfo{author}{\bibfnamefont{P.~S.} \bibnamefont{Dodds}} \bibnamefont{and}
  \bibinfo{author}{\bibfnamefont{C.~M.} \bibnamefont{Danforth}},
  \bibinfo{journal}{Journal of Happiness Studies}  (\bibinfo{year}{2009}),
  \bibinfo{note}{doi:10.1007/s10902-009-9150-9}.

\bibitem[{mec()}]{mechturk}
\bibinfo{note}{Amazon's Mechanical Turk service. Available at
  \url{https://www.mturk.com/}. Accessed October 24, 2011.}

\bibitem[{\citenamefont{Edgeworth}(1881)}]{edgeworth1881a}
\bibinfo{author}{\bibfnamefont{F.~Y.} \bibnamefont{Edgeworth}},
  \emph{\bibinfo{title}{Mathematical Physics: An Essay into the Application of
  Mathematics to Moral Sciences}} (\bibinfo{publisher}{Kegan Paul},
  \bibinfo{address}{London, UK}, \bibinfo{year}{1881}).

\bibitem[{\citenamefont{Killingsworth and Gilbert}(2010)}]{killingsworth2010a}
\bibinfo{author}{\bibfnamefont{M.~A.} \bibnamefont{Killingsworth}}
  \bibnamefont{and} \bibinfo{author}{\bibfnamefont{D.~T.}
  \bibnamefont{Gilbert}}, \bibinfo{journal}{Science Magazine}
  \textbf{\bibinfo{volume}{330}}, \bibinfo{pages}{932} (\bibinfo{year}{2010}).

\bibitem[{\citenamefont{Kahneman and Riis}(2005)}]{kahneman2005a}
\bibinfo{author}{\bibfnamefont{D.}~\bibnamefont{Kahneman}} \bibnamefont{and}
  \bibinfo{author}{\bibfnamefont{J.}~\bibnamefont{Riis}}, in
  \emph{\bibinfo{booktitle}{The science of well-being}}, edited by
  \bibinfo{editor}{\bibfnamefont{F.~A.} \bibnamefont{Huppert}},
  \bibinfo{editor}{\bibfnamefont{N.}~\bibnamefont{Baylis}}, \bibnamefont{and}
  \bibinfo{editor}{\bibfnamefont{B.}~\bibnamefont{Keverne}}
  (\bibinfo{publisher}{Oxford University Press}, \bibinfo{address}{Oxford, UK},
  \bibinfo{year}{2005}), pp. \bibinfo{pages}{285--304}.

\bibitem[{\citenamefont{Fowler and Christakis}(2008)}]{fowler2008a}
\bibinfo{author}{\bibfnamefont{J.~H.} \bibnamefont{Fowler}} \bibnamefont{and}
  \bibinfo{author}{\bibfnamefont{N.~A.} \bibnamefont{Christakis}},
  \bibinfo{journal}{BMJ} \textbf{\bibinfo{volume}{337}},
  \bibinfo{pages}{article \#2338} (\bibinfo{year}{2008}).

\bibitem[{\citenamefont{Jost}(2006)}]{jost2006a}
\bibinfo{author}{\bibfnamefont{L.}~\bibnamefont{Jost}},
  \bibinfo{journal}{Oikos} \textbf{\bibinfo{volume}{113}}, \bibinfo{pages}{363}
  (\bibinfo{year}{2006}).

\bibitem[{gal()}]{gallup}
\bibinfo{note}{Gallup Healthways Well-Being Index. Available at
  \url{http://www.well-beingindex.com/}. Accessed October 24, 2011.}

\bibitem[{fac()}]{facebookgnh}
\bibinfo{note}{Facebook Gross National Happiness. Available at
  \url{http://apps.facebook.com/usa_gnh/}. Accessed October 24, 2011}.

\bibitem[{\citenamefont{Kramer}(2010)}]{kramer2010a}
\bibinfo{author}{\bibfnamefont{A.~D.~I.} \bibnamefont{Kramer}}, in
  \emph{\bibinfo{booktitle}{Proceedings of the 28th international conference on
  Human factors in computing systems}} (\bibinfo{publisher}{ACM},
  \bibinfo{address}{New York, NY}, \bibinfo{year}{2010}), CHI '10, pp.
  \bibinfo{pages}{287--290}.

\bibitem[{\citenamefont{O'Connor et~al.}(2010)\citenamefont{O'Connor,
  Balasubramanyan, Routledge, and Smith}}]{oconnor2010a}
\bibinfo{author}{\bibfnamefont{B.}~\bibnamefont{O'Connor}},
  \bibinfo{author}{\bibfnamefont{R.}~\bibnamefont{Balasubramanyan}},
  \bibinfo{author}{\bibfnamefont{B.}~\bibnamefont{Routledge}},
  \bibnamefont{and} \bibinfo{author}{\bibfnamefont{N.}~\bibnamefont{Smith}}, in
  \emph{\bibinfo{booktitle}{Proceedings of 4th International AAAI Conference on
  Weblogs and Social Media (ICWSM-2010)}} (\bibinfo{year}{2010}).

\bibitem[{\citenamefont{Mitrovi\'{c} et~al.}()\citenamefont{Mitrovi\'{c},
  Paltoglou, and Tadi\'{c}}}]{mitrovic2010a}
\bibinfo{author}{\bibfnamefont{M.}~\bibnamefont{Mitrovi\'{c}}},
  \bibinfo{author}{\bibfnamefont{G.}~\bibnamefont{Paltoglou}},
  \bibnamefont{and}
  \bibinfo{author}{\bibfnamefont{B.}~\bibnamefont{Tadi\'{c}}},
  \bibinfo{note}{available at \url{http://arxiv.org/abs/1011.6268}. Accessed
  October 24, 2011.}

\bibitem[{\citenamefont{Golder and Macy}(2011)}]{golder2011a}
\bibinfo{author}{\bibfnamefont{S.~A.} \bibnamefont{Golder}} \bibnamefont{and}
  \bibinfo{author}{\bibfnamefont{M.~W.} \bibnamefont{Macy}},
  \bibinfo{journal}{Science Magazine} \textbf{\bibinfo{volume}{333}},
  \bibinfo{pages}{1878} (\bibinfo{year}{2011}).

\bibitem[{\citenamefont{Bollen et~al.}(2011{\natexlab{a}})\citenamefont{Bollen,
  Pepe, and Mao}}]{bollen2011c}
\bibinfo{author}{\bibfnamefont{J.}~\bibnamefont{Bollen}},
  \bibinfo{author}{\bibfnamefont{A.}~\bibnamefont{Pepe}}, \bibnamefont{and}
  \bibinfo{author}{\bibfnamefont{H.}~\bibnamefont{Mao}}, in
  \emph{\bibinfo{booktitle}{In Proceedings of the Fifth International AAAI
  Conference on Weblogs and Social Media (ICWSM)}}
  (\bibinfo{address}{Barcelona, Spain}, \bibinfo{year}{2011}{\natexlab{a}}).

\bibitem[{\citenamefont{Bollen et~al.}(2011{\natexlab{b}})\citenamefont{Bollen,
  Mao, and Zeng}}]{bollen2011b}
\bibinfo{author}{\bibfnamefont{J.}~\bibnamefont{Bollen}},
  \bibinfo{author}{\bibfnamefont{H.}~\bibnamefont{Mao}}, \bibnamefont{and}
  \bibinfo{author}{\bibfnamefont{X.-J.} \bibnamefont{Zeng}},
  \bibinfo{journal}{Journal of Computational Science}
  \textbf{\bibinfo{volume}{2}}, \bibinfo{pages}{1}
  (\bibinfo{year}{2011}{\natexlab{b}}).

\bibitem[{\citenamefont{Rentfrow et~al.}(2008)\citenamefont{Rentfrow, Gosling,
  and Potter}}]{rentfrow2008a}
\bibinfo{author}{\bibfnamefont{P.~J.} \bibnamefont{Rentfrow}},
  \bibinfo{author}{\bibfnamefont{S.~D.} \bibnamefont{Gosling}},
  \bibnamefont{and} \bibinfo{author}{\bibfnamefont{J.}~\bibnamefont{Potter}},
  \bibinfo{journal}{Perspectives on Psychological Science}
  \textbf{\bibinfo{volume}{3}}, \bibinfo{pages}{339} (\bibinfo{year}{2008}).

\bibitem[{\citenamefont{Balog et~al.}(2006)\citenamefont{Balog, Mishne, and
  de~Rijke}}]{balog2006b}
\bibinfo{author}{\bibfnamefont{K.}~\bibnamefont{Balog}},
  \bibinfo{author}{\bibfnamefont{G.}~\bibnamefont{Mishne}}, \bibnamefont{and}
  \bibinfo{author}{\bibfnamefont{M.}~\bibnamefont{de~Rijke}},
  \bibinfo{journal}{11th Meeting of the European Chapter of the Association for
  Computational Linguistics}  (\bibinfo{year}{2006}).

\bibitem[{\citenamefont{Balog and de~Rijke}(2006)}]{balog2006a}
\bibinfo{author}{\bibfnamefont{K.}~\bibnamefont{Balog}} \bibnamefont{and}
  \bibinfo{author}{\bibfnamefont{M.}~\bibnamefont{de~Rijke}},
  \bibinfo{journal}{3rd Annual Workshop on the Weblogging Ecosystem}
  (\bibinfo{year}{2006}).

\bibitem[{\citenamefont{Mishne and de~Rijke}(2005)}]{mishne2005a}
\bibinfo{author}{\bibfnamefont{G.}~\bibnamefont{Mishne}} \bibnamefont{and}
  \bibinfo{author}{\bibfnamefont{M.}~\bibnamefont{de~Rijke}},
  \bibinfo{journal}{AAAI 2006 Spring Symposium on Computational Approaches to
  Analysing Weblogs}  (\bibinfo{year}{2005}).

\bibitem[{\citenamefont{Mishne}(2005)}]{mishne2005b}
\bibinfo{author}{\bibfnamefont{G.}~\bibnamefont{Mishne}},
  \bibinfo{journal}{Style2005, 1st Workshop on Stylistic Analysis of Text for
  Information Access}  (\bibinfo{year}{2005}).

\bibitem[{\citenamefont{Mishne and de~Rijke}(2006{\natexlab{a}})}]{mishne2006f}
\bibinfo{author}{\bibfnamefont{G.}~\bibnamefont{Mishne}} \bibnamefont{and}
  \bibinfo{author}{\bibfnamefont{M.}~\bibnamefont{de~Rijke}},
  \bibinfo{journal}{AAAI 2006 Spring Symposium on Computational Approaches to
  Analysing Weblogs}  (\bibinfo{year}{2006}{\natexlab{a}}).

\bibitem[{\citenamefont{Mihalcea and Liu}(2010)}]{mihalcea2006a}
\bibinfo{author}{\bibfnamefont{R.}~\bibnamefont{Mihalcea}} \bibnamefont{and}
  \bibinfo{author}{\bibfnamefont{H.}~\bibnamefont{Liu}}, in
  \emph{\bibinfo{booktitle}{AAAI 2006 Spring Symposium on Computational
  Approaches to Analysing Weblogs}} (\bibinfo{publisher}{AAAI Press},
  \bibinfo{year}{2010}), AAAI-CAAW 2006, pp. \bibinfo{pages}{591--600}.

\bibitem[{gig()}]{gigatweet}
\bibinfo{note}{Gigatweet. Available at
  \url{http://www.gigatweeter.com/counter}. Accessed November 6, 2010.}

\bibitem[{twi({\natexlab{a}})}]{twitterapi}
\bibinfo{note}{Twitter API. Available at \url{http://dev.twitter.com/}.
  Accessed October 24, 2011.}

\bibitem[{twi({\natexlab{b}})}]{twitterdemog}
\bibinfo{note}{Inside Twitter: An In-Depth Look Inside the Twitter World,
  Sysmos Resource Library. Available at
  \url{http://www.sysomos.com/insidetwitter/}. Accessed August 1, 2010.}

\bibitem[{\citenamefont{Fox et~al.}(2006)\citenamefont{Fox, Zickuhr, and
  Smith}}]{fox2009a}
\bibinfo{author}{\bibfnamefont{S.}~\bibnamefont{Fox}},
  \bibinfo{author}{\bibfnamefont{K.}~\bibnamefont{Zickuhr}}, \bibnamefont{and}
  \bibinfo{author}{\bibfnamefont{A.}~\bibnamefont{Smith}}, \bibinfo{type}{Tech.
  Rep.}, \bibinfo{institution}{Pew Internet \& American Life Project}
  (\bibinfo{year}{2006}), \bibinfo{note}{accessed August 1, 2011},
  \urlprefix\url{http://www.pewinternet.org/Reports/2009/17-Twitter-and-Status-Updating-Fall-2009.aspx}.

\bibitem[{\citenamefont{Smith}(2011)}]{smith2011a}
\bibinfo{author}{\bibfnamefont{A.}~\bibnamefont{Smith}}, \bibinfo{type}{Tech.
  Rep.}, \bibinfo{institution}{Pew Internet \& American Life Project}
  (\bibinfo{year}{2011}), \bibinfo{note}{accessed August 1, 2011},
  \urlprefix\url{http://www.pewinternet.org/Reports/2011/Twitter-Update-2011.aspx}.

\bibitem[{twi({\natexlab{c}})}]{twitterusers}
\bibinfo{note}{Inside Twitter: An In-Depth Look at the 5\% of Most Active
  Users, Sysmos Resource Library. Available at
  \url{http://sysomos.com/insidetwitter/mostactiveusers}. Accessed August 1,
  2010.}

\bibitem[{\citenamefont{Kwak et~al.}(2010)\citenamefont{Kwak, Lee, Park, and
  Moon}}]{kwak2010a}
\bibinfo{author}{\bibfnamefont{H.}~\bibnamefont{Kwak}},
  \bibinfo{author}{\bibfnamefont{C.}~\bibnamefont{Lee}},
  \bibinfo{author}{\bibfnamefont{H.}~\bibnamefont{Park}}, \bibnamefont{and}
  \bibinfo{author}{\bibfnamefont{S.}~\bibnamefont{Moon}}, in
  \emph{\bibinfo{booktitle}{Proceedings of the 19th international conference on
  World wide web}} (\bibinfo{publisher}{ACM}, \bibinfo{address}{New York, NY,
  USA}, \bibinfo{year}{2010}), WWW '10, pp. \bibinfo{pages}{591--600}, ISBN
  \bibinfo{isbn}{978-1-60558-799-8}.

\bibitem[{con()}]{congresstweets-nytimes}
\bibinfo{note}{``Library of Congress will save tweets,'' New York Times.
  Available at
  \url{http://www.nytimes.com/2010/04/15/technology/15twitter.html}. Accessed
  April 15, 2011.}

\bibitem[{\citenamefont{Sandhaus}(2008)}]{nytimescorpus2008a}
\bibinfo{author}{\bibfnamefont{E.}~\bibnamefont{Sandhaus}},
  \emph{\bibinfo{title}{The {N}ew {Y}ork {T}imes {A}nnotated {C}orpus}},
  \bibinfo{howpublished}{{L}inguistic Data Consortium, Philadelphia}
  (\bibinfo{year}{2008}).

\bibitem[{\citenamefont{Zipf}(1949)}]{zipf1949a}
\bibinfo{author}{\bibfnamefont{G.~K.} \bibnamefont{Zipf}},
  \emph{\bibinfo{title}{Human Behaviour and the Principle of Least-Effort}}
  (\bibinfo{publisher}{Addison-Wesley}, \bibinfo{address}{Cambridge, MA},
  \bibinfo{year}{1949}).

\bibitem[{\citenamefont{Bradley and Lang}(1999)}]{bradley1999a}
\bibinfo{author}{\bibfnamefont{M.}~\bibnamefont{Bradley}} \bibnamefont{and}
  \bibinfo{author}{\bibfnamefont{P.}~\bibnamefont{Lang}},
  \bibinfo{type}{Technical report C-1}, \bibinfo{institution}{University of
  Florida}, \bibinfo{address}{Gainesville, FL} (\bibinfo{year}{1999}).

\bibitem[{\citenamefont{Redondo et~al.}(August 2007)\citenamefont{Redondo,
  Fraga, Padron, and Comesana}}]{redondo2007a}
\bibinfo{author}{\bibfnamefont{J.}~\bibnamefont{Redondo}},
  \bibinfo{author}{\bibfnamefont{I.}~\bibnamefont{Fraga}},
  \bibinfo{author}{\bibfnamefont{I.}~\bibnamefont{Padron}}, \bibnamefont{and}
  \bibinfo{author}{\bibfnamefont{M.}~\bibnamefont{Comesana}},
  \bibinfo{journal}{Behavior Research Methods} \textbf{\bibinfo{volume}{39}},
  \bibinfo{pages}{600} (\bibinfo{year}{August 2007}).

\bibitem[{\citenamefont{Osgood et~al.}(1957)\citenamefont{Osgood, Suci, and
  Tannenbaum}}]{osgood1957a}
\bibinfo{author}{\bibfnamefont{C.}~\bibnamefont{Osgood}},
  \bibinfo{author}{\bibfnamefont{G.}~\bibnamefont{Suci}}, \bibnamefont{and}
  \bibinfo{author}{\bibfnamefont{P.}~\bibnamefont{Tannenbaum}},
  \emph{\bibinfo{title}{The Measurement of Meaning}}
  (\bibinfo{publisher}{University of Illinois}, \bibinfo{address}{Urbana, IL},
  \bibinfo{year}{1957}).

\bibitem[{\citenamefont{Wilbur and Sirotkin}(1992)}]{wilbur1992a}
\bibinfo{author}{\bibfnamefont{W.~J.} \bibnamefont{Wilbur}} \bibnamefont{and}
  \bibinfo{author}{\bibfnamefont{K.}~\bibnamefont{Sirotkin}},
  \bibinfo{journal}{Journal of Information Science}
  \textbf{\bibinfo{volume}{18}}, \bibinfo{pages}{45} (\bibinfo{year}{1992}).

\bibitem[{\citenamefont{Kloumann et~al.}(2011)\citenamefont{Kloumann, Danforth,
  Harris, Bliss, and Dodds}}]{kloumann2011a}
\bibinfo{author}{\bibfnamefont{I.~M.} \bibnamefont{Kloumann}},
  \bibinfo{author}{\bibfnamefont{C.~M.} \bibnamefont{Danforth}},
  \bibinfo{author}{\bibfnamefont{K.~D.} \bibnamefont{Harris}},
  \bibinfo{author}{\bibfnamefont{C.~A.} \bibnamefont{Bliss}}, \bibnamefont{and}
  \bibinfo{author}{\bibfnamefont{P.~S.} \bibnamefont{Dodds}}
  (\bibinfo{year}{2011}), \bibinfo{note}{available at
  \url{http://arxiv.org/abs/1108.5192}. Accessed October 24, 2011}.

\bibitem[{\citenamefont{Dodds et~al.}(2011)\citenamefont{Dodds, Harris,
  Kloumann, Bliss, and Danforth}}]{dodds2011a}
\bibinfo{author}{\bibfnamefont{P.~S.} \bibnamefont{Dodds}},
  \bibinfo{author}{\bibfnamefont{K.~D.} \bibnamefont{Harris}},
  \bibinfo{author}{\bibfnamefont{I.~M.} \bibnamefont{Kloumann}},
  \bibinfo{author}{\bibfnamefont{C.~A.} \bibnamefont{Bliss}}, \bibnamefont{and}
  \bibinfo{author}{\bibfnamefont{C.~M.} \bibnamefont{Danforth}}
  (\bibinfo{year}{2011}), \bibinfo{note}{draft version. Available at
  \url{http://arxiv.org/abs/1101.5120v3}. Accessed October 24, 2011}.

\bibitem[{\citenamefont{Lee}(2004)}]{lee2004a}
\bibinfo{author}{\bibfnamefont{L.}~\bibnamefont{Lee}}, in
  \emph{\bibinfo{booktitle}{Computer Science: Reflections on the Field,
  Reflections from the Field}}, edited by
  \bibinfo{editor}{\bibfnamefont{C.}~\bibnamefont{on~the Fundamentals~of
  Computer Science:~Challenges}}, \bibinfo{editor}{\bibfnamefont{C.~S.}
  \bibnamefont{Opportunities}}, \bibnamefont{and}
  \bibinfo{editor}{\bibfnamefont{N.~R.~C.}
  \bibnamefont{Telecommunications~Board}} (\bibinfo{publisher}{The National
  Academies Press}, \bibinfo{year}{2004}), pp. \bibinfo{pages}{111--118}.

\bibitem[{\citenamefont{Choi et~al.}(2005)\citenamefont{Choi, Cardie, Riloff,
  and Patwardhan}}]{choi2005a}
\bibinfo{author}{\bibfnamefont{Y.}~\bibnamefont{Choi}},
  \bibinfo{author}{\bibfnamefont{C.}~\bibnamefont{Cardie}},
  \bibinfo{author}{\bibfnamefont{E.}~\bibnamefont{Riloff}}, \bibnamefont{and}
  \bibinfo{author}{\bibfnamefont{S.}~\bibnamefont{Patwardhan}},
  \bibinfo{journal}{Proceedings of Human Language Technology
  Conference/Conference on Empirical Methods in Natural Language Processing
  (HLT/EMNLP 2005)}  (\bibinfo{year}{2005}).

\bibitem[{\citenamefont{Choi et~al.}(2006)\citenamefont{Choi, Breck, and
  Cardie}}]{choi2006a}
\bibinfo{author}{\bibfnamefont{Y.}~\bibnamefont{Choi}},
  \bibinfo{author}{\bibfnamefont{E.}~\bibnamefont{Breck}}, \bibnamefont{and}
  \bibinfo{author}{\bibfnamefont{C.}~\bibnamefont{Cardie}},
  \bibinfo{journal}{Conference on Empirical Methods in Natural Language
  Processiong (EMNLP-2006)}  (\bibinfo{year}{2006}).

\bibitem[{\citenamefont{List}(2006)}]{list2006a}
\bibinfo{author}{\bibfnamefont{J.~A.} \bibnamefont{List}},
  \bibinfo{journal}{Advances in Economic Analysis \& Policy}
  \textbf{\bibinfo{volume}{6}}, \bibinfo{pages}{Article 8}
  (\bibinfo{year}{2006}).

\bibitem[{\citenamefont{Skitka and Sargis}(2006)}]{skitka2006a}
\bibinfo{author}{\bibfnamefont{L.~J.} \bibnamefont{Skitka}} \bibnamefont{and}
  \bibinfo{author}{\bibfnamefont{E.~G.} \bibnamefont{Sargis}},
  \bibinfo{journal}{Annu. Rev. Psychol.} \textbf{\bibinfo{volume}{57}},
  \bibinfo{pages}{529} (\bibinfo{year}{2006}).

\bibitem[{\citenamefont{Kahneman et~al.}(2006)\citenamefont{Kahneman, Krueger,
  Schkade, Schwarz, and Stone}}]{kahneman2006a}
\bibinfo{author}{\bibfnamefont{D.}~\bibnamefont{Kahneman}},
  \bibinfo{author}{\bibfnamefont{A.~B.} \bibnamefont{Krueger}},
  \bibinfo{author}{\bibfnamefont{D.}~\bibnamefont{Schkade}},
  \bibinfo{author}{\bibfnamefont{N.}~\bibnamefont{Schwarz}}, \bibnamefont{and}
  \bibinfo{author}{\bibfnamefont{A.~A.} \bibnamefont{Stone}},
  \bibinfo{journal}{Science} \textbf{\bibinfo{volume}{312}},
  \bibinfo{pages}{1908} (\bibinfo{year}{2006}).

\bibitem[{\citenamefont{Simpson}(1949)}]{simpson1949a}
\bibinfo{author}{\bibfnamefont{E.~H.} \bibnamefont{Simpson}},
  \bibinfo{journal}{Nature} \textbf{\bibinfo{volume}{163}},
  \bibinfo{pages}{688} (\bibinfo{year}{1949}).

\bibitem[{\citenamefont{Mishne and de~Rijke}(2006{\natexlab{b}})}]{mishne2006a}
\bibinfo{author}{\bibfnamefont{G.}~\bibnamefont{Mishne}} \bibnamefont{and}
  \bibinfo{author}{\bibfnamefont{M.}~\bibnamefont{de~Rijke}},
  \bibinfo{journal}{5th International Conference on Natural Language
  Processing}  (\bibinfo{year}{2006}{\natexlab{b}}).

\bibitem[{\citenamefont{Harris and Kamvar}(2009)}]{harris2009a}
\bibinfo{author}{\bibfnamefont{J.}~\bibnamefont{Harris}} \bibnamefont{and}
  \bibinfo{author}{\bibfnamefont{S.}~\bibnamefont{Kamvar}},
  \emph{\bibinfo{title}{We Feel Fine: An Almanac of Human Emotion}}
  (\bibinfo{publisher}{Scribner}, \bibinfo{address}{New York, NY},
  \bibinfo{year}{2009}).

\bibitem[{nor()}]{northeasterntwittermood}
\bibinfo{note}{Pulse of the nation: U.S.~Mood Throughout the Day inferred from
  Twitter. Available at
  \url{http://www.ccs.neu.edu/home/amislove/twittermood/}. Accessed October 24,
  2011.}

\bibitem[{\citenamefont{Stone et~al.}(2006)\citenamefont{Stone, Schwartz,
  Schkade, Schwarz, Krueger, and Kahneman}}]{stone2006a}
\bibinfo{author}{\bibfnamefont{A.~A.} \bibnamefont{Stone}},
  \bibinfo{author}{\bibfnamefont{J.~E.} \bibnamefont{Schwartz}},
  \bibinfo{author}{\bibfnamefont{D.}~\bibnamefont{Schkade}},
  \bibinfo{author}{\bibfnamefont{N.}~\bibnamefont{Schwarz}},
  \bibinfo{author}{\bibfnamefont{A.}~\bibnamefont{Krueger}}, \bibnamefont{and}
  \bibinfo{author}{\bibfnamefont{D.}~\bibnamefont{Kahneman}},
  \bibinfo{journal}{Emotion} \textbf{\bibinfo{volume}{6}}, \bibinfo{pages}{139}
  (\bibinfo{year}{2006}).

\bibitem[{\citenamefont{Bollen et~al.}(2011{\natexlab{c}})\citenamefont{Bollen,
  Goncalves, Ruan, and Mao}}]{bollen2011a}
\bibinfo{author}{\bibfnamefont{J.}~\bibnamefont{Bollen}},
  \bibinfo{author}{\bibfnamefont{B.}~\bibnamefont{Goncalves}},
  \bibinfo{author}{\bibfnamefont{G.}~\bibnamefont{Ruan}}, \bibnamefont{and}
  \bibinfo{author}{\bibfnamefont{H.}~\bibnamefont{Mao}},
  \bibinfo{journal}{Artificial Life} \textbf{\bibinfo{volume}{17}},
  \bibinfo{pages}{237} (\bibinfo{year}{2011}{\natexlab{c}}).

\bibitem[{\citenamefont{Shannon}(1948)}]{shannon1948a}
\bibinfo{author}{\bibfnamefont{C.~E.} \bibnamefont{Shannon}},
  \bibinfo{journal}{The Bell System Tech. J.} \textbf{\bibinfo{volume}{27}},
  \bibinfo{pages}{379} (\bibinfo{year}{1948}).

\end{thebibliography}
\end{document}